\documentclass[twocolumn]{aa}

\usepackage{graphicx}
\usepackage{amsmath,amsfonts,amssymb}
\usepackage{txfonts}
\usepackage{color}
\usepackage{natbib}
\usepackage{float}
\usepackage{dblfloatfix}
\usepackage{afterpage}
\usepackage{ifthen}
\usepackage[morefloats=12]{morefloats}
\usepackage{placeins}
\usepackage{multicol}
\bibpunct{(}{)}{;}{a}{}{,}
\usepackage[switch]{lineno}
\definecolor{linkcolor}{rgb}{0.6,0,0}
\definecolor{citecolor}{rgb}{0,0,0.75}
\definecolor{urlcolor}{rgb}{0.12,0.46,0.7}
\usepackage[breaklinks, colorlinks, urlcolor=urlcolor,
    linkcolor=linkcolor,citecolor=citecolor,pdfencoding=auto]{hyperref}
\hypersetup{linktocpage}
\usepackage{bold-extra}
\usepackage{csquotes}
\usepackage{breakurl}
\usepackage{lipsum}
\usepackage{mathtools}
\usepackage{cuted}
\usepackage{upgreek}
\usepackage[utf8]{inputenc}
\usepackage{eurosym}
\DeclareUnicodeCharacter{20AC}{\euro}

\def\setsymbol#1#2{\expandafter\def\csname #1\endcsname{#2}}
\def\getsymbol#1{\csname #1\endcsname}

\def\Planck{\textit{Planck}}

\newbox\tablebox    \newdimen\tablewidth
\def\leaderfil{\leaders\hbox to 5pt{\hss.\hss}\hfil}
\def\endPlancktablewide{\tablewidth=\textwidth 
    $$\hss\copy\tablebox\hss$$
    \vskip-\lastskip\vskip -2pt}
\def\tablenote#1 #2\par{\begingroup \parindent=0.8em
    \abovedisplayshortskip=0pt\belowdisplayshortskip=0pt
    \noindent
    $$\hss\vbox{\hsize\tablewidth \hangindent=\parindent \hangafter=1 \noindent
    \hbox to \parindent{$^#1$\hss}\strut#2\strut\par}\hss$$
    \endgroup}
\def\doubleline{\vskip 3pt\hrule \vskip 1.5pt \hrule \vskip 5pt}

\def\L2{\ifmmode L_2\else $L_2$\fi}

\def\DeltaT{\ifmmode \Delta T\else $\Delta T$\fi}
\def\deltat{\ifmmode \Delta t\else $\Delta t$\fi}
\def\fknee{\ifmmode f_{\rm knee}\else $f_{\rm knee}$\fi}
\def\Fmax{\ifmmode F_{\rm max}\else $F_{\rm max}$\fi}
\def\solar{\ifmmode{\rm M}_{\mathord\odot}\else${\rm M}_{\mathord\odot}$\fi}
\def\Msolar{\ifmmode{\rm M}_{\mathord\odot}\else${\rm M}_{\mathord\odot}$\fi}
\def\Lsolar{\ifmmode{\rm L}_{\mathord\odot}\else${\rm L}_{\mathord\odot}$\fi}
\def\inv{\ifmmode^{-1}\else$^{-1}$\fi}
\def\mo{\ifmmode^{-1}\else$^{-1}$\fi}
\def\sup#1{\ifmmode ^{\rm #1}\else $^{\rm #1}$\fi}
\def\expo#1{\ifmmode \times 10^{#1}\else $\times 10^{#1}$\fi}
\def\,{\thinspace}
\def\lsim{\mathrel{\raise .4ex\hbox{\rlap{$<$}\lower 1.2ex\hbox{$\sim$}}}}
\def\gsim{\mathrel{\raise .4ex\hbox{\rlap{$>$}\lower 1.2ex\hbox{$\sim$}}}}

\def\simprop{\mathrel{\raise .4ex\hbox{\rlap{$\propto$}\lower 1.2ex\hbox{$\sim$}}}}
\def\deg{\ifmmode^\circ\else$^\circ$\fi}
\def\pdeg{\ifmmode $\setbox0=\hbox{$^{\circ}$}\rlap{\hskip.11\wd0 .}$^{\circ}
          \else \setbox0=\hbox{$^{\circ}$}\rlap{\hskip.11\wd0 .}$^{\circ}$\fi}
\def\arcs{\ifmmode {^{\scriptstyle\prime\prime}}
          \else $^{\scriptstyle\prime\prime}$\fi}
\def\arcm{\ifmmode {^{\scriptstyle\prime}}
          \else $^{\scriptstyle\prime}$\fi}
\newdimen\sa  \newdimen\sb
\def\parcs{\sa=.07em \sb=.03em
     \ifmmode \hbox{\rlap{.}}^{\scriptstyle\prime\kern -\sb\prime}\hbox{\kern -\sa}
     \else \rlap{.}$^{\scriptstyle\prime\kern -\sb\prime}$\kern -\sa\fi}
\def\parcm{\sa=.08em \sb=.03em
     \ifmmode \hbox{\rlap{.}\kern\sa}^{\scriptstyle\prime}\hbox{\kern-\sb}
     \else \rlap{.}\kern\sa$^{\scriptstyle\prime}$\kern-\sb\fi}
\def\ra[#1 #2 #3.#4]{#1\sup{h}#2\sup{m}#3\sup{s}\llap.#4}
\def\dec[#1 #2 #3.#4]{#1\deg#2\arcm#3\arcs\llap.#4}
\def\deco[#1 #2 #3]{#1\deg#2\arcm#3\arcs}
\def\rra[#1 #2]{#1\sup{h}#2\sup{m}}

\def\dots{\relax\ifmmode \ldots\else $\ldots$\fi}
\def\WHzsr{\ifmmode $W\,Hz\mo\,sr\mo$\else W\,Hz\mo\,sr\mo\fi}
\def\mHz{\ifmmode $\,mHz$\else \,mHz\fi}
\def\GHz{\ifmmode $\,GHz$\else \,GHz\fi}
\def\mKs{\ifmmode $\,mK\,s$^{1/2}\else \,mK\,s$^{1/2}$\fi}
\def\muKs{\ifmmode \,\mu$K\,s$^{1/2}\else \,$\mu$K\,s$^{1/2}$\fi}
\def\muKRJs{\ifmmode \,\mu$K$_{\rm RJ}$\,s$^{1/2}\else \,$\mu$K$_{\rm RJ}$\,s$^{1/2}$\fi}
\def\muKHz{\ifmmode \,\mu$K\,Hz$^{-1/2}\else \,$\mu$K\,Hz$^{-1/2}$\fi}
\def\MJysr{\ifmmode \,$MJy\,sr\mo$\else \,MJy\,sr\mo\fi}
\def\MJysrmK{\ifmmode \,$MJy\,sr\mo$\,mK$_{\rm CMB}\mo\else \,MJy\,sr\mo\,mK$_{\rm CMB}\mo$\fi}
\def\microns{\ifmmode \,\mu$m$\else \,$\mu$m\fi}

\def\muK{\ifmmode \,\mu$K$\else \,$\mu$\hbox{K}\fi}
\def\microK{\ifmmode \,\mu$K$\else \,$\mu$\hbox{K}\fi}
\def\muW{\ifmmode \,\mu$W$\else \,$\mu$\hbox{W}\fi}
\def\kms{\ifmmode $\,km\,s$^{-1}\else \,km\,s$^{-1}$\fi}
\def\kmsMpc{\ifmmode $\,\kms\,Mpc\mo$\else \,\kms\,Mpc\mo\fi}
\providecommand{\sorthelp}[1]{}

\def\WMAP{\emph{WMAP}}
\def\COBE{\emph{COBE}}

\def\commander{\texttt{Commander}}
\def\commanderone{\texttt{Commander1}}
\def\commandertwo{\texttt{Commander2}}
\def\commanderthree{\texttt{Commander3}}

\makeatletter
\ifcase \@ptsize \relax
  \newcommand{\miniscule}{\@setfontsize\miniscule{5}{6}}
\or
  \newcommand{\miniscule}{\@setfontsize\miniscule{6}{7}}
\or
  \newcommand{\miniscule}{\@setfontsize\miniscule{6}{7}}
\fi
\makeatother

\newcommand{\gray}[0]{\color{gray}}

\renewcommand{\d}[0]{\vec{d}}
\renewcommand{\t}[0]{\vec{t}}
\newcommand{\A}[0]{\tens{A}}

\newcommand{\V}[0]{\tens{V}}

\newcommand{\n}[0]{\vec{n}}
\newcommand{\x}[0]{\vec{x}}

\definecolor{orange}{RGB}{255,127,0}

\newcommand{\s}[0]{\vec{s}}
\renewcommand{\a}[0]{\vec{a}}
\newcommand{\m}[0]{\vec{m}}

\newcommand{\F}[0]{\tens{F}}
\newcommand{\B}[0]{\tens{B}}
\newcommand{\T}[0]{\tens{T}}

\renewcommand{\L}[0]{\tens{L}}
\newcommand{\g}[0]{\vec{g}}
\renewcommand{\b}[0]{\vec{b}}
\newcommand{\N}[0]{\tens{N}}
\newcommand{\Z}[0]{\tens{Z}}
\renewcommand{\C}[0]{\tens{C}}
\newcommand{\I}[0]{\tens{I}}
\newcommand{\M}[0]{\tens{M}}

\renewcommand{\S}[0]{\tens{S}}
\renewcommand{\r}[0]{\vec{r}}

\renewcommand{\v}[0]{\vec{v}}
\renewcommand{\P}[0]{\tens{P}}
\newcommand{\R}[0]{\tens{R}}
\renewcommand{\G}[0]{\tens{G}}

\newcommand{\Dbp}[0]{\Delta_{\mathrm{bp}}}

\newcommand{\Te}[0]{T_{e}}
\newcommand{\EM}[0]{{\rm EM}}

\newcommand{\BP}{\textsc{BeyondPlanck}}
\newcommand{\cosmoglobe}{\textsc{Cosmoglobe}}

\newcommand{\npipe}[0]{\texttt{NPIPE}}
\newcommand{\sroll}[0]{\texttt{SROLL}}
\newcommand{\srollTwo}[0]{\texttt{SROLL2}}
\newcommand{\HEALPix}[0]{\texttt{HEALPix}}
\newcommand{\e}{\mathrm e}

\def\Tcmb{\ifmmode T_\mathrm{CMB}\else $T_{\mathrm{CMB}}$\fi}
\def\Tdust{\ifmmode T_\mathrm{d}\else $T_{\mathrm{d}}$\fi}
\def\scmb{\ifmmode s_\mathrm{CMB}\else $s_{\mathrm{CMB}}$\fi}
\def\squad{\ifmmode s_\mathrm{quad}\else $s_{\mathrm{quad}}$\fi}
\def\ssynch{\ifmmode s_\mathrm{s}\else $s_\mathrm{s}$\fi}
\def\sdust{\ifmmode s_\mathrm{d}\else $s_{\mathrm{d}}$\fi}
\def\ssdust{\ifmmode s_\mathrm{sd}\else $s_{\mathrm{sd}}$\fi}
\def\same{\ifmmode s_\mathrm{ame}\else $s_{\mathrm{ame}}$\fi}
\def\ssrc{\ifmmode s_\mathrm{src}\else $s_{\mathrm{src}}$\fi}
\def\sco{\ifmmode s_\mathrm{CO}\else $s_{\mathrm{CO}}$\fi}
\def\sff{\ifmmode s_\mathrm{ff}\else $s_{\mathrm{ff}}$\fi}
\def\gff{\ifmmode g_\mathrm{ff}\else $g_{\mathrm{ff}}$\fi}
\def\fsynch{\ifmmode f_\mathrm{s}\else $f_{\mathrm{s}}$\fi}
\def\fsd{\ifmmode f_\mathrm{sd}\else $f_{\mathrm{sd}}$\fi}
\def\fame{\ifmmode f_\mathrm{ame}\else $f_{\mathrm{ame}}$\fi}
\def\alphasrc{\ifmmode \alpha_\mathrm{src}\else $\alpha_{\mathrm{src}}$\fi}
\def\bdust{\ifmmode \beta_\mathrm{d}\else $\beta_{\mathrm{d}}$\fi}
\def\bsynch{\ifmmode \beta_\mathrm{s}\else $\beta_{\mathrm{s}}$\fi}
\def\bsun{\ifmmode \beta_\mathrm{sun}\else $\beta_{\mathrm{sun}}$\fi}
\def\nuzeros{\ifmmode \nu_{0,\mathrm{s}}\else $\nu_{0,\mathrm{s}}$\fi}
\def\nuzeroff{\ifmmode \nu_{0,\mathrm{ff}}\else $\nu_{0,\mathrm{ff}}$\fi}
\def\nuzerod{\ifmmode \nu_{0,\mathrm{d}}\else $\nu_{0,\mathrm{d}}$\fi}
\def\nuzeroame{\ifmmode \nu_{0,\mathrm{ame}}\else $\nu_{0,\mathrm{ame}}$\fi}
\def\nuzerosd{\ifmmode \nu_{0,\mathrm{}}\else $\nu_{0,\mathrm{sd}}$\fi}
\def\nuzerosrc{\ifmmode \nu_{0,\mathrm{src}}\else $\nu_{0,\mathrm{src}}$\fi}
\def\nup{\ifmmode \nu_{\mathrm{p}}\else $\nu_{\mathrm{p}}$\fi}
\def\alphasd{\ifmmode \alpha_{\mathrm{sd}}\else $\alpha_{\mathrm{sd}}$\fi}
\def\Te{\ifmmode T_{\mathrm{e}}\else $T_{\mathrm{e}}$\fi}
\def\kB{\ifmmode k_\mathrm{B}\else $k_{\mathrm{B}}$\fi}

\def\inv{^{-1}}

\begin{document}

\title{\bfseries{\scshape{BeyondPlanck}} I. Global Bayesian analysis of the \Planck\ Low Frequency Instrument data }
\newcommand{\oslo}[0]{1}
\newcommand{\milanoA}[0]{2}
\newcommand{\milanoB}[0]{3}
\newcommand{\milanoC}[0]{4}
\newcommand{\triesteB}[0]{5}
\newcommand{\planetek}[0]{6}
\newcommand{\princeton}[0]{7}
\newcommand{\jpl}[0]{8}
\newcommand{\helsinkiA}[0]{9}
\newcommand{\helsinkiB}[0]{10}
\newcommand{\nersc}[0]{11}
\newcommand{\haverford}[0]{12}
\newcommand{\mpa}[0]{13}
\newcommand{\triesteA}[0]{14}
\author{\small
BeyondPlanck Collaboration\thanks{Corresponding author: H.~K.~Eriksen; \url{h.k.k.eriksen@astro.uio.no}}:
K.~J.~Andersen\inst{\oslo}
\and
\textcolor{black}{R.~Aurlien}\inst{\oslo}
\and
\textcolor{black}{R.~Banerji}\inst{\oslo}
\and
\textcolor{black}{A.~Basyrov}\inst{\oslo}
\and
M.~Bersanelli\inst{\milanoA, \milanoB, \milanoC}
\and
S.~Bertocco\inst{\triesteB}
\and
M.~Brilenkov\inst{\oslo}
\and
M.~Carbone\inst{\planetek}
\and
L.~P.~L.~Colombo\inst{\milanoA,\milanoC}
\and
H.~K.~Eriksen\inst{\oslo}
\and
J.~R.~Eskilt\inst{\oslo}
\and
\textcolor{black}{M.~K.~Foss}\inst{\oslo}
\and
C.~Franceschet\inst{\milanoA,\milanoC}
\and
\textcolor{black}{U.~Fuskeland}\inst{\oslo}
\and
S.~Galeotta\inst{\triesteB}
\and
M.~Galloway\inst{\oslo}
\and
S.~Gerakakis\inst{\planetek}
\and
E.~Gjerl{\o}w\inst{\oslo}
\and
\textcolor{black}{B.~Hensley}\inst{\princeton}
\and
\textcolor{black}{D.~Herman}\inst{\oslo}
\and
M.~Iacobellis\inst{\planetek}
\and
M.~Ieronymaki\inst{\planetek}
\and
\textcolor{black}{H.~T.~Ihle}\inst{\oslo}
\and
J.~B.~Jewell\inst{\jpl}
\and
\textcolor{black}{A.~Karakci}\inst{\oslo}
\and
E.~Keih\"{a}nen\inst{\helsinkiA, \helsinkiB}
\and
R.~Keskitalo\inst{\nersc}
\and
J.~G.~S.~Lunde\inst{\oslo}
\and
G.~Maggio\inst{\triesteB}
\and
D.~Maino\inst{\milanoA, \milanoB, \milanoC}
\and
M.~Maris\inst{\triesteB}
\and
A.~Mennella\inst{\milanoA, \milanoB, \milanoC}
\and
S.~Paradiso\inst{\milanoA, \milanoC}
\and
B.~Partridge\inst{\haverford}
\and
M.~Reinecke\inst{\mpa}
\and
M.~San\inst{\oslo}
\and
N.-O.~Stutzer\inst{\oslo}
\and
A.-S.~Suur-Uski\inst{\helsinkiA, \helsinkiB}
\and
T.~L.~Svalheim\inst{\oslo}
\and
D.~Tavagnacco\inst{\triesteB, \triesteA}
\and
H.~Thommesen\inst{\oslo}
\and
D.~J.~Watts\inst{\oslo}
\and
I.~K.~Wehus\inst{\oslo}
\and
A.~Zacchei\inst{\triesteB}
}
\institute{\small
Institute of Theoretical Astrophysics, University of Oslo, Blindern, Oslo, Norway\goodbreak
\and
Dipartimento di Fisica, Universit\`{a} degli Studi di Milano, Via Celoria, 16, Milano, Italy\goodbreak
\and
INAF-IASF Milano, Via E. Bassini 15, Milano, Italy\goodbreak
\and
INFN, Sezione di Milano, Via Celoria 16, Milano, Italy\goodbreak
\and
INAF - Osservatorio Astronomico di Trieste, Via G.B. Tiepolo 11, Trieste, Italy\goodbreak
\and
Planetek Hellas, Leoforos Kifisias 44, Marousi 151 25, Greece\goodbreak
\and
Department of Astrophysical Sciences, Princeton University, Princeton, NJ 08544,
U.S.A.\goodbreak
\and
Jet Propulsion Laboratory, California Institute of Technology, 4800 Oak Grove Drive, Pasadena, California, U.S.A.\goodbreak
\and
Department of Physics, Gustaf H\"{a}llstr\"{o}min katu 2, University of Helsinki, Helsinki, Finland\goodbreak
\and
Helsinki Institute of Physics, Gustaf H\"{a}llstr\"{o}min katu 2, University of Helsinki, Helsinki, Finland\goodbreak
\and
Computational Cosmology Center, Lawrence Berkeley National Laboratory, Berkeley, California, U.S.A.\goodbreak
\and
Haverford College Astronomy Department, 370 Lancaster Avenue,
Haverford, Pennsylvania, U.S.A.\goodbreak
\and
Max-Planck-Institut f\"{u}r Astrophysik, Karl-Schwarzschild-Str. 1, 85741 Garching, Germany\goodbreak
\and
Dipartimento di Fisica, Universit\`{a} degli Studi di Trieste, via A. Valerio 2, Trieste, Italy\goodbreak
}

\authorrunning{BeyondPlanck Collaboration}
\titlerunning{Global Bayesian analysis of \Planck\ LFI}

\abstract{We describe the \BP\ project in terms of motivation,
  methodology and main products, and provide a guide to a set of
  companion papers that describe each result in fuller
  detail. Building directly on experience from ESA's \emph{Planck}
  mission, we implement a complete end-to-end Bayesian analysis
  framework for the \emph{Planck} Low Frequency Instrument (LFI)
  observations. The primary product is a full joint posterior
  distribution $P(\omega \mid \d)$, where $\omega$ represents the set
  of all free instrumental (gain, correlated noise, bandpass etc.),
  astrophysical (synchrotron, free-free, thermal dust emission etc.),
  and cosmological (CMB map, power spectrum etc.) parameters. Some
  notable advantages of this approach compared to a traditional
  pipeline procedure are seamless end-to-end propagation of
  uncertainties; accurate modeling of both astrophysical and
  instrumental effects in the most natural basis for each uncertain
  quantity; optimized computational costs with little or no need for
  intermediate human interaction between various analysis steps; and a
  complete overview of the entire analysis process within one single
  framework. As a practical demonstration of this framework, we focus
  in particular on low-$\ell$ CMB polarization reconstruction with
  \Planck\ LFI. In this process, we identify several important new
  effects that have not been accounted for in previous pipelines,
  including gain over-smoothing and time-variable and non-$1/f$
  correlated noise in the 30 and 44\,GHz channels. Modelling and
  mitigating both previously known and newly discovered systematic
  effects, we find that all results are consistent with the
  $\Lambda$CDM model, and we constrain the reionization optical depth
  to $\tau=0.066\pm0.013$, with a low-resolution CMB-based $\chi^2$
  probability-to-exceed of 32\,\%. This uncertainty is about 30\,\%
  larger than the official pipelines, arising from taking into account
  a more complete instrumental model. The marginal CMB Solar dipole
  amplitude is $3362.7\pm1.4\muK$, where the error bar is derived
  directly from the posterior distribution without the need of any
  \emph{ad-hoc} instrumental corrections. We are currently not aware
  of any significant unmodelled systematic effects remaining in the
  \Planck\ LFI data, and, for the first time, the 44\,GHz channel is
  fully exploited in the current analysis. We argue that this
  framework can play a central role in the analysis of many current
  and future high-sensitivity CMB experiments, including
  \emph{LiteBIRD}, and it will serve as the computational foundation
  of the emerging community-wide \textsc{Cosmoglobe} effort, which
  aims to combine state-of-the-art radio, microwave, and submillimeter
  data sets into one global astrophysical model. All software is made
  publicly available under an OpenSource license, and both codes and
  products may be obtained through
  \url{http://beyondplanck.science}. }


\maketitle

\tableofcontents


\section{Introduction}
\label{sec:introduction}

\subsection{CMB cosmology}

According to the current cosmological concordance model, the
observable universe came into existence some 13.8 billion years ago in
a process often referred to as the Big Bang. While the physical laws
underpinning this singular event remain unknown, it is a testament to
the success of modern cosmology that physicists today are able to
measure and model the evolution and energy content of the universe to
exquisite precision, starting only a fraction of a second after the
Big Bang.

Among the most important cosmological observables is the cosmic
microwave background (CMB), first detected by \citet{penzias:1965}.
This leftover heat from the Big Bang fills the entire
universe, and may today be observed as a nearly isotropic
blackbody signal with a temperature of 2.7255~K
\citep{fixsen2009}. Initially, CMB photons were
trapped locally within a dense electron--proton plasma by Thomson
scattering. However, once the mean plasma temperature fell below
3000~K as the Universe expanded, electrons and protons combined into neutral hydrogen atoms,
and the photons were free to move throughout the entire observable
universe, with almost no further scattering. This event took place some
380\,000 years after the Big Bang, at a time often referred to as the
\emph{recombination epoch}. To any observer, the
resulting photons appear to come from a so-called
\emph{last-scattering surface}, a sphere corresponding to a light 
travel distance of just under the entire history of the universe.

While the CMB field is very nearly isotropic, it does exhibit small
spatial variations at the $\mathcal{O}(30~\mu\mathrm{K})$ level
\citep[e.g.,][and references therein]{hu:2002}. These fluctuations are
produced primarily by variations in the local gravitational potential,
temperature, density, and velocity at the last-scattering surface.
Smaller amplitude fluctuations arise from various secondary
interactions taking place after the photons leave the last-scattering
surface, for instance through gravitational lensing or Thomson
scattering in the hot, ionized medium in clusters of galaxies. It is
precisely by measuring and modelling all these small variations that
cosmologists are able to decipher the history of the universe in ever
greater detail. The current best-fit cosmological model derived from
this work is often referred to as \emph{the $\mathit\Lambda$CDM
  model}, which posits that the universe is isotropic and homogeneous
on large scales; that it started in a hot Big Bang; that it underwent
a brief period of exponential expansion called inflation that seeded
the universe with Gaussian random density fluctuations drawn from a
scale-invariant power spectrum; and that the energy contents of the
universe comprise 4.9\,\% baryonic matter, 26.5\,\% cold dark
matter, and 68.5\,\% dark energy \citep{planck2016-l06}. Flat spatial
curvature is also frequently assumed.

\begin{figure}[t]
  \center
  \includegraphics[width=\linewidth]{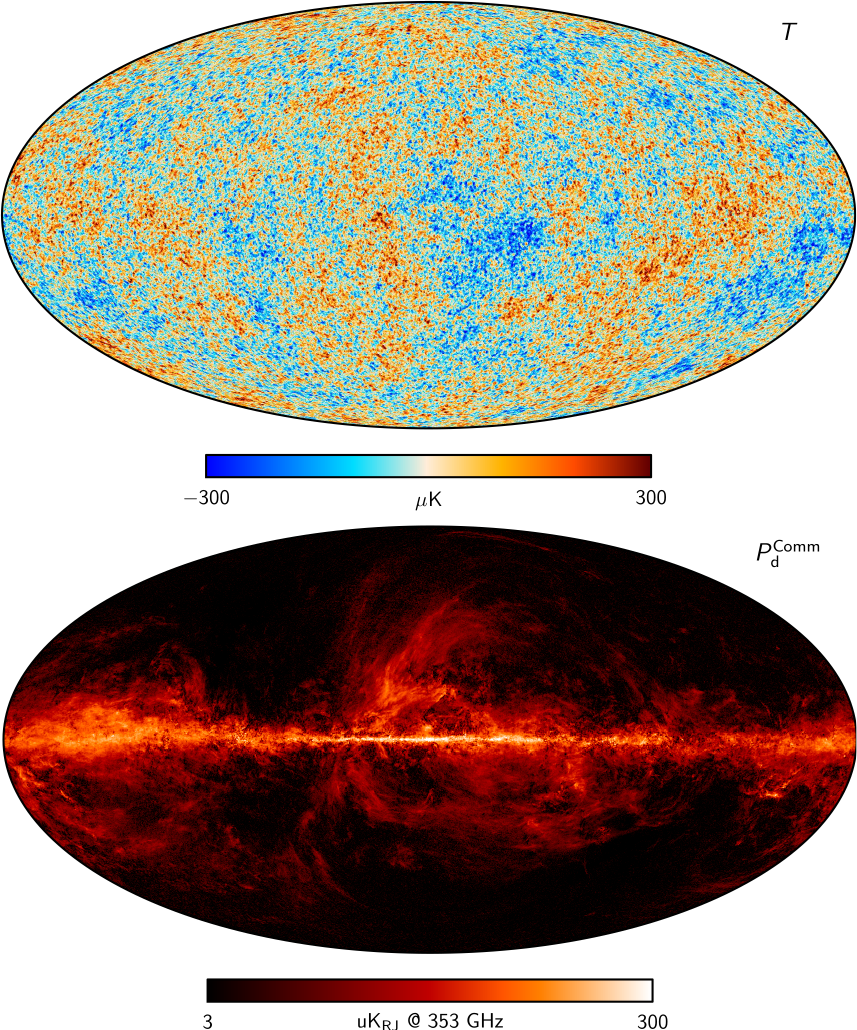}
  \caption{\Planck\ 2018 \commander\ CMB temperature (\emph{top
      panel}) and thermal dust polarization amplitude (\emph{bottom
      panel}) maps. Reproductions from \citet{planck2016-l04}.    }\label{fig:planck_maps}
\end{figure}

The rich cosmological information embedded in the CMB is not, however,
easy to extract.  Even the most dominant physical effects produce only
temperature fluctuations at the $10^{-5}$--$10^{-4}$ level in the
CMB.  A primary goal for next-generation CMB experiments is the
detection of primordial gravitational waves through the subtle
polarization they imprint on the CMB \citep[e.g.,][and references
  therein]{kamionkowski:2016}.  These so-called $B$-modes are likely
to have an amplitude no larger than 30\,nK, or a relative amplitude
smaller than $\mathcal{O}(10^{-8})$.

The fact that current CMB observations reach the $\mathrm{\mu K}$ level in the
face of instrument noise and systematics is a testament to the effort
of many scientists and engineers in this field, and to the time and
money they have spent.  Here, we list only a few of the results of
five decades of observational milestones.  NASA's \emph{COBE} mission
produced the first highly accurate measurement of the thermal spectrum
of the CMB \citep{mather:1994} and the first detection of large scale
fluctuations in the CMB \citep{smoot:1992}. The first high-fidelity
map of smaller scale CMB fluctuations was made by the BOOMERanG team
\citep{debernardis:2000}, and the first detection of polarized
fluctuations by DASI \citep{kovac:2002}.  These are among the more
than 50 past and present CMB projects, all of which have contributed to
technological innovations or scientific breakthroughs.

Two space missions, however, are primarily responsible for today's 
cosmological concordance model. They are NASA's \emph{Wilkinson Microwave
 Anisotropy Probe} (\WMAP; \citealp{bennett2012}) and ESA's
\emph{Planck} \citep{planck2016-l01} satellite missions. \emph{WMAP}
was the first experiment to take full advantage of the exquisite
thermal stability at Earth's second Lagrange point (L2), and observed
the CMB sky for nine years (2001--2010) in five frequency bands
(23--94~GHz) with precision unprecedented at the time. 

\subsection{\Planck}

The state-of-the-art in all-sky CMB observations as of 2022 is defined
by ESA's cosmology flagship mission called
\Planck\ \citep{planck2016-l01}. \Planck\ observed the CMB sky for
four years (2009--2013) in nine frequency bands (30--857~GHz), with
three times higher angular resolution and ten times higher sensitivity
than \WMAP. Its original design goal was to measure the primary CMB
temperature fluctuations with a precision limited only by fundamental
physical processes, including cosmic variance, not by instrumental
sensitivity \citep{planck2005-bluebook}.

\Planck\ comprised two separate instruments within a common focal plane. One was 
the \emph{Low Frequency Instrument} (LFI; \citealp{planck2016-l02}),
which employed coherent High Electron Mobility Transistor (HEMT)
radiometers with center frequencies near 30, 44 and 70~GHz,
 each with a fractional bandwidth of roughly 20\,\%. The other
was the \emph{High Frequency Instrument} (HFI;
\citealp{planck2016-l03}), which employed spider-web and polarization
sensitive bolometers with center frequencies of 100, 143, 217, 353,
545 and 857~GHz, each with a fractional bandwidth of 25\,\%.

Two different detector technologies were required to span
\emph{Planck}'s frequency range.  The use of two very different
detector technologies also provided crucial cross-checks against
subtle instrumental errors. \emph{Planck}'s wide frequency range fully
covered most of the spectrum of a 2.7255 K blackbody, but more
crucially allowed for the removal of contaminating foreground signals
\citep[e.g.,][]{leach2008}.  These arise from synchrotron emission
from relativistic electrons moving in the magnetic field of the
Galaxy, thermal emission from warm Galactic dust and bremsstrahlung
emission from ionized gas, as well as microwave emission from
extra-galactic sources.  This list is not exhaustive; but each
mechanism for foreground emission has a unique spatial distribution on
the sky and a unique, non-blackbody spectrum which allows it to be
distinguished from the CMB.  The preferred method for separating
cosmological fluctuations in the CMB from astrophysical foreground
signals is to map the sky at multiple frequencies, and then perform a
joint fit to this set of maps while taking into account the particular
spatial and spectral behaviour of each foreground.  These
considerations drove the design of
\Planck\ \citep{planck2005-bluebook}.  The capability to detect
polarized signals was added at the seven lowest frequency bands, from
30 to 353 GHz. Figure~\ref{fig:planck_maps} shows the CMB temperature
fluctuation and the polarized thermal dust emission maps as derived
from \Planck\ observations, which rank among the most iconic results
from the experiment.

The \Planck\ satellite was launched to L2 on May 14th 2009 and
deactivated on October 23rd 2013; it thus completed in total almost
4.5 years of observations \citep{planck2016-l01}. Unlike the case for
\WMAP, both \Planck\ instruments were cryogenically cooled. The last
18 months of operation included only LFI science measurements, as HFI
exhausted its cooling capacity in January 2012.

The first \Planck\ data release (denoted either ``PR1'' or 2013 here;
\citealp{planck2013-p01}) took place in March 2013, and was based on
the first 15.5~months of data, covering the full sky twice. By and
large, these measurements confirmed the cosmological model presented
by \WMAP\ and other previous experiments, but with significantly
higher signal-to-noise ratio. This higher sensitivity also supported
several truly groundbreaking results, two of which were a $25\sigma$
detection of gravitational lensing of CMB anisotropies
\citep{planck2013-p12}, and a revolutionary new image of polarized
thermal dust emission in the Milky Way \citep{planck2013-p06b}.

The 2013 release, however, did not include any CMB polarization
results. In addition, the initial angular power spectrum of CMB
anisotropies exhibited a $\sim$$2\,\%$ shift in amplitude compared to
the earlier \WMAP\ power spectrum \citep{planck2013-p08}. Both of
these issues had a common origin, namely incompletely controlled
systematic errors arising from instrumental effects
\citep{planck2014-a01}.  As noted earlier, CMB observations are not
easy: even small errors in assumptions made about foregrounds or
instrumental behaviour can have dramatic effects on the recovered CMB
signal.  Examples of instrumental effects include: uncertainties in
the beam shape and far sidelobes; mis-estimation of the frequency
response of detectors, which can introduce temperature to polarization
leakage; unaccounted-for non-linearity in the analog-to-digital
converters (ADCs) used in each detector chain; and uncertainties in
the polarization properties of detectors.

The \Planck\ team grappled with all of these, as well as uncertainties
in foreground contamination, in the years between 2013 and the release
of the final \Planck\ results in 2020 \citep{planck2016-l01}.  Very
substantial investments of time and money were made to develop
increasingly accurate models of the two \Planck\ instruments; these
allowed for more precise and robust science results. We emphasize that
the official LFI and HFI pipelines evolved step-by-step in the
post-launch period as instrument-specific effects emerged due to
increased calibration accuracy. \BP\ builds on all this accumulated
experience in implementing a global approach to the data analysis
problem.

A major milestone in this iterative process was the second
\Planck\ data release (``PR2'' or 2015; \citealp{planck2014-a01}),
which for the first time included the full set of
\Planck\ observations (50 months of LFI data and 27 months of HFI
data). At this point, the polarization properties of both the LFI and
HFI instruments were sufficiently well understood to allow for a
direct measurement of CMB polarization on intermediate and small
angular scales \citep{planck2014-a13}.  For HFI, however, accurate
large-scale polarization was still out of reach due to systematic
errors, and only LFI provided such constraints. The original power
spectrum discrepancy relative to \WMAP\ was tracked down to
inaccuracies in the calibration procedure and reference dipole values
used for the \Planck\ 2013 analysis, and these were subsequently
corrected in the 2015 release. With this second data release,
\Planck\ finally fulfilled its promise of measuring the primary CMB
temperature fluctuations to the limits set by astrophysical and
cosmological effects \citep{planck2014-a01}.  

\subsection{Large-scale CMB polarization, the reionization optical depth, and systematic errors}
\label{sec:tau_intro}

\Planck\ analysis continued beyond 2015, with a particular emphasis on
reducing large-scale polarization systematics
\citep{planck2016-l01}. Both the importance and difficulty of this
specific issue may be summarized in terms of the reionization optical depth, $\tau$
\citep[e.g.,][]{planck2014-a25}. This parameter is directly related to
the epoch during which the first stars
were born, often called the \emph{epoch of reionization}
\citep[e.g.,][and references therein]{loeb:2001}. According to
detailed measurements of the abundance of neutral hydrogen in the
universe from quasar spectra (the so-called ``Lyman alpha forest'';
\citealp{gunn:1965}), this event cannot have happened later than about
1 billion years after the Big Bang, corresponding to an optical depth
of $\tau\gtrsim0.048$. However, an independent measurement of $\tau$
may also be derived through CMB observations, by noting that the first
stars or galaxies ionized their surrounding medium, and thereby
released large numbers of free electrons off which CMB photons could
scatter. Detailed models predict a CMB polarization signal with an
amplitude of about $0.5~\mu\textrm{K}$ on angular scales larger
than $10^{\circ}$ \citep[e.g.,][and references
  therein]{alvarez:2006}.

While the scientific potential in
establishing robust large-scale polarization measurements is very
high, potentially pinpointing a critical epoch in the history of the
universe, the technical challenges are massive. The expected curl-free
$E$-mode polarization signal is only about 1\,\% of the corresponding
CMB temperature fluctuations, and the signal is only clearly visible on large
angular scales. Among all parameters in the cosmological concordance
model, the reionization optical depth is the most susceptible to
systematic errors, and for this reason it is often adopted as a
monitor for residual errors.

To illustrate the difficulties associated with measuring $\tau$, it is
interesting to consider its value as reported in the literature as a
function of time. The first CMB constraint was reported in the
first-year \WMAP\ release, which claimed $\tau=0.17\pm0.04$
corresponding to a reionization epoch of $t_{\mathrm{r}} =
180^{+220}_{-80}\,\textrm{Myr}$ \citep{kogut2003}. Such an early
reionization epoch imposed strong limits on galaxy formation
processes, and was not immediately compatible with standard
theories. However, this preliminary measurement was based on the
cross-correlation between temperature and polarization fluctuations
for which uncertainties and degeneracies are large. Furthermore, it
also did not account for bias introduced by foreground emission.

After adding more data, and, critically, allowing more time for
understanding the data and controlling systematic errors, the 3-year
\WMAP\ data release resulted in a significantly revised estimate of
$\tau=0.089\pm0.03$, nearly doubling the time allowed for structure
formation \citep{page2007}. This estimate was derived directly from
polarization-only measurements, and included proper foreground
corrections. Based on further improvements and additional data, the
reported 5-year \WMAP\ posterior mean value was $\tau=0.085\pm0.016$
\citep{komatsu2009}, while in the 7-year release it was
$\tau=0.088\pm0.015$ \citep{larson2010}, before finally settling on
$\tau=0.089\pm0.014$ in the 9-year release \citep{hinshaw2012}. This
represented the state-of-the-art before \Planck\ in terms of
large-scale CMB polarization measurements.

As already mentioned, no CMB polarization measurements were included
in the first \Planck\ 2013 release \citep{planck2013-p01}. However,
from temperature measurements alone, the best-fit optical depth was
constrained to $\tau=0.097\pm0.038$, in seemingly excellent
agreement with the final \WMAP\ polarization results
\citep{planck2013-p11}. Then, in the \Planck\ 2015 release, the LFI
data allowed the first independent constraint on large-scale CMB
polarization since \WMAP\ \citep{planck2014-a13}. At the same time,
the HFI polarization observations provided new and powerful
constraints on Galactic polarized thermal dust \citep{planck2014-a12},
to which the \WMAP\ experiment was only marginally sensitive. The
combination of LFI CMB and HFI thermal dust polarization measurements
alone resulted in $\tau=0.064^{+0.022}_{-0.023}$, or $1.1\sigma$ lower
than the 9-year \WMAP\ value. Furthermore, when combining the
\WMAP\ large-scale polarization CMB data with the same HFI polarization
foreground data, the best-fit value was $\tau=0.067\pm0.013$, in full agreement
with LFI.

The HFI large-scale CMB polarization data were not considered
sufficiently mature for scientific analysis until 2016, when new
calibration, mapmaking, and simulation procedures had been implemented
in a code called \sroll\ \citep{Planck_PIP_XLVIII}. Taking advantage
of these new developments, and leveraging the higher statistical power
of the HFI data, the reported estimate of the reionization optical
depth was adjusted further down by HFI to $\tau=0.055\pm0.009$. In
parallel, the LFI procedure was improved by merging calibration and
component separation into one framework.  Combined, these new analysis
procedures formed the basis for the third official
\Planck\ release \citep{planck2016-l01}, for which a final value of
$\tau=0.053\pm0.009$ was reported. The good agreement with the lower
limit imposed by quasar measurements, $\tau > 0.048$, implies both
that reionization by the first generation of stars occurred relatively
late, and that we can pin down the epoch of reionization with
precision.

While a stable and internally consistent $\Lambda$CDM model, including
$\tau$, had emerged by the official end of the \Planck\ consortium in
2018, one could still see clear signatures of residual systematics
present in various subsets of the data. For HFI, several internal
cross-correlations did not agree with each other to statistical
precision \citep{planck2016-l03}. For LFI the 44\,GHz channel failed
internal null tests \citep{planck2016-l02}, and there were clear
discrepancies between the raw frequency maps as seen by LFI and
\WMAP\ \citep{planck2016-l04}, indicating that there were still issues
to be resolved within either LFI or \WMAP, or both.

The last effort of the \Planck\ collaboration to resolve these
questions was organized within \Planck\ Release 4 (PR4; also sometimes referred to as ``Data Release 4'' --- DR4 --- or simply \npipe\footnote{This name is short for ``NERSC
  pipeline'', a name deriving from the computer facilities at which it
  is executed, namely the National Energy Research Scientific
  Computing Center (NERSC).}; \citealp{planck2020-LVII}). One
unique feature of this pipeline was its ability to analyze both LFI
and HFI jointly within the same framework. Combining some of the most
powerful features from each of the instrument analysis pipelines, this
approach led to further reduction of systematic errors in both data
sets. The resulting best-fit estimate of the reionization optical
depth from DR4 reads $\tau=0.058\pm0.006$ \citep{tristram:2020}.

An independent initiative to improve the \Planck\ processing was
\srollTwo\ \citep{delouis:2019}, which was a direct continuation of the
HFI \sroll\ effort \citep{planck2016-l03}. A defining feature of this
approach is improved ADC corrections, which in particular leads to
more robust large-scale polarization estimates. From the
\srollTwo\ polarization analysis alone, the current best-fit estimate
of the reionization optical depth is $\tau=0.0566^{+0.0053}_{-0.0062}$
\citep{pagano:2020}.

A second independent initiative is called \BP, and this is the primary
focus of the current paper and suite of companion papers. The scope of
this project is significantly different than the previous efforts, as
\BP\ aims at building a complete integrated end-to-end analysis
pipeline for current and future CMB experiments. The current work
focuses in particular on the \Planck\ LFI data set, although
significant effort is spent ensuring that the tools are generalizable
to other experiments. Indeed, one example of this is already presented
within the current project in the form of a preliminary application to
\WMAP\ \citep{bennett2012,bp17}.

Because instrumental systematics and residual foreground contamination
have such a dramatic impact on the large-scale CMB polarization
estimates, in this paper we will use the reionization optical depth as
a direct demonstration of the \BP\ framework, and our ultimate
scientific goal is to estimate the posterior distribution $P(\tau\mid
\d)$ from \Planck\ LFI and \WMAP\ observations, $\d$. The posterior
summarizes our knowledge about $\tau$ in the form of a probability
distribution, and we will estimate $P(\tau\mid \d)$ within a strict
Bayesian framework, with as few approximations and little data
selection as possible. We will avoid the use of cross-spectrum
techniques, which frequently are used to reduce the sensitivity of the
final products to instrumental systematics
\citep[e.g.,][]{planck2016-l05}. In this project, we aim to do the
opposite, and \emph{highlight} the impact of residual systematics,
such that, if needed, they can be addressed at a lower level of the
analysis. As such, internal consistency, goodness-of-fit and $\chi^2$
tests will play critical roles.

\subsection{Lessons learned from \Planck}
\label{sec:lessons}

To understand the background, historical context, and motivation for
the \BP\ program, it is useful to revisit the ``Lessons learned from
\Planck,''\footnote{\href{URL}{https://www.cosmos.esa.int/web/planck/lessons-learned}}
as compiled by the \Planck\ consortium in 2016. In Section~9.6
(``Understanding the data'') one can read the following:

\begin{quotation}
\emph{In a project like \Planck, ``understanding the data'' is
  certainly the most significant driver of the quality of the final
  products and science it can produce. This activity must be at the
  core of the data processing. It covers a lot of ground --
  photometry, optical response, time response, calibration, systematic
  effects, etc. -- all interlinked issues that can be diagnosed at
  many different levels in the data processing pipelines, from raw
  data streams to finished maps and scientific products.}

\emph{(\ldots) In the early phases of \Planck, much of the strategy was based on
  separating the various elements of the problem into independent
  parts. This was adequate for a first treatment of the data.
  However, as the quality of the data improved, it became harder to
  find and analyse subtler non-ideal effects, and to do so required a
  more integrated approach, where a variety of effects were treated
  simultaneously.}

\emph{(\ldots) An example is the influence of foregrounds on
  calibration: initially model foreground templates were used to
  isolate the CMB dipole signal (the calibrator), but in later stages
  the template had to be iterated within the calibration pipeline to
  include and self-consistently reduce the effects of polarization,
  sidelobes, dipoles, etc.}

\emph{(\ldots) As understanding of the data progresses, analysis --
  and the teams doing it -- need to become more and more integrated,
  pulling in parts of the pipeline which initially could be separated
  out.}
\end{quotation}

As described in these paragraphs, the analysis approach adopted by
\Planck\ became gradually more and more integrated as the effective
sensitivity of the data set improved through more refined analysis,
and new systematic effects were uncovered. Indeed, only toward the end
of the \Planck\ mission period did it become evident that the single
most limiting factor for the overall analysis was neither instrumental
systematics nor astrophysical foregrounds as such, but rather the
\emph{interplay} between the two. Intuitively speaking, the problem
may be summarized as follows: \emph{One cannot robustly characterize
  the astrophysical sky without knowing the properties of the
  instrument, and one cannot characterize the instrument without
  knowing the properties of the astrophysical sky.}  The calibration
and component separation procedures are intimately tied together. By
the time this issue was fully understood, there were neither
sufficient resources nor time to redesign a complete \Planck\ analysis
pipeline from bottom-up. An important organizational goal of the
\BP\ program has therefore been to provide a financial structure that
allows the team to consolidate this experience into practical computer
code, and make this publicly available to the general community.

\subsection{The next frontier: Primordial gravitational waves}
\label{sec:frontier}

While a statistically coherent analysis of existing data is
undoubtedly both interesting and useful in its own right, it is
important to emphasize that none of the developments detailed in this
work are likely to impact the overall cosmological concordance model
to any significant degree. Indeed, looking at the big picture, the
cosmological model has been remarkably stable even before \WMAP\ and
\Planck\ provided their high-precision measurements; see, e.g.,
\citet{wang:2003} for a summary of pre-\WMAP\ measurements and
constraints. The main achievement of \WMAP\ and \Planck\ has been to
refine this model to the level at which cosmology now is a
high-precision science within which competing theoretical models can
be tested and rejected at high statistical significance.

\Planck\ has for all practical purposes completed the study of primary
CMB temperature fluctuations.  Currently, however, another frontier is
driving the CMB field, namely the search for primordial gravitational
waves created during inflation. These are predicted to exist by
most inflationary theories, although their predicted amplitudes can vary
by many orders of magnitude, depending on the precise details of the
assumed inflationary model \citep[e.g.,][]{kamionkowski:2016}. Typically, this amplitude is quantified in
terms of the tensor-to-scalar ratio, $r$, which measures the ratio in
fluctuation power attributable to gravitational waves and ordinary
density perturbations, respectively.

If such gravitational waves do exist, one generically expects a
specific imprint in the CMB polarization field in the form of a
large-scale ``divergence-free'' or $B$-mode polarization signal. The
observational challenges associated with gravitational wave detection
are essentially the same as those for measuring $\tau$.  However, the
state-of-the-art upper limits on the tensor-to-scalar ratio are
$r<0.036$ \citep{bicep2021} and $r<0.032$ \citep{tristram:2021}, both
at 95\,\% confidence, which immediately implies that the $B$-mode
signal must be more than one order of magnitude smaller than the
$E$-mode signal, and thus no more than a few tens of nK in amplitude.

With such a small target amplitude, it is safe to assume that an
integrated analysis approach will no longer be optional for future CMB
missions, but rather a strict prerequisite. Establishing both the
experience and appropriate code required to implement such an approach
for future CMB missions is a main long-term scientific motivation
for the \BP\ program; current experiments such as \Planck\ and
\WMAP\ provide real-world test-beds that ensure that the \BP\ approach
is both realistic and practical.

\subsection{The \BP\ program}
\label{sec:bp_program}

We are now ready to formulate the main goal of the
\BP\ program:
\begin{quotation}
  \emph{\BP\ aims to implement and apply a single statistically
    coherent analysis pipeline to \Planck\ and other CMB data sets,
    processing raw uncalibrated time-ordered data into final
    astrophysical component maps, angular power spectra, and
    cosmological parameters within one single code.}
\end{quotation}
Important secondary goals include
\begin{enumerate}
\item to model and propagate instrumental uncertainties from raw
  time-ordered data into final high-level \Planck\ LFI scientific results;
\item to provide a computationally convenient interface to the raw
  \Planck\ LFI data that can be accessed and extended by external users;
\item to develop a framework that allows joint analysis of \Planck\ LFI with
  other data sets; and
\item to prepare for next-generation CMB experiments, in particular
  those aiming to detect primordial gravitational waves through their
  imprint on large-scale polarization of the CMB.
\end{enumerate}
The ``\BP'' name serves as a reminder that this work builds directly
on several decades of \Planck\ efforts and experience, while at the
same time highlights the fact that it aims to apply the
\Planck\ methodology to data sets beyond \Planck, both archival and
future.

Clearly, this is a very ambitious program that will require long-term
and dedicated support. The first stage of the program, which is
reported in the current suite of papers, has been funded within an
EU-based Horizon 2020 action called ``Leadership in Enabling and
Industrial Technologies'' (LEIT), as well as through various
individual grants. This funding only covers end-to-end analysis of the
\Planck\ LFI data, which is smaller in volume than HFI data, and therefore
serves as a convenient real-world test case for development purposes,
while still representing a very important scientific data set in its
own right.

As detailed in the H2020 LEIT contract, the \BP\ program started on
March 1st 2018, and ended formally on November 30th 2020; the total
duration of the funded program was thus strictly limited to two years
and nine months. During this period, large amounts of software,
products and documentation had to be written from scratch. Indeed, a
first fully operational pipeline was completed as late as June 2020,
and the first data release took place during an online conference on
November 20-22 2020; the conference was remote due to COVID-19 travel
restrictions. However, as discussed during that conference, the first
\BP\ data products were significantly affected by important unresolved
systematic effects, including a notable power excess in the Southern
Galactic hemisphere. Understanding those effects has been the main
focus of the collaboration since that time, and, as will be detailed
in the following and in a suite of companion papers, we now finally
believe that all previously reported issues have been successfully
understood and resolved.

\subsection{From \BP\ to \textsc{Cosmoglobe}}
\label{sec:cosmoglobe}

The amount of funding and human effort spent on mapping the sky at
radio, microwave, and submillimeter frequencies during the last few
decades has truly been massive \citep{bp05}, and an experiment like
\Planck\ alone represents an investment of about \euro 700\,M for the
spacecraft, payload, launch, and operations. The cost of \WMAP\ was
about \$150\,M, while a typical ground-based or sub-orbital CMB
experiment costs about 10\,M euros or dollars; LiteBIRD's cost cap is
\$300\,M, while the future CMB-S4 project is likely to cost about
\$500\,M.

Optimally exploiting these expensive data is clearly of the essence
for the field as whole. In particular, it is important to realize that
each experiment is unique in its own way, and teaches us something new
about the astrophysical sky. Some experiments have very broad
frequency coverage, while others have very high angular resolution;
some provide absolutely calibrated data, while others have very low
amounts of correlated noise. Some are designed for intensity
observations, while others are optimized for polarization
measurements. At the same time, each experiment invariably has its own
``blind-spots'' or degeneracies. For instance, both \Planck\ and
\WMAP\ have their own so-called ``poorly measured modes'', which
correspond to sky modes that are not well measured by their respective
scanning strategies \citep{planck2016-l01,bennett2012,bp17}.

The ultimate solution to breaking such degeneracies is through joint
analysis of complementary data sets; \WMAP\ can be used to constrain
the largest polarization modes and thereby improve \Planck's gain
model \citep{planck2016-l02}, while \Planck\ can be used to constrain
\WMAP's transmission imbalance parameters
\citep{jarosik2007}. Similarly, for future LiteBIRD or CMB-S4
analyses, \Planck\ and \WMAP\ must be used as models of the
temperature sky, while the new experiments will provide an entirely
new view of the polarization sky. And once those new polarization
measurements become available, it will be essential to re-analyze the
\Planck\ and \WMAP\ measurements, taking into account the new
polarization information, to further reduce their systematic
uncertainties.

In general, the most cost-efficient and productive way for the field
as a whole to make progress in the future is through joint and global
analysis of all available experiments.
\textsc{Cosmoglobe}\footnote{\url{https://cosmoglobe.uio.no}}
represents an Open Science platform for this work, which main goal is
to establish a global model of the radio, microwave, and submillimeter
sky through joint analysis of all available state-of-the-art data
sets. This project builds directly on the current \BP\ efforts, and
will use the computational machinery developed in the following as the
starting point for its multi-experiment analysis. All interested
parties and collaborations are highly encouraged to participate in the
\textsc{Cosmoglobe} effort.

\section{\BP\ analysis strategy and organization}
\label{sec:overview}

\begin{table*}[t]
  \begingroup
  \newdimen\tblskip \tblskip=5pt
  \caption{
    Overview of \BP\ and preliminary \cosmoglobe\ papers.
  }
  \label{tab:papers}
  \nointerlineskip
  \vskip -3mm
  \footnotesize
  \setbox\tablebox=\vbox{
    \newdimen\digitwidth
    \setbox0=\hbox{\rm 0}
    \digitwidth=\wd0
    \catcode`*=\active
    \def*{\kern\digitwidth}
    \newdimen\signwidth
    \setbox0=\hbox{-}
    \signwidth=\wd0
    \catcode`!=\active
    \def!{\kern\signwidth}
 \halign{
      \hbox to 5.5cm{#\leaderfil}\tabskip 1em&
      #\hfil \tabskip 0pt\cr
    \noalign{\doubleline}
      \omit\textsc{Reference}\hfil&
      \omit\textsc{Title}\hfil\cr
      \noalign{\vskip 4pt\hrule\vskip 4pt}
      \omit\parbox{3cm}{\textit{Pipeline}}\cr
      \noalign{\vskip 2pt}
      \hspace{3mm}\citet{bp01}&{I. Global Bayesian analysis of the \Planck\ Low Frequency
      Instrument data}\hfil\cr
      \hspace{3mm}\citet{bp02}&{II. CMB mapmaking through Gibbs sampling}
      \cr      
      \hspace{3mm}\citet{bp03}&{III. \commanderthree} \cr
      \hspace{3mm}\citet{bp04}&{IV. On end-to-end simulations in CMB analysis --- Bayesian versus frequentist statistics}  \cr
      \noalign{\vskip 5pt\hrule\vskip 5pt}
      \multispan2{\textit{Instrument characterization}}\hfil\cr
      \noalign{\vskip 2pt}
      \hspace{3mm}\citet{bp25}&{V. Minimal ADC Corrections for Planck LFI} \cr
      \hspace{3mm}\citet{bp06}&{VI. Noise characterization and modelling} \cr
      \hspace{3mm}\citet{bp07}&{VII. Bayesian estimation of gain and absolute calibration for CMB experiments} \cr
      \hspace{3mm}\citet{bp08}&{VIII. Efficient Sidelobe Convolution and Correction through Spin Harmonics} \cr
      \hspace{3mm}\citet{bp09}&{IX. Bandpass and beam leakage corrections} \cr
      \noalign{\vskip 5pt\hrule\vskip 5pt}
      \multispan2\textit{Cosmological and astrophysical
          results}\hfil\cr
      \noalign{\vskip 2pt}
      \hspace{3mm}\citet{bp10}&{X. \Planck\ LFI frequency maps with sample-based error propagation} \cr
      \hspace{3mm}\citet{bp11}&{XI. Bayesian CMB analysis with sample-based end-to-end error propagation}  \cr
      \hspace{3mm}\citet{bp12}&{XII. Cosmological parameter estimation with end-to-end error propagation}  \cr
      \hspace{3mm}\citet{bp13}&{XIII. Intensity foregrounds, degeneracies and priors}  \cr
      \hspace{3mm}\citet{bp14}&{XIV. Polarized foreground emission between 30 and 70\,GHz }  \cr
      \hspace{3mm}\citet{bp15}&{XV. Limits on Large-Scale Polarized Anomalous Microwave Emission from \Planck\ LFI and \WMAP}  \cr
      \noalign{\vskip 5pt\hrule\vskip 5pt}
      \multispan2{\cosmoglobe}\hfil\cr
      \noalign{\vskip 2pt}
      \hspace{3mm}\citet{bp05}&{From \BP\ to \cosmoglobe: Open Science, reproducibility, and data longevity} \cr
      \hspace{3mm}\citet{bp17}&{From \BP\ to \cosmoglobe: Preliminary WMAP Q-band analysis} \cr
      \noalign{\vskip 4pt\hrule\vskip 5pt} } }
  \endPlancktablewide \endgroup
\end{table*}

\subsection{End-to-end Bayesian CMB analysis}

Recognizing the lessons learned from \Planck\ as summarized in
Sect.~\ref{sec:lessons}, the defining design philosophy of \BP\ is
tight integration of all steps from raw time-ordered data processing
to high-level cosmological parameter estimation. Traditionally, this
process has been carried out in a series of weakly connected steps,
pipelining independent executables with or without human
intervention. Some steps have mostly relied on frequentist statistics,
employing forward simulations to propagate uncertainties, while other
steps have adopted a Bayesian approach, using the posterior
distribution to quantify uncertainties. For instance, traditional
mapmaking is a typical example of the former
\citep[e.g.,][]{ashdown2007b}, while cosmological parameter estimation
is a typical example of the latter \citep[e.g.,][]{cosmomc}; for component
separation purposes, both approaches have been explored in the
literature \citep[e.g.,][]{planck2014-a10}.

\BP\ is the first real-world CMB analysis pipeline to adopt an
end-to-end Bayesian approach. This solution was in fact first proposed
by \citet{jewell2004}. However, it took more than 15 years of computational and
algorithmic developments to actually make it feasible.

Perhaps the single most important advantage of a uniform Bayesian
approach is that it allows seamless propagation of uncertainties
within a well-established statistical framework. This aspect will become critically
important for future experiments, as demonstrated by \Planck. For most
CMB experiments prior to \Planck, the dominant source of uncertainty
was noise; for most CMB experiments after \Planck, the dominant source
of uncertainty will be instrumental systematics, foreground
contamination, and the interplay between the two.  As a logical
consequence of this fact, \BP\ adopts a consistent statistical
framework that integrates detailed error propagation as a foundational
feature.

The Bayesian approach also has several notable advantages in terms of
intuition and transparency. In particular, the most critical step for
any Bayesian analysis is the definition of the data model. This may
often be described in terms of a handful of equations, and these
equations subsequently serve as a road-map for the entire
analysis. While the complexity of the numerical implementation may
vary from model to model, the posterior distribution itself has a very
intuitive and direct interpretation.

At a practical level, integrating the entire pipeline into a single
computational code also has significant advantages in terms of net
computational speed and resources. Not only are slow disk operations
reduced to a minimum by performing all operations within one single
code, but more importantly, all intermediate human interactions are
eliminated from the process. This both saves significant amounts of
human time required for ``code shepherding'' and file transfers, and
it significantly reduces the risk of human errors. Thus human
resources are saved that can be better spent on fundamental modelling
aspects.

A fourth significant advantage of end-to-end integration is increased
transparency of implicit and explicit priors. For a distributed
analysis process, it is critically important to communicate all
assumptions made in each step to avoid errors, while in an integrated
approach internal inconsistencies become much more visible; there are
simply fewer opportunities for misunderstandings to propagate
undetected throughout an integrated analysis pipeline.

\subsection{\commander}
\label{sec:commander}

We adopt
\commandertwo\ \citep{eriksen:2004,eriksen2008,seljebotn:2019}, a
well-established Bayesian CMB Gibbs sampler developed for \Planck, as
the starting point of our pipeline. As demonstrated in
\citet{planck2016-l04}, this code already supports Bayesian
multi-resolution component separation, which is precisely the
operation that connects low-level mapmaking to high-level cosmological
parameter estimation. A main implementational goal for \BP\ is thus to
extend this framework to incorporate Bayesian calibration and
mapmaking, as well as to connect component separation and cosmological
parameter estimation.

We will refer to three different versions of the \commander\ code in
the following. \commanderone\ refers to the original implementation
described by \citet{eriksen:2004,eriksen2008}, which at the beginning
of the \BP\ project represented the most mature version in terms of
foreground spectral parameter fitting. However, a major limitation of
that code is a requirement of common angular resolution among all data
sets. \commandertwo\ removes this limitation through explicit beam
convolution for each frequency map during component separation, as
detailed by \citet{seljebotn:2019}, and thereby allows for full
resolution analysis of the \Planck\ data. Due to the much higher
computational cost associated with increased angular resolution, the
development of \commandertwo\ required a re-implementation of the
original algebra from scratch, adopting a much more fine-grained
parallelization strategy than \commanderone.

\commanderthree\ \citep{bp03} refers to the time-domain version of the
algorithm, as developed in \BP, and is a direct generalization and
extension of \commandertwo\ in terms of code implementation. As a
result, \commandertwo\ is no longer an independent code, but we will
still refer to it in cases where it might be convenient for
pedagogical purposes to distinguish between multi-resolution component
separation in the pixel-domain versus the time-domain. All
\commander\ source codes are available under a GNU Public Library
(GPL) OpenSource
license.\footnote{\url{https://github.com/Cosmoglobe/Commander}}

\subsection{Paper organization}
\label{sec:papers}

The \BP\ methodology and results are described in a suite of companion
papers, as listed in Table~\ref{tab:papers}. The present paper
provides a broad overview in terms of motivation, algorithms, and main
results. However, it is not intended to be comprehensive, and specific
details are deferred to the relevant companion papers.

The remaining papers may be divided into four main categories, namely
1) pipeline papers; 2) instrument characterization papers; 3)
cosmological and astrophysical results papers; and 4) joint
\BP--Cosmoglobe papers.  The first category of papers provides a
comprehensive overview of the current implementation of the
\BP\ pipeline, at a level that is hopefully sufficiently detailed to
allow external users to understand intuitively its statistical and
computational basis, what assumptions it relies on, and what its
limitations are.  The ultimate goal of these papers is that external
users should be able to repeat and extend the work that is presented
here.

The second category of papers address the various relevant
instrumental parameters required to process the raw time-ordered data
into sky maps. These include noise characterization, gain estimation,
sidelobe corrections, and bandpass and beam mismatch modelling. Each
paper aims both to provide an intuitive understanding of the effect in
question, and to show how it impacts the final results.  These papers
also demonstrate how to quantitatively model each instrumental effect,
and how to propagate uncertainties into other parameters. Particular
emphasis is placed on building intuition regarding leading internal
parameter degeneracies, both among the various instrumental parameters
and with astrophysical and cosmological parameters.

The third category of papers present the main scientific results in
terms of frequency and component maps, as well as angular power
spectra and cosmological parameters. Consistency between the
\BP\ products and non-\Planck\ sets is also considered in this
category of papers.

The fourth category includes papers that aim to generalize the
\BP\ data model to other data sets within the
\cosmoglobe\ framework. One worked example is provided in the form of
a preliminary \WMAP\ $Q$-band reanalysis \citep{bp17}.

We note that, in the spirit of reproducibility and accessibility
\citep{bp05}, a significant emphasis is put on intuition and
background throughout the \BP\ papers. The paper suite is intended to
be largely self-contained, and detailed knowledge of the
\Planck\ publication list is not an assumed prerequisite. As such, a
substantial amount of review material is included, both in terms of
general background material and algorithmic details. The style of the
papers is consciously tuned toward Ph.D.\ students and early
postdoctoral fellows, rather than seasoned CMB experts.

\section{Parameterizing the microwave sky}
\label{sec:sky}

As already noted, the single most important component in any Bayesian
analysis is the parametric model that is fitted to the data. In our
case, this model consists of both astrophysical and instrumental
components. In this section we consider the cosmological and
astrophysical parameters, before introducing the instrumental
parameters in the next section.

\subsection{Conventions: Stokes parameters, pixelization, spherical harmonics, and units}
\label{sec:units}

In order to characterize each astrophysical component quantitatively,
we need to introduce some general notation and conventions. First,
each astrophysical component will be described in terms of three
Stokes parameters, namely intensity (denoted either $I$ or $T$) and
two linear polarizations (denoted $Q$ and $U$). We will ignore
circular polarization ($V$) for now, but we note that this may be
added in future work.

To discretize the Stokes parameters on the sphere, we adopt the
\HEALPix\ pixelization\footnote{\url{http://healpix.jpl.nasa.gov}}
\citep{gorski2005}. This pixelization has several highly desirable
properties, including equal-area pixels and support for fast spherical
harmonics transforms, and is now effectively a standard in modern CMB
analysis. The \HEALPix\ pixel resolution is controlled through a
parameter called $N_{\mathrm{side}}$, and the total number of pixels
on the sky is $N_{\mathrm{pix}}=12N_{\mathrm{side}}^2$. We organize
the Stokes parameters into vectors of length $3N_{\mathrm{pix}}$,
simply by stacking $\{T,Q,U\}$ into a map vector $\s(\hat{n})$, where
$\hat{n}$ is a unit direction vector.

Unless otherwise noted, we define the Stokes parameters with respect
to Galactic coordinates. We adopt the cosmological convention for the
polarization angle, $\gamma$, in which $\gamma=0$ for vectors pointing
north and increases westward. This is opposite to the IAU convention
used in most other fields of astronomy, in which $\gamma$ increases
eastward. To convert from one convention to the other, one must
multiply Stokes $U$ by $-1$.

The Stokes polarization parameters $Q$ and $U$ form a spin-2 field,
which intuitively may be interpreted as a ``headless vector
field''. In contrast, the intensity $T$ is a spin-0 field, and does
not change under rotations. Thus, when rotating Stokes parameters by
an angle $\alpha$, the transformed Stokes parameters are
\begin{equation}
  \left[
    \begin{array}{c}
      T' \\
      Q' \\
      U'
    \end{array}
    \right]
  =
  \left[
    \begin{array}{ccc}
      1 & 0 & 0 \\
      0 & \cos 2\alpha &  -\sin 2\alpha \\
      0 & \sin 2\alpha &  \phantom{-}\cos 2\alpha
    \end{array}
    \right]  
  \left[
    \begin{array}{c}
      T \\
      Q \\
      U
    \end{array}
    \right].
\end{equation}

As described by \citet{zaldarriaga1997}, the polarization Stokes
parameters may be expanded into spherical harmonics through the
following relations,
\begin{align}
  T(\hat{n}) &=
  \sum_{\ell=0}^{\ell_{\mathrm{max}}} \sum_{m=-\ell}^{\ell} a_{\ell m} 
  Y_{\ell m} (\hat{n})\\
  (Q\pm \mathrm iU)(\hat{n}) &=
  \sum_{\ell=2}^{\ell_{\mathrm{max}}} \sum_{m=-\ell}^{\ell} {_{\pm2}}a_{\ell m} 
    \,{_{\pm2}}Y_{\ell m} (\hat{n}),
\end{align}
where $_ka_{\ell m}$ are called (spin-$k$) spherical harmonic
coefficients. The polarization coefficients are often combined
algebraically into $E$ and $B$ coefficients,
\begin{align}
  a_{\ell m}^E &= -\frac{1}{2}\left(_{2}a_{\ell m} + {_{-2}}a_{\ell m}
  \right)\\
  a_{\ell m}^B &= \frac{\mathrm i}{2}\left(_{2}a_{\ell m} - {_{-2}}a_{\ell m} \right),
\end{align}
which each form a spin-0 field, fully analogous to the intensity
$T$.

From the spherical harmonic coefficients we may compute the observed
angular power spectrum as
\begin{align}
\sigma_{\ell}^{XY} = \frac{1}{2\ell+1} \sum_{m=-\ell}^{\ell} \left(a_{\ell}^X\right)^*
a_{\ell m}^Y,
\label{eq:sigmal}
\end{align}
where $\{X,Y\} \in \{T,E,B\}$. These quantify the strength of
fluctuations at a given multipole $\ell$ as directly measured from
some sky map. In addition, we define the ensemble-averaged power
spectrum as
\begin{equation}
C_{\ell}^{XY} \equiv \left<\left(a_{\ell}^X\right)^* a_{\ell m}^Y \right> = \left<
\sigma_{\ell}^{XY} \right>,
\end{equation}
where brackets indicate an average over statistical realizations. This
function is thus independent of the observed sky, and only depends on
the model that describes the field in question. 

Finally, each sky map $\s$ must be quantified in terms of a physical
unit. In the following work, we will encounter many different
conventions for this, depending on the particular application in
question. However, three conventions are more common than others, and
we limit our discussion here to these special cases.

The first measure is \emph{surface brightness per solid angle}, which
simply measures the amount of energy emitted by some source per
surface area, per frequency interval,  per sky solid angle. This is often
measured in units of ${\textrm{MJy}\,\textrm{sr}^{-1} \equiv
10^{-20}\,\textrm{W}\,\textrm{m}^{-2}\,\textrm{Hz}^{-1}\textrm{sr}^{-1}}$,
and it quantifies the specific intensity $I_{\nu}$ of a given source as a
function of wavelength, $\nu$.

The second measure we will use is \emph{thermodynamic temperature}. In
this case, we identify the intensity with that emitted by a blackbody
source with temperature $T$,
\begin{equation}
  I_{\nu} = B_{\nu}(T) =
  \frac{2h\nu^3}{c^2}\frac{1}{\e^{\frac{h\nu}{kT}}-1},
  \label{eq:T_thermo}
\end{equation}
where $h$ is Planck's constant, $c$ is the speed of light, and $k$ is
the Boltzmann constant. This measure is particularly useful for CMB
applications, because the CMB is itself a near-perfect blackbody, and
a single temperature $T(\hat{n})$ therefore uniquely specifies its
intensity at any wavelength at a given position. The unit for
thermodynamic temperature is denoted $\textrm{K}_{\mathrm{CMB}}$ or
simply $\mathrm{K}$. 

Our third and final measure is the \emph{brightness temperature} or
\emph{Rayleigh-Jeans temperature}, $T_{\mathrm{RJ}}$. This is defined
by the long wavelength limit ($h\nu \ll kT$) of
Eq.~\eqref{eq:T_thermo}, such that
\begin{equation}
  I_{\nu} = \frac{2\nu^2kT_{\mathrm{RJ}}}{c^2}.
  \label{eq:Krj}
\end{equation}

While the thermodynamic temperature is convenient to describe the CMB,
most astrophysical foreground signals have a non-blackbody nature, and
are more naturally quantified in terms of brightness temperature. In
particular, while the spectral energy density of many foregrounds can
span many tens of orders of magnitude when expressed in
$\mathrm{K_{CMB}}$, they are usually limited to a few orders of
magnitude when expressed in either \MJysr\ or
$\textrm{K}_{\mathrm{RJ}}$. To avoid numerical problems, all
astrophysical components are therefore expressed in units of
$\textrm{K}_{\mathrm{RJ}}$ internally in \commander, and only
converted to the respective natural unit before outputting results to
disk. Monochromatic conversion between $\mathrm{K}_{\mathrm{RJ}}$ and
\MJysr\ is performed through Eq.~\eqref{eq:Krj}, while monochromatic
conversion between $\mathrm{K}_{\mathrm{RJ}}$ and
$\mathrm{K}_{\mathrm{CMB}}$ is given by
\begin{equation}
  \Delta T_{\mathrm{CMB}} = \frac{\left(\e^{x}-1\right)^2}{x^2\e^{x}}
  T_{\mathrm{RJ}},
  \label{eq:rj2cmb}
\end{equation}
where $x=h\nu/kT_{0}$, and $T_0 = 2.7255\,\mathrm{K}$ is the mean CMB
temperature \citep{fixsen2009}. Note that this conversion applies only
to small temperature variations around the CMB mean value, $\Delta T
\equiv T - T_0$, which is precisely the form of most CMB temperature
maps in common use today.

We are now ready to write down parametric models for each of the main
astrophysical components that are relevant for the \Planck\ frequency
range. Each component will be described in terms of a spectral energy
density (SED) in brightness temperature units, and, in some cases, in terms of an
angular power spectrum or some other similar spatial coherence
measure.

\subsection{Cosmic microwave background anisotropies}
\label{sec:cmb}

We start our survey with the CMB component, which is the
scientifically most important one for \Planck. For this, we first
define $\s^{\mathrm{CMB}}$ to be a $3N_{\mathrm{pix}}$ sky vector of
CMB Stokes parameters as described above. Second, we assume that the CMB
SED may be approximated as a blackbody. As such, its brightness
temperature SED is given by Eq.~\eqref{eq:rj2cmb}, 
\begin{equation}
  \s_{\mathrm{RJ}}^{\mathrm{CMB}}(\nu) \propto
  \frac{x^2\e^x}{\left(\e^{x}-1\right)^2} \s^{\mathrm{CMB}},
\end{equation}
where $x=h\nu/kT_0$. (Note that we define the effective SED only up to
a normalization constant, as we will typically parameterize each
component in terms of an amplitude map at a given reference frequency
times the SED normalized to unity at the reference; any normalization
factor is therefore accounted for in the amplitude coefficient.)

In addition to the pure cosmological blackbody SED, the CMB component
exhibits a relativistic correction due to our motion with respect to
the CMB monopole. This effect is frequency dependent, and is primarily
dominated by quadrupole correction \citep{Notari:2015}. Unlike the
official \Planck\ sky maps \citep{planck2016-l02,npipe}, we do
\emph{not} subtract this component from the frequency maps
\citep{bp10}, but rather include it as part of the
signal model \citep{bp13}.

For component separation purposes, these are the \emph{only} assumption
we make regarding the CMB. However, for cosmological parameter
estimation purposes, we make two important additional assumptions,
namely that the CMB temperature flucutations are both Gaussian
distributed and statistically isotropic. The
assumption of Gaussianity determines the conditional probability
distribution for the CMB signal,
\begin{equation}
  P(\s\mid C_{\ell}) \propto \frac{\e^{-\frac{1}{2}\s^t
      \S^{-1} \s}}{\sqrt{|\S|}},
  \label{eq:Pcmb}
\end{equation}
where $\S$ is the covariance matrix of the CMB fluctuation field, and
we have dropped the ``CMB'' superscript for convenience. The assumption
of statistical isotropy implies that $\S$ is fully specified in terms
of the angular power spectrum,
\begin{equation}
  \S^{XY}_{\ell m, \ell' m'} \equiv \left<\left(a^X_{\ell}\right)^*
  a^{Y}_{\ell' m'}\right> = C^{XY}_{\ell m} \delta_{\ell
    \ell'}\delta_{m m'}.
\end{equation}
For practical parameter estimation purposes, both of these assumptions
have been shown to be excellent approximations to the true CMB sky
\citep[see, e.g.,][and references
  therein]{planck2016-l07,planck2016-l09}.

The connection to cosmological parameters, such as the Hubble constant
$H_0$ or the reionization optical depth $\tau$, is made through
cosmological Boltzmann codes, such as \texttt{CMBfast}
\citep{seljak:1996} or \texttt{CAMB} \citep{Lewis:1999bs}. These
deterministically calculate the ensemble-averaged CMB power spectrum
based on well-understood physics given some specific set of
cosmological parameters, $\xi$. However, this calculation is only
straightforward going from $\xi$ to $C_{\ell}$; it is highly
nontrivial to go directly from $C_{\ell}$ to $\xi$. Instead, Markov
Chain Monte Carlo (MCMC) methods such as \texttt{CosmoMC}
\citep{cosmomc} are typically employed to perform the inversion, in
which a series of parameter combinations are proposed and rejected or
accepted, ultimately resulting in a set of parameter samples that
jointly represents the final parameter posterior distribution. As
described in Sect.~\ref{sec:bp_program}, the goal of the \BP\ program
is to implement a similar MCMC method that accounts for the entire
process from raw time-ordered data to final cosmological parameters
with full Bayesian end-to-end error propagation.

\subsection{Galactic foreground emission}
\label{sec:gal_fg}

The second most important class of sky emission components consists of diffuse
Galactic foregrounds. These all originate from within the Milky Way,
and are due to particles (electrons, ions, dust, etc.) associated with
various processes such as star formation or supernova
explosions. Furthermore, these particles all interact with the same
magnetic field, and as a result they produce correlated polarized
emission. In this section, we provide a brief survey of each of the
main physical emission mechanisms, with a particular focus on
parametric models.

\subsubsection{Synchrotron emission}
\label{sec:synchrotron}

At low microwave frequencies, synchrotron emission dominates the radio
sky. This emission is mostly due to relativistic electrons ejected
from supernova, spiralling in the magnetic field of the Milky Way. CMB
observations are typically made at frequencies in the range of tens or
hundreds of GHz, and at these frequencies, the synchrotron SED falls
rapidly with increasing frequency. Indeed, detailed models and observations both
suggest that the effective spectrum may be closely approximated by a
power-law at frequencies higher than a few gigahertz, with some
evidence for possible curvature. In this work, we therefore follow
\citet{kogut:2012}, and adopt a general SED model of the form
\begin{equation}
  \s_{\mathrm{RJ}}^{\mathrm{s}}(\nu) \propto
  \left(\frac{\nu}{\nu_{0,\mathrm{s}}} \right)^{\beta + C\ln \nu/\nu_{0,\mathrm{s}}},
\end{equation}
where $\nu_{0,\mathrm{s}}$ is a reference frequency, $\beta$ is a
power-law index, and $C$ is a curvature parameter. However, in most
cases we set $C=0$, as the signal-to-noise ratio for this parameter is
very low with the limited data set considered in this work \citep{bp14}.

When the local magnetic field is highly structured, synchrotron
emission can be highly polarized, with a theoretical maximum
polarization fraction of $p=75\,\%$. In practice, this value is
decreased due to line-of-sight and volume integration effects, and
according to \Planck\ and \WMAP, more typical values are
$\lesssim15\,\%$ at high Galactic latitudes, with extreme cases
reaching 30--50\,\% only in a few large-scale supernova remnants that,
when projected on the sky, take the form of so-called ``Galactic
spurs'' \citep{planck2014-a31}.

At low frequencies, polarized synchrotron emission is also
significantly affected by Faraday rotation \citep[e.g.,][and
  references therein]{beck:2013}. This effect is caused by circular
birefringence, i.e., left- and right-handed circular polarized
emission travel at different speeds through a magnetic field embedded
in an ionized medium, resulting in a net rotation of the polarization
angle of linearly polarized emission. The polarization angle rotation
is proportional to the magnetic field strength as well as to the
square of the wavelength of the emission. Numerically, the rotation
angle is typically a few degrees at 23\,GHz at low Galactic latitudes
\citep{Carretti:2019, fuskeland:2019}, and we neglect it for
the higher-frequency \Planck\ and \WMAP\ surveys considered here.

\subsubsection{Free-free emission}
\label{sec:freefree}

Free-free emission (or \emph{bremsstrahlung}) arises primarily from free
electrons scattering off protons without being captured, and emitting a
photon in the process. Since free electrons only exist in appreciable
amounts when the temperature of the medium is comparable to the
hydrogen binding energy, corresponding to $10^3-10^4\,\mathrm{K}$,
free-free emission predominantly traces hot \ion{H}{ii} regions and, as such,
active star forming regions. Free-free emission is particularly
important for CMB experiments because it is the only foreground
component that is non-negligible at all frequencies between 1 and
1000~GHz, and it is therefore particularly sensitive to degeneracies
with respect to both the CMB and other foreground components.

The free-free SED depends primarily on the number of free protons and electrons
along the line of sight, which typically is quantified in terms of the
\emph{emission measure} (EM), i.e., the integral of the square
electron density along the line of sight,
\begin{equation}
\mathrm{EM} \equiv \int_{0}^{\infty} n_e^2\, \mathrm dl,
\end{equation}
where the number densities of free protons and electrons are assumed to be equal.
The conventional unit adopted for the EM is
pc~cm$^{-6}$, and typical values for the Milky Way range between 0 and
1000 \citep{planck2014-a12}.

Assuming local thermodynamic equilibrium and first considering an
optically thick medium, the free-free SED is determined by a blackbody
spectrum given its electron temperature, $\Te$, alone. Since the optical
depth drops rapidly with increasing frequency, however, free-free emission in
astrophysical contexts and at CMB frequencies is optically thin. Hence,
the effective SED can be expressed as
\begin{equation}
  \s^{\mathrm{ff}}_{\mathrm{RJ}}(\nu) = \Te \,(1-\e^{-\tau}).
  \label{eq:ff_sed}
\end{equation}
As shown by \citet{dickinson2003} and \citet{draine2011}, $\tau$ may be very well
approximated by
\begin{equation}
    \tau = 0.05468 \cdot \Te^{-3/2} \cdot \nu_{9}^{-2} \cdot \EM \cdot g_\mathrm{ff},
\end{equation}
where
\begin{equation}
g_{\mathrm{ff}} = \log\left\{\exp\left[5.960
  -\sqrt{3}/\pi\log(\nu_9 \cdot T_{4}^{-3/2})\right] + \e\right\}
\end{equation}
is called the Gaunt factor, and $\nu_9$ and $T_{4}$ are the frequency
and the electron temperature measured in units of GHz and
$10^4\,\mathrm{K}$, respectively. 

This SED is a nonlinear function of EM and $T_{e}$. A complete
free-free model therefore corresponds to a complicated probability
distribution with expensive special-purpose sampling algorithms, as
for instance employed in \citet{planck2014-a11}. In this work, we
instead adopt a simpler linearized version of Eq.~\eqref{eq:ff_sed} that
is only strictly valid in the optically thin case, $\tau \ll 1$, namely
\begin{equation}
  \s^{\mathrm{ff}}_{\mathrm{RJ}}(\nu) \propto\frac{g_{\mathrm{ff}}(T_e)}{\nu^2},
  \label{eq:ff_sed_linear}
\end{equation}
and we correspondingly quantify the free-free amplitude in terms of
the observed signal at a given reference frequency in
$\mu\mathrm{K}_{\mathrm{RJ}}$, as opposed to the full nonlinear EM
parameter described above.

There is essentially no effective alignment mechanism for thermal
electrons in a hot medium, and large-scale free-free emission is
therefore expected to be nearly unpolarized. The main exception to
this are sharp edges around hot \ion{H}{ii} regions, which do
introduce a preferred direction in the emission geometry. However,
even these are only expected to be mildly polarized, and over large
angular scales, the net polarization fraction is expected to be well
below 1\,\% \citep[see discussion in][]{keating1998}. In the
following, we thus assume that free-free emission is completely
unpolarized.

\subsubsection{Thermal dust emission}
\label{sec:dust}

The interstellar medium (ISM) is filled not only with hydrogen and
electrons, but also with tiny dust grains ranging in diameter from less 
than a nanometer (i.e., a few atoms across) to roughly a micron (i.e.,
thousands of atoms across). Dust grains typically condense
from stellar outflows and ejecta, and so dust abundance is correlated
with star formation. Newly-formed dust is rapidly mixed in the dynamic,
turbulent ISM, where it undergoes significant processing. Dust is
therefore ubiquitous in the Galaxy, found wherever there is interstellar
gas. 

It is known from spectroscopic features that dust is made from, at
minimum, silicate and carbonaceous materials. However, the precise
grain composition is likely to vary with local environment. Dust
grains are heated by ambient stellar radiation, and large grains
accounting for the bulk of the dust mass equilibriate to a
steady-state temperature ranging between 10 and 30\,K. This energy is
thermally re-emitted with a peak wavelength in the sub-mm frequency
range, typically between 1000 and 3000\,GHz. Since these grains are
inefficient radiators at longer wavelengths, the thermal dust SED
falls rapidly at frequencies below the peak, where CMB observations
are typically carried out.  The varied composition and geometry of ISM
dust particles makes the thermal dust SED significantly more
complicated to model from first principles, when compared to the
free-free emission described above; for recent examples of such
modelling efforts, see, e.g., \citet{guillet2018} and
\citet{Hensley2022}.

In practice, simpler fitting formulae are
therefore usually adopted for practical analyses, and one particularly
popular class of models is the so-called modified blackbody
spectrum, which in intensity units reads
\begin{equation}
I^{\mathrm{d}}_{\nu} \propto \tau \nu^{\beta_{\mathrm{d}}} B_{\nu}(T_{\mathrm{d}}).
\end{equation}
This function is simply a blackbody spectrum with temperature
$T_{\mathrm{d}}$, modulated by a power-law having index
$\beta_{\mathrm{d}}$. In physical terms, this corresponds to dust having
an opacity that scales as $\nu^\beta_{\mathrm{d}}$, a reasonable approximation
for wavelengths longer than $\sim20\,\mu$m \citep{Hensley2020}.

The amplitude is, as for free-free emission,
given by the optical depth, $\tau$, which depends directly on the
surface density of particles along the line of sight. Typical
numerical values for these three parameters are $\tau \sim 10^{-6}$, 
$\beta_{\mathrm{d}}\sim 1.6$, and
$T_{\mathrm{d}}\sim20\,$K. Intuitively speaking, $\beta_{\mathrm{d}}$
determines the slope (or first derivative in log-log space)
of the SED below 200~GHz, while $T_{\mathrm{d}}$ determines the SED
peak position, and second derivative at lower frequencies. However, we
will model thermal dust emission in terms of brightness
temperature, and in these units the effective SED may be written in
the form
\begin{equation}
\s^{\mathrm{d}}_{\mathrm{RJ}}(\nu) \propto
\frac{\nu^{\beta_{\mathrm{d}}+1}}{\e^{h\nu/kT_{\mathrm{d}}}-1}. 
\end{equation}

Interaction with gas and radiation torques up grains, and they tend to rotate
about their axis of greatest moment of inertia, i.e., their short axis.
Dust grains having unpaired electrons can develop a non-zero
magnetic moment anti-parallel to their angular velocity through the Barnett effect
\citep{Dolginov1976}.
Dissipative processes act to align the rotation axis with the local magnetic field.
For a more detailed discussion of grain alignment, see \citet{Andersson2015}.

The preferential alignment of the short axes of grains with the local magnetic field
leads to significant net polarization from the ensemble of grains. Thermal dust polarization 
fractions as large as 20\,\% are found using the high frequency \Planck\ polarization
measurements \citep{planck2016-l11A}. We therefore include all three
Stokes parameters in our thermal dust model. At the same time, we note
that the highest polarization-sensitive \Planck\ frequency channel is
353~GHz, and this does not provide sufficient frequency range to allow
an independent estimate of the thermal dust temperature in polarization. We therefore assume
the same $T_{\mathrm{d}}$ for intensity and polarization, while
$\beta_{\mathrm{d}}$ is allowed to be different.

\subsubsection{Spinning dust (or anomalous microwave) emission}
\label{sec:ame}

Dust grains rotate with rotational kinetic energy of order the thermal
energy in the ambient gas. Consequently, sub-nanometer grains can achieve
rotational frequencies of tens of GHz. If these grains possess an electric dipole
moment, as generally expected for particles of this size \citep{MaciaEscatllar2020}, 
this rotation produces emission in the microwave frequency
range, as first predicted theoretically by \citet{erickson:1957}, and
described quantitatively by \citet{draine:1998}. The spinning dust
mechanism currently provides the most popular theoretical explanation
for so-called ``anomalous microwave emission'' (AME) observed around
20~GHz in CMB surveys, as first identified and named by
\citet{leitch:1997}.

In this work, we will adopt a spinning dust model for this component,
starting from an SED template, $\s^{\mathrm{sd}}_0(\nu)$, computed
with the \texttt{SpDust2} code \citep{ali-haimoud:2009,
  ali-haimoud:2010, silsbee:2011} for environmental parameters typifying
the Cold Neutral Medium. This
spectrum is intrinsically computed in intensity units, in which it
peaks at 30~GHz. After converting to brightness temperature by scaling
with $\nu^{-2}$, as given by Eq.~\eqref{eq:Krj}, the peak shifts to
17.4~GHz, and the overall spectrum is less than 1\,\% of its peak
value at frequencies below 1.3~GHz or above 66~GHz. To fit this SED
model to the data, we follow \citet{bennett2012}, and introduce a peak
position parameter, $\nu_{\mathrm{p}}$, that shifts the spectrum
rigidly in $\log\nu$--$\log\s$ space,
\begin{equation}
\s^{\mathrm{sd}}_{\mathrm{RJ}}(\nu) \propto
\nu^{-2}\,\s^{\mathrm{sd}}_0\left(\nu \cdot
\frac{30.0\,\mathrm{GHz}}{\nu_p}\right)
\label{eq:spindust}
\end{equation}

We note, however, that this emission component is associated with
large uncertainties, both in terms of the physical mechanism that is
actually responsible for the observed emission, and in terms of
detailed modelling within the chosen paradigm. In general, we assume
this component to be unpolarized, and we adopt the spinning dust model
in Eq.~\eqref{eq:spindust} for the AME component.

Despite sensitive searches in individual objects \citep{QUIJOTE_I_2015, QUIJOTE_II_2016}
and over large sky areas \citep{macellari2011},
polarization has not been detected in the AME. In principle, AME could be highly polarized if
small spinning grains are efficiently aligned. Theoretical estimates of the alignment efficiency
of ultrasmall grains vary widely, with predicted AME polarization fractions ranging 
from $\lesssim 1\%$ \citep{hoang2013} to 
completely negligible \citep{draine2016}. We perform a detailed study of AME
polarization in \citet{bp15}, but assume it to be unpolarized in all other analysis.

\subsubsection{Carbon monoxide emission}
\label{sec:co}

In the same way that rotating dust particles can emit radio emission,
so can molecules with a non-zero electric dipole moment. One
particularly important example of such molecules is carbon monoxide
(CO), which resides primarily in dense clouds where it is shielded from
destruction by UV radiation. The most common isotopologe of CO is
$^{12}\mathrm{C}^{16}\mathrm{O}$ (abbreviated $^{12}\mathrm{CO}$),
which is typically 10--100 times more abundant than
$^{13}\mathrm{C}^{16}\mathrm{O}$ (abbreviated $^{13}\mathrm{CO}$)
\citep{szucs:2014}.

For a simple system such as CO, quantum mechanical effects are highly
significant. In particular, only very specific rotational states are
allowed by quantization of angular momentum. Let us denote the masses
of the two atoms by $m_{\mathrm{C}}$ and $m_{\mathrm{O}}$,
respectively, and the corresponding atomic distances from their center
of mass by $r_{\mathrm{C}}$ and $r_{\mathrm{O}}$. We also define
$r_{\mathrm{CO}}=r_{\mathrm{C}}+r_{\mathrm{O}}$ to be the effective
atom size and $m_{\mathrm{CO}} =
m_{\mathrm{C}}m_{\mathrm{O}}/(m_{\mathrm{C}}+m_{\mathrm{O}})$ its
reduced mass.

With this notation, the moment of inertia of the CO molecule is $I =
m_{\mathrm{C}}^{\phantom 2}r_{\mathrm{C}}^2 + m_{\mathrm{O}}^{\phantom 2}r_{\mathrm{O}}^2$. The
corresponding eigenvalues of the Schr\"odinger equation are given by
\begin{equation}
  E_{\mathrm{rot}} = \frac{J(J+1)\hbar^2}{2I},
\end{equation}
where $J=0,1,\ldots$ is the angular momentum quantum number. Quantum
mechanically allowed energy changes are given by $\Delta J=\pm1$, and
each such transition either absorbs or emits a photon with wavelength
\begin{equation}
  \nu_0 = \frac{\Delta E_{\mathrm{rot}}}{h} = \frac{\hbar J}{2\pi I}
	= \frac{\hbar J}{2\pi m_\mathrm{CO}^{\phantom{2}} r_{\mathrm{CO}}^2}, \quad J = 1, 2, \ldots
\end{equation}
For the $^{12}\mathrm{CO}$ $J$=1$\leftarrow$0 transition, one finds
$\nu_0=115.27$~GHz, while for the $^{13}\mathrm{CO}$ $J$=1$\leftarrow$0
transition, it is $\nu_0=110.20$~GHz. Higher-order transitions, such
as $J$=2$\leftarrow$1, are very
nearly multiples of these frequencies.

The width of CO lines is small compared to the broad \Planck\ bandpasses, 
and so we model the corresponding SED by a delta function at the
respective frequency,
\begin{equation}
\s^{\mathrm{CO}}_{\mathrm{RJ}}(\nu) \propto
\delta(\nu-\nu_0).
\label{eq:CO}
\end{equation}
We note that specific intensity units are not appropriate for
CO emission since all of the energy is being emitted in a narrow
spectral range. Instead, CO
emission is conventionally quantified in terms of
$\mathrm{K}\,\mathrm{km}\,\mathrm{s}^{-1}$. Because the central frequency
is known from theory, emission away from line center can be attributed
to radial motion with a definite mapping between frequency and velocity.
The line intensity in brightness temperature units is integrated over
all velocities, yielding $\mathrm{K}\,\mathrm{km}\,\mathrm{s}^{-1}$.

CO emission is expected to be only weakly polarized, with a
polarization fraction around 1\,\% in molecular clouds
\citep{greaves:1999}. Detecting such low levels of true polarization
is challenging with the currently available \Planck\ data, primarily
due to instrumental temperature-to-polarization leakage. For now, we
assume CO line emission to be fully unpolarized, but note that this 
is likely to change in future analysis.

Finally, we note that although the base CO frequencies do not lie
within the \Planck\ LFI frequency bands themselves, CO emission is
nevertheless indirectly important for LFI because of its prevalence in
the HFI channels, and these are in turn critical to model thermal dust
emission for LFI.

\subsection{Extra-galactic compact sources}
\label{sec:pointsources}

In addition to the Galactic foreground emission mechanisms discussed
in Sect.~\ref{sec:gal_fg}, several extra-galactic effects are also
important for CMB frequencies, including cosmic infrared background
(CIB) radiation \citep{hauser:2001} and the Sunyaev-Zeldovich effect
\citep{sunyaev:1972}, and there are also contributions arising from
within the Solar system, namely zodiacal light emission
\citep{kelsall1998,planck2014-a12,san:2022}. However, the \Planck\ LFI
data have a very low signal-to-noise ratio to any of these effects,
and we therefore ignore these in the following.

For LFI frequencies, the most important class of extra-galactic
components are compact radio sources. All the emission mechanisms
listed above operate in external galaxies, but the radio source
population is dominated by active galactic nuclei (AGN). Radio
emission from AGN is largely synchrotron, and comes from either the
galactic nucleus itself or from jets and lobes associated with the
nucleus.  While the morphology of individual sources may be
complicated, few are resolved by most CMB experiments and hence can be
treated as ``point'' sources. Thus, while individual components of an
AGN may exhibit polarized microwave emission, the emission from an unresolved 
source as a whole is rarely strongly polarized; typical polarization
fractions are a few percent \citep{datta2019}. 

AGN have a wide range
of SEDs, but most AGN spectra at CMB frequencies can be adequately
modeled by a simple power law with a spectral index determined primarily
by the energy spectrum of the relativistic electrons generating the
synchrotron emission. The spectral indices (in brightness) typically
range from 0 to $-1$, and most fall in a narrower range of $-0.5$ to $-0.7$.
Hence we adopt a simple power law SED as our model for radio sources,
and fit for the amplitude and spectral index of the radio source
contribution, 
\begin{equation}
\s^{\mathrm{src}}_{\mathrm{RJ}}(\nu) \propto \left(\frac{\nu}{\nu_{0,\mathrm{src}}} \right)^{\alpha-2}
\label{eq:ptsrc}
\end{equation}
Here $\nu_{\mathrm{src}}$ is a fixed reference frequency, and $\alpha$
is the spectral index defined in intensity units; the conversion
between intensity and brightness temperature is proportional to
$\nu^2$. 

As we move to higher CMB frequencies, or to more sensitive
experiments, the counts of extra-galactic sources begin to include dusty
galaxies.  These objects emit modified blackbody radiation, like
Galactic dust, but typically at higher temperatures. Emission from
this class of objects is included in the cosmic infrared background
discussed below.

Unlike the dusty galaxies, which tend to be clustered, synchrotron-dominated
radio sources are quite randomly distributed on the sky, and hence have a flat 
angular power spectrum. On the other hand, the emission of synchrotron dominated 
sources is frequently variable, on time scales ranging from days to
years. Time variability is not accounted for in the current model, and
variable sources are therefore likely to leave residuals in the final
maps. For this reason, we will apply a dedicated point source mask
during the final CMB parameter estimation, to minimize contamination
in the final cosmological parameters.

\subsection{Default sky model}
\label{sec:default_sky_model}

Based on the above survey, and unless specified otherwise, the default
\BP\ astrophysical sky model (in brightness temperature units) reads as follows,
\begin{align}
  \vec{s}_{\mathrm{RJ}} &= \left(\vec{a}_{\mathrm{CMB}}+\vec{a}_{\mathrm{quad}}(\nu)\right) \frac{x^2 e^x}{(e^x -1)^2}+\label{eq:cmb_astsky}\\
  &+ \vec{a}_{\mathrm{s}} \left(\frac{\nu}{\nuzeros}\right)^{\bsynch} + \label{eq:synch_astsky}\\
  &+ \vec{a}_{\mathrm{ff}} \left(\frac{\nuzeroff}{\nu}\right)^2 \frac{g_{\mathrm{ff}}(\nu;\Te) }{g_{\mathrm{ff}}(\nuzeroff;\Te)} +\label{eq:ff_astsky}\\
  &+ \vec{a}_{\mathrm{ame}} \left(\frac{\nuzeroame}{\nu}\right)^2 \frac{f_{\mathrm{ame}} \left(\nu\cdot \frac{30.0\,\mathrm{GHz}}{\nup}\right)}{f_{\mathrm{ame}} \left(\nuzeroame\cdot \frac{30.0\,\mathrm{GHz}}{\nup}\right)}+\label{eq:ame_astsky}  \\ 
  &+ \vec{a}_{\mathrm{d}} \left(\frac{\nu}{\nuzerod}\right)^{\bdust+1} \frac{e^{h\nuzerod/\kB\Tdust}-1}{e^{h\nu/\kB\Tdust}-1}+ \label{eq:dust_astsky}\\
  &+ U_{\mathrm{mJy}} \sum_{j=1}^{N_{\mathrm{src}}} \vec{a}_{j,\mathrm{src}} \left(\frac{\nu}{\nuzerosrc}\right)^{\alpha_{j,\mathrm{src}}-2}, \label{eq:sum_ptsrc}
\end{align}
where $x=h\nu/kT_0$ and $\nu_{0,i}$ is the reference frequency for
component $i$; $U_{\mathrm{mJy}}$ is a conversion factor between flux
density in milli-jansky and brightness temperature. Thus, $\a_i$ is
the amplitude of component $i$ in units of
$\mu\mathrm{K}_{\mathrm{RJ}}$, as observed at a monochromatic
frequency $\nu_{0,i}$.
The sum in line~\ref{eq:sum_ptsrc} runs over all sources brighter than
some flux threshold as defined by an external source catalog, and both
the amplitude and spectral index are fitted individually per
source. We adopt the same catalog as \citet{planck2016-l04}, which is
hybrid of the AT20G \citep{murphy2010}, GB6 \citep{gregory1996}, NVSS
\citep{condon1998} and PCCS2 \citep{planck2014-a35} catalogs
comprising a total of 12\,192 individual sources.


Only $\{\s_{\mathrm{RJ}}, \a_{\mathrm{CMB}}, \a_{\mathrm{s}},
\a_{\mathrm{d}}\}$ are assumed to be polarized in this model, and these
comprise 3-component vectors including Stokes $T$, $Q$, and $U$
parameters. The remaining amplitudes parameters, $\{\a_{\mathrm{ff}},
\a_{\mathrm{AME}}, \a^j_{\mathrm{src}}\}$, are assumed unpolarized, and have vanishing
Stokes $Q$ and $U$ parameters.

For algorithmic reasons, we distinguish between linear and nonlinear
parameters. The former group includes $\{\a_{\mathrm{CMB}},
\a_{\mathrm{s}}, \a_{\mathrm{ff}}, \a_{\mathrm{AME}}, \a_{\mathrm{d}},
\a_{\mathrm{src}}\}$, collectively denoted $\a$; as
described in Sect.~\ref{sec:sigamp}, this set of parameters may be
estimated jointly and efficiently through a multivariate Gaussian
Monte Carlo sampler. In contrast, the nonlinear parameters include
$\{\beta_{\mathrm{s}}, T_e, \nu_{p}, \beta_{\mathrm{d}},
T_{\mathrm{d}}, \beta_{\mathrm{src}}\}$, and these must be estimated
independently and with computationally far more expensive algorithms;
see Sect.~\ref{sec:beta} for specific details. In practice, we fit
individual compact source amplitudes jointly with the corresponding
spectral indices using a general sampling algorithm, since these are
much more correlated with these than with any of the diffuse component
parameters. 

\section{Instrument characterization}
\label{sec:instrument}

We now turn to the second half of the parametric model
employed in the \BP\ analysis, which describes the instrument used to
collect the measurements. So that the \BP\ analysis may freely be used by
others, we aim to
keep the presentation and notation as general as possible, and only
introduce \BP\ and LFI-specific notation where strictly necessary. We
start our discussion by first defining an ideal detector response
model, and then increase the level of realism step-by-step, until we
reach the final instrument model.

\subsection{Ideal instrument model}
\label{sec:ideal_model}

Let us first consider an ideal monochromatic detector observing at
frequency $\nu$ a stationary sky signal with local Stokes parameters
$\{T,Q,U\}$ at Galactic coordinates $(l,b)$ and polarization angle
$\psi$.  We also initially assume infinite angular resolution. In this 
ideal case, the signal recorded by the detector as a function of time $t$ 
may be written as
\begin{equation}
  d(t) = g(t)\,\left[T + Q\,\cos2\psi + U\,\sin2\psi\right]
  + n(t),
  \label{eq:ideal_model}
\end{equation}
where $g$ is a multiplicative factor called the gain, which converts
between physical signal units (which in our case will be
$\mathrm{K}_{\mathrm{CMB}}$) and digitized instrumental detector units
(which in our case will be V), and $n$ denotes instrumental
noise.

In order to obtain data that may be processed on a computer, it is
necessary to discretize the measurements by averaging over some
(short) time period, $\Delta t$. For most CMB experiments, typical
samples rates are between 10 and 200~Hz. A single recorded datum,
$d_t$, thus corresponds to the detector output averaged over a period 
typically between 0.005 and 0.1~s.

For an ideal detector, the noise may be approximated as Gaussian and
uncorrelated in time, and, as such, its variance decreases
proportionally to $1/\Delta t$. We define the standard deviation of a
single time sample to be $\sigma_0$.

A CMB experiment scans the sky according to some scanning strategy,
$p(t) = [l(t),b(t),\psi(t)]$, while continuously recording the signal $d_t$. To
describe this behaviour in a convenient notation, we first discretize the
sky as described in Sect.~\ref{sec:units}, $\s = \s_p$, and then
re-write Eq.~\eqref{eq:ideal_model} in vector form as follows,
\begin{equation}
  \d = \G\, \P\s + \n,
  \label{eq:ideal_model2}
\end{equation}
where $\d = [d_1, d_2, \ldots, d_{n_{\mathrm{TOD}}}]^t$ and $\n =
[n_1, n_2, \ldots, n_{n_{\mathrm{TOD}}}]^t$ are time-domain vectors of
length $N_{\mathrm{TOD}}$, and $\G$ is a diagonal
$N_{\mathrm{TOD}}\times N_{\mathrm{TOD}}$ matrix with $g_t$ on the
diagonal. The scanning strategy is encoded in an $N_{\mathrm{TOD}}
\times 3N_{\mathrm{pix}}$ matrix that contains $(1, \cos 2\psi, \sin
2\psi)$ in the columns that correspond to pixel $p$ that happens to be
observed at time $t$, and zero elsewhere, i.e.,
\begin{equation}
  {\miniscule
\P = 
  \left[
    \begin{array}{ccccccccccc}
      0 & 1 & 0 & \ldots & 0 & \cos 2\psi_1 & 0 & \ldots & 0 & \sin 2\psi_1 & 0 \\
      1 & 0 & 0 & \ldots & \cos 2\psi_2 & 0 & 0 & \ldots & \sin 2\psi_2 & 0 &  0 \\
      \vdots & \vdots & \vdots & \vdots & \vdots & \vdots & \vdots & \vdots & \vdots & \vdots & \vdots \\
      0 & 0 & 1 & \ldots & 0 & 0 & \cos 2\psi_1  & \ldots &
      0 & 0 & \sin 2\psi_1
    \end{array}
    \right]}.
  \label{eq:pointmat}
\end{equation}
This matrix is called the pointing matrix.\footnote{Only the
  nonzero entries need to be stored in the pointing matrix, and
  memory requirements are therefore manageable.} Correspondingly, the
sky vector consists of the three pixelized Stokes parameter maps
stacked into a single vector, $\s = [T_1, \ldots,
  T_{N_{\mathrm{pix}}}, Q_1, \ldots, Q_{N_{\mathrm{pix}}}, U_1,
  \ldots, U_{N_{\mathrm{pix}}}]^t$.

Equation~\eqref{eq:ideal_model2} describes an ideal instrument that
cannot be realized in actual hardware. The remainder of this section
is therefore dedicated to generalizing this equation to the point that
it actually does serve as a useful model for real-world CMB
experiments.

\subsection{Spectral response, bandpass averaging, and unit conversion}
\label{sec:bandpass}

The first generalization we will consider is the assumption of
monochromaticity. No real detector can measure a single frequency
signal, but it is instead sensitive to a range of frequencies. This
sensitivity is described by a so-called \emph{bandpass profile} or
\emph{spectral transmission}, $\tau(\nu)$, which quantifies how much
of the radiation at a given frequency is actually recorded by the
detector. We define $\tau$ to be normalized to unity when integrated across
all frequencies. Adopting brightness temperature units for all
quantities (i.e., $\tau$, $\d$, and the monochromatic sky signal,
$\s(\nu)$), the data model in Eq.~\eqref{eq:ideal_model2} generalizes to
\begin{equation}
  \d = \G \P\int \s(\nu)\tau(\nu)\,\mathrm d\nu + \n,
  \label{eq:bp}
\end{equation}
after taking into account the bandpass effect.\footnote{Note that many
  experiments, including \Planck\ HFI, defines the bandpass profile in
  intensity units rather than brightness temperature units, and in
  this case an additional factor of $2h\nu^2/c^2$ must be included in
  the integral, as given by Eq.~\eqref{eq:Krj}; see
  \citet{planck2013-p03d} for details.}

However, most data sets are not provided in terms of brightness
temperature units, but more often in either thermodynamic temperature
or intensity units. As described in detail in
\citet{planck2013-p03d}, in order to convert from unit
convention $X_i$ to unit convention $X_j$, one must multiply with a
unit conversion factor that is given by
\begin{equation}
U_{ij} = \frac{\int \tau(\nu)\frac{dI_{\nu}}{dX_i}\, \mathrm d\nu}{\int
  \tau(\nu)\frac{dI_{\nu}}{dX_j}\,\mathrm d\nu},
\end{equation}
where $dI_{\nu}/dX_i$ is the intensity derivative expressed in unit convention
$X_i$. In particular, the conversion factors from brightness
temperature to thermodynamic temperature and intensity units are given
by
\begin{align}
U_{\mathrm{K}_{\mathrm{RJ}}\rightarrow \mathrm{K}_{\mathrm{CMB}}} &= \frac{\int
  \tau(\nu)\,\frac{2k\nu^2}{c^2}\,\mathrm d\nu}{\int \tau(\nu)\,b'_{\nu}\,
  \mathrm d\nu}\\
U_{\mathrm{K}_{\mathrm{RJ}}\rightarrow \mathrm{MJy}\,\mathrm{sr}^{-1}} &= \frac{\int
  \tau(\nu)\,\frac{2k\nu^2}{c^2}\,\mathrm d\nu}{\int
    \tau(\nu)\,\frac{\nu_c}{\nu}\,\mathrm d\nu},
\end{align}
where
\begin{equation}
  b'_{\nu} = \left.\frac{\partial B(T,\nu)}{\partial T}\right|_{T=T_0}
\end{equation}
is the derivative of the blackbody function with respect to
temperature, evaluated at the CMB temperature $T_0$, and $\nu_c$ is an
arbitrary selected reference frequency for the channel in
question. For other conversions, including to
$\mathrm{K}\,\mathrm{km}\,\mathrm{s}^{-1}$ and the SZ $y$-parameter,
we refer the interested reader to \citet{planck2013-p03d}. Taking into
account both bandpass integration and unit conversion, the instrument
model reads
\begin{equation}
  \d = U \G\P\int \s(\nu)\tau(\nu)\,\mathrm d\nu + \n.
  \label{eq:unit}
\end{equation}

We aim to constrain $\s$ given $\d$. It is
therefore important to be able to quickly evaluate the integral and
unit conversion factors in Eq.~\eqref{eq:unit}. With this in mind, we
consider signal component $i$ as defined by the sky model in
Sect.~\ref{sec:default_sky_model}, and write it in the general form
$\s_i(\nu) = \a_i\,f_i(\nu; \nu_0, \beta)$, where $\a_i$ is the linear
amplitude relative to some reference frequency, $\nu_{0,i}$, and
$f_i(\nu; \beta)$ is the frequency scaling from that reference
frequency to an arbitrary frequency $\nu$, which depends on some set
of spectral parameters $\beta$. The total signal measured by detector
$j$ may then be written as
\begin{equation}
  \s^j = \sum_{i=1}^{N_{\mathrm{comp}}}\a_i\, \left[\,U_j \int f_i(\nu; \beta)\,
    \tau_j(\nu)\,\mathrm d\nu\right] \equiv
  \sum_{i=1}^{N_{\mathrm{comp}}}\M^j_i\,\a_i = \M^j\,\a,
  \label{eq:mixmat}
\end{equation}
where $\M^j_i$ is called the \emph{mixing matrix}. In order to take
into account bandpass integration and unit conversion, the idealized data
model in Eq.~\eqref{eq:ideal_model2} must be generalized as
follows,
\begin{equation}
  \d = \G\P\M\a + \n.
  \label{eq:ideal_model_bp}
\end{equation}

It is evident that $\M$ depends only on the spectral parameters
$\beta$ and the bandpass $\tau$, but not the amplitudes. Since most
signal components are parameterized with limited number of spectral
parameters (see Sect.~\ref{sec:sky}), and these parameters are
typically also fairly uniform on the sky, it is possible to pre-compute
accurate lookup tables for $\M$ for each component and detector. In
our current code, we adopt (bi-)cubic splines with regular grids for
these lookup tables, and the computational cost of performing a full
bandpass integral is thus equal to that of a simple polynomial
evaluation.

\subsubsection{Bandpass uncertainties and corrections}

\begin{figure}[t]
  \center
  \includegraphics[width=\linewidth]{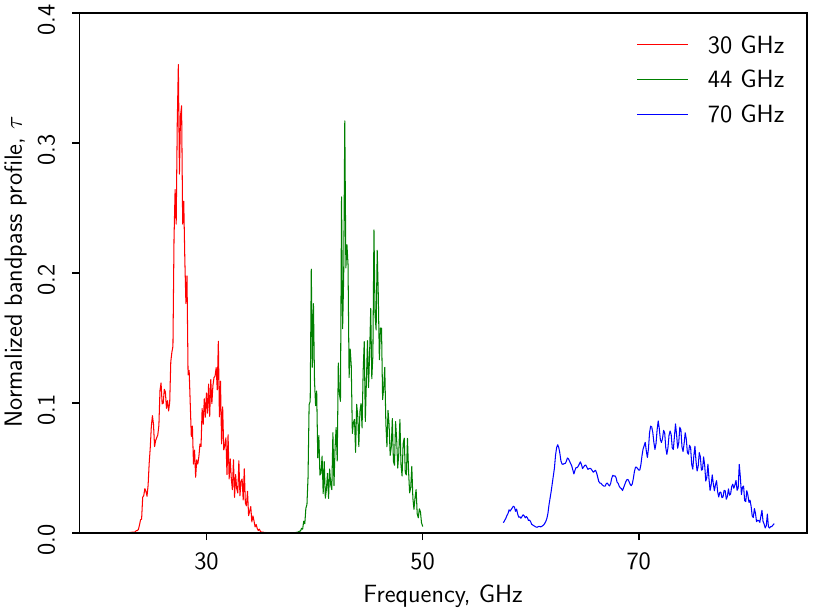}
  \caption{Detector averaged bandpass profiles, $\tau$, for the three
    \Planck\ LFI frequency channels. }
  \label{fig:bp_LFI}
\end{figure}

While the bandpass integral described by Eq.~\eqref{eq:bp} may look
simple enough at first glance, it does lead to a wide variety of
important complications in practice. The most important among these is
the fact that the exact shape of the bandpass profile itself is
unknown. In particular, it is highly nontrivial to measure $\tau$
accurately in a laboratory for a combined multi-component instrument,
and it is obviously impossible to do so after commissioning for
satellite missions.

As a concrete real-world illustration of this, Fig.~\ref{fig:bp_LFI}
shows the laboratory-determined (normalized) bandpass profiles after
averaging over all radiometers for a given LFI channel. First, we see
that the profiles for both 44 and 70\,GHz are truncated, and therefore
significant response is likely present outside the measured
range. Second, for all three channels we see notable small scale
ripples, which are due to standing waves. These may be due to real
standing waves within the optical assembly of the LFI instrument
itself; but some part of them may also be due to standing waves in the
test equipment used to make the measurements. These artefacts were
already noted by \citet{zonca2009}, but they were never actually
corrected during the course of the official processing; this is now
done as part of the current reprocessing, as described by
\citet{bp09}. In addition to these visually obvious effects, there may
also be systematic errors in the actual shape, for instance in the
form of a smooth slope across the bands, or in the precise position of
the peaks within the band.

As described in Sect.~\ref{sec:gain}, the CMB dipole serves as our primary
calibrator for \BP, following both \WMAP\ and the official
\Planck\ pipelines. Because the CMB SED very closely follows a
blackbody spectrum, which translates into a frequency independent
scaling in thermodynamic units, the precise shape of the bandpass is
irrelevant for the CMB component. Instead, errors in the bandpass
shape effectively translate into incorrectly estimated foreground
components, and introduce inaccuracies in the relative foreground
SEDs between different frequency channels. In turn, foreground errors
can affect the CMB reconstruction.

To account for the uncertainties noted above, we introduce one or more
free parameters that can modify the bandpass shape, and allow the data
to inform us about, and hence mitigate, potential inaccuracies in the
laboratory bandpass measurements. The simplest and most common model
we adopt is a simple linear shift, $\Delta_{\mathrm{bp}}$, in
frequency space,
\begin{equation}
  \tau(\nu) = \tau_0(\nu+\Delta_{\mathrm{bp}}),
  \label{eq:bp_shift}
\end{equation}
where $\tau_0$ is the default laboratory measurement. One value of
$\Delta_{\mathrm{bp}}^i$ is allowed per radiometer $i$, but (in most
cases) either with the prior that $\sum_i \Delta_{\mathrm{bp}}^i = 0$,
or that one particular channel is held fixed. Without any priors, the
bandpass parameters are fully degenerate with the spectral parameters
$\beta$ of the foreground model, and no stable solution can be
found. Various choices of both bandpass models and priors are
considered by \citet{bp09}. In general, we note that the impact of
$\Delta_{\mathrm{bp}}$ is essentially to scale the amplitude of
foregrounds, while leaving the CMB unchanged. At CMB dominated
frequency channels, the bandpass shift is therefore non-degenerate
with respect to the gain, while at foreground-dominated channels, it
is virtually impossible to distinguish between a bandpass error and a
gain error.

In addition to this fundamental uncertainty in the bandpass profile
for each detector, we note, first, that different detectors within the
same frequency band observe different sky signals, and if not properly
accounted for, this can create so-called bandpass mismatch errors in
co-added frequency maps (see Sect.~\ref{sec:leakmaps}). Second, as
discussed in the next section, the instrumental beam is also
intrinsically frequency dependent, with an angular resolution of the
main beam that is inversely proportional to the frequency for
diffraction-limited observations, as is the case for LFI.  In
addition, far sidelobes can vary rapidly with frequency through
complicated diffraction patterns. Unless properly accounted for, all
these effects can potentially compromise final estimates. In \BP\ we
account for sidelobes as modelled by the \Planck\ team
\citep{planck2014-a05}, but we do not explore uncertainties in the
beam model itself.

\subsection{Beam and pixel window convolution}
\label{sec:beam}

\begin{figure}[t]
  \center
  \includegraphics[width=\linewidth]{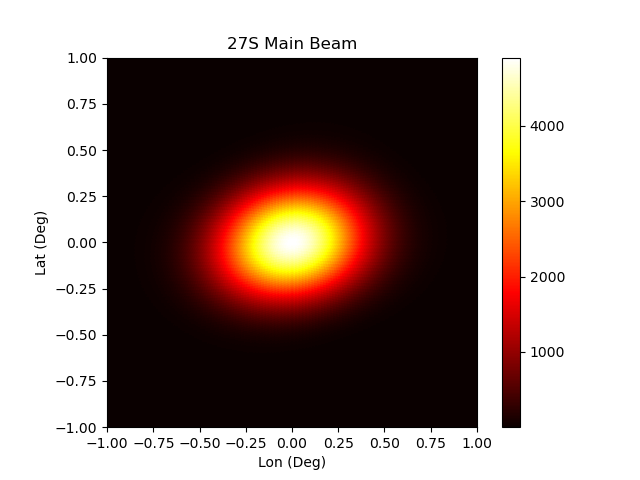}\\
  \includegraphics[width=\linewidth]{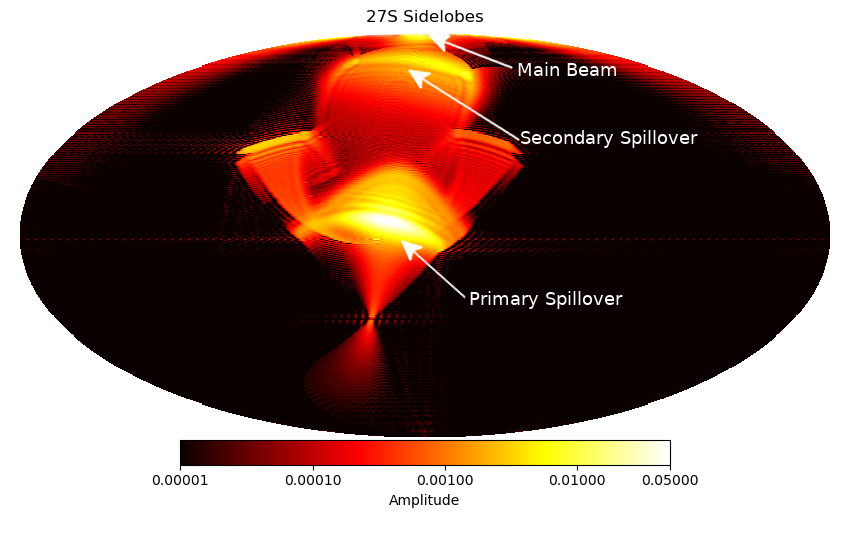}
  \caption{(\emph{Top}:) Scanning beam (or main beam) of the 30\,GHz
    LFI 27S radiometer in local telescope coordinates, i.e., the
    instantaneous spatial sensitivity to a point source
    centered at the beam maximum.  {\emph{Bottom}:} Corresponding
    $4\pi$ beam map, oriented such that the main beam is located on
    the north pole. The main \Planck\ far sidelobes are caused by
    spillover from (i.e., diffraction around) the primary and
    secondary mirrors. The beams are normalized such that their
    combined integral over the full sky equals unity. }
  \label{fig:beam_LFI}
\end{figure}

\begin{figure}[t]
  \center
  \includegraphics[width=\linewidth]{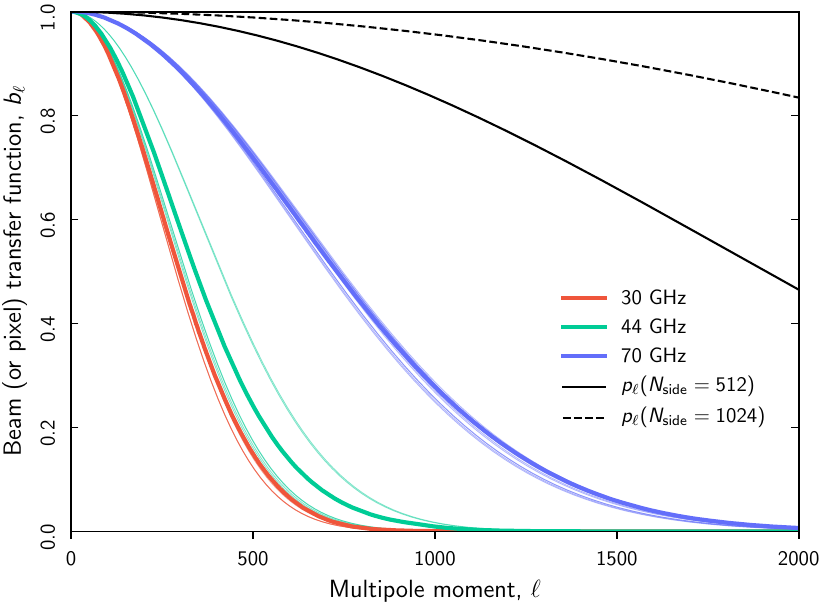}
  \caption{Azimuthally symmetric and normalized beam transfer
    functions, $b_{\ell}$ for each co-added \Planck\ LFI frequency
    channel (thick colored lines) and for each radiometer (thin
    colored lines). The former include the effects of non-Gaussian
    tails, while the latter are strictly Gaussian
    approximations. Black curves show the \HEALPix\ pixel window
    functions for $N_{\mathrm{side}}=512$ and 1024, respectively.}
  \label{fig:bl_LFI}
\end{figure}

In the same way that no real detector can measure the signal from only
a single monochromatic frequency, no real detector can measure the
signal from a single point on the sky. Rather, each detector is
associated with a so-called ``point spread function'' (PSF) or
``beam'', $\b(\hat{n})$, that characterizes its instantaneous spatial
sensitivity. Conventionally, we define $\b(\hat{n})$ to be normalized
by setting the full-sky integral equal to unity, and to be oriented such that
its maximum value is centered on the north pole.

The recorded value of the sky signal, $s^{\mathrm{beam}}_t$, as seen
through the beam at time $t$ is then given as the convolution of $\b$
and $\s$,
\begin{equation}
  s^{\mathrm{beam}}_t = \int_{4\pi} \b_t(\hat{n})\s(\hat{n})
  \,\mathrm d\Omega,
  \label{eq:beam_conv}
\end{equation}
where $\b_t(\hat{n})=\R_t(\hat{n},\hat{n}')\b(\hat{n}')$, and $\R_t$
is a time-dependent rotation matrix that rotates the beam as given by
the scanning strategy of the instrument. Since convolution is a linear
operation, we may define a matrix operator, $\B$, such that
$\s^{\mathrm{beam}} = \B\s$, and the data model in
Eq.~\eqref{eq:ideal_model_bp} may therefore be generalized further into
its final form,
\begin{equation}
  \d = \G\P\B\M\a + \n,
  \label{eq:ideal_model_final}
\end{equation}
where the position of the operator is defined by noting that the beam
only acts on the true sky signal, and not on instrumental effects such
as gain or noise.

Noting that modern CMB maps typically comprise up to several hundred
million pixels, Eq.~\eqref{eq:beam_conv} is prohibitively expensive to
evaluate directly in pixel space. Instead, we take advantage of the
convolution theorem, which states that any convolution in pixel space
may be expressed as a multiplication in harmonic space, and vice
versa. As first demonstrated by \citet{wandelt:2001}, and later
optimized by \citet{prezeau:2010}, Eq.~\eqref{eq:beam_conv} may be
computed efficiently through reduced Wigner matrices, reducing the
cost by a factor of $\mathcal{O}(\sqrt{N_{\mathrm{pix}}})$ per
evaluation for a general $\b$. A particularly computationally
convenient formulation of this algorithm is presented by \citet{bp08}
as part of the current \BP\ work.

Another substantial saving can be made if we additionally assume that
$\b$ is azimuthally symmetric. In that case, the spherical harmonics
expansion of $\b$ is independent of $m$, and may be expressed in terms
of its Legendre transform, $b_{\ell}$. The full convolution may (by
the convolution theorem) in this case be written as\footnote{This expression
  applies to temperature convolution; polarization convolution is
  notationally slightly more involved, but mathematically fully
  analogous.}
\begin{equation}
s^{\mathrm{beam}}(\hat{n}) =
\sum_{\ell=0}^{\ell_{\mathrm{max}}}\sum_{m=-\ell}^{\ell} b_{\ell}
s_{\ell m} Y_{\ell m}(\hat{n}),
\label{eq:conv_symm}
\end{equation}
where $s_{\ell m}$ are the spherical harmonics coefficients of
$\s$. Often, $b_{\ell}$ is referred to as the beam transfer function.

Note that the bandlimit, $\ell_{\mathrm{max}}$, in
Eq.~\eqref{eq:conv_symm} should be selected sufficiently large that
$b_{\ell} \approx 0$ as compared to the noise level of the
instrument. Conversely, if a too low value of $\ell_{\mathrm{max}}$ is
adopted for analysis, the most notable artifacts arising from the
convolution is ringing around bright point sources, resulting from
premature harmonics truncation.

Note also that $s^{\mathrm{beam}}(\hat{n})$ in Eq.~\eqref{eq:conv_symm}
is written as a function
of position rather than time in the above expression, which is only
possible in the case of an azimuthally symmetric beam. To obtain the
time-dependent signal, one simply reads off the value of
$s^{\mathrm{beam}}(\hat{n})$ given by the beam center position at
time $t$. In this approximation, a full real-space convolution may be
carried out at the cost of only two spherical harmonics transforms. 

As discussed in Sect.~\ref{sec:units}, all CMB sky maps are pixelized
in order to allow for efficient analysis on a computer. Such pixelization
corresponds to an additional smoothing operation of the true sky
signal that can be approximated with a top-hat convolution kernel of a
given pixel size. For \HEALPix, the effect of this kernel in harmonic
space is described in terms of a pixel window function, $p_{\ell}$,
that is provided with the library. Implementationally, it is often
convenient to redefine $b_{\ell} \rightarrow b_{\ell}p_{\ell}$
internally in computer codes, as the beam and pixel window affect the
signal in the same way, and accounting for the pixel window can
therefore usually be done with no additional computational cost
compared to beam convolution.

In Euclidean space, the Nyquist theorem assures that any bandwidth
limited signal may be reconstructed exactly with at least two samples
per bandwidth. No corresponding exact theorem exists on the
sphere. Instead, a rough rule of thumb for smooth spherical fields is
to allow for at least two or three pixels per beam width. Likewise, no
exact multipole bandlimit exists for given a \HEALPix\ pixelization;
however, numerical experiments suggest that multipoles above $\ell
\gtrsim 2.5N_{\mathrm{side}}$ are poorly resolved on the \HEALPix\ 
grid. Combined, these rules of thumb provide useful relationships
between a given beam width and the corresponding appropriate values of
$N_{\mathrm{side}}$ and $\ell_{\mathrm{max}}$. 

Figure~\ref{fig:beam_LFI} shows the beam of the \Planck\ 27S
radiometer \citep{planck2013-p02d}. The bottom panel shows the full
$4\pi$ beam, while the top panel shows a zoom-in on the north
pole. Clearly, this beam pattern is not azimuthally
symmetric. However, in this respect it is useful to distinguish
between the \emph{main beam}, which is highlighted in the top panel,
and the \emph{sidelobes}, which are highlighted in the bottom
panel. Furthermore, since convolution is a linear operation,
contributions from the main beam and sidelobes may be computed
separately.

The sidelobes are caused by optical imperfections,
typically by diffraction around the main optical elements. In the case
of \Planck, these are the primary and secondary mirrors (see
Fig.~\ref{fig:beam_LFI}). As such, the resulting beam structures tend
to be highly frequency dependent, and also cover large angular
scales. While they clearly cannot be described as azimuthally
symmetric in any meaningful way, they are associated with relatively
modest bandlimits, $\ell_{\mathrm{max}}$, and this leads to acceptable
computational costs for treating this component.

The main beam, on the other hand, can often be described reasonably
well as azimuthally symmetric, when centered on the north (or south)
pole. Of course, the LFI 27S beam shown in the top panel of
Fig.~\ref{fig:beam_LFI} exhibits a substantial ellipticity of
$\epsilon \approx 1.3$, but this instantaneous beam profile is at
least partially symmetrized by averaging due to the scanning strategy. 
The remaining effects of beam asymmetries may be accounted for, at least
in terms of power spectrum bias, by adjusting the transfer function
$b_{\ell}$ through simulations, as described by, e.g.,
\citet{mitra2010}.

For simplicity or because of low signal-to-noise, the beam profile is also
sometimes approximated in terms of a two-dimensional Gaussian with
some full-width-half-maximum (FWHM), or $\sigma_{\mathrm{FWHM}}$, in the
following expressed in radians. In the Gaussian case, one can derive an
explicit expression for the beam transfer function in the form
\begin{equation}
  b_{\ell} = \e^{-\frac{1}{2}\ell(\ell+1)\frac{\sigma_{\mathrm{FWHM}}^2}{8\ln 2}},
\end{equation}
where the factor $8\ln 2$ simply accounts for the conversion between
the square of the FWHM and the variance for a Gaussian.

Figure~\ref{fig:bl_LFI} compares the azimuthally symmetric beam
transfer functions of the three \Planck\ LFI channels, co-added over
all radiometers, as well as the Gaussian approximations to the
individual radiometer beam transfer functions. For reference, we also
show the \HEALPix\ window transfer functions for $N_{\mathrm{side}}=512$ and
1024, which are the typical pixelizations used for LFI and \WMAP\
analysis.

We see that the general azimuthal approximations tend to have slightly
heavier tails than the Gaussian approximations, and this is important
to account for when estimating the CMB power spectrum, $C_{\ell}$. At
the same time, we also see that for applications for which only
percent-level accuracy is required, the Gaussian approximations may
very well be sufficient. In the following analyses, we will adopt the
general azimuthally symmetric approximations for co-added frequency
maps, which will be used for component separation and CMB estimation
purposes, but Gaussian approximations for radiometer-specific signal
modelling during time-domain processing, where the signal-to-noise
ratio per sample is low, and sub-percent precision is irrelevant. The
reason for the latter approximation is simply that the
\Planck\ collaboration only provides FWHM estimates for individual
radiometers, not full transfer functions.

In the current work, we assume that the transfer functions provided by
the \Planck\ collaboration are exact, and do not assign dedicated
stochastic parameters to them. This is neither a realistic description, 
nor a testament to the accuracy of the provided
products, but only a statement of currently limited human resources; a
high-priority task for future work is to implement full support for
dynamic beam modelling and error propagation. As presented in this
work, however, beam convolution is assumed to be a fully deterministic
operation, dependent on officially available beam characterizations
alone.

\subsection{Gain and analog-to-digital conversion}
\label{sec:gain}

While the instrument model in Eq.~\eqref{eq:ideal_model_final} is
structurally complete in terms of components, we still need to
introduce a few generalizations before we can apply it to our
data. The first regards the gain $g$, simply by reemphasizing that
this should be interpreted as a truly time-dependent object, $g_t$.

To understand why this is the case, it is useful to consider its
origin and physical interpretation, and to focus the discussion we
will consider the special case of a perfect total-power receiver. The
output voltage of such a device is given by
\begin{equation}
  P = GkT_{\mathrm{sys}}\Delta\nu,
  \label{eq:power}
\end{equation}
where $G$ is a unit-less gain factor, and $\Delta\nu$ is the width of
the bandpass. The system temperature is defined as
$T_{\mathrm{sys}}=T_{\mathrm{ant}}+T_{\mathrm{recv}}$, where
$T_{\mathrm{ant}}=T_{\mathrm{CMB}} + T_{\mathrm{fg}}$ is the antenna temperature,
and $T_{\mathrm{recv}}$ is the receiver temperature; the latter
essentially defines the intrinsic noise level of the receiver.

For a \Planck\ LFI 30~GHz radiometer, the bandwidth is 6~GHz, and
the receiver temperature is typically 10\,K. The antenna temperature
is dominated by the CMB temperature, $T_{\mathrm{CMB}}=2.7$\,K, as other sky
components typically only make up a few mK at
most. Assuming, therefore, a system temperature of about 13\,K,
Eq.~\eqref{eq:power} predicts that the power measured by this device is
$P=1.1\,$pW or $P=-90$~dBm,\footnote{The unit dBm measures power
  ratios, $x$, in decibel relative to 1\,mW, i.e., $x = 10\,\log_{10}
  \frac{P}{1\,\mathrm{mW}}$.} assuming no amplification
($G=1$). However, current microwave detectors are typically only able
to reliably record power levels larger than $P\gtrsim-30\,$dBm. For
this reason, the signal level must be actively amplified by a factor
of 60\,dB or more between the optical assembly and the detector. For
\Planck\ LFI, such amplification is achieved through the use of
high-electron-mobility transistors (HEMTs).

HEMTs provide high gain factors, while adding only very low levels of
additional noise to the data. However, they are not perfectly stable
in time. Rather, their effective gains exhibit time-dependent drifts
with typical overall variations at the $\mathcal{O}(10^{-6})$ level,
and correlations in time that are often well described by a so-called
$1/f$ spectrum (see Sect.~\ref{sec:noise}). Unless explicitly
accounted for in the model, these time-dependent gain fluctuations can
and will bias the derived sky model.

The gain defined by our original instrument model in
Eq.~\eqref{eq:ideal_model_final}, denoted $\G$, is in principle the same
gain as in Eq.~\eqref{eq:power}, but with two important
differences. First, while $G$ is defined as a pure power
amplification, and therefore unit-less, $\G$ takes into account the
end-to-end conversion from a raw sky signal to final recorded data
values. As such, $\G$ has units of $\mathrm{V}\,\mathrm{K}^{-1}$, in
order to be dimensionally correct.

As far as measuring the actual gain from a given experiment at high
precision is concerned, an important practical aspect is suppressing
measurement noise through various gain smoothing algorithms. The
official \Planck\ processing employed a simple boxcar average
algorithm for this purpose \citep{planck2014-a03,planck2016-l06}. As
shown by \citet{bp07}, this method can lead to striping in the final
map by averaging over real variations. The current \BP\ processing
therefore replaces the boxcar algorithm with a tuned Wiener filter
which allows variations on much finer time-scales, and this greatly
suppresses gain-induced striping. In fact, this modification
represents the single most important algorithmic improvement in the
\BP\ analysis in terms of improving data quality, and it was a key
step in understanding long-standing issues regarding the LFI 44\,GHz
channel \citep{planck2016-l02,planck2016-l05}.

Second, $\G$ additionally takes into account the digitization process
that converts analog signals to digital bits stored on a
computer. This process takes place in a so-called analog-to-digital
converter (ADC). An ideal ADC is perfectly linear. Unfortunately, many
real-world ADCs exhibit important imperfections, for instance in the
form of smooth nonlinear conversion within given signal ranges, or,
as for LFI, sharp jumps at specific signal or bit values.

Overall, ADC errors are indistinguishable from gain fluctuations in
terms of their direct impact on the recorded data. However, there is
one critical difference between the two effects: While gain
fluctuations are stochastic and random in time, and do not correlate
with the sky signal, ADC errors are perfectly reproducible, and depend
directly on the sky signal. Consequently, while the archetypical
signature of unmitigated gain fluctuations are coherent stripes or
large-scale features in the final sky maps, the corresponding unique
signature of unmitigated ADC errors is an asymmetry in the amplitude
of the CMB dipole along its positive and negative directions. This
effect can be used to characterize and mitigate ADC non-linearity, as
done both for \Planck\ LFI and HFI
\citep{planck2016-l02,planck2016-l03,planck2020-LVII}. For a
discussion of ADC corrections within the \BP\ framework, see
\citet{bp25}.

\subsection{Instrumental noise}
\label{sec:noise}

We complete our review of the instrument model by considering the
properties of the instrumental noise, $\n$. This component may be
decomposed into two main contributions, called correlated and white
noise,
\begin{equation}
  \n = \n_{\mathrm{corr}} + \n_{\mathrm{wn}}.
  \label{eq:noise}
\end{equation}
Both terms may be approximated as Gaussian, but they have different
covariances.

The dominant physical source of white noise is Johnson (or thermal)
noise, typically excited by thermal electron motions within the
electric radiometer circuits. This noise is temperature dependent, and
cryogenic cooling is usually required to achieve sufficient
sensitivity. The dominant source of the correlated noise term are
rapid gain fluctuations modulating the system temperature,
$T_{\mathrm{sys}}$, as discussed in Sect.~\ref{sec:gain}.

Based on this decomposition, the standard deviation of the total
instrumental noise term for a sample of duration $\Delta t$ (i.e.,
$\sigma_0$ in Eq.~\eqref{eq:ideal_model2}) may be estimated through
the so-called radiometer equation,
\begin{equation}
\sigma_0 = T_{\mathrm{sys}} \sqrt{\frac{1}{\Delta\nu \Delta t} +
  \left(\frac{\Delta g}{g}\right)^2}.
\end{equation}
Here, $\Delta g$ is the root-mean-square gain variation over $\Delta
t$, and $\Delta \nu$ is as usual the receiver bandwidth. Intuitively
speaking, this equation summarizes the following facts. First, the
noise level is proportional to the system temperature, in recognition
of the fact that Johnson noise scales with temperature. Second, the
white noise term is inversely proportional to the square root of both
bandwidth and integration time; this is simply by virtue of collecting
more photons, and noting that Gaussian errors add in quadrature. Third
and finally, the correlated noise component is proportional to the
overall gain fluctuation level. Typical values of $\sigma_0$ for the
LFI radiometers range between 600 and 1700~$\mu\mathrm{K}$ per sample
in temperature units, or between 50 and 200$~\mu\mathrm{V}$ in
detector units. If $N_{\mathrm{obs}}$ independent observations are
made of the same sky pixel $p$, then the effective noise of the
corresponding pixel integrates down roughly as $\sigma_p = \sigma_0 /
\sqrt{N_{\mathrm{obs}}}$.

The different correlation structures of the white and correlated noise
terms are most conveniently described in frequency domain through the
noise power spectrum density (PSD), $P_n(f) = \left<|n_f|^2\right>$,
where $n_f$ are the Fourier coefficients of $\n_t$. This PSD is often
modelled in terms of a so-called $1/f$ profile, which takes the form
\begin{equation}
  P_n(f) = \sigma_0^2 \left[1 +
    \left(\frac{f}{f_{\mathrm{k}}}\right)^{\alpha} \right].
  \label{eq:n_psd}
\end{equation}
Here, $f_\mathrm{k}$ is the knee frequency at which the variance of
the correlated noise equals that of the white noise, and $\alpha$ is
the slope of the spectrum at low frequencies. Typical best-fit values
for LFI radiometers are $f_{\mathrm{k}}\approx10$~mHz and
$\alpha\approx-1$. However, this model is obviously only approximate;
if for no other reasons, the real spectrum has to flatten at low
frequencies by energy considerations, whereas the power predicted by
this model would approach infinity at low frequencies.

\begin{figure}
        \begin{center}
                \includegraphics[width=\linewidth]{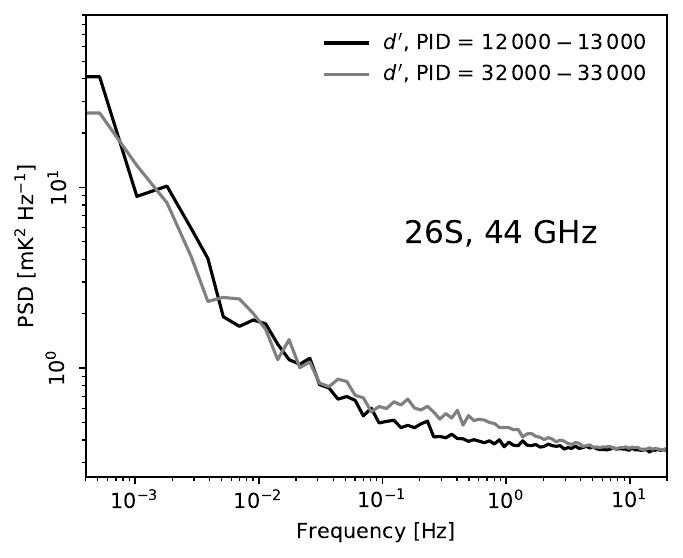}
        \end{center}
        \caption{PSD of signal subtracted data from radiometer 26S, averaged over ten PIDs (at intervals of 100 PIDs) in the ranges 
    12\,000--13\,000 (black) and 32\,000--33\,000 (grey). We see that there is
    significantly more power in the frequency range 0.1--10\,Hz in the
    later period. For further discussion, see \citet{bp06}.
                \label{fig:ps_compare_26S}}
\end{figure}

An important novel result that emerged from the current
\BP\ reprocessing is that the correlated noise for the LFI 30 and
44\,GHz in fact cannot be adequatly described by this simple $1/f$
model \citep{bp06}. Rather, for most detectors we find a significant
noise excess for frequencies between 0.01 and 1\,Hz that may
be described, at least phenomenologically, through the addition of a
log-normal power term of the form,
\begin{equation}
  P(f) = A_\mathrm{p}
    \exp\left[-\frac{1}{2}\left(\frac{\log_{10}f - \log_{10} f_\mathrm{p}}{\sigma_\mathrm{dex}}\right)^2\right],
    \label{eq:1fmodel_lognorm}
\end{equation} 
where $A_{\mathrm{p}}$ is a freely fitted amplitude, and
$f_\mathrm{p}$ represents the peak frequency of the log-normal
term. An example of this power excess for a 44\,GHz radiometer is
shown in Fig.~\ref{fig:ps_compare_26S}; the black line shows the
measured noise power spectrum for a period of time in which the power
excess is negligible, while the gray line shows the same for a period
in which it is large. Properly modelling this term yields a
significantly better $\chi^2$, in particular on angular scales that
are important for large-scale polarization measurements; for further
discussion, see \citet{bp06}.

\section{Data}
\label{sec:data}

The instrument discussion has until this point for the most part been
kept general and applicable to a wide range of different data sets. In
this section, we specialize our discussion to \Planck\ LFI. As
discussed in Sect.~\ref{sec:overview}, only this data set will be
considered in the time-domain, while external data sets will be
considered in the form of processed pixelized maps.

We note that the minimal sky model summarized in
Sect.~\ref{sec:default_sky_model} includes seven distinct
astrophysical components, three polarized and four
unpolarized. Considering that there are only three LFI frequency
channels, we immediately recognize that the LFI data must be augmented
with at least four external frequency channels, just in order to make
the model minimally constrained. In the default analysis
configuration, we therefore include select observations also from
\Planck\ HFI \citep{planck2020-LVII} and \WMAP\ \citep{bennett2012}, as well as
from some ground-based surveys. In this section, we provide a brief
overview of these data sets, and refer the interested reader to the
respective papers for full details.

The precise combination of data sets used in any particular
\BP\ analysis will depend on the goal of the respective
application. For instance, the main scientific goal of the current
paper is to introduce the concept of Bayesian end-to-end CMB analysis,
and provide a first demonstration of this framework as applied to the
LFI observations. Consequently, we here only include a minimal set of
external observations, allowing LFI to play the dominant role, in
particular with respect to CMB constraints. Specifically, in this
paper we include only
\begin{itemize}
\item \Planck\ 857~GHz to constrain thermal dust emission in intensity;
\item \Planck\ 353~GHz in polarization to constrain polarized thermal
  dust emission;
\item \WMAP\ 33, 41, and 61~GHz (called \emph{Ka}, \emph{Q} and \emph{V}-bands,
  respectively) in intensity at full angular resolution to constrain
  free-free emission and AME;
\item the same \WMAP\ channels in polarization to increase the signal-to-noise
  ratio of polarized synchrotron emission, but only at low angular
  resolution, where a full noise covariance matrix is available; and
\item Haslam 408~MHz \citep{haslam1982} to constrain synchrotron emission in intensity.
\end{itemize}
That is, we include neither intermediate HFI channels nor the
\WMAP\ \emph K-band (23~GHz) channel, because of their higher
signal-to-noise ratio relative to the LFI channels. The \WMAP\ \emph W-band
is excluded because of known systematics effects \citep{bennett2012},
and it does not have particularly unique features with respect to the
signal model that are not already covered by other data sets.

We also note that \citet{bp13}, \citet{bp14}, and \citet{bp15} focus
on general foreground constraints, and these papers therefore also
consider additional channels. The ultimate long-term goal of the
global Bayesian CMB analysis program in general is of course to
integrate as many data sets as possible into a single coherent sky
model, and thereby produce the strongest possible constraints on the
true astrophysical sky. One leading example of such an effort is the
\textsc{Cosmoglobe}\footnote{\href{url}{http://cosmoglobe.uio.no}}
project, which specifically aims to combine many state-of-the-art
experiments with the ones listed above, including CHIPASS
\citep{calabretta:2014}, \emph{COBE}-DIRBE \citep{hauser:1998} and
FIRAS \citep{mather:1994}, PASIPHAE \citep{tassis:2018}, SPIDER
\citep{gualtieri:2018}, and many more. The \BP\ methodology presented
here represents an ideal statistical framework for performing such
global data integration.

\begin{figure}[t]
	\center
	\includegraphics[width=\linewidth]{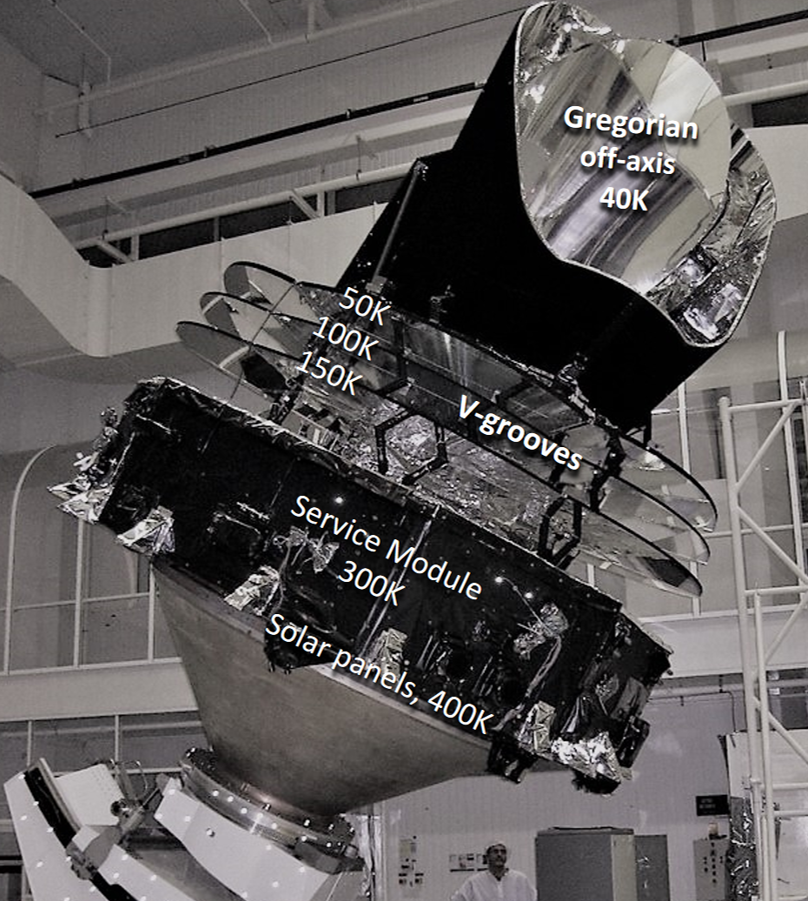}
	\caption{Flight model of the \Planck\ spacecraft. The satellite size is about $4.2 \times 4.2$\,m, and its mass at launch was 1950 kg. \Planck\ was launched on May 14, 2009, and operated for 4.4 years from a Lissajous orbit around the Lagrangian point L2 of the Sun--Earth system. Shown are the approximate temperatures of different critical parts of the satellite during nominal operation in space (see \citealp{planck2011-1.3}).}
	\label{fig:PLANCK-PICTURE}
\end{figure}

\subsection{LFI instrument overview}
\subsubsection{Instrument configuration}

We now provide a synthetic description of the LFI instrument
configuration, which directly impacts the structure of the LFI data
and the potential systematic effects addressed in the \BP\ 
analysis. For more details on the LFI instrument, its ground
calibration and in-flight performance, see
\cite{bersanelli2010}, \citet{planck2011-1.4}, and references therein; the
overall LFI programme is described by \citet{mandolesi2010}.

The heart of the LFI instrument is an array of 22 differential
receivers based on high-electron-mobility transistor (HEMT) low noise
amplifiers. The instrument operates in three frequency bands,
nominally centred at 30, 44 and 70~GHz, with angular resolutions of
about $32\arcm$, $28\arcm$, and $13\arcm$ FWHM, respectively.  The front end of the
receivers is cooled to 20\,K, which dramatically reduces the noise
temperature of the HEMT amplifiers and of the overall system.  In each
receiver, the signal coming from different directions of the sky,
intercepted by the telescope as the satellite spins, is compared to a
stable internal blackbody reference load at 4\,K. It is this
differential scheme that allows the LFI to achieve its excellent
stability.

\begin{figure}[t]
	\center
	\includegraphics[width=\linewidth]{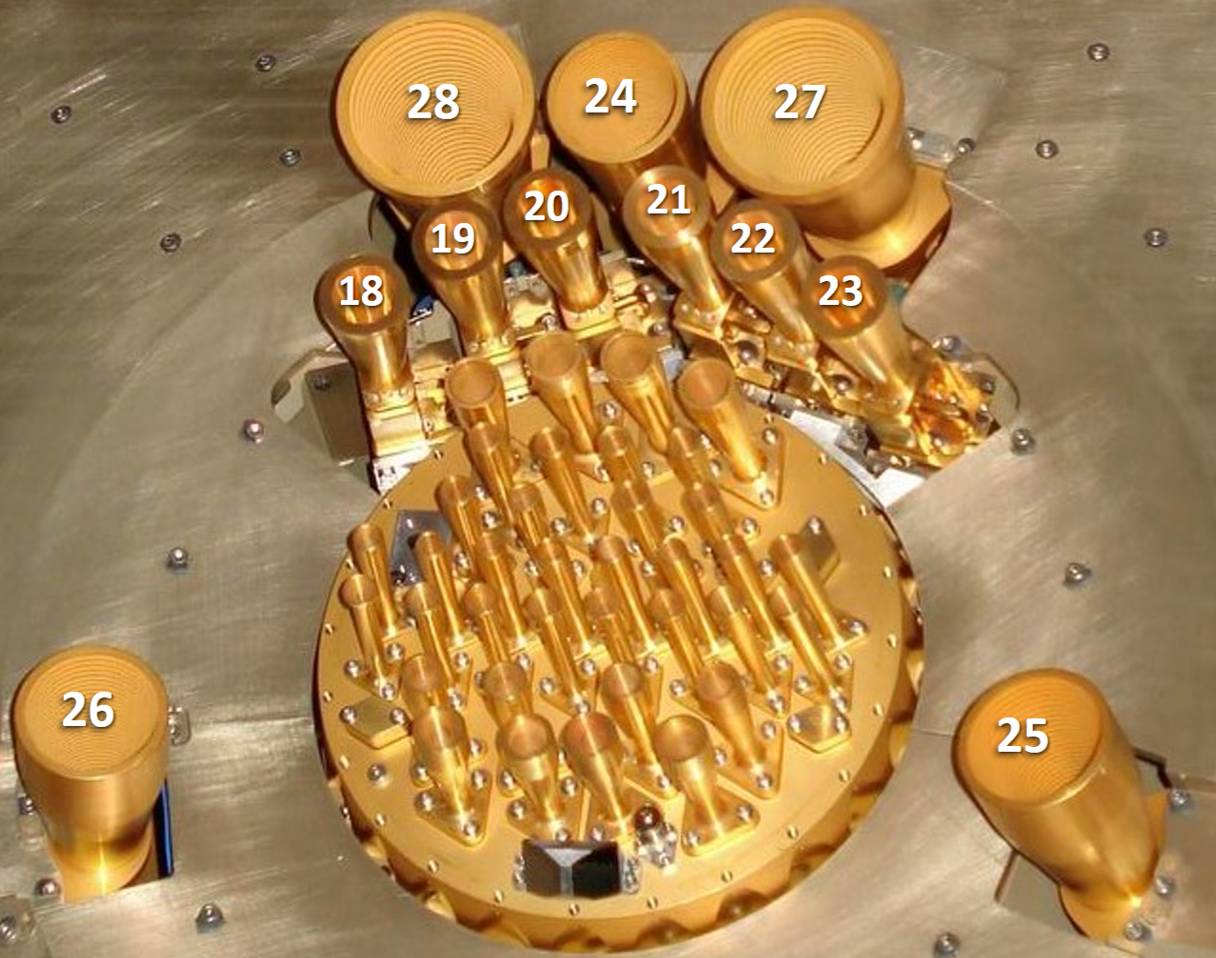}
	\caption{Top view of the \Planck\ focal plane. The central array contains
    the HFI feed-horns, cooled to 4\,K, feeding bolometric detectors cooled to
    0.1\,K. The LFI horns in the outer part of the array are labelled with
    numbers; they are cooled to 20\,K. The LFI horn numbers 18--23, 24--26, and
    27--28 correspond to the 70~GHz, 44~GHz, and 30~GHz channels, respectively. }
	\label{fig:FOCAL-PLANE}
\end{figure}

Radiation from the sky is coupled to 11 corrugated feed horns, shown in
Fig.~\ref{fig:FOCAL-PLANE}. Each horn is followed by an orthomode
transducer (OMT), which splits the incoming radiation into two
perpendicular linear polarizations that propagate through two
independent differential radiometers; see
Fig.~\ref{fig:RADIOMETER-SCHEME}. The OMT provides exquisite
polarization purity, with typical isolation of ${<-30~\mathrm{dB}}$. Each
radiometer pair has a front-end module (FEM), cooled to 20\,K, and a
back-end module (BEM), operated at 300\,K. The FEM is connected to the
BEM by four composite wave-guides (two for each radiometer), thermally
coupled to the three \Planck\ V-groove radiators to minimize parasitic
heat transfer to the cold focal plane (see
Fig.~\ref{fig:PLANCK-PICTURE}). The cryogenically cooled front-end
modules include the first stage HEMT amplifiers and the differencing
system, while the back-end modules provide further radio frequency
amplification. Detection is made via two square-law detector diodes for
each radiometer.

After detection, an analog circuit in the data acquisition electronics
is used to adjust the offset to obtain a nearly null DC output
voltage, and a programmable gain is applied on-board to match the signal level
to the analog-to-digital converter (ADC) input range. After the ADC,
data are digitally down-sampled, re-sampled to match beam resolution
($>3$ samples per beam), compressed, and assembled into telemetry
packets, which are then downlinked to the ground station.

\subsubsection{Stabilization}

\begin{figure}[t]
	\center
	\includegraphics[width=\linewidth]{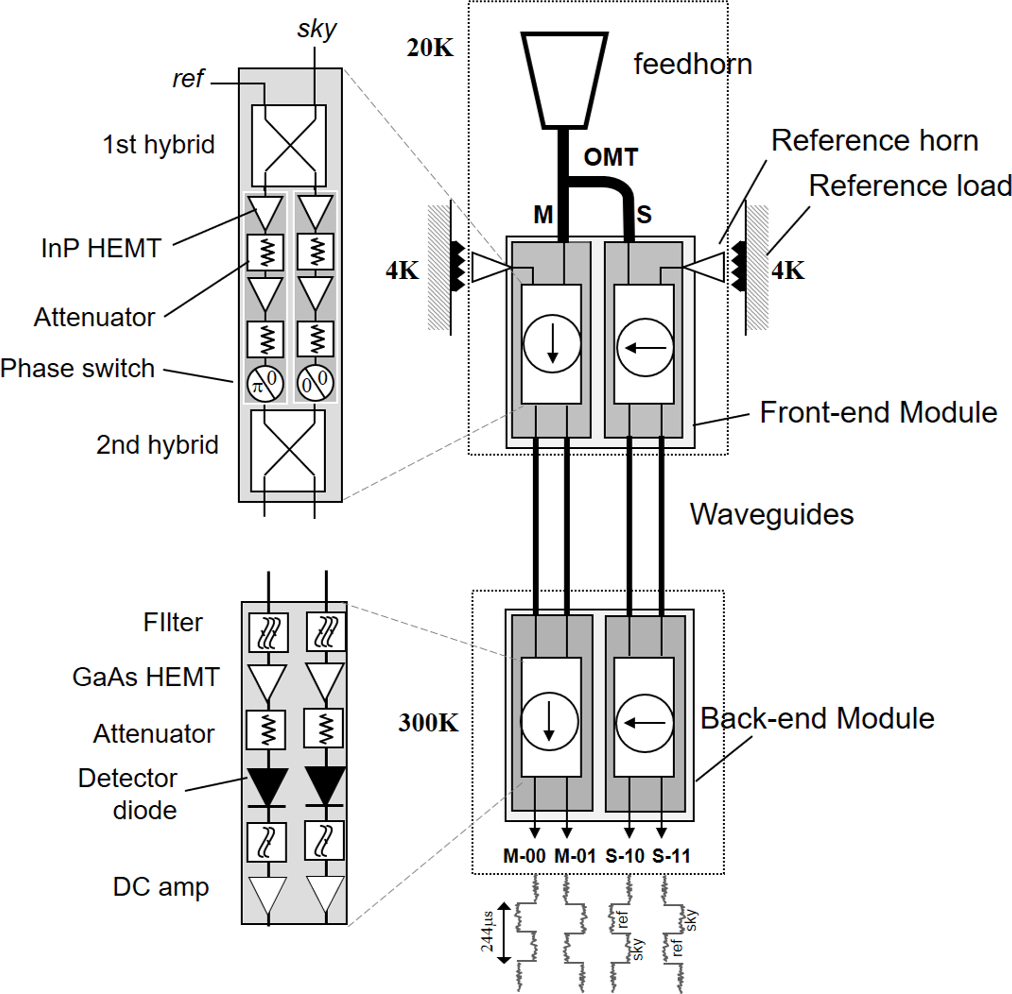}
	\caption{Schematic of an LFI radiometer chain, consisting of a feedhorn and OMT, and the two associated radiometers, each feeding two diode detectors. The insets show details of a front-end module (top) and of a back end module (bottom). }
	\label{fig:RADIOMETER-SCHEME}
\end{figure}

Cryogenic HEMT amplifiers exhibit excellent low-noise performance, but
are affected by significant instability in terms of gain and
noise-temperature fluctuations, typically modelled in terms of a $1/f$
spectrum as discussed in Sect.~\ref{sec:noise}. The LFI system is
designed to efficiently reject such fluctuations in the radiometer
response. The main differential process responsible for radiometer
stabilization takes place in the front-end modules. The signals from
the sky and 4\,K reference load are injected into a hybrid coupler,
which splits the two signals, and redirects them to both of its output
ports (see inset of Fig.~\ref{fig:RADIOMETER-SCHEME}). Then the two
mixed signals are amplified by $\sim$30\,dB by the two amplifier
chains. Thus, any fluctuation in the FEM amplifiers affects both the
sky and the reference load components in exactly the same way. After
amplification, a second hybrid coupler reconstructs the sky and
reference components, which now contain the same fluctuations. Then
the signals are transported by the wave-guides in the warm back-end
modules, where they are further amplified and detected by the
diodes. Finally, when taking the difference between the two diodes,
the FEM fluctuations cancel out. This ``pseudo-correlation'' scheme
reduces front-end fluctuations by a factor of $\mathcal{O}(10^{3})$.

However, instabilities downstream of the FEMs, particularly those
originating in the back-end amplifiers and in the detector diodes,
would still affect the measurements. For this reason, a further level
of stabilization is built into the LFI design. A phase shifter,
alternating between 0\deg\ and 180\deg\ at a frequency of 4096\,Hz,
is applied in one of the two amplification chains within the front-end
modules, as shown in Fig.~\ref{fig:RADIOMETER-SCHEME}. In this way, the DC
output from each diode rapidly alternates the sky and reference
signals, with opposite phase in the two detectors. By taking the
difference between time-averaged values of sky and reference, any
residual fluctuations on time scales longer than ${\sim (1/4096)\,\mathrm s=
0.244\,\mathrm{ms}}$ are removed.

Of course, any non-ideality in the receiver components will introduce
some level of residual fluctuations. Further strategies to suppress
remaining instabilities and potential systematics introduced by the
receiver are described below.


\subsubsection{LFI signal model}
\label{sec:lfi_signal_model}

Based on the above description, the differential power output for each
of the four diodes associated with a feedhorn can be written as the following
special case of Eq.~\eqref{eq:power},
\begin{equation}
  P_{{\rm out},0}^{\rm diode} = a\, G_{\rm tot}\,k\,\Delta\nu \left[ T_{\rm sky} + T_{\rm noise} - r\left( T_{\rm ref} + T_{\rm noise}\right) \right],
  \label{eq_p0}
\end{equation}
where $G_{\rm tot}$ is the total gain, $k$ is the Boltzmann constant,
$\Delta\nu$ the receiver bandwidth, and $a$ is the diode constant.
$T_{\rm sky}$ and $T_{\rm ref}$ are the sky and reference load antenna
temperatures at the inputs of the first hybrid and $T_{\rm noise}$ is
the receiver noise temperature, averaged over an appropriate
integration time. The gain modulation factor, $r$, is a factor of order
unity ($0.8 < r < 1.0$ depending on channel) used to balance in
software the temperature offset between the sky and reference load
signals. This has the important effect of minimising the residual
1/$f$ noise and other non-idealities in the differential data stream.
In the DPC analysis \citep{planck2016-l02} $r$ was determined using
the approximation
\begin{equation}
  r = \frac{T_{\rm sky} + T_{\rm noise}}{T_{\rm ref} + T_{\rm noise}} \approx \frac{V_{\rm sky}}{V_{\rm ref}},
  \label{eq_r}
\end{equation}
and we adopt the same procedure without modification in \BP\ for
now. However, we do note that the \npipe\ analysis pipeline implements
an alternative approach in which $V_{\mathrm{ref}}$ is low-pass
filtered prior to differencing, and this reduces the amount of high
frequency noise in the final maps. Future \BP\ versions can and should
implement a similar solution.

Although somewhat counter-intuitive, the sensitivity of the LFI
radiometers does not depend significantly on the absolute temperature
of the reference load.  In fact, to first order, the white noise
spectral density at the output of each diode is given by
\begin{equation}
  \Delta T_0^{\rm diode} = \frac{2\,(T_{\rm sky}+T_{\rm noise})}{\sqrt{\Delta\nu}}.
  \label{eq_deltat_diode_ideal}
\end{equation}
However, a large imbalance between $T_{\rm sky}$ and $T_{\rm ref}$
would have the effect of amplifying residual fluctuations in the
differential signal. For this reason the LFI reference loads are
cooled to about $4$\,K, exploiting the HFI pre-cooling stage.

The above description holds for the ideal case where all front-end
amplifiers and phase switches have perfectly balanced properties.  In
presence of some level of mismatch, the separation of the sky and
reference load signals after the second hybrid is not perfect and the
outputs are slightly mixed. If the front-end imbalance is small,
Eq.~\eqref{eq_deltat_diode_ideal} may be written as
\begin{equation}
  \left(\Delta T^{\rm diode}\right)^2 \approx  \left(\Delta T_0^{\rm diode}\right)^2
  \left( 1 \pm \frac{\epsilon_{A_1}-\epsilon_{A_2}}{2} + \alpha\epsilon_{T_{\rm n}}\right),
  \label{eq_deltat_diode_nonideal}
\end{equation}
where $\epsilon_{T_{\rm n}}$ is the imbalance in front end noise
temperature between the two radiometer arms, and $\epsilon_{A_1}$ and
$\epsilon_{A_2}$ are the imbalance in signal attenuation in the two
states of the phase switch. Eq.~\eqref{eq_deltat_diode_nonideal}
shows that the output is identical for the two diodes apart from the
sign of the term $(\epsilon_{A_1}-\epsilon_{A_2})/2$, representing the
phase switch amplitude imbalance. For this reason, the LFI scientific
data streams are obtained by averaging the voltage outputs from the
two diodes in each radiometer,
\begin{equation}
  V_{\rm out}^{\rm rad} = w_1 V_{\rm out}^{\rm diode\, 1} + w_2 V_{\rm out}^{\rm diode\, 2},
  \label{eq_vout_radiometer}
\end{equation}
where $w_1$ and $w_2$ are inverse-variance weights calculated from the
data.  Thus, the diode-diode anti-correlation is cancelled, and the
radiometer white noise becomes
\begin{equation}
  \Delta T^{\rm rad} \approx  \frac{\Delta T_0^{\rm diode}}{\sqrt{2}}
  \left( 1  + \alpha\epsilon_{T_{\rm n}}\right)^{1/2}.
  \label{eq_deltat_rad_nonideal}
\end{equation}
In Eqs.~\eqref{eq_deltat_diode_nonideal} and
\eqref{eq_deltat_rad_nonideal}, $\epsilon\ll 1$, while $\alpha$ is a
term of order unity defined by a combination of the input signals and
noise temperature of the radiometer; for details, see Eq.~(8) in
\citet{planck2011-1.4}.

In the current \BP\ processing, we follow the LFI DPC procedure for
all these steps. Future versions of the framework may also account for
these pre-processing steps, and jointly estimate $r$, $\alpha$,
$\epsilon_i$, and $w_i$, but this is left for future work, simply due
to the strong time limitations of the current project (see
Sect.~\ref{sec:bp_program}).

\subsubsection{Naming convention}

As described in the previous section, LFI has 11 horns and
associated OMTs, FEMs and BEMs; 22 radiometers (two for each horn);
and a total of 44~detectors (two for each radiometer). For historical
reasons, the 11~horns are labelled by numbers from 18 to 28 as shown
in Fig.~\ref{fig:FOCAL-PLANE}.

The radiometers associated with each horn are labelled as ``M'' or
``S'' depending on the arm of the OMT they are connected to (``Main''
or ``Side'', as shown in Fig.~\ref{fig:RADIOMETER-SCHEME}). Each
radiometer has two output diodes that are labelled with binary codes
``00'', ``01'' (radiometer M) and ``10'', ``11'' (radiometer S), so
that the four outputs of each radiometer pair can be named with the
following sequence; M-00, M-01, S-10, S-11.

As the telescope scans, the observed region of the sky sweeps across
the focal plane in the horizontal direction as appearing in
Fig.~\ref{fig:FOCAL-PLANE}. Since the reconstruction of the
polarization information requires at least two horns, every pair of
horns aligned in the scan direction are oriented such that their
linear polarizations are rotated by 45\deg\ from each other (with the
exception of LFI-24, which is an unpaired 44~GHz horn). Thus, LFI can
produce independent polarization measurements from the ``horn pairs''
18--23, 19--22, 20--21 (at 70~GHz); 25--26 (at 44~GHz); and 27--28 (at
30~GHz).


\subsection{Implementation details}

Since the \BP\ project aims to establish an open-source, reproducible
and externally extendable analysis framework, it is no longer possible
to rely on direct access to the existing LFI-DPC database, which both
employs proprietary software and runs on one specific computer. To
circumvent this issue, we convert the LFI TOD into a convenient
\texttt{HDF5} format \citep{bp03} that may be accessed using publicly
available tools. This, however, does lead to some adjustments in the
scientific pre-processing pipeline, which now uses this new interface.
At the same time, we have converted the scientific pipeline to
\texttt{C++11}, and a number of optimizations are applied at the same
time, exploiting the new possibilities given by that language.

\subsubsection{Unprocessed Level-1 data}

The extraction of time-ordered Level-1 data from the LFI-DPC database
and the conversion to \texttt{HDF5} format only need to be performed once
within the LFI-DPC environment. We create one file for each LFI horn for
each Operational Day, i.e., the time between two consecutive daily
telecommunication periods. The extracted file contains sky, reference
load and quality flags for each of the diodes of the horn and timing
information, including On-Board Time, Spacecraft Event Time (SCET) and
Modified Julian Date (MJD). It also contains attitude information that
is critical for the analysis; Pointing Period ID (PID); start and end
time of each Pointing Period; end time of the maneuver of each
Pointing Period; and number of data samples.

To optimize the computational time of Level-2 processing, various
deterministic operations are implemented during extraction. For
instance, missing data are added back into the time streams and
flagged as bad data; this ensures that all the timelines for each
frequency are of the same length. Also, planet transits are flagged,
and instrumental flags are added to the extracted data.
		
\subsubsection{Level-2 data pre-processing}

In the DPC pipeline, the main pre-processing of the LFI data occurs at
the Level-2 stage (see \citealp{planck2016-l02} and references
therein).  The same is true in the \BP\ framework, although this is
now implemented as an integrated component in the full algorithm. The
first step in this process is to correct the low-level diode
observations for ADC non-linearities.

The analog signal from each detector is processed by an
analog-to-digital converter (ADC), which ideally provides a digitized
output exactly proportional to the applied voltage.  If the voltage
step sizes between successive binary outputs of the ADC are not
constant, then the ADC introduces a nonlinear response that leads to
calibration errors.  In differential measurements such as those of
LFI, small localized distortions due to ADC non-linearity can have a
significant impact, since the calibration reconstruction depends on
the gradient of the ADC response curve at the point at which the
differential measurements are made.

A non-linearity of the ADC produces a variation in the white noise
level of a detector which does not correspond to a variation in the
input voltage level, as one would expect if the effect were due to a
gain shift.  This subtle effect was observed in some of the LFI
radiometer data for the first time in flight, where drops of a few
percent were observed in the voltage white noise but not in the output
level over periods of few weeks \citep{planck2013-p02a}.  Because of
their lower detector voltages, the 44\,GHz channels showed the
strongest effect, reaching levels of 3 to 5\,\%.  The typical
amplitude of the region where the non-linearity occurs is on the order
of 1\,mV, corresponding to about three bits in the ADC.

The ADC non-linearity effect has been characterised from flight data
and removed from the data streams.  The correct response curves is
reconstructed by tracking how the noise amplitude varies with the
apparent detector voltage in the TOD.  Under the assumption that the
radiometers are stable, the intrinsic white noise is taken to be
constant, so any voltage variations are taken to be due to a
combination of gain drift and ADC effects.  A mathematical model of
the effect and the details of the correction method are described in
Appendix~A of \citet{planck2013-p02a}. In \BP, we adopt a very similar
algorithm, but with one notable improvement: While the DPC algorithm
allow for one correction per narrow voltage bin across the full
observed range, the \BP\ ADC model instead introduces a
low-dimensional parametric model for each ADC glitch. As such, the
\BP\ model has many fewer degrees of freedom, and there is
correspondingly less risk of over-fitting non-ADC-related artefacts;
for further details, see \citet{bp25}.

The next low-level step is to compute a gain modulation factor, $r$,
from the data streams, and apply this to minimize $1/f$ noise as given
by Eq.~\eqref{eq_r}. The outputs from the two detector diodes of each
radiometer are then combined with appropriate noise weights, to remove
the effect of phase switch mismatch, as given by
Eq.~\eqref{eq_vout_radiometer}. For now, we adopt the same weights as
in the official processing. However, we note that future work should
aim to implement a frequency-dependent weighting scheme as introduced
by \Planck\ PR4, which leads to about 8\,\% lower white noise
\citep{npipe}.

In the current \BP\ implementation, neither the ADC correction nor
diode weighting algorithms depend on the assumed sky model, and there
is therefore no feedback from higher-level analysis steps to these
steps. Accordingly, these low-level operations are performed as
one-time pre-processing steps, saving both CPU time and RAM. Only the
co-added and ADC-corrected TOD are actually stored in memory during
the main processing.

\subsubsection{1\,Hz spike correction}

The output signal of the LFI receivers exhibits a set of narrow spikes
at 1\,Hz and harmonics with different amplitude and shape for each
detector.  These subtle artifacts are due to a common-mode additive
effect caused by interference between scientific and housekeeping data
in the analog circuits of the LFI data acquisition electronics.  The
spikes are present at some level in the output from all detectors, but
affect the 44\,GHz data most strongly because of the low voltage and
high post-detection gain values in that channel. The spikes are
nearly identical in sky and reference load samples, and therefore 
almost completely removed by the LFI differencing scheme. However, a
residual effect remains in the differenced data, which needs to be
carefully considered in the data processing.

These features are synchronous with the On-Board time, with no
measurable change in phase over the entire survey, allowing
construction of a piecewise-continuous template by stacking the data
for a given detector onto a one second interval, and fitting a single
overall amplitude. In the DPC analysis the spikes were deemed to
produce negligible effects in the 30 and 70 GHz channels, and were
removed only from the 44\,GHz time-ordered data via template fitting,
while the \Planck\ PR4 processing corrected all three channels. One
notable difference between the two algorithms is that while the DPC
analysis created the template shape by binning the TOD in time-domain,
the PR4 analysis instead fitted a series of Fourier modes.

The BeyondPlanck 1\,Hz correction algorithm combines aspects from both
the DPC and PR4 analysis, and also introduces one new feature. First,
and similar to the DPC analysis, we build the template shapes in
time-domain, and fit one single overall amplitude across the full
mission. Second, and in contrast to the DPC method, we do correct all
three channels, similar to PR4. However, and unlike both the DPC and
PR4, we perform the 1\,Hz correction on the \emph{co-added} time
streams. The main advantage of this is that the precise binning used
to construct the shape template does not need to be aligned between
diodes to avoid edge effects, while the main disadvantage is that the
ADC correction makes the 1\,Hz effect in principle
time-dependent. However, this latter issue a second-order correction,
and, as shown by \citet{bp10}, the full magnitude of the 1\,Hz
correction is anyway several orders of magnitude lower than any other
systematic effect, and typically amounts to only a $\sim\,$0.3\muK
correction in map domain; a second-order ADC correction with respect
to this is entirely negligible. 

\subsection{Pixel-domain data}

In addition to time-domain LFI data, we consider several external data
sets in the pixel domain, as described in the introduction to this
section, simply in order to be able to constrain the full
astrophysical sky model as defined in
Sect.~\ref{sec:default_sky_model}.

\subsubsection{\Planck\ HFI data}

The first external data set we consider is \Planck\ HFI, primarily in
order to constrain thermal dust emission in the LFI frequencies. The
HFI measurements were taken during the first 29 months of
\Planck\ observations, from August 2009 until January 2011, at which
time the helium coolant was depleted. The HFI instrument includes a
total of six frequency bands, centered on 100, 143, 217, 353, 545, and
857\,GHz, respectively. The first four channels are polarized, while
the latter two are (at least nominally) only sensitive to intensity.

While LFI employs coherent radiometers and HEMTs for signal detection,
HFI employs bolometers. One important difference between these two
detector types is that while the former records both the phase and the
amplitude of the incoming electric field, the latter is sensitive only
to the amplitude. In practice, this difference translates into
different sensitivity as a function of frequency, as well as different
instrumental systematics. Generally speaking, bolometers have lower
noise levels than coherent radiometers over relevant CMB frequencies,
but they also tend to be more susceptible to various systematic
errors. For instance, for the LFI 70\,GHz radiometers the noise
equivalent temperature\footnote{The noise equivalent temperature (NET)
  represents the noise standard deviation, $\sigma_0$, expressed in
  thermodynamic units of $\mu\mathrm{K}_{\mathrm{CMB}}$ with an
  integration time of $\Delta t=1$\,s.} is
$152\,\mu\mathrm{K}_{\mathrm{CMB}}\,\mathrm{s}^{-1/2}$
\citep{planck2014-a03}, while it for the HFI 143~GHz bolometers is
$57.5\,\mu\mathrm{K}_{\mathrm{CMB}}\,\mathrm{s}^{-1/2}$
\citep{planck2014-a08}. At the same time, the size of CMB detectors
typically scales with wavelength, and it is therefore possible to fit
a larger number of high frequency detectors than low-frequency
detectors into the same focal plane area. In sum, HFI nominally has
more than six times higher sensitivity than LFI with respect to CMB
fluctuations, as measured in terms of white noise alone. However, a
non-negligible fraction of this sensitivity advantage is lost because
of higher sensitivity to cosmic rays, ADC non-linearities, and
long-duration bolometer time constants \citep{planck2016-l03}.

Several different HFI analysis pipelines were developed within the
nominal \Planck\ collaboration period, as detailed by
\citet{planck2013-p03}, \citet{planck2014-a08}, and
\citet{planck2016-l03}. The two most recent and advanced efforts are
summarized in terms of the \srollTwo\ \citep{delouis:2019} and
\npipe\ \citep{planck2020-LVII} pipelines. For \BP, we adopt by default the
\npipe\ processing as our HFI data set, which is the most recent among
the various available options. However, we note that most analyses here
will only consider the highest frequency channels (857\,GHz in
temperature and 353\,GHz in polarization), in order to constrain
thermal dust emission, and the precise details of the HFI processing
are largely irrelevant for these purposes.

The HFI data are pre-processed as follows before integration into the
\BP\ pipeline. First, we note that the HFI frequency channels have
angular resolutions ranging between 9.7~arcmin at 100~GHz and
4.4~arcmin at 857~GHz. The natural \HEALPix\ pixel resolution for HFI
is thus either $N_{\mathrm{side}}=2048$ or 4096. While our
computational codes do support full resolution analysis, such high
resolution is computationally wasteful for the purposes of LFI
analysis. We therefore smooth the HFI maps to a common angular
resolution of $10\arcm$ FWHM (which is still smaller than the
$14\arcm$ beam of the 70~GHz channel), and we re-pixelize each map at
$N_{\mathrm{side}}=1024$. Overall, this reduces both CPU and memory
requirements for the component separation phase of the algorithm by
about one order of magnitude. Second, we subtract estimates of both
zodiacal light and the kinematic CMB quadrupole from each sky map
prior to analysis, following \citet{planck2020-LVII}, but exclude the
quadrupole from the astrophysical sky model for these channels.

\subsubsection{\emph{Wilkinson Microwave Anisotropy Probe}}

Second, we consider observations from the \emph{Wilkinson Microwave
  Anisotropy Probe} (\WMAP; \citealp{bennett2012}), primarily in order
to constrain synchrotron, free-free, and anomalous microwave
emission. \WMAP\ was funded by the National Aeronautics and Space
Administration (NASA), and operated for 9 years between 2001 and
2010. \WMAP\ observed the microwave sky in five frequency bands,
centered on 23, 33, 41, 61, and 94\,GHz, with an angular resolution
varying from $53\arcm$ at 23\,GHz to $13\arcm$ at 94\,GHz, and with
sensitivities that range between 0.8 and 1.6\,mK\,s$^{-1/2}$.

Like LFI, the \WMAP\ detectors are based on coherent HEMT
technology. However, there are (at least) two critical differences
between the practical implementation of the two experiments. First,
while the LFI detectors measure the difference between the sky signal
in a single direction and that from an internal 4\,K reference load,
the \WMAP\ detection chain is intrinsically differential. That is,
each radiometer is coupled to two independent feedhorns that are
separated by an angle of $141^{\circ}$ on the sky, and each TOD sample
is given by the difference between the signals recorded by those two
horns. For this reason, each \WMAP\ channel is often referred to as a
``differencing assembly'' (DA), rather than a radiometer. Second,
while the basic \Planck\ scanning strategy is fixed by its single
reaction wheel, supporting smooth rotation only around a single axis,
the \WMAP\ satellite carried three orthogonal reaction wheels that
allow for much more tightly interconnected scanning strategies. In
sum, these differences lead to independent instrumental systematics
between the two instruments and consequently to different strategies
to minimise their impact.  The two data sets are thus complementary,
and can be used to break each other's internal degeneracies.

As discussed above, we will in this paper only use enough external
data to break parameter degeneracies that cannot be resolved by
\Planck\ LFI alone, thereby leaving enough room to allow this data set
to provide the main CMB constraints. Therefore, we include in the
following only the \WMAP\ channels between 33 and 61\,GHz. In
intensity, we use the \WMAP\ 9-year full-resolution maps with a
diagonal noise covariance matrix, while in polarization we use the
low-resolution maps with full noise covariance. No pre-processing is
applied to any \WMAP\ data before integration into the \BP\ pipeline.

\subsubsection{Low-frequency surveys}

As discussed by \citet{planck2014-a12}, because of the roughly similar
shapes of the synchrotron, free-free and AME SEDs between 20 and
70\,GHz, \Planck\ and \WMAP\ are not able to resolve these components
on their own. Rather, it is critically important to complement these
data with at least one low-frequency survey in order to establish a
statistically non-degenerate model.

In \BP, we follow \citet{planck2014-a12}, and include the celebrated
408\,MHz survey by \citet{haslam1982}. Although this is widely
believed to suffer more from instrumental systematic errors than
comparable recent surveys, such as S-PASS \citep{Carretti:2019} or
C-BASS \citep{king2014}, it also has the distinct advantages of both
being publicly available and covering the full sky. This full-sky
coverage was achieved by combining observations taken by the Jodrell
Bank MkI 76\,m telescope, the Bonn 100\,m telescope, and the Parkes
64\,m telescope during the 1960's and 1970's. A second advantage is
its very low frequency, which allows for a very clean separation of
synchrotron emission, with only a minor additional contribution from
free-free emission.

We adopt the reprocessed version of the Haslam map that was presented
by \citet{remazeilles2014} for our analyses, and, following
\citet{planck2014-a12}, we model the uncertainty of this map with a
uniform standard deviation of 0.8\,K per pixel, added in quadrature to
1\,\% of the amplitude in that pixel. Finally, we adopt the monopole and
dipole corrections presented by \citet{wehus2014} to fix the largest
angular scales.


\begin{figure*}[t]
	\center
	\includegraphics[width=\linewidth]{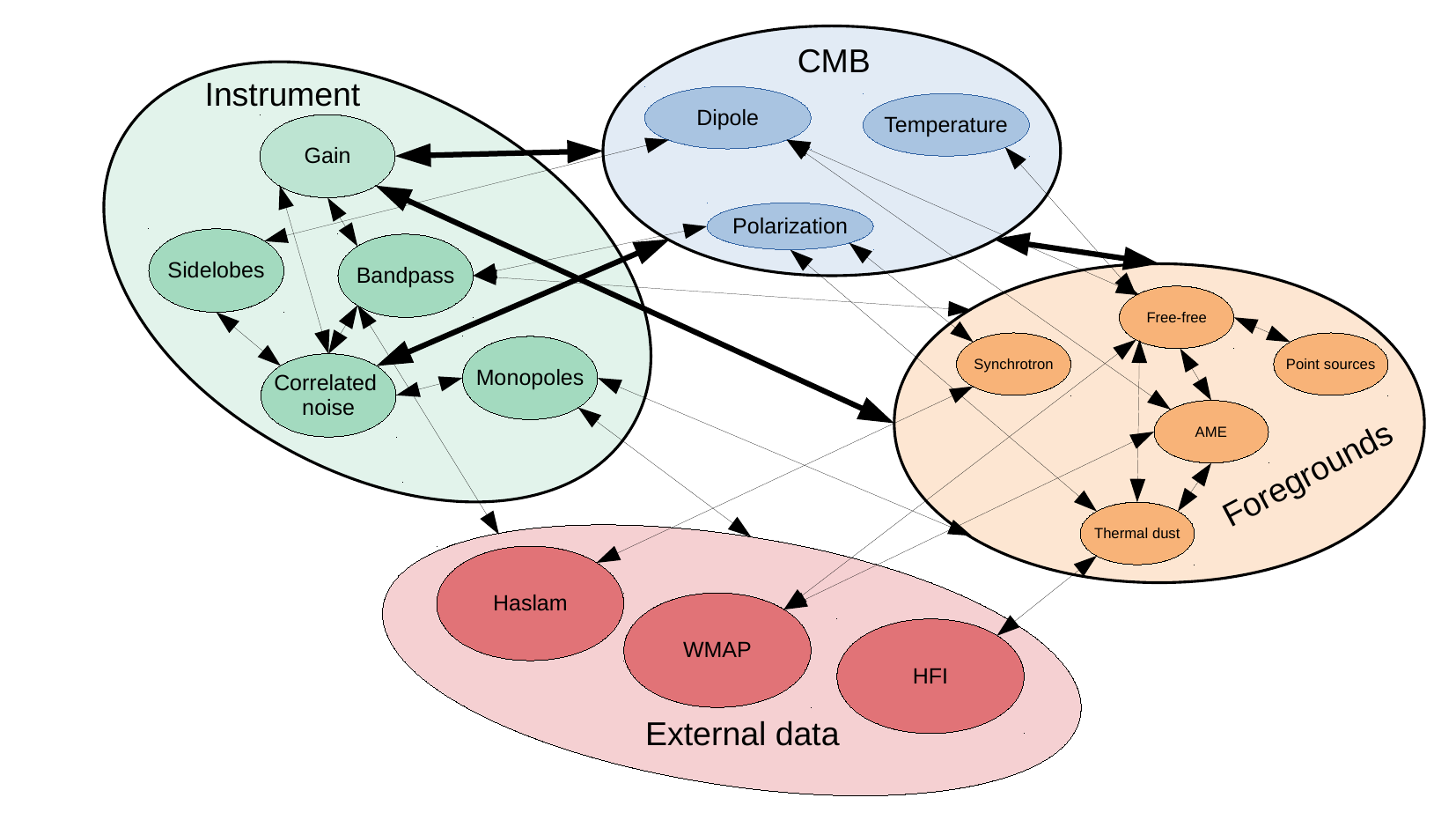}
	\caption{Schematic overview of the primary parameters and
          external data sets considered in the current \BP\ analysis
          and their inter-dependencies. This chart is intended to
          visualize the deeply integrated nature of a modern CMB
          analysis problem; changing any one of these parameter can
          lead to significant changes in a wide range of other
          parameters, and tracing these joint uncertainties is
          critically important for high-precision experiments. }
	\label{fig:dependencies}
\end{figure*}

\section{Bayesian analysis and MCMC sampling}

We have now defined an effective parametric model of the astrophysical
sky in Sect.~\ref{sec:default_sky_model}, and an effective instrument
model in Eq.~\eqref{eq:ideal_model_final}, and we seek to constrain
these models using the data summarized in Sect.~\ref{sec:data}. Let us
for convenience denote the combined set of all free parameters by
$\omega$, such that $\omega\equiv\{g,\Dbp,\n_{\mathrm{corr}}, \a_i,
\beta_i, C_{\ell},\ldots\}$. In \BP, we choose to work within the
well-established Bayesian framework, and as such, our main goal is to
estimate the posterior distribution, $P(\omega\mid \d)$, where $\d$
denotes all available data, both in the form of time-ordered LFI
observations and pre-pixelized external sky maps.

Clearly, this distribution involves billions of non-Gaussian and
highly correlated parameters. Figure~\ref{fig:dependencies} is an
informal attempt to visualize some of the main degeneracies of this
distribution. Thick arrows indicate particularly strong correlations,
while thin arrows indicate weaker ones. This chart is just intended to be a
rough illustration, based on our practical experience, rather than a
formal posterior exploration, and so it is obviously
incomplete. Still, it may serve as useful reminder for new readers
about how individual parameters affect other parts of the system. To
consider one specific example, the gain has a direct and strong impact
on both the CMB and foreground maps by virtue of multiplying the TOD,
and this impact goes both ways; if the current CMB or foreground
parameters are biased, then the estimated gains will also be
biased. The same observations also hold with respect to the correlated
noise and bandpasses, although at a lower level. On the other hand,
the gains are only weakly dependent on the monopoles or sidelobes. The
sidelobes do affect the CMB dipole, however, which is a critically
important component for the gain estimation, and so there is a
second-order dependency. Similar observations hold for most
other parameters; the distribution is tightly integrated, and
each parameter affects a wide range of the full model, either directly
or indirectly. This integrated nature of the full posterior
distribution emphasizes the importance of global end-to-end analysis
with full propagation of uncertainties, as implemented in the
following.

To start our formal exploration of this full posterior distribution,
we write down Bayes' theorem,
\begin{equation}
  P(\omega\mid \d) = \frac{P(\d\mid \omega)P(\omega)}{P(\d)} \propto
  \mathcal{L}(\omega)P(\omega),
  \label{eq:jointpost}
\end{equation}
where $P(\d\mid \omega)\equiv\mathcal{L}(\omega)$ is called the
likelihood; $P(\omega)$ is called the prior; and $P(\d)$ is a
normalization factor usually referred to as the ``evidence''. By
virtue of being independent of $\omega$, the evidence is irrelevant
for parameter estimation purposes, and we ignore it in the current work,
although we note that it is important for model selection
applications.

For a one-, two-, or three-dimensional parametric model, the simplest
way to numerically evaluate the posterior distribution is often to
compute the right-hand side of Eq.~\eqref{eq:jointpost} over some grid
in $\omega$. However, this approach quickly becomes computationally
expensive in higher-dimensional parameter spaces, since the number of
grid points grows exponentially with the number of parameters. For
models with more than three parameters, it is common practice to
resort to Markov Chain Monte Carlo (MCMC) sampling techniques rather
than grid techniques. The main advantage of these techniques is that
computing resources are mostly spent on exploring the peak of the
posterior, which is the region in parameter space that actually
matters for final parameter estimates. In contrast, gridding
techniques spend most of their time evaluating probability densities
that are statistically equivalent to zero. In this section, we will
briefly review three particularly important examples of such MCMC
sampling techniques, as they play a fundamental role in the
\BP\ pipeline.

\subsection{Metropolis sampling}
\label{sec:metropolis}

By far the most commonly applied, and widely known, MCMC algorithm is
the Metropolis sampler \citep{metropolis:1953}. Let $\omega_i$ denote
the $i$th sample in a Markov chain,\footnote{A Markov chain is a
  stochastic sequence of parameter states, $\{\omega_i\}$, in which
  $\omega_i$ only depends on $\omega_{i-1}$, but not earlier states.}
and $T(\omega_{i+1}\mid \omega_i)$ be a stochastic transition probability
density for $\omega_{i+1}$ that depends on $\omega_i$, but not on
earlier states. Assume further that $T$ is symmetric, such that
$T(\omega_{i+1}\mid \omega_i) = T(\omega_i\mid \omega_{i+1})$. The most
typical example of such a transition rule is a Gaussian distribution
with mean equal to $\omega_i$ and with some predefined standard
deviation (or ``step size''), $\sigma$.

With these definitions, the Metropolis sampling algorithm can be
summarized in terms of the following steps:
\begin{enumerate}
  \item Initialize the chain at some arbitrary parameter set,
    $\omega_0$.
  \item Draw a random proposal\footnote{The symbol ``$\leftarrow$''
    indicates setting the symbol on the left-hand side equal to a
    sample drawn from the distribution on the right-hand side.} for the next
    sample based on the transition rule, i.e., $\omega_{i+1}
    \leftarrow T(\omega_{i+1}\mid \omega_i)$.
  \item Compute the acceptance probability, $a$, defined by
    \begin{equation}
      a = \min\left(1,
      \frac{P(\omega_{i+1}\mid \d)}{P(\omega_i\mid \d)}\right)
      \label{eq:metropolis}
    \end{equation}
  \item Draw a random number, $\eta$, from a uniform distribution,
    $U[0,1]$. Accept the proposal if $\eta < a$; otherwise, set
        ${\omega_{i+1} = \omega_i}$.
  \item Repeat steps 2--4 until convergence.
\end{enumerate}

The critical component in this algorithm is the acceptance rule in
Eq.~\eqref{eq:metropolis}. On the one hand, this rule ensures that the
chain is systematically pushed toward the posterior maximum by always
accepting proposals that are more likely than the previous step. In
this sense, the Metropolis sampler can be considered a nonlinear
optimization algorithm that performs a random walk in the
multidimensional parameter space. However, unlike most standard
optimization algorithms, the method also does allow samples with lower
probability density than the previous state. In particular, by
accepting samples with a probability given by the relative posterior
ratio of the two samples, one can show that the time spent at a
given differential parameter volume is proportional to the underlying distribution
density at that state. Thus, the multidimensional histogram of MC
samples produced with this algorithm converges to $P(\omega\mid \d)$ in
the limit of an infinite number of samples.

\subsection{Metropolis-Hastings sampling}

We note that there is no reference to the proposal distribution $T$ in
the Metropolis acceptance probability as defined by
Eq.~\eqref{eq:metropolis}. This is because we have explicitly assumed
that $T$ is symmetric. If we were to choose an asymmetric transition
distribution, this equation would no longer hold, as proposals within
the heavier tail would be systematically proposed more often than
proposals within the lighter tail, and this would overall bias the
chain. 

For asymmetric transition distributions, we need to replace
Eq.~\eqref{eq:metropolis} with
\begin{equation}
  a = \min\left(1,
  \frac{P(\omega_{i+1}\mid \d)}{P(\omega_i\mid \d)}\frac{T(\omega_{i}\mid
    \omega_{i+1})}{T(\omega_{i+1}\mid \omega_i)}\right),
  \label{eq:metropolis-hastings}  
\end{equation}
as shown by \citet{hastings:1970}.  Without further changes, the
algorithm in Sect.~\ref{sec:metropolis} is then valid for arbitrary
distributions $T$, and the algorithm is in this case called
Metropolis-Hastings sampling.

\subsection{Gibbs sampling}
\label{sec:gibbs}

While the Metropolis and Metropolis-Hastings samplers are prevalent in
modern Bayesian analysis applications, they do require a well-tuned
proposal distribution $T$ in order to be computationally efficient. If
the step size is too small, it takes a prohibitive number of proposals
to move from one tail of the distribution to another, whereas if the
step size is too large, then all proposals are in effect rejected by
the acceptance rate. The latter issue is particularly critical in
high-dimensional spaces, and for this reason Metropolis-type samplers
are usually only applied to moderately high-dimensional parameter
spaces, for instance 20 or 50 dimensions. For millions of dimensions,
traditional non-guided Metropolis sampling becomes entirely intractable.

In order to achieve acceptable efficiencies in such cases, one must
typically exploit additional information within the transition
probability. For instance, the Hamiltonian sampler exploits the
derivative of the posterior distribution to establish proposals
\citep[e.g,][]{liu:2008}, while the Langevin Monte Carlo algorithm can
also incorporate second-order derivatives \citep{girolami:2011}.

Another effective way of improving computing efficiency is to decompose
complicated high-dimensional joint distributions into its various
conditional distributions, a process that is called Gibbs sampling
\citep{geman:1984}. In this case, one exploits the shape of the
posterior distribution itself to make proposals, but only in the form
of conditionals. To illustrate the process, let us for the sake of
notational simplicity consider a two-dimensional distribution
$P(\alpha, \beta)$. In that case, the Gibbs sampling transition
probability takes the form
\begin{equation}
  T_{\mathrm{Gibbs}}(\alpha_{i+1}, \beta_{i+1}\mid \alpha_i, \beta_i) =
  P(\alpha_{i+1}\mid \beta_{i})\,\delta(\beta_{i+1}-\beta_i),
  \label{eq:gibbs_prop}
\end{equation}
where $\delta(x)$ denotes the Dirac delta function, which vanishes for
$x\ne 0$, but has a unit integral. The $\delta$ function in
Eq.~\eqref{eq:gibbs_prop} ensures that $\beta_{i+1}=\beta_i$, i.e., that
$\beta$ is kept fixed.

\definecolor{gray}{rgb}{0.75, 0.75, 0.75}

This is an asymmetric proposal distribution, and the corresponding
acceptance probability is therefore given by inserting
Eq.~\eqref{eq:gibbs_prop} into the Metropolis-Hastings rule in
Eq.~\eqref{eq:metropolis-hastings}:
\begin{align}
  a &= \frac{P(\alpha_{i+1}, \beta_{i+1})}{P(\alpha_{i}, \beta_{i})}
  \frac{T_{\mathrm{Gibbs}}(\omega_{i}\mid
    \omega_{i+1})}{T_{\mathrm{Gibbs}}(\omega_{i+1}\mid \omega_i)}\\
  &= 
  \frac{P(\alpha_{i+1}, \beta_{i+1})}{P(\alpha_{i}, \beta_{i})}
  \frac{P(\alpha_{i}\mid
    \beta_{i+1})\,\delta(\beta_{i}-\beta_{i+1})}{P(\alpha_{i+1}\mid \beta_{i})\,\delta(\beta_{i+1}-\beta_i)}\\
  &= 
  \frac{P(\alpha_{i+1}, \beta_{i})}{P(\alpha_{i}, \beta_{i})}
  \frac{P(\alpha_{i}\mid \beta_{i})}{P(\alpha_{i+1}\mid \beta_{i})}\quad \quad\quad\quad\quad{\color{gray}\beta_{i+1}=\beta_i} \\
  &= 
  \frac{P(\alpha_{i+1}\mid \beta_i)\,P(\beta_{i})}{P(\alpha_{i}\mid \beta_{i})\,P(\beta_i)}
  \frac{P(\alpha_{i}\mid \beta_{i})}{P(\alpha_{i+1}\mid \beta_{i})}\quad {\color{gray} P(\alpha,\beta) = P(\alpha\mid\beta)P(\beta)}\\
  &= 1,
\end{align}
where we have used the definitions of both
conditional\footnote{Definition of a conditional distribution: $P(\alpha\mid \beta) \equiv P(\alpha,\beta)/P(\beta)$}
and marginal\footnote{Definition of a marginal distribution: $P(\beta) \equiv \int P(\alpha, \beta)\,\mathrm
  d\alpha$} distributions; the equations marked in gray indicate which
relation is used in a given step. From this calculation, we see that
when proposing samples from a conditional distribution within a larger
global joint distribution, the Metropolis-Hastings acceptance rate is
always unity. Consequently, there is no need to even compute it, and
this can save large amounts of computing time for complex
distributions. However, one does of course have to propose from the
proper conditional distribution for this result to hold.

It is also important to note that only a sub-space of the full
distribution is explored within a single Markov step with this
algorithm. To explore the full distribution, it is therefore necessary
to iterate through all possible conditionals, and allow changes in all
dimensions. Note, however, that there are no restrictions in terms of
order in which the conditionals are explored. Any combination of
sampling steps is valid, as long as all dimensions are explored
sufficiently to reach convergence.

The Gibbs sampling algorithm forms the main computational framework of
the \BP\ analysis pipeline. However, within this larger framework a
large variety of different samplers are employed in order to explore
the various conditionals. For convenience, Appendix~\ref{app:samplers}
provides a summary of the most important samplers, while specific
implementation details are deferred to the individual companion
papers.

We conclude this section by noting that Gibbs sampling only works well
for uncorrelated and weakly degenerate distributions. For strongly
degenerate distributions, the number of Gibbs iterations required to
explore the full distribution becomes prohibitive, as the algorithm
only allows parameter moves parallel to coordinate axes. In such
cases, it is usually necessary either to reparametrize the model in
terms of less degenerate parameters; or, if possible, sample the
degenerate parameters jointly. A commonly used solution in that respect
is to exploit the identity $P(\alpha,\beta) = P(\alpha\mid
\beta)P(\beta)$, which tells us that a joint sample may be established
by first sampling $\beta$ from its \emph{marginal} distribution, and
then $\alpha$ from the corresponding conditional distribution as
before. The marginal sampling step ensures the Markov chain
correlation length becomes unity. This method is used in several places
in the \BP\ Gibbs chain, for instance for the combination of
instrumental gain and correlated noise \citep{bp07}, and for the
combination of astrophysical component amplitudes and spectral
parameters in intensity \citep{bp13}, both of which are internally
strongly correlated.

\section{Global model specification}
\label{sec:model}

The previous section provides a very general overview of our analysis
strategy. In this section, we provide a detailed specification of the
parametric \BP\ model that is appropriate for actual implementation
and processing.

\subsection{Global parametric model}
\label{sec:unimodel}

Following the general model introduced in
Sects.~\ref{sec:sky}--\ref{sec:instrument}, we adopt the following
time-ordered data model,
\begin{equation}
  \begin{split}
    d_{j,t} = g_{j,t}&\P_{tp,j}\left[\B^{\mathrm{symm}}_{pp',j}\sum_{c}
      \M_{cj}(\beta_{p'}, \Dbp^{j})a^c_{p'}  + \B^{4\pi}_{j,t}\s^{\mathrm{orb}}_{j}  
      + \B^{\mathrm{asymm}}_{j,t} \s^{\mathrm{fsl}}_{t} \right]
    +   \label{eq:todmodel} \\
    + &a_{\mathrm{1Hz}}\s^{\mathrm{1Hz}}_{j} + n^{\mathrm{corr}}_{j,t} + n^{\mathrm{w}}_{j,t}.
  \end{split}
\end{equation}
Here $j$ represents a radiometer label, $t$ indicates a single
time sample, $p$ denotes a single pixel on the sky, and $c$ represents
one single astrophysical signal component. Further,
\begin{itemize}
\item $d_{j,t}$ denotes the measured data value in units of V;
\item $g_{j,t}$ denotes the instrumental gain in units of V\,K$_{\mathrm{cmb}}^{-1}$
\item $\P_{tp,j}$ is the $N_{\mathrm{TOD}}\times 3N_{\mathrm{pix}}$
  pointing matrix defined in Eq.~\eqref{eq:pointmat}, where $\psi$ is
  the polarization angle of the respective detector with respect to
  the local meridian;
\item $\B_{j}$ denotes the beam convolution in
  Eq.~\eqref{eq:beam_conv} in the form of a matrix operator, either
  for the (symmetric) main beam, the (asymmetric) far sidelobes, or
  the full $4\pi$ beam; note that for computational efficiency reasons
  we only take into account beam asymmetries for the sidelobes and
  orbital dipole in this paper;
\item $\M_{cj}(\beta_{p}, \Dbp)$ denotes element $(c,j)$ of an
  $N_{\mathrm{comp}}\times N_{\mathrm{comp}}$ mixing matrix defined in
  Eq.~\eqref{eq:mixmat}, describing the amplitude of component $c$ as
  seen by radiometer $j$ relative to some reference frequency $j_0$
  when assuming some set of bandpass correction parameters $\Dbp$;
\item $a^c_{p}$ is the amplitude of component $c$ in pixel $p$,
  measured at the same reference frequency as the mixing matrix $\M$,
  and expressed in brightness temperature units;
\item $\s^{\mathrm{orb}}_{j}$ is the orbital CMB dipole signal in units
  of K$_{\mathrm{cmb}}$, including relativistic quadrupole
  corrections;
\item $\s^{\mathrm{1Hz}}_{j}$ represents the electronic 1\,Hz spike
  correction;
\item $\s^{\mathrm{fsl}}_{j}$ denotes the contribution from far
  sidelobes, also in units of K$_{\mathrm{cmb}}$;
\item $n^{\mathrm{corr}}_{j,t}$ denotes correlated instrumental noise,
  as defined by Eqs.~\eqref{eq:noise} and \eqref{eq:n_psd}; and
\item $n^{\mathrm{w}}_{j,t}$ is uncorrelated (white) instrumental
  noise.
\end{itemize}
For notational convenience, we also define
\begin{equation}
    \s^{\mathrm{sky}}_{j} = \sum_{c} \M_{cj}(\beta, \Dbp^{j})\a^c
\end{equation}
to be the sky model for detector $j$ without beam convolution, but
integrated over the bandpass.

For external data sets, which are defined in terms of pre-pixelized
maps, this model simplifies to
\begin{equation}
  d_{j,p} = g_{j}\B^{\mathrm{symm}}_{pp',j}\sum_{c} \M_{cj}(\beta_{p'},
  \Dbp^{j})a^c_{p'} + n^{\mathrm{w}}_{j,p},
  \label{eq:pixmodel}
\end{equation}
which is identical to the \commandertwo\ data model considered by
\citet{seljebotn:2019}.

The free parameters in Eq.~\eqref{eq:todmodel} are
$\{\g,\Dbp,\n_{\mathrm{corr}}, \a, \beta\}$. All other
quantities are either provided as intrinsic parts of the original data
sets (e.g., the pointing matrix, the beam profile, and the orbital
dipole), or given as a deterministic function of already available
parameters (e.g., the mixing matrix and the far sidelobe
component). The only exception to this is the white noise component,
which is neither fitted explicitly nor given by prior knowledge, but
is simply left as a stochastic uncertainty in the model.

In addition to the parameters defined by Eq.~\eqref{eq:todmodel}, our
model includes a handful of parameters that describe the statistical
properties of the stochastic random fields included in the
model. Specifically, we associate each of the astrophysical component
maps $\a^c$ with a covariance matrix $\S^c$, which in most cases is
assumed to be statistically isotropic. Expanding $a^c_p = \sum_{\ell
  m} a^{c}_{\ell m} Y_{\ell}(p)$ into spherical harmonics, this matrix may
then be written as
\begin{equation}
S^{c}_{\ell m, \ell' m'} \equiv \left< a^{c}_{\ell m} a^{c,*}_{\ell' m'}\right> = C^{c}_{\ell} \delta_{\ell\ell'} \delta_{mm'},
\end{equation}
where $C^{c}_{\ell}$ denotes the angular power spectrum of component
$c$. (Here we have for notational simplicity assumed that the
component in question is unpolarized; the appropriate generalization
to polarization is straightforward, and will be discussed in Sect.~\ref{sec:powspec}.)
This power spectrum is a stochastic parameter on the same footing as
$\a$ or $\beta$, and may as such be included in the model fitted to
the data. Alternatively, the power spectrum may be modelled in terms
some smaller set of parameters, $\xi$, through some deterministic
function $C_{\ell}(\xi)$, in which case $\xi$ is the set of
stochastic parameters included in the model. For notational
simplicity, we will only include the power spectrum in the various
posterior distributions below, but we note that $C_\ell$ may be
replaced with $\xi$ without loss of generality.

Finally, similar considerations hold for the two noise
components. First, the white noise component is assumed to be
piece-wise stationary and Gaussian distributed with vanishing mean and
a covariance matrix equal to $\N^{\mathrm{w}}_{tt'}=\sigma_0^2
\delta_{tt'}$. In the following, we will assume the stationary period
to be given by PIDs, and $\sigma_0$ will be fitted independently for
each period. Second, the correlated noise component is also assumed to
be piece-wise stationary and Gaussian distributed with zero mean, but
with a nontrivial covariance structure in time. Traditionally, this
has been described by the $1/f$ model in Eq.~\eqref{eq:n_psd} for
\Planck\ LFI, but as discussed by \citet{bp06}, the \BP\ processing
has identified an additional component at intermediate frequencies in
the 30 and 44\,GHz channels between 0.01 and 1\,Hz. In this current
analysis, this is modelled in terms of a log-normal contribution, and the full noise power spectral density (PSD) reads
\begin{equation}
        P(f) = \sigma_0^2\left[1 +
          \left(\frac{f}{f_\mathrm{knee}}\right)^\alpha\right] +
        A_\mathrm{p} \exp\left[-\frac{1}{2}\left(\frac{\log_{10}f -
            \log_{10}
            f_\mathrm{p}}{\sigma_\mathrm{dex}}\right)^2\right],
        \label{eq:1fmodel_lognorm}
\end{equation} 
where $A_\mathrm{p}$ is the peak amplitude of the additional term,
$f_\mathrm{p}$ isthe peak frequency, and $\sigma_\mathrm{dex}$ is the
width; for now, the latter two are fixed at fiducial values, but this
may change in future analyses.  With this approximation, the total
noise PSD for the 30 and 44\,GHz channels is modelled in terms of a
total of four free parameters, namely the white noise level
$\sigma_0$, a knee frequency $f_{\mathrm{knee}}$, a low-frequency
spectral slope $\alpha$, and the amplitude of the log-normal
contribution. We denote the spectral noise parameters collectively as
$\xi_n$. For the 70\,GHz channel, the log-normal term is omitted.

So far, the discussion has been kept general, aiming to fit all
necessary parameters into one succinct and computationally convenient
framework. However, at this point it is useful to remind ourselves
that one of the astrophysical component carries particular importance
in this work, namely the CMB. This component is accommodated in
Eq.~\eqref{eq:todmodel} in the form of $\a = \a^{\mathrm{cmb}}$ and
$\M^{\mathrm{cmb}} = 1$ in thermodynamic temperature units, with an
angular CMB power spectrum defined as $C_{\ell} = \left<
|a^{\mathrm{cmb}}|^2\right>$. Computing $P(C_{\ell}\mid \d)$ (or
$P(\xi\mid \d)$, where $\xi$ represents a set of cosmological parameters)
properly marginalized over all relevant astrophysical and instrumental
parameters, is the single most important scientific goal of the
current algorithm.

In summary, the total set of free stochastic parameters adopted in
this work is
$\omega\equiv\{\g,\Dbp,\n_{\mathrm{corr}},\xi_{n},\a,\beta,C_{\ell}\}$,
where each symbol collectively represents a larger set of individual
parameters, typically depending on radiometer, time, pixel, or
component. For notational convenience, we will usually suppress
individual indices, unless explicitly required for context. Likewise,
we also note that in most cases, each of the parameters and quantities
discussed above is associated with its own technicalities, which have
been omitted in the above discussion. Such details will be provided in
dedicated companion papers, with appropriate references given where
appropriate. Finally, a full specification of the astrophysical
component model considered in this analysis is provided in
Sect.~\ref{sec:default_sky_model}.

\subsection{Deterministic quantities}
\label{sec:derquant}

Before considering the posterior distribution $P(\omega\mid \d)$, it is
useful to introduce some extra notation regarding various quantities
that may either be derived deterministically from ancillary
information or from other parameters in our model. These quantities
are not stochastic variables in their own right within our model, and
are as such not associated with independent degrees of freedom, but
they are simply computationally convenient auxiliary variables.

\subsubsection{Frequency maps and leakage corrections}
\label{sec:freqmaps}

Our first derived quantity are frequency maps, which we
will denote $\m_{\nu}$. In our framework, frequency maps are not
stochastic parameters, but instead they represent a deterministic
compression of the full data set from time-ordered data into sky
pixels, conditioning on any parameter or quantity that is not
stationary, such as the gain, correlated noise, and the orbital
dipole.

In order to construct frequency sky maps, we start by computing the
following \emph{residual calibrated TOD} for each detector,
\begin{equation}
  r^{(0)}_{j,t} = \frac{d_{j,t}- n^{\mathrm{corr}}_{j,t}- s^{\mathrm{1Hz}}_{j,t}}{g_{t,j}} - \left(s^{\mathrm{orb}}_{j,t}  
  + s^{\mathrm{fsl}}_{j,t}\right).
  \label{eq:res_bin1}
\end{equation}
According to Eq.~\eqref{eq:todmodel}, $r_{j,t}$ now contains only
stationary sky signal and white noise, given the current estimates of
all other parameters.

In principle, $r^{(0)}_{j,t}$ could be individually binned into a
pixelized map for each radiometer $j$ given the pointing information
in $P^{j}_{tp}$. Unfortunately, due to the poor cross-linking
properties of the \Planck\ scanning strategy, it is very difficult to
solve for three independent Stokes parameters per pixel based on only
information from a single radiometer. In practice, four radiometers
are required in order to obtain well-conditioned maps with robust
statistical properties. In the following we will mostly consider
full-frequency maps, combining all four, six and twelve LFI
radiometers into respective 30, 44 and 70\,GHz maps.

Unfortunately, combining multiple radiometers into a single pixelized
map carries its own complications. Since each radiometer has its own
separate bandpass and beam profile, the observed sky will appear
slightly different for each radiometer. However, when creating a
single joint frequency map, only one single value per pixel is
allowed. Any deviation from this mean value will be interpreted within
the data model as either correlated or white noise, and consequently
be filtered according to $\xi_n$ or down-weighted according to
$\sigma_0$ during processing, or be split among the various other free
parameters, including the CMB map. This typically gives rise to
artifacts proportional to the total signal amplitude, but modulated by
the scanning strategy of the instrument. These effects are often
referred to as bandpass or beam mismatch contamination,
respectively. Informally speaking, this is also often referred to as
``temperature-to-polarization leakage,'' in recognition of the fact
that the temperature signal is orders of magnitude brighter than the
polarization signal, and therefore even a small bandpass or beam
difference can induce a large spurious polarization signal.

Fortunately, with the model described above, which includes a full and
explicit model of the astrophysical sky signal as part of its
parameter space, it is possible to correct for such leakages. As
described by \citet{bp09}, we adopt a very straightforward approach by
simply subtracting a correction from each detector TOD, prior to map
binning, of the form
\begin{equation}
\delta s^{\mathrm{leak}}_{j, t} = \P^{j}_{tp}\B_{pp'}^j\left(s^{\mathrm{sky}}_{jp'} - \left<s^{\mathrm{sky}}_{jp'}\right>\right),
\label{eq:leak}
\end{equation}
where $s^{\mathrm{sky}}_{j}$ denotes the sky model as seen by detector
$j$, accounting for separate bandpass profiles, and angle brackets
indicate an average over all radiometers included in the map. For
computational efficiency reasons, the beam is here approximated as
azimuthally symmetric, which allows the average over detector
indicated by brackets in the equation to be performed
pixel-by-pixel. However, since $\delta \s^{\mathrm{leak}}$ is already a
difference between two very similar sky models with slightly different
bandpasses, the error due to asymmetric beams is a second-order
effect, and completely negligible compared to instrumental noise.

In order to correct for bandpass and beam leakage effects, we modify
Eq.~\eqref{eq:res_bin1} accordingly,
\begin{equation}
  r_{j,t} = \frac{d_{j,t}- n^{\mathrm{corr}}_{j,t} - s^{\mathrm{1Hz}}_{j,t}}{g_{t,j}} - \left(s^{\mathrm{orb}}_{j,t}  
  + s^{\mathrm{fsl}}_{j,t} + \delta s^{\mathrm{leak}}_{j, t}\right).
\end{equation}
After applying this correction, all detector TODs exhibit the same net
sky signal, up to the accuracy of the instrument model, which itself
is sampled over within the Markov chain. At the same time, the mean
signal is not affected by this correction, independent of the accuracy
of the instrument model, as ${\left<\delta s^{\mathrm{leak}}\right> =
0}$ when averaged over all detectors.

With calibrated and cleaned TOD ready at hand which contain
exclusively equalized signal and white noise for each detector,
optimal mapmaking is performed simply by solving the corresponding
normal equations pixel-by-pixel (see, e.g.,
Appendix~\ref{sec:gauss_highdim} or \citealp{ashdown:2007}),
\begin{equation}
\left(\sum_{j \in \nu} \P_j^t (\N^{\mathrm{w}}_{j})^{-1} \P_j\right) \m_{\nu} =
\sum_j \P_j^t (\N^{\mathrm{w}}_j)^{-1}\d_j.
\label{eq:binmap}
\end{equation}
For our pointing matrix definition and white noise covariance matrix,
this equation may for a single pixel be written explicitly as
{\scriptsize
\begin{equation}
  \left[\begin{array}{ccc}
      \sum \frac{1}{\sigma_{0,j}^2} & \sum \frac{\cos
        2\psi_{j,t}}{\sigma_{0,j}^2} & \sum \frac{\sin
        2\psi_{j,t}}{\sigma_{0,j}^2} \\
            \sum \frac{\cos 2\psi_{j,t}}{\sigma_{0,j}^2} & \sum \frac{\cos^2
        2\psi_{j,t}}{\sigma_{0,j}^2} & \sum \frac{\cos 2\psi_{j,t} \sin
              2\psi_{j,t}}{\sigma_{0,j}^2} \\
                  \sum \frac{\sin 2\psi_{j,t}}{\sigma_{0,j}^2} & \sum
                  \frac{\sin 2\psi_{j,t} \cos
        2\psi_{j,t}}{\sigma_{0,j}^2} & \sum \frac{\sin^2
        2\psi_{j,t}}{\sigma_{0,j}^2}
    \end{array}\right]
  \left[\begin{array}{c}
      T \\ Q \\ U
    \end{array}\right]
  =
  \left[\begin{array}{c}
      \sum \frac{d_j}{\sigma_{0,j}^2} \\ \sum \frac{d_j\cos
        2\psi_{j,t}}{\sigma_{0,j}^2} \\ \sum \frac{d_j\sin
        2\psi_{j,t}}{\sigma_{0,j}^2}
    \end{array}\right],
  \label{eq:binmap2}
\end{equation}}
where the sums run over both detector $j$ and all time samples $t$ that
point toward pixel $p$. The associated inverse white noise pixel-pixel
covariance matrix, $\N^{-1}_{pp'}$, is given simply by the inverse
of the matrix on the left-hand side of Eq.~\eqref{eq:binmap}.

It is important to note that the frequency maps defined by
Eq.~\eqref{eq:binmap} have a slightly different statistical
interpretation than those delivered by earlier CMB analysis pipelines,
for instance from the \Planck\ DPCs or \WMAP\ science team. With our
definition, $\m_{\nu}$ represents one possible realization of the
frequency sky map \emph{assuming perfect knowledge about the
  correlated noise, gain, bandpass, leakage effects}, among others; the only
unmitigated stochastic quantity is instrumental white noise. The
uncertainties due to all those other effects are instead accounted for
by the fact that we produce an entire ensemble, $\m_{\nu}^i$, each
with different combinations of systematic effects. For full error
propagation, it is thus important to analyze the full set of available
frequency maps, not just one single realization. In contrast,
traditional frequency maps represent an approximation to the overall
maximum likelihood solution, and error propagation can only be
achieved through analysis of end-to-end simulations.

We conclude this section by emphasizing that $\s^{\mathrm{leak}}$ as
defined above is not a separate stochastic parameter within our
model. It neither increases the total uncertainty in the system, nor
does it induce new parameter degeneracies; it is a simple
deterministic correction that removes a known bias in co-added
frequency maps.

\subsubsection{Spurious leakage maps}
\label{sec:leakmaps}

The correction for spurious leakages from bandpass and beam mismatch
defined in Eq.~\eqref{eq:leak} is only exact to the extent that the
assumed bandpass and beam profiles are accurate. In order to monitor
the efficiency of the leakage correction, it is therefore useful to
establish a dedicated goodness-of-fit statistic for this correction.
For this purpose, we adopt the ``spurious map'' approach pioneered by
\citet{page2007}, and later adapted within various pipelines,
including \citet{planck2016-l02} and \citet{planck2020-LVII}.

The central idea underlying this approach is to modify the pointing
matrix to allow for a set of additional temperature maps, each
corresponding to the difference between the temperature sky as seen by
radiometer $j$ and the temperature sky as seen by the mean of the
detectors at that frequency. However, to prevent the linear mapmaking
equation from becoming degenerate, one can at most include
$N_{\mathrm{det}}-1$ such spurious maps for a configuration involving
$N_{\mathrm{det}}$ detectors. Thus, we generalize the pointing model
for a single observation in terms of the Stokes parameters and
spurious maps as follows,
\begin{equation}
s_{j} = T + Q\cos2\psi_j + U\sin2\psi_j +
\sum_{i=1}^{N_{\mathrm{det}}-1}S_i \,\delta_{ij}. 
\end{equation}
Given this definition, the mapmaking equation in Eq.~\eqref{eq:binmap}
generalizes straightforwardly, and for the special case of three
detectors, the contribution of a single sample from detector $j$ takes
the schematic form
{\fontsize{5.3}{4}\selectfont
\begin{equation}
  \left[\begin{array}{ccccc}
      1 & \cos 2\psi & \sin 2\psi & \,\delta_{1j} & \,\delta_{2,j} \\
      \cos 2\psi & \cos^2 2\psi & \cos 2\psi \sin 2\psi & \cos 2\psi\,
      \,\delta_{1j} & \cos 2\psi\,\delta_{2,j} \\
      \sin 2\psi & \sin 2\psi\cos 2\psi & \sin^2 2\psi & \sin 2\psi\,
      \,\delta_{1j} & \cos 2\psi\sin 2\psi\,\delta_{2,j} \\
      \,\delta_{1j} & \cos 2\psi\,\delta_{1i} & \sin 2\psi\,\delta_{1j} & \,\delta_{1j} & 0 \\
      \,\delta_{2j} & \cos 2\psi\,\delta_{2i} & \sin 2\psi\,\delta_{2j} & 0 & \,\delta_{2j}
    \end{array}\right]
  \left[\begin{array}{c}
      T \\ Q \\ U \\ S_1 \\ S_2
    \end{array}\right]
  =
  \left[\begin{array}{c}
      d \\ d\cos 2\psi  \\ d\sin 2\psi  \\ d\,\delta_{1j} \\ d\,\delta_{2j}
    \end{array}\right].
  \label{eq:Smap}
\end{equation}
}\normalfont

For \WMAP, it is in fact possible to solve this equation
pixel-by-pixel, due to the highly interconnected \WMAP\ scanning
strategy \citep{page2007}.  The resulting Stokes parameter maps solved
jointly with $S$ were therefore released as primary mission products
\citep{bennett2012}. Unfortunately, the same is not possible for
\Planck\ without inducing an unacceptable increase in the overall
noise level, as the coupling matrix in Eq.~\eqref{eq:Smap} is poorly
conditioned over most of the sky. However, the resulting $S$ maps are
still very useful for monitoring purposes, and we will in fact use
these maps to optimize a small number of bandpass parameters, for
which a high level of noise is of no concern; see \citet{bp09} and
Sects.~\ref{sec:bandpass} and \ref{sec:bp} for further details.

\subsubsection{Orbital dipole}
\label{sec:orbital}

The third derived quantity we will need is the orbital dipole,
$s^{\mathrm{orb}}_{j,t}$. Including a relativistic quadrupole
correction, this has a closed form as given by
\begin{equation}
s^{\mathrm{orb}}_{j,t} =
\frac{T_{\mathrm{CMB}}}{c}\left(\,\v_{\mathrm{sat}} \cdot
\hat{\n}_{j,t} + q (\v_{\mathrm{sat}} \cdot
\hat{\n}_{j,t})^2\right),
\end{equation}
where
\begin{equation}
  q = \frac{x(\e^{2x}+1)}{\e^{2x}-1}; \quad x = \frac{h\nu}{2kT_{\mathrm{CMB}}}
\end{equation}
is the frequency dependency of the relativistic quadrupole term. The
CMB temperature is in our analysis fixed to
${T_{\mathrm{CMB}}=2.7255\,\textrm{K}}$, following
\citet{fixsen2009}. Finally, $c$ is the speed of light, $h$ is
Planck's constant, $k$ is Boltzmann's constant, $\v_{\mathrm{sat}}$ is
the satellite velocity, and $\hat{\n}_{j,t}$ is the pointing vector of
detector $j$ at time $t$. The satellite velocity is known with an
absolute precision better than $1\,\textrm{cm}\,\textrm{s}^{-1}$
\citep{planck2020-LVII}. An efficient convolution algorithm for this component
that takes into account the full $4\pi$ beam is described by
\citet{bp08}.

It is important to note the critical role of this particular signal
term. Depending only on the velocity of the satellite (which is known to
exceedingly high precision) and the CMB temperature (which is known to
a precision of 0.02\,\%; \citealp{fixsen2009}), it provides the best
absolute calibration source in microwave astronomy, if not all of
astronomy. For \BP, as for both \Planck\ and \WMAP, this signal is
therefore used to determine the overall absolute calibration of
the entire data set.

\subsubsection{Far sidelobe corrections}
\label{sec:fsl}

The last derived quantity we will need at this stage is the far
sidelobe correction, $\s^{\mathrm{fsl}}$, as defined in
Sect.~\ref{sec:beam}. As shown by \citet{planck2014-a05}, the
\Planck\ LFI optics have several significant sidelobes at large angles
from the optical axis.  The most important is due to spillover around
the main reflector, and located about $85^{\circ}$ from the main
beam. The second most important is due to spillover around the
secondary reflector, and located about $20^{\circ}$ from the main
beam. To account for these, we convolve the parametric sky model with
the (near-)$4\pi$ beam profile, $\B$, of each radiometer (regions
closer than $5^{\circ}$ from the main beam are excluded),
\begin{equation}
  s^{\mathrm{fsl}}_{j,t} = \int_{4\pi} \left[\R(\Omega_t)\B(\Omega) \right]\s^{\mathrm{sky}}_j(\Omega)
  \,\mathrm d\Omega,
\end{equation}
where $\R(\Omega_t)$ is a rotation matrix that rotates the beam as
specified by the satellite pointing at time $t$.  To evaluate this
integral, we employ an algorithm that is algebraically equivalent to
the \texttt{conviqt} approach described by \citet{prezeau:2010}, but
implemented in terms of spin harmonics, as described by
\citet{bp08}.

We stress, however, that uncertainties in the far-sidelobe model are
not yet accounted for, and this represents a significant model
uncertainty in the current analysis. Generalizing the parametric model
in Eq.~\eqref{eq:todmodel} to allow for new beam-related degrees of
freedom is an important goal for future algorithm development.

\section{The \BP\ Gibbs sampler}
\label{sec:posterior}

\subsection{Global posterior distribution}

Given the global parametric model defined in Sect.~\ref{sec:unimodel},
and the ancillary quantities summarized in Sect.~\ref{sec:derquant},
we are now finally ready to consider the full global \BP\ posterior
distribution, $P(\omega\mid \d)$, and describe the computational
algorithms required to map it out. In practice, this entails writing
down explicit expressions for the likelihood and priors in
Eq.~\eqref{eq:jointpost}, as well as specifying an explicit Gibbs chain
that is able to explore the posterior distribution efficiently.

Starting with the likelihood, $\mathcal{L}(\omega)$, we first note
that the data model defined in
Eqs.~\eqref{eq:todmodel}--\eqref{eq:pixmodel} is given as a linear sum of
various components, all of which are specified precisely in terms of
our free parameters $\omega$. This applies even to the correlated
noise component, $\n^{\mathrm{corr}}$, which for the purposes of the
likelihood is fully equivalent to any of the other physical
components. As such, we may symbolically write $\d =
\s^{\mathrm{tot}}(\omega) + \n^{\mathrm{w}}$, where
$\s^{\mathrm{tot}}(\omega)$ is the sum of all model components in
Eq.~\eqref{eq:todmodel}, whether they have a cosmological, astrophysical
or instrumental origin. With this notation, we immediately see that
\begin{equation}
P(\d\mid \omega) \propto P(\n^{\mathrm{w}}\mid \omega) \propto \exp\left({-\frac{1}{2}\left(\frac{\d-\s^{\mathrm{tot}}(\omega)}{\sigma_0}\right)^2}\right),
\end{equation}
since $\n^{\mathrm{w}} = \d - \s^{\mathrm{tot}}(\omega)$, 
$P(\n^{\mathrm{w}}) \propto N(0,\sigma^2)$, and $\s^{\mathrm{tot}}$ is
deterministically given by $\omega$.

Next, the prior $P(\omega)$ should encapsulate all our prior knowledge
about any of the model parameters. For instance, we may use this term
to introduce information regarding the instrumental gain from
temperature measurements of the 4\,K load onboard the
\Planck\ satellite during the calibration stage; or we can use it to
impose prior knowledge regarding the CIB zero-level amplitude at each
frequency during component separation; or we may introduce a prior on
the Hubble constant during cosmological parameter estimation; or we
may use it to regularize posterior volume effects through the
application of a Jeffreys ignorance prior \citep{jeffreys:1946}. A
detailed breakdown of the priors used in this particular analysis will
be presented in association with the respective steps.

\subsection{Overview of Gibbs chain}
\label{sec:gibbschain}

As already discussed, the posterior distribution defined by
Eq.~\ref{eq:jointpost} involves millions of tightly correlated and
non-Gaussian parameters, and it is clearly unfeasible to optimize or
sample from it directly. We therefore resort to the Gibbs sampling
algorithm described in Sect.~\ref{sec:gibbs}: We compute a Markov
chain of correlated samples by initializing on some arbitrary
parameter combination, $\omega_0$, and then iteratively sample from
each conditional distribution from the full distribution. In
practice, most runs are initialized on the outcome of an earlier
analysis, in order to save burn-in time.

The \BP\ Gibbs chain may be written schematically as follows,
\begin{alignat}{11}
\g &\,\leftarrow P(\g&\,\mid &\,\d,&\, & &\,\xi_n, &\,\a^{\mathrm{1Hz}}, &\,\Dbp, &\,\a, &\,\beta, &\,C_{\ell})\\
\n_{\mathrm{corr}} &\,\leftarrow P(\n_{\mathrm{corr}}&\,\mid &\,\d, &\,\g, &\,&\,\xi_n,
&\,\a^{\mathrm{1Hz}}, &\,\Dbp, &\,\a, &\,\beta, &\,C_{\ell})\\
\xi_n &\,\leftarrow P(\xi_n&\,\mid &\,\d, &\,\g, &\,\n_{\mathrm{corr}}, &\,
&\,\a^{\mathrm{1Hz}}, &\,\Dbp, &\,\a, &\,\beta, &\,C_{\ell})\\
\a^{\mathrm{1Hz}} &\,\leftarrow P(\a^{\mathrm{1Hz}}&\,\mid &\,\d,
&\,\g, &\,\n_{\mathrm{corr}}, &\,\xi_n, &\,
&\,\Dbp, &\,\a, &\,\beta, &\,C_{\ell})\\
\Dbp &\,\leftarrow P(\Dbp&\,\mid &\,\d, &\,\g, &\,\n_{\mathrm{corr}}, &\,\xi_n,
&\,\a^{\mathrm{1Hz}}, &\,&\,\a, &\,\beta, &\,C_{\ell})\\
\beta &\,\leftarrow P(\beta&\,\mid &\,\d, &\,\g, &\,\n_{\mathrm{corr}}, &\,\xi_n,
&\,\a^{\mathrm{1Hz}}, &\,\Dbp, & &\,&\,C_{\ell})\\
\a &\,\leftarrow P(\a&\,\mid &\,\d, &\,\g, &\,\n_{\mathrm{corr}}, &\,\xi_n,
&\,\a^{\mathrm{1Hz}}, &\,\Dbp, &\,&\,\beta, &\,C_{\ell})\\
C_{\ell} &\,\leftarrow P(C_{\ell}&\,\mid &\,\d, &\,\g, &\,\n_{\mathrm{corr}}, &\,\xi_n,
&\,\a^{\mathrm{1Hz}}, &\,\Dbp, &\,\a, &\,\beta&\,\phantom{C_{\ell}})&,
\end{alignat}
where the conditional variables have been vertically aligned for
clarity only.  As usual, the symbol $\leftarrow$ means setting the
variable on the left-hand side equal to a sample from the distribution
on the right-hand side. For convenience, in the following we also
define the notation ``$\omega\setminus \xi$'' to imply the set of
parameters in $\omega$ except $\xi$.

Note that the first conditional in this Gibbs chain, ${P(\g\mid
  \d,\ldots)}$ represents a marginal distribution with respect to
$\n_{\mathrm{corr}}$. As such, $\g$ and $\n_{\mathrm{corr}}$ are in
effect sampled jointly in the \BP\ Gibbs chain \citep{bp07,bp06},
using the properties discussed in Sect.~\ref{sec:gibbs}. The reason
for this choice is that these two parameters are particularly strongly
degenerate, and joint sampling therefore leads to a much shorter
overall correlation length than strict Gibbs sampling. This outweighs
by far the somewhat higher computational cost per iteration that is
required for sampling the gain from its marginal distribution.

The same method is applied when sampling astrophysical component
parameters, $\a$ and $\beta$ in the case of intensity maps. In this case, we first
sample $\beta$ marginalized over $\a$, and then $\a$ conditionally on
$\beta$ \citep{bp13}. Since $\a$ is a set of linear parameters, the
integral over $\a$ may be computed analytically, as first exploited
for CMB component separation purposes in the \texttt{Miramare} code
\citep{stompor:2008,stivoli:2010}. For polarization, we still sample
$\beta$ conditionally on $\a$, as described by \citet{bp14}, because
the low-resolution \WMAP\ data with full covariance matrix prohibits
smoothing to a common angular resolution, as needed for the marginal
sampling approach.

We will now describe each of these distributions in turn, with the
main goal being to build intuition regarding each distribution. For
specific implementational details we refer the interested reader to
companion papers.

At this point, we note that if a joint maximum likelihood estimate is
required as opposed to a sample set, the same methodology applies as
described below, with the exception that one should then maximize each
conditional, rather than sample from it. The algorithm then becomes
equivalent to a (slow but computationally convenient) steepest descent
nonlinear optimizer. In our codes, we have implemented support for
both modes of operation.

\subsection{Specification of conditional sampling steps}
\label{sec:conditionals}

\subsubsection{Gain and calibration sampling}
\label{sec:gain}

We start our review of the various Gibbs sampling steps with the gain,
$g_t$. In this paper, we only summarize the main central steps, and we
refer the interested reader to \citet{bp07} for full algorithmic
details.

The gain is among the most critical parameters in our model in terms
of the overall resulting data quality, and even relative errors at the
$\mathcal{O}(10^{-4})$ level are highly significant. At the same time,
it is also one of the parameters we have the least prior information
about, as it is entirely specific for each individual instrument. To
estimate the gain robustly, we therefore exploit the following
observations: First, we note that the orbital CMB dipole (see
Sect.~\ref{sec:orbital}) depends only the satellite velocity, which is
known to a precision of $10^{-6}$ \citep{godard2009}, and
the CMB monopole value, which is known to a precision of 0.02\,\%
\citep{fixsen2009}. The orbital dipole therefore by far provides the
most robust constraints on the mean calibration.\footnote{The term
  ``calibration'' refers in this paper to the time average of the
  gain.} However, since the Earth's orbital velocity is
30\,km\,s$^{-1}$ and the CMB monopole is 2.7255\,K, the absolute
amplitude of the orbital dipole is only 270\,\muK, which is small
compared to typical signal and noise variations. As a result, the
orbital dipole is not strong enough to directly determine the gain
alone on short time scales.

In contrast, the amplitude of the solar CMB dipole is 3\,mK, about ten
times brighter than the orbital dipole. Of course, the true solar CMB
dipole parameters are unknown, and must be estimated jointly with
everything else; but we do know that all detectors observe the
\emph{same} solar dipole. We also know that its frequency spectrum is
given by a perfect blackbody with temperature $T_{\mathrm{CMB}}$. Together, these
two facts provide a strong handle on relative gain variations, both
between detectors and in time.

\begin{figure}[t]
  \center
  \includegraphics[width=\linewidth]{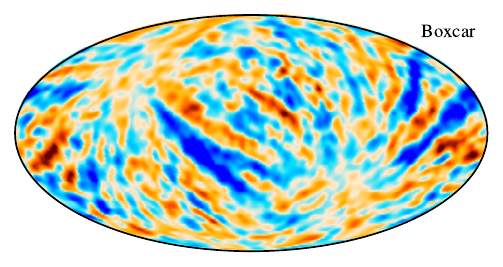}\\\vspace*{-3mm}
  \includegraphics[width=\linewidth]{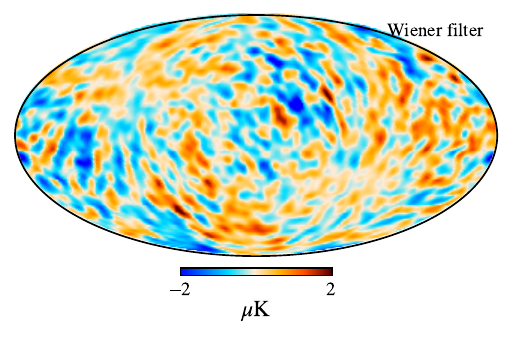}
  \caption{Random correlated noise Stokes $Q$ Gibbs sample,
    $\n^{\mathrm{corr}}$, for the 44\,GHz frequency channel, smoothed
    to an effective angular resolution of $5^{\circ}$ FWHM. The top
    figure shows the map resulting from a boxcar smoothed gain solution,
    whereas the bottom figure is the map which results from smoothing
    the gain solution with a Wiener filter. Reproduced from \citet{bp07}  }\label{fig:corrstripes}
\end{figure}

\begin{figure}[t]
  \center
  \includegraphics[width=\linewidth]{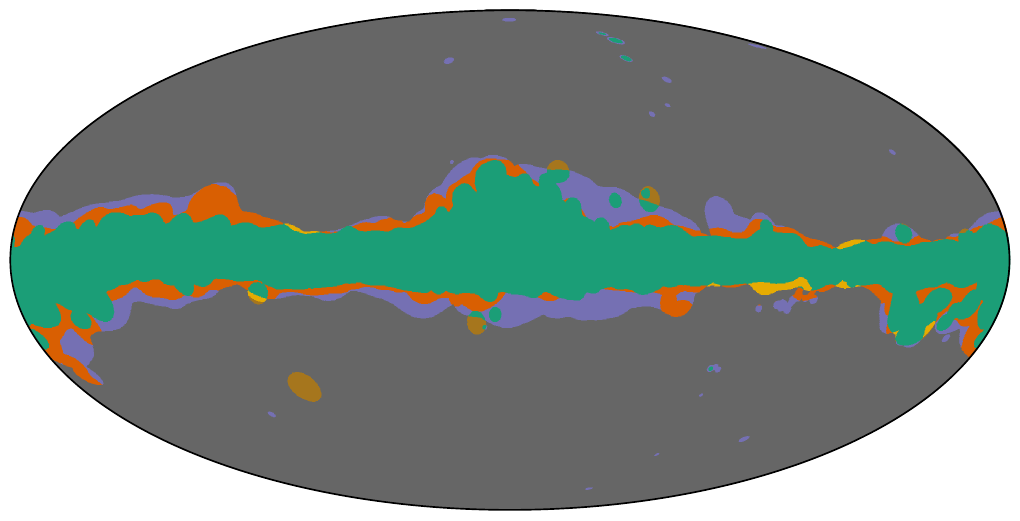}
  \caption{Processing masks used in gain and correlated noise
    estimation for each frequency channel. The allowed 30\,GHz sky
    fraction (blue) is 0.73, the 44\,GHz sky fraction (green) is 0.81,
    and the 70\,GHz sky fraction (red) is 0.77. Reproduced from
    \citet{bp07}.}
  \label{fig:procmask}
\end{figure}

First, we note that the fundamental data model in
Eq.~\eqref{eq:todmodel} may for each detector be written in the form
\begin{equation}
  d_{t} = g_{t} s^{\mathrm{tot}}_{t} + s^{\mathrm{1Hz}}_{t} + n^{\mathrm{corr}}_{t} + n^{\mathrm{w}}_{t} 
  \label{eq:gainmodel}
\end{equation}
where
\begin{equation}
  s^{\mathrm{tot}}_t = \P_{tp}\left[\B_{pp'}^{\mathrm{symm}}s^{\mathrm{sky}}_{p'} + \B_{pp'}^{\mathrm{asymm}}\left(s^{\mathrm{orb}}_{t}  
  + s^{\mathrm{fsl}}_{t}\right)\right]
  \label{eq:stot}
\end{equation}
is the total sum of all true astrophysical signals entering through
the optical assembly of the instrument. Noting that both the gain and
correlated noise \emph{a priori} are unknown quantities with no and
weak priors, respectively, it is clear from the structure of
Eq.~\eqref{eq:gainmodel} that these two parameters are highly
degenerate: significant variations in $\g$ can be accounted for by
adjusting $\n_{\mathrm{corr}}$ with only a small penalty in terms of
total goodness-of-fit through the noise power spectrum, $\xi$.

Recognizing the importance of this degeneracy, and the resulting
sensitivity to potential modelling errors, we decompose the full
time-dependent gain function into three components, and sample each of
these with a special-purpose sampler. Specifically, we write the full
gain for detector $i$ in the form ${g^i_t = g_0 + \Delta g^i + \delta
  g^i_t}$, where $g_0$ is the gain averaged both over all detectors
within a given frequency map and over time; $\Delta g^i$ is the mean
gain differential for detector $i$ averaged over time, with the
additional requirement that $\sum_i \Delta g^i = 0$; and $\delta
g^i_t$ represents the time dependence in gain, with the additional
requirement that $\sum_t g^i_t = 0$ for each $i$. In addition, when
sampling the these gain parameters, we marginalize over the correlated
noise component, as discussed in Sect.~\ref{sec:gain}, in order to
minimize the Gibbs chain correlation length. In total, the data model
used for gain sampling therefore reads
\begin{equation}
  d^i_{t} = (g_0 + \Delta g^i + \delta g_t^i) s^{\mathrm{tot}}_{t} + s^{\mathrm{1Hz}}_t + n^{\mathrm{tot}}_{t},
  \label{eq:gainmodel2}
\end{equation}
where $n^{\mathrm{tot}}_{t} = n^{\mathrm{corr}}_{t} +
n^{\mathrm{wn}}_{t}$ is the total noise contribution with a full
covariance matrix given by $\xi^n$.

Formally speaking, the statistically optimal sampling algorithm for
any of the three gain parameters is given by correlating the full sky
signal with the observed data. In effect, this direct approach was
adopted by the LFI DPC pipeline \citep{planck2016-l02}. A significant
disadvantage of this direct approach, however, is sensitivity to
foreground and bandpass mismatch errors. Instead, we adopt the
following three-step approach, which is structurally similar to the
\npipe\ algorithm \citep{planck2020-LVII}.

First, we sample $g_0$ with the orbital CMB dipole alone as
calibration source, based on the residual
\begin{equation}
  r^i_t \equiv d^i_{t} - \hat{g} (s^{\mathrm{tot}}_{t} -
  s^{\mathrm{orb}}) - (\Delta \hat{g}^i + \delta \hat{g}_t^i) s^{\mathrm{orb}}_{t} - s^{\mathrm{1Hz}}_t
 =  g_0 s^{\mathrm{orb}}_{t} + n^{\mathrm{tot}}_{t},
\end{equation}
where the symbol ``\char`\^'' denotes the respective estimate from the
previous Gibbs iteration. Noting that $n^{\mathrm{tot}}$ is a Gaussian
field with covariance $\N$ and $\s^{\mathrm{orb}}$ is a deterministic
quantity, the sampling algorithm for $g_0$ is given by that of a
univariate Gaussian distribution as described in
Appendix~\ref{sec:gauss_lowdim},
\begin{equation}
  g_0 =
  \frac{\sum_i\s_i^{\mathrm{orb},t}\N_i^{-1}\r_i^{\mathrm{orb}}}{\sum_i\s_i^{\mathrm{orb},t}\N_i^{-1}\s_i^{\mathrm{orb}}}
  +
  \frac{\eta}{\sqrt{\sum_i\s_i^{\mathrm{orb},t}\N_i^{-1}\s_i^{\mathrm{orb}}}},
  \label{eq:g0}
\end{equation}
where the sum runs over all detectors in the frequency map, and $\eta
\sim N(0,1)$.  We note that this expression represents a formal
violation of the Gibbs chain, since $g_0$ in reality affects both
$\s^{\mathrm{tot}}$ and $\s^{\mathrm{orb}}$. By fitting only to
$\s^{\mathrm{orb}}$ we effectively achieve robustness with respect to
modelling errors at the cost of increased statistical
uncertainty.

Note that the noise covariance matrices in Eq.~\eqref{eq:g0} include
both correlated and white noise contributions; this is algebraically
equivalent to marginalizing over $\n_{\mathrm{corr}}$ as described
above. In constrast, in a classic Gibbs sampling algorithm we would
subtract $\n_{\mathrm{corr}}$ from $r^i$ in Eq.~\eqref{eq:g0_res}, and
then only include the white noise component in $\N_i$. This, however,
would lead to a much longer Markov correlation length, since the joint
uncertainties between $g_0$ and $\n_{\mathrm{corr}}$ then would have
to be explored by sampling, rather than algebraically.

Second, we sample the detector dependent calibration factors, $\Delta
g_i$, based on the residual
\begin{equation}
  r^i_t \equiv d^i_{t} - (g_0 + \delta \hat{g}_t^i) s^{\mathrm{tot}}_{t}  - s^{\mathrm{1Hz}}_t
  =  \Delta g_i s^{\mathrm{tot}}_{t} + n^{\mathrm{tot}}_{t},
  \label{eq:g0_res}
\end{equation}
for each detector, which now includes contributions from both the
solar CMB dipole and astrophysical foregrounds, and therefore supports
a significantly higher signal-to-noise ratio than the orbital dipole
alone. At the same time, we impose the additional linear constraint that
\begin{equation}
  \sum \Delta g_i = 0,
\end{equation}
such that possible contamination in this step does not affect the
absolute mean calibration of the full frequency channel. The total
system may be solved using the Lagrange multiplier technique
\citep[e.g.,][]{bertsekas:1996} with a Lagrangian of the form
\begin{equation}
\mathcal{L}(\Delta g_i, \lambda) = \sum_i \left(\r^{i}-\Delta g_i
\s^{\mathrm{tot}}\right)^t \N_i^{-1} \left(\r^{i}-\Delta g_i \s^{\mathrm{tot}}\right) +
\lambda \sum_i \Delta g_i,
\end{equation}
where $\lambda$ is a Lagrange multiplier. The maximum posterior
solution is found by solving the linear equations resulting from setting
$\partial \mathcal{L}/\partial x_i = 0$ with $x_i = \{\Delta g_i,
\lambda\}$.

Third and finally, the time-dependent gain fluctuations are sampled
starting from the residual 
\begin{equation}
  r^i_t \equiv d^i_{t} - \left(g_0 + \Delta g_i\right) s^{\mathrm{tot}}_{t} - s^{\mathrm{1Hz}}_t
 =  \delta g_i s^{\mathrm{tot}}_{t} + n^{\mathrm{tot}}_{t},
\end{equation}
where $\delta g^i_t$ is assumed to be constant within each PID period,
but free to vary between consecutive PIDs. Unfortuately, the noise on
the estimated gain per PID is very large, and in practice, these
estimates must be smoothed. For a fully stationary system, the
mathematically optimal approach to do so would be to sample $\delta
g^i_t$ from a multivariate Gaussian distribution, taking into account
their known $1/f$ power dependency and associated temporal
correlations. This would correspond to an optimal Wiener filter, and
the optimal sampling algorithm takes the following form \citep{bp07},
\newcommand{\tG}[0]{\tens{G}}
\begin{equation}
    (\tG^{-1} + \N^{-1})\delta \hat{\g} = \N_i^{-1}\r_i + \N_i^{-1/2}\vec\eta_1 + \tG^{-1/2}\vec\eta_2.
    \label{eq:delta_gain_sampling_with_prior}
\end{equation}
In this expression, $\tG$ is the covariance matrix of the Gaussian
prior on $\delta\g_i$, $N_i$ is the noise covariance of the individual
gain measurements, and $\eta_1$ and $\eta_2$ are two independent
vectors drawn from a normal distribution with unity
variance.

Motivated by the fact that gain fluctuations originate from the same
$1/f$ fluctuations as correlated noise, we adopt a similar functional
form in Fourier space for $\tG$ as the correlated noise component,
\begin{equation}
    \tG(f) = \sigma_0^2\left(\frac{f}{f_0}\right)^\alpha,
\end{equation}
where $\alpha$ and $\sigma_0$ are parameters to be determined, while
$f_0$ is just a reference frequency. These two parameters effectively
determine the smoothness of the resulting solution; for instance, the
lower the value of $\sigma_0$, the smoother the final gain estimate
will be. In principle, these parameters could be fitted directly from
the data, fully analogous to the correlated noise parameters discussed
in the next section. However, the time-dependent gain shows clear
evidence of non-thermal variations, for instance in the form of
sharp jumps due to abrupt changes in the thermal or electronic
environment. To avoid over-smoothing these, we instead choose values
that impose less temporal smoothing than dictated by the stationary
$1/f$ behaviour; for further details, see \citet{bp07}.

For comparison, the official \Planck\ LFI DPC processing adopted a
boxcar average algorithm for smoothing the gain
\citep{planck2014-a03}, effectively averaging over a gliding window in
time. The main difference between these two algorithms is illustrated
in Fig.~\ref{fig:corrstripes}, which compares the correlated noise
components that results from the two methods: The boxcar averaging
introduces significant striping by over-smoothing real gain
fluctuations, which in turn biases the estimated CMB Solar dipole and
Galactic foregrounds for the relevant time periods. The novel
Wiener-filter smoothing algorithm presented by \citet{bp07} represents
one of the most important individual algorithmic improvements in
\BP\ with respect to the official \Planck\ processing, and it was key
to understanding the long-standing issues with the 44\,GHz frequency
channel \citep{planck2016-l02}. 

To prevent foreground modelling errors from affecting the various gain
estimates, we apply the processing masks indicated in gray in
Fig.~\ref{fig:procmask} in each of the above equations. Any sample
that falls within the masked region is omitted from the corresponding
inner product, and does not contribute to the overall estimate. The
same applies to any sample that may be flagged by the instrument
database. Removing individual samples, however, does introduce a
slight computational complication because of the
$\N_{\mathrm{tot}}=\N_{\mathrm{corr}}+\N_{\mathrm{wn}}$ operator,
which denotes a dense noise covariance matrix that includes both
correlated and white noise. Application of this operator at full
temporal TOD resolution is computationally expensive. However, we note
that since the gain is defined only by a single value per PID,
small-scale fluctuations can be averaged down with minimal loss of
information in all the above equations. We therefore down-sample each
time-ordered data object to 1\,Hz before evaluating the above
equations, and this reduces the overall computational cost for gain
sampling by almost two orders of magnitude; see \citet{bp07} and \citet{bp06} for
further details.

\subsubsection{Correlated noise sampling}
\label{sec:ncorr}

Since the gain is sampled from a marginal distribution with respect to
correlated noise, not a conditional distribution, it is essential to
sample the correlated noise immediately following the gain; otherwise
the Gibbs chain would end up in an internally inconsistent
state. However, as far as the actual sampling algorithm for the
correlated noise is concerned, this is a normal conditional with
respect to the gain, akin to any other standard Gibbs step, and was
first described in a CMB setting by \citet{wehus:2012}. The same
algorithm has now also, for the first time, been used to solve the CMB
mapmaking problem by \citet{bp02}.

To derive the appropriate sampling equation for $\n^{\mathrm{corr}}$,
we return to the full data model in Eq.~\eqref{eq:todmodel}, and note
that it may be written on the form
\begin{equation}
  r_{t} \equiv d_{t} - g_{t} s^{\mathrm{tot}}_{t} - s^{\mathrm{1Hz}}_t =
  n^{\mathrm{corr}}_{t} + n^{\mathrm{w}}_{t},
  \label{eq:ncorr_model}
\end{equation}
where $s^{\mathrm{tot}}_{t}$ is defined in Eq.~\eqref{eq:stot}. As
discussed in Sect.~\ref{sec:unimodel}, $\n^{\mathrm{corr}}$ is assumed
to be Gaussian distributed with zero mean and covariance
$\N^{\mathrm{corr}}$, while the white noise term is uncorrelated
Gaussian with variance $\sigma_0^2$. Eq.~\eqref{eq:ncorr_model}
therefore also describes a correlated Gaussian distribution, and the
sampling equation is in this case given by Eq.~\eqref{eq:multigauss}
with a template matrix $\T=1$, a signal covariance matrix $\S =
\N^{\mathrm{corr}}$, a noise covariance matrix $\N = \N^{\mathrm{w}}$,
and data $\d=\r$.

Let us first consider the ideal case of a single PID with no missing
data due to either instrument flags or processing mask. In that
case, Eq.~\eqref{eq:multigauss} can be solved very conveniently in
the Fourier domain, and the appropriate sampling equation for the $k$th
Fourier mode reads
\begin{equation}
n_k = \frac{r_k + \eta_1\sigma_0^2/\sqrt{P_k^{\mathrm{corr}}} + \eta_2\sigma_0
  }{1 + \sigma_0^2/P^{\mathrm{corr}}_k}.
\label{eq:ncorr_samp1}
\end{equation}
For this case, we note that the computational cost is equivalent to
two Fourier transforms of the full time-ordered data.

As usual, the first term in the numerator of Eq.~\eqref{eq:ncorr_samp1}
is simply a Wiener filtered version of the residual, $\r$. As such, it
represents a biased estimate of $\n^{\mathrm{corr}}$, with a noise
suppression factor given by the relative inverse signal-to-noise
ratio, $P^{\mathrm{corr}}_k/\sigma_0^2$. The two last terms are
stochastic random fluctuations that ensure that the resulting sample
has the appropriate covariance structure.

Equation~\eqref{eq:ncorr_samp1} only applies to data with no missing
time samples, as the Fourier transforms require continuous inputs. In
practice, however, we drop all samples that are removed by either the
instrument flags or by the processing mask shown in
Fig.~\ref{fig:procmask}. In this case, the optimal solution is
therefore given by Eq.~\eqref{eq:multigauss}, where rows and columns
corresponding to masked samples are set to zero in $\N^{-1} $. The resulting
equation is therefore solved efficiently by a Conjugate Gradient (CG)
technique, as described by \citet{bp02}. As reported by \citet{bp03},
and summarized in Table~\ref{tab:resources}, this particular step
accounts for about 40\,\% of the total computational cost of the
\BP\ Gibbs sampler, and it is as such by far the most expensive single
component in the entire analysis.

In the current framework, correlated noise estimation plays the role
of the traditional CMB mapmaking problem with correlated noise in a
traditional pipeline. In this respect, it is worth noting that the
correlated noise sample is constructed based on the signal-subtracted
data, $r$, alone. Under the assumption of a perfect signal model,
inaccuracies in the correlated noise model can therefore not introduce
any signal bias. Using the analogy of traditional destriping codes
\citep[e.g.,][]{Maino1999,Keihanen2004, Keihanen2005,Keihanen2010}, the signal
subtraction plays the same role in the Gibbs sampling approach as the
projection operator $\Z = \I - \P(\P^t \N^{-1}\P)^{-1}\P^t\N^{-1}$
does for destriping, shielding any stationary signal from the noise
filter. The main fundamental difference between the two approaches
lies in the fact that while the traditional destriper only exploits
information from a single frequency channel at any given time, the
Gibbs sampling approach simultaneously exploits information from all
frequencies to construct a joint signal model, which then is used to
shield the signal during correlated noise estimation. The Gibbs
sampling approach is thus mathematically equivalent to destriping all
frequencies at once. The effect of this global correlated noise
estimation will become evident later, in the form of lower correlated
noise residuals in the joint approach.

Second, it is important to note that the correlated noise solution
resulting from Eq.~\eqref{eq:ncorr_samp1} is moderately robust against
model errors, whether they are due to foreground modelling errors or
inaccuracies in the bandpass or beam profile. The reason is simply
that Eq.~\eqref{eq:ncorr_samp1} is a Wiener filter, and therefore has
greatly suppressed power in any frequency mode for which
$P^{\mathrm{corr}}_k \ll \sigma_0^2$. Intuitively, this means that any
feature that cannot be readily identified in the raw time-ordered data
as compared with $\sigma_0$, will only be weakly affected by the
correlated noise component. Minor errors in the signal model, beam or
bandpass profiles are therefore mostly negligible.

There are, however, two important exceptions to this general
rule. First, some point sources, such as Tau A or the Galactic center,
are sufficiently bright that uncertainties in the beam or foreground
model can be large compared to the white noise level. If so, the
resulting errors will typically translate into bright stripes passing
through the respective source, extending along the scanning path of
the satellite. To avoid this, it is critically important to mask all
bright sources as part of the processing mask, and replace those
regions with a proper constrained realization as described above. 

The second important exception is the CMB dipole. This signal is both
bright, with a peak-to-peak amplitude of about 3\,mK, and concentrated
at a very low frequency that corresponds to the satellite spin rate of
$1/60\,$Hz. This is typically comparable to (or lower than) the
correlated noise knee frequencies \citep{planck2016-l02}. Furthermore,
the ring-based \Planck\ scanning strategy provides minimal modulation
of the dipole signal on frequencies beyond the spin frequency. The
combination of these facts leads to a strong degeneracy between the
CMB dipole parameters, the time-dependent gain, and the correlated
noise. Indeed, experience shows that \Planck\ is, for all practical
purposes, unable to break this degeneracy through statistical power
alone. Instead, various strong priors are typically imposed to
regularize these degeneracies. For instance, the LFI DPC processing
impose the requirement that $\m_{\mathrm{D}}\cdot\m=0$, where
$\m_{\mathrm{D}}$ is a map of the CMB dipole and $\m$ is the sky map;
this effectively leaves the full instrumental noise component aligned
with the CMB dipole in the final sky map
\citep{planck2014-a06}. Additionally, the LFI pipeline makes no
explicit corrections for bandpass mismatch during gain
calibration. For the HFI 2018 DPC processing, the dominant assumption
is that the gain is fully independent of time, and the only source of
apparent gain fluctuations are ADC non-linearities
\citep{planck2016-l03}. For \npipe, two important assumptions are that
polarized foregrounds at frequencies between 44 and 217\,GHz may be
fully modelled in terms of the signal observed by 30 and 353\,GHz, and
that CMB polarization may be ignored during calibration
\citep{planck2020-LVII}. Obviously, none of these assumptions are formally
correct, and they will necessarily lead to systematic biases at some
level.

In \BP, we adopt a different approach to the problem, by actually
exploiting information beyond \Planck. Specifically, as described in
Sect.~\ref{sec:data}, we will in the following perform a joint
analysis of \WMAP\ and \Planck\ observations, and thereby take
advantage of information in one experiment to break degeneracies in
the other. Most notably, the \WMAP\ scanning strategy covers 70\,\% of
the sky every hour, as compared to less than 1\,\% per hour for
\Planck. This strategy is thus obviously better suited for measuring
the very largest angular scales on the sky, despite higher white
noise. On the other hand, the differential structure of the
\WMAP\ differencing assemblies leads to particularly large
uncertainties for some specific modes, including $E_{\ell=5}$ and
$B_{\ell=3}$ \citep{jarosik2010}. In \BP\ we therefore choose to
combine \Planck\ and \WMAP\ data while taking into account the full
covariance information of each experiment, and thereby optimally
leverage the individual strengths of each experiment. Still, we
emphasize the importance of visually inspecting binned sky maps of
$\n_{\mathrm{corr}}$ for dipole-like residuals, which is the
archetypical signature of calibration errors; such residuals may occur
if the assumed signal model is inadequate for the data set in
question.

\subsubsection{Noise PSD sampling}
\label{sec:ncorr_psd}

The third conditional distribution, $P(\xi^{n}\,\mid \,\d,
\omega\setminus \xi^{n})$, in the \BP\ Gibbs chain describes the noise
power spectrum density parameters, $P_k$ and $\sigma_0$, collectively
denoted $\xi^{n}$. In the following, we will make the assumptions that
$\xi^n$ is constant within each PID and uncorrelated between PIDs.
Being closely connected to the previous sampling step, the following
procedure was also first presented for CMB applications by
\citet{wehus:2012}.

To sample from $P(\xi^{n}\,\mid \,\d, \omega\setminus \xi^{n})$, we
recall that $\n_{\mathrm{corr}}\sim
N(0,\N_{\mathrm{corr}})$. Therefore,
\begin{align}
  P(\xi\mid \,\d, \omega\setminus \xi^{n}) &\propto P(\xi\mid \,\n_{\mathrm{corr}}) \\
  &\propto \frac{\e^{-\frac{1}{2}\n_{\mathrm{corr}}^t\N_{\mathrm{corr}}^{-1}\n_{\mathrm{corr}}}}{\sqrt{|\N_{\mathrm{corr}}|}},
  \label{eq:psd}
\end{align}
where $\N_{\mathrm{corr}}=\N_{\mathrm{corr}}(\xi^n)$. To sample from
this distribution, we could for instance run a Metropolis sampler over
$\xi^n$, using Eq.~\eqref{eq:psd} to define the acceptance
probability.  However, at this stage we introduce an approximation to
the exact solution, trading a small amount of statistical optimality
for increased robustness to modelling errors and minimal parameter
degeneracies. Specifically, we decouple the white noise variance from
the correlated noise model simply by defining
\begin{equation}
\sigma_0^2 \equiv \frac{\mathrm{Var}(r_{t}-r_{t-1})}{2},
\end{equation}
where we define
\begin{equation}
r_t \equiv d_t - g s_t^{\mathrm{tot}} - n_t^{\mathrm{corr}} - s^{\mathrm{1Hz}}_t
\end{equation}
to be the residual time stream after subtracting both the current
total sky signal and correlated noise estimates. Thus, we take the
variance of the difference between any two neighboring residual
samples to be our white noise variance. On the one hand, this
represents the single most robust estimate of the total white noise
level one can form from a finite data set. On the other hand, it is of
course only an approximation to the true white noise level, since the
correlated noise component may also happen to include a flat and
non-negligible power spectrum density at the highest frequency
mode. This situation typically arises more often for bolometers (as
for instance employed by the \Planck\ HFI detectors) than for coherent
detectors (as employed by the \Planck\ LFI detectors and considered
here), but the principle is the same both cases.

Thus, we define the component of the correlated noise at half the
Nyquist frequency to be part of the white noise, and the correlated
noise is consequently defined as the difference between the total
noise and the white noise. For error propagation into other parameters
in the model, only the sum of the two components is significant. This
split is thus essentially just a computational short-cut that
eliminates internal degeneracies between the two noise components, and
maximizes the relative contribution of the white noise component. This
has two main numerical advantages. First, noting that white noise
marginalization is performed algebraically, while correlated noise
marginalization is done through sampling, a high relative white noise
fraction leads to a shorter overall Markov chain correlation length
for all steps in the algorithm. Second, by fixing the white noise
level, we break degeneracies within the $\xi^{n}$ parameters, which
otherwise lead a very long correlation length between $\sigma_0$,
$\alpha$, $f_{\mathrm{knee}}$ and $A_p$ (see Sect.~\ref{sec:noise}),
making convergence assessment difficult.

Given this definition of the white noise variance, the correlated
noise level may now be sampled from Eq.~\eqref{eq:psd} by fixing
$N^{\mathrm{w}}$. Specifically, as discussed by \citet{bp06}, the
conditional posterior may be written in Fourier space as
\begin{equation}
  -\ln P(\xi^n) = \sum_{f>0}
  \left[\frac{|n_f^\mathrm{corr}|^2}{P_{\mathrm{corr}}(f)}
    + \ln{P_{\mathrm{corr}}(f)}\right],
  \label{eq:logP_S}
\end{equation}
up to an irrelevant constant, where $n_f^\mathrm{corr}$ are the
Fourier coefficients of the correlated noise estimate,
$\n_{\mathrm{corr}}$, and $P_{\mathrm{corr}}(f)$ represent all
non-white contributions to the noise PSD. We sample from this
distribution with a simple inversion sampler (see
Appendix~\ref{sec:inversion}), iteratively Gibbs sampling over
$\alpha$, $f_{\mathrm{knee}}$, and $A_\mathrm{p}$. Masking and
in-painting is handled by the $\n_{\mathrm{corr}}$ sampling step
described in Sect.~\ref{sec:ncorr}.

\subsubsection{Bandpass sampling}
\label{sec:bp}  

Next, we consider the bandpass correction conditional distribution,
$P(\Delta_{\mathrm{bp}}\mid \d,\omega\setminus\Delta_{\mathrm{bp}})$,
and in the following we will consider the most basic form of bandpass
correction, namely a linear shift as defined by
Eq.~\eqref{eq:bp_shift}; see \citet{bp09} for further details.

Similar to the gain case, we find it useful to decompose the full
bandpass shift for detector $j$ as follows,
\begin{equation}
  \Delta_{\mathrm{bp}}^j = \bar{\Delta}_{\mathrm{bp}} + \delta_{\mathrm{bp}}^j.
\end{equation}
Here the first term is the average over all radiometers within a given
frequency channel and the second term is constrained by $\sum_j
\delta_{\mathrm{bp}}^j = 0$. The motivation for this decomposition is
that the two terms impact the data in qualitatively different
ways. The average bandpass shift, $\bar{\Delta}_{\mathrm{bp}}$, change
the overall effective frequency of the full frequency channel, and is
as such highly degenerate with the foreground SED parameters; a given
bandpass frequency shift may often be counteracted by adjusting the
values of the synchrotron or thermal dust spectral indices. This mean
bandpass shift does not in general change the polarization properties
of the resulting frequency map. The relative bandpass corrections,
however, have a strong impact in terms of polarization through
temperature-to-polarization leakage, as discussed in
Sect.~\ref{sec:bandpass} and by \citet{bp09}.

For this reason, we have implemented two different sampling algorithms
for these parameters. First, the mean bandpass correction is sampled
with the full time-domain residual on the form
\begin{align}
    \r_j &= \d_{j} - \n^{\mathrm{corr}}_{j} - \G_j \P_{j}\B_{j}^{\mathrm{asymm}}\left(\s^{\mathrm{orb}}_{j}  
      + \s^{\mathrm{fsl}}_{j} \right) - \s^{\mathrm{1Hz}}_j \\ &= \G_{j}\P_{j}\B_{j}^{\mathrm{symm}}\sum_{c}
      \M_{cj}(\beta, \Dbp^{j})\a^c  + \n^{\mathrm{w}}_{j}.
      \label{eq:res_bp}
\end{align}
Clearly, this residual is highly nonlinear in $\Delta_{\mathrm{bp}}$,
and no analytic distribution or sampler exist. We therefore once again
resort to the Metropolis sampler described in
Sect.~\ref{sec:metropolis}. Specifically, we propose small variations
to the current mean bandpass shift (while keeping the relative
differences between radiometers fixed); we compute the resulting
mixing matrices $\M$ and sky maps for the new proposals; and we
finally then apply the Metropolis acceptance rule as given by the
resulting $\chi^2$. Only samples within the small processing mask in
Fig.~\ref{fig:procmask} are included in the $\chi^2$. Since mixing
matrix updates are computationally expensive, bandpass corrections are
among of the most difficult parameters to explore within the entire
model. However, as discussed by \citet{bp09}, the degeneracies between
CMB, free-free, AME and $\bar{\Delta}_{\mathrm{bp}}$ are too strong to
support a robust determination of $\bar{\Delta}_{\mathrm{bp}}$ when
including only LFI and \WMAP\ data. In the final \BP\ production runs,
we therefore adopt priors based on the \Planck\ 2015 analysis
\citep{planck2014-a12}, which used HFI data to break these
degeneracies. In practice, we only apply an overall mean correction of
0.24\,GHz to the 30\,GHz channel, and no mean corrections to the 44 and 70\,GHz
channels. In future analyses also including the full HFI set, these
priors will obviously be removed.

For the relative bandpass corrections, $\delta_{\mathrm{bp}}$, we
adopt an alternative approach that is specifically tuned to increase
robustness in the final polarization maps. Specifically, after
proposing changes to each of the detector-specific bandpasses (under
the constraint that their sum vanishes), we compute the resulting
$IQUS$ map that was defined in Eq.~\eqref{eq:Smap} for both the old and
new parameter values. Next, we define a special purpose $\chi^2$ of
the form
\begin{equation}
  \chi^2 = \sum_{j=1}^{N_{\mathrm{det}}-1} \sum_p \left(\frac{S_j(p)}{\sigma_{j}(p)}\right)^2,
\end{equation}
where $S_j(p)$ is the spurious map corresponding to radiometer $j$ in
pixel $p$, and $\sigma_j(p)$ is the associated uncertainty resulting
from the $IQUS$ solution. This $\chi^2$ defines the Metropolis
acceptance probability as follows,
\begin{equation}
a = \min\left(1,\e^{-\frac{1}{2}(\chi^2_{\mathrm{prop}}-\chi^2_{\mathrm{i-1}})}\right),
\end{equation}
where $\chi^2_{\mathrm{prop}}$ and $\chi^2_{\mathrm{i-1}}$ are the
$\chi^2$'s of the proposed and previous parameter states, respectively.

Overall, this approach builds on the same fundamental ideas as the
$IQUS$ approach pioneered by \WMAP\ \citep{page2007}, but using vastly
fewer free parameters: Rather than fitting one free parameter per
pixel, this algorithm introduces only one additional free parameter
per radiometer. To achieve acceptable precision, it instead uses the
current foreground model to predict the resulting corrections in each
pixel. Thus, while the direct $IQUS$ method is not applicable for
\Planck\ due to its poorly interconnected scanning strategy, our
approach implements the same basic idea but without excessively
increasing the overall white noise level of the final maps. For
further discussion of the method, we refer the interested reader to
\citet{bp09}.

\subsubsection{Diffuse component spectral parameter sampling}
\label{sec:beta}  

The fifth conditional distribution in the \BP\ Gibbs chain concerns
the foreground SED parameters, $P(\beta\mid \d,
\omega\setminus\beta)$. Noting that the linear amplitudes $\a$ and
spectral parameters $\beta$ are in general highly degenerate for high
signal-to-noise models, we employ the same computational method for
intensity sampling as for the gain and correlated noise, and sample
these jointly. In practice, this is achieved by first sampling $\beta$
from the marginal probability distribution with respect to $\a$, and
then $\a$ conditionally on $\beta$. For specific details regarding the
following algorithm, we refer the interested reader to
\citet{bp13}. For polarization, we employ a standard Metropolis
sampler that is conditional on the foreground amplitudes; see
\citet{bp14} for details. 

For CMB component separation applications, the two-step marginal
sampling approach was first described by \citet{stompor:2008} and
later implemented in the \texttt{Miramare} code by \citet{stivoli:2010}. To see
how their methodology connects with our notation, as defined by
Eq.~\eqref{eq:todmodel}, we can write the relevant residual in the
following form,
\begin{align}
    \nonumber
    \r_j &= \left(\d_j - \n_j^{\mathrm{corr}} - \s_j^{\mathrm{1Hz}}\right)/g_j -
\left(\s_j^{\mathrm{orb}} + \s_j^{\mathrm{sl}}\right)
    \\
    &= g_j\P_j\B_j\s_j^{\mathrm{sky}}(\beta) +
\n_j^{\mathrm{w}}.
\label{eq:beta_res}
\end{align}
The left-hand side in this equation is identical to the residual in
Eq.~\eqref{eq:res_bin1}, which is the input to the binned mapmaker
defined by Eq.~\eqref{eq:binmap}. Under the assumption of azimuthally
symmetric beams,\footnote{In the current \BP\ implementation, we assume
  azimuthally symmetric beams for all component separation steps,
  following all previous CMB analysis pipelines.} $\B_j$, this
expression may therefore be rewritten in terms of binned sky maps on
the form
\begin{equation}
  \m_{\nu} = \A_{\nu}(\beta)\a +\n^{\mathrm{w}}_{\nu},
  \label{eq:beta_data}
\end{equation}
where $\A(\beta) \equiv \B_{\nu}\M_{\nu}(\beta)$ is an effective
mixing matrix that accounts for both beam convolution and
astrophysical component SEDs. Given this expression, the marginal
log-posterior for $\beta$ then reads \citep{stompor:2008}
\begin{equation}
-2\ln P(\beta\mid \m) = \sum_{\nu}\left(\A_{\nu}^t\N_{\nu}^{-1}\m_{\nu}\right)^t
\left(\A_{\nu}^t\N_{\nu}^{-1}\A_{\nu}\right)^{-1}\left(\A_{\nu}^t\N_{\nu}^{-1}\m_{\nu}\right).
\label{eq:beta_marg}
\end{equation}
However, the derivation of this expression relies on an assumption of
identical beam responses across all frequency channels, and it is
therefore necessary to smooth all input maps to a common angular
resolution before evaluating this expression. We therefore use
this expression only for intensity sampling, coupled to a tuned Metropolis
sampler.

For polarization, we employ a likelihood given by the original
residual defined by Eq.~\eqref{eq:beta_data}, 
\begin{equation}
  -2\ln P(\beta\mid \m, \a) = \sum_{\nu}\left(\frac{\m_{\nu} - \A_{\nu}(\beta)\a}{\sigma_{\nu}(p)}\right)^2
\label{eq:beta_cond}
\end{equation}
where $\sigma_{\nu}(p)$ is the standard deviation map of channel
$\nu$.
When estimating the spectral index of synchrotron emission, we
partition the sky into four large disjoint regions, and sample one
constant value of $\beta_{\mathrm{s}}$ per region, while still
allowing for smooth transitioning between regions. Sky partitioning
allows us both to tune the signal-to-noise ratio per fitted parameter,
and also to reduce the overall computational cost. All other free
spectral parameters are fitted using a single constant value across
the full sky. For both temperature and polarization, we employ tuned
Metropolis samplers to explore the posterior distribution
\citep{bp13,bp14}.

Finally, we note that even with low-dimensional spectral parameter
models, it is useful to impose additional priors on $\beta$ to
stabilize the fits. Specifically, we consider two types of priors in
the following. First, in order to be able to pre-compute efficient
mixing matrix lookup tables for each parameter, we impose a hard
uniform prior on each parameter as discussed in
Sect.~\ref{sec:powspec}. Second, we impose informative Gaussian priors
on $\beta$, with parameters informed from the literature; see
\citet{bp13} and \citet{bp14} for further details.

\subsubsection{Diffuse component amplitude sampling}
\label{sec:sigamp}  

Since we sample $\beta$ from a marginal distribution with respect to
$\a$ for the intensity case, we must also sample $P(\a\mid \d,
\omega\setminus\a)$ directly following $\beta$. The relevant data
model for $\a$ is (similar to $\beta$) given by
Eq.~\eqref{eq:beta_data}, but this time interpreted as a function of
$\a$ instead of $\beta$. As applied to CMB estimation, this model was
first introduced into the CMB literature by
\citet{jewell2004}, 
\citet{wandelt2004},
and \citet{eriksen:2004}, and later generalized to
joint CMB power spectrum estimation and astrophysical component
separation by \citet{eriksen2008}. With the uniformized notation
defined above, the same formalism applies both to CMB and diffuse
astrophysical foregrounds, just with different parametric forms for
the mixing matrices, $\M$, signal covariance matrices, $\S$, and
optional priors.

Noting that $\n^{\mathrm{w}}_{\nu}$ represents Gaussian white noise
and $\sum_{\nu}\B_{\nu}\M_{\nu}$ is a deterministic linear operation
given $\omega\setminus\a$, the appropriate sampling equation for $\a$
is yet again given by the multivariate Gaussian sampler in
Eq.~\eqref{eq:multigauss} with a template matrix $\T = \sum_{\nu}\B_{\nu}\M_{\nu}$,
i.e.,
\begin{equation}
    \begin{split}
\biggl(\S^{-1} +
\sum_{\nu}\M^t_{\nu}\B^t_{\nu}&\N_{\nu}^{-1}\B_{\nu}\M_{\nu}\biggr)\,\a
= \\\sum_\nu\M_\nu^t\B_{\nu}^t\N_\nu^{-1}\m_{\nu} + &\S^{-1}\mu + \sum_{\nu}\M_{\nu}^t\B_{\nu}^t\N_{\nu}^{-1/2}\eta_{\nu} +
\S^{-1/2}\eta_{0}.
\label{eq:wiener}
    \end{split}
\end{equation}
Here we have included the signal covariance matrix, $\S=\S(C_{\ell})$,
which is a prior that depends on the angular power spectrum of the
respective component. If no spatial prior is desired, $\S^{-1}$ may
simply be set to zero. 

Equation~\eqref{eq:wiener} arguably represents the single most
challenging step in the entire \BP\ analysis pipeline in terms of
computational complexity. Fortunately, an efficient iterative solver
was recently developed by \citet{seljebotn:2019} for precisely this
equation, and this algorithm forms the computational engine of
\commandertwo\ (see Sect.~\ref{sec:commander}). The main new idea in
that work is the use of a pseudo-inverse preconditioner coupled to a
Conjugate Gradient (CG) solver that easily supports multi-resolution
observations, as required for Eq.~\eqref{eq:wiener}. For specific
details, we refer the interested reader to \citet{seljebotn:2019}.

Computationally speaking, the main complicating factor associated with
Eq.~\eqref{eq:wiener} is the application of an analysis mask. For CMB
likelihood estimation purposes, it is necessary to exclude pixels with
particularly bright astrophysical foregrounds by setting
$\N_{\nu}^{-1}=\tens 0$, in order not to contaminate the resulting CMB
map. Unfortunately, this makes the coefficient matrix on the left-hand
side of Eq.~\eqref{eq:wiener} poorly conditioned, and the resulting CG
search expensive. At the same time, we are also scientifically
interested in the properties of astrophysical foregrounds inside the Galactic
mask, and simply discarding all this useful information is clearly
undesirable.

Rather than directly applying a processing mask, we therefore instead
choose to solve Eq.~\eqref{eq:wiener} twice. First, within the main
Gibbs loop (as defined in Sect.~\ref{sec:gibbschain}) we solve
Eq.~\eqref{eq:wiener} imposing neither a spatial prior on the CMB
component, nor an analysis mask. In this configuration the CG search
converges typically within $\mathcal{O}(10^2)$ iterations, which
corresponds to a computational cost that is smaller than the TOD
processing steps by one order of magnitude \citep{bp03}. The resulting
CMB sky map samples correspond to prior-free, full-sky CMB maps,
similar to those produced by classic component separation algorithms;
see, e.g., \citet{planck2014-a11} and \citet{planck2016-l04}.

However, in order to produce the clean full-sky CMB map and power
spectrum samples that are required for high-resolution CMB likelihood
estimation purposes (see Sect.~\ref{sec:powspec} and \citealp{bp11}),
we additionally solve Eq.~\eqref{eq:wiener} with $\S^{-1}$ and a mask,
but \emph{condition} on all non-CMB parameters. Statistically
speaking, this is equivalent to writing the full joint posterior
distribution in Eq.~\eqref{eq:jointpost} in the form
\begin{equation}
  P(\a^\mathrm{CMB}, \omega\setminus\a^{\mathrm{CMB}}\mid \d) =
  P(\a^\mathrm{CMB}\mid \d, \omega\setminus\a^{\mathrm{CMB}})
  P(\omega\setminus\a^{\mathrm{CMB}}\mid \d),
  \label{eq:cmb_cond}
\end{equation}
and using the first main Gibbs loop to draw samples from the second
factor on the right-hand side, and the second solution of
Eq.~\eqref{eq:wiener} to sample from the first factor.

Formally speaking, we note that this approach is only approximate,
since $C_{\ell}$ should in principle also be conditioned upon in the
second factor in Eq.~\eqref{eq:cmb_cond}. The penalty of not doing so
is slightly more noise in the non-CMB parameters, since the prior-free
CMB sky map sample is less smooth than it is with the prior. However,
the practical benefits gained by separating the TOD processing steps
from the CMB likelihood estimation step more than outweighs a small
increase in statistical uncertainties for several reasons: 1) it
greatly reduces overall computational costs for the joint Gibbs chain;
2) it allows CMB estimation from individual frequency channels or
channel combinations; and 3) it allows rapid exploration of different
analysis masks and/or cosmological models without having to rerun the
costly TOD processing steps. Thus, this split plays the same role in
the \BP\ pipeline as the split between mapmaking and likelihood
estimation does in a traditional CMB analysis pipeline.

We employ a similar method also for low-resolution likelihood analysis,
and re-sample CMB multipoles below $\ell\le64$, while conditioning on
all higher multipole CMB modes and other parameters. In this case, we
do not impose the $C_{\ell}$ prior term, but rather set $\S^{-1}=\tens 0$ as
in the original analysis. This allows us to generate tens of thousands
of low-resolution samples at a greatly reduced computational cost, and
derive a well-converged brute-force low-$\ell$ likelihood from a
relatively limited number of full-scale samples. For further details,
see Sect.~\ref{sec:cmb_params}, \citet{bp11}, and \citet{bp12}.

For two of the astrophysical foregrounds, namely free-free emission
and AME, we use informative priors to stabilize the model
\citep{bp13}. For free-free emission, we adopt the \Planck\ 2015 model
\citep{planck2014-a12} as a spatial template for the prior mean, while
the AME prior is based on the \Planck\ HFI 857\,GHz map, but with a
free scaling factor, under the assumption that the AME surface
brightness correlates strongly with thermal dust emission
\citep{planck2014-a12}. In both cases, the signal covariance matrices
are empirically tuned to allow sufficient variations to statistically
fit the data, while at the same time not introducing too many
unconstrained degrees-of-freedom.\footnote{Note that $\S$ plays a
  fully analoguous role in a multivariate Gaussian prior as the usual
  standard deviation in a univariate Gaussian prior, and can be used
  to adjust the strength of the prior.}

\subsubsection{Compact source sampling}
\label{sec:ptsrc}  

The two previous sections described sampling algorithms for diffuse
components (such as CMB, synchrotron or thermal dust emission) in
terms of their amplitude and SED parameters. These algorithms are
strongly tuned toward global modelling in terms of spherical harmonics
expansions through the use of computationally efficient spherical
harmonics transforms. However, as discussed in
Sect.~\ref{sec:pointsources}, a multitude of compact objects also
scatters the sky, and some of these are extremely bright. Formally
speaking, these may of course also be described in terms of a
spherical harmonics decomposition, since the instrumental beam ensures
that they are indeed bandwidth limited in the observed data. However,
in practice this would require an extremely high bandwidth limit for the
diffuse components, and this is therefore impractical because of the
high associated computational costs.

Instead, we follow \citet{planck2016-l04}, and individually model the
brightest compact sources based on a pre-existing catalog of object
candidates. Each source candidate is mathematically modelled spatially
as a delta function convolved with the instrumental beam evaluated at
the source location, and with a power-law SED given by an amplitude,
$\a^{\mathrm{src}}$, and a spectral index, $\alpha$. For
\Planck\ frequencies, we take into account the full asymmetric beam
profiles as evaluated with \texttt{FEBeCOP} \citep{mitra2010}, while for
non-\Planck\ frequency maps, we adopt azimuthally symmetric beams.

The conditional posterior for the $i$th compact object is given by
subtracting all diffuse components and all other compact objects from
the map-based data model in Eq.~\eqref{eq:beta_data}, such that the
effective residual at frequency $\nu$ reads
\begin{equation}
  \r_{i} = \m_{\nu} - \sum_{c\ne i} \B_{\nu}\M^c_{\nu}\a_c,
\end{equation}
where $c$ runs both over all diffuse components and all compact
objects except the $i$'th source. The likelihood then takes the form
\begin{equation}
  -2\ln P(a_i, \alpha_i\mid \m, \omega\setminus \{a_i, \alpha_i\}) =
  \sum_{\nu}\left(\frac{\m_{\nu} - U_\nu
    a_i\left(\frac{\nu}{\nu_{\mathrm{ptsrc}}}\right)^{\alpha-2}_i \t_{\nu}^i}{\sigma_{\nu}(p)}\right)^2,
\label{eq:lnL_likelihood}
\end{equation}
where $\nu_{\mathrm{ptsrc}}$ is the reference frequency adopted for
the point source component, $\t_\nu^i$ is the spatial (pre-computed)
beam template for the current source, and $U_\nu$ is the unit
conversion factor for frequency $\nu$. (As usual, bandpass integration
is suppressed in the notation for readability, but is of course taken
into account in the actual calculations, as described in
Sect.~\ref{sec:bandpass}.) 

In addition, we impose a Gaussian prior on the spectral index of
$P(\alpha) = N(-0.1,0.3^2)$, motivated by \citet{bennett2012}, and a
positivity prior on the amplitude, $a_i \ge 0$.

The full conditional posterior is sampled using a Metropolis sampler
for $(a_i, \alpha_i)$, running 100 MCMC steps for each source, while
completing 3 full scans through the full source set per full Gibbs
cycle. This step represents a relatively minor computational cost, due
to extensive pre-computation of both effective beam and bandpass
profiles.

\subsubsection{$C_{\ell}$ and cosmological parameter sampling}
\label{sec:powspec}  

The final conditional distribution in the \BP\ Gibbs chain concerns
the angular power spectrum, $C_{\ell}$, of each component, possibly as
parameterized in terms of a smaller number of general parameters. In
the following, we will actually apply this only to the angular CMB
power spectrum, but we note that the formalism applies without changes to
any other statistically isotropic component, for instance the CIB.

Before we start the discussion, we remind the reader that, as
mentioned in Sect.~\ref{sec:sigamp}, we apply three different
sampling steps for the CMB amplitude map:
\begin{enumerate}
\item full-resolution solution of Eq.~\eqref{eq:wiener} with no
  spatial CMB prior, $\S_{\mathrm{CMB}}^{-1} = \tens 0$; the resulting
  samples are primarily used for CMB prior-free component separation
  and deriving unbiased frequency maps, but not directly for
  cosmological parameter estimation;
\item low-resolution solution of Eq.~\eqref{eq:wiener} with no spatial
  CMB prior,\footnote{In practice, we do formally apply a prior also
    in this case, but with a sufficiently large numerical value that
    $\S_{\mathrm{CMB}}^{-1}\approx0$.} $\S_{\mathrm{CMB}}^{-1} = \tens
  0$, but only including multipoles ${\ell \le 64}$, and conditioning
  on all other parameters; typically, 50 low-resolution samples are
  drawn based on each high-resolution sample. These samples form the
  basis for the low-$\ell$ temperature-plus-polarization CMB
  likelihood described below.
\item full-resolution solution of Eq.~\eqref{eq:wiener} with a spatial
  CMB prior, $\S_{\mathrm{CMB}}^{-1} \ne \tens 0$, where $C_{\ell}$ is
  sampled with an inverse Wishart sampler as summarized below. The
  resulting samples form the basis for our high-$\ell$ temperature
  likelihood.
\end{enumerate}
In practice, the first step is run together with the full Gibbs
analysis, including both TOD and component separation steps, while the
other two are performed by re-running the code after the main run has
been completed. From the point of view of CMB estimation alone, the
primary purpose of the main Gibbs run is thus to derive an ensemble of
frequency maps and corresponding astrophysical sky models, that later
can be re-sampled with respect to CMB parameters.

\paragraph{Low-resolution temperature-plus-polarization likelihood}
From step 2 above, we typically have a sample set of
$\mathcal{O}(10^4)$ CMB-only samples, each corresponding to one
possible combination of TOD, foreground and high-$\ell$ CMB
parameters. Clearly, the information contained in this sample set may
be combined into an effective CMB likelihood in many different ways,
each with its own algorithmic advantages and disadvantages. For
instance, they could form the basis of a highly robust cross-spectrum
estimator, by analysing two halves of the data set at a time, and
cross-correlating the resulting CMB map; for a recent example of such
cross-spectrum approach applied to the \Planck\ data, see, e.g.,
\citet{planck2016-l05}.

However, since our main goal of this work is to introduce a
statistically well-motivated end-to-end Bayesian approach, we prefer to stay as
close as possible to the exact Bayesian solution. And, practically
speaking, that corresponds most closely to a Gaussian multivariate
distribution on the form,
\begin{equation}
P(C_{\ell}\mid \hat{\s}_{\mathrm{CMB}}) \propto
\frac{\e^{-\frac{1}{2}\hat{\s}^t_{\mathrm{CMB}}\left(\S(C_{\ell}) +
  \N\right)^{-1}\hat{\s}_{\mathrm{CMB}}}}{\sqrt{|\S(C_{\ell})+\N|}},
\label{eq:cmb_bf}
\end{equation}
where $\hat{\s}_{\mathrm{CMB}}$ represents a CMB-plus-noise map and $\N$ is its
corresponding effective noise covariance map.\footnote{We note that this
  expression does not correspond to the exact Bayesian solution,
  strictly speaking, because the true uncertainty of a given pixel may
  be non-Gaussian due to the presence of both foregrounds and TOD
  corrections. To account for this, cosmological parameters should
  ideally be sampled within the full-resolution Gibbs chain, for
  instance using the algorithms proposed by \citet{racine:2016}; this,
  however, is left for future work, and we adopt a Gaussian
  approximation for now.} Since we at this point have access to a
full ensemble of low-resolution CMB samples that span the full allowed
posterior volume, we may estimate these quantities as 
\begin{align}
  \hat{\s}_{\mathrm{CMB}} &= \left<\s_{\mathrm{CMB}}^i\right>\\
  \N &=
  \left<(\s^i_{\mathrm{CMB}}-\hat{\s}_{\mathrm{CMB}})(\s^i_{\mathrm{CMB}}-\hat{\s}_{\mathrm{CMB}})^t\right>,
  \label{eq:cmb_lowl_input}
\end{align}
where brackets indicate average over the sample set. In the limit of
an infinite number of samples, these quantities will converge to the
Gaussian approximation of the full pixel-based CMB posterior. The
resulting covariance matrices are shown and discussed by \citet{bp11}.

This approach is conceptually very similar to that adopted by both the
\Planck\ LFI DPC \citet{planck2016-l05} and the \WMAP\ science team
\citet{hinshaw2012} for low-$\ell$ likelihood estimation, both of
which rely on brute-force likelihood estimation according to
Eq.~\eqref{eq:cmb_bf}. However, there is one critically important
difference: with our approach, all sources of uncertainty that are
sampled over in the Gibbs chain with $\omega$ are seamlessly
propagated to the CMB likelihood, including gain and bandpass
uncertainties; foreground uncertainties; correlated noise etc. For the
traditional approaches, typically only correlated noise and overall
calibration is accounted for in the covariance matrix.

An important question regarding the practicality of
Eq.~\eqref{eq:cmb_bf} is how many samples are required for
convergence. As discussed by \citet{sellentin2016}, an absolute
minimum criterion for a sampled $n\times n$ covariance matrix simply
to be invertible is that $N_{\mathrm{samp}} > n$. However, this is by
no means sufficient to obtain a \emph{robust} estimate, and, more
typically, numerical experiments indicate that many times this is
required for matrices of moderate size and relatively weak
correlations; the precise value, however, is something that must be
tested on a case-by-case matrix.

In any case, since we have a relatively limited number of samples
available, it is of great interest to compress the relevant
information in $\hat{\s}_{\mathrm{CMB}}$ into as few spatial modes as
possible, while still retaining the lion's share of its full
information content. With this in mind, we note that the main
scientific target for low-$\ell$ likelihood estimation for \Planck\ is
the reionization optical depth, $\tau$. In this case, $\tau$
typically only depends on the first 6 or 8 multipoles, because of the
limited sensitivity of the instrument \citep{planck2016-l05}. As such,
a first natural compression is to retain only modes with $\ell \le 8$,
which corresponds to a total of $3(\ell_{\mathrm{max}}+1)^2\approx
240$ modes. However, many of these modes fall within a typical
analysis mask \citep{bp11}, and therefore carry no statistical weight
in the final answer.

One particularly convenient method of isolating the actually useful
modes is through Karhunen-Lo\`eve compression, as discussed by
\citet{tegmark1997} and \citet{gjerlow2015}. This approach essentially
corresponds to retaining the eigenvectors of $\S+\N$ with the highest
eigenvalues, where $\S$ is evaluated for a typical model of
interest. Adopting the notation of \citet{gjerlow2015}, we organize
the eigenmodes with eigenvalues higher than some user-specified
threshold row-by-row into a projection operator, $\P$, and apply this
to the CMB samples derived above. The compressed data and covariance
matrix then reads
\begin{align}
  \tilde{\s}_{\mathrm{CMB}} &= \P\hat{\s}_{\mathrm{CMB}}\\
  \tilde{\N} &= \P\N\P^t \\
  \tilde{\S} &= \P\S\P^t.
\end{align}
Adopting a multipole threshold of $\ell_{\mathrm{max}}=8$ and a
signal-to-noise threshold of $10^{-6}$ typically leaves around 170
spatial modes in the full data set, for which we that convergence is
typically reached with about 100\,000 fast samples, corresponding to
2000 full samples including all systematic effects; see
\citet{bp12}. The computational cost of a single likelihood evaluation
is also correspondingly reduced because of this compression, and only
takes a few hundredths of a second.

\paragraph{High-resolution Blackwell-Rao estimator}

The above estimator can only be employed at low angular resolution
because of its strong dependence on the size of the covariance
matrix. For high angular resolution analysis, we use another
well-established solution, namely the Blackwell-Rao estimator
\citep{chu2005}, which works very well for high signal-to-noise
data. In practice, we only use this for temperature analysis in the
current paper, since the signal-to-noise ratio for high-$\ell$
polarization is very low with only LFI and \WMAP\ data. However, we
keep the following presentation general, such that it can be used for
both temperature and polarization analysis for other experiments.

To derive the appropriate sampling algorithm for ${P(C_{\ell}\mid \d,
  \omega\setminus C_{\ell})}$ from first principles, we first note
that $P(C_{\ell}\mid \d, \omega\setminus C_{\ell}) = P(C_{\ell}\mid
\a^{\mathrm{CMB}})$; if the true CMB map, $\s_{\mathrm{CMB}}$, is
perfectly known, then no further knowledge regarding the measured data
can possibly provide more useful information about the angular CMB power
spectrum, $C_{\ell}$. Second, as discussed in
Sect.~\ref{sec:cmb}, we assume that the CMB fluctuation field is
isotropic and Gaussian distributed, and its probability distribution
is therefore given by Eq.~\eqref{eq:Pcmb}. Noting that individual
$a_{\ell m}$'s are statistically independent by the assumption of
isotropy, we can write \citep{wandelt2004}
\begin{align}
  P(C_{\ell}\mid \a^{\mathrm{CMB}}) &\propto
  P(\a^{\mathrm{CMB}}\mid C_{\ell})P(C_{\ell}) \\
  &= \prod_{m=-{\ell}}^{\ell} \frac{\e^{-\frac{1}{2}\a_{\ell
        m}^{\dagger} \C_{\ell}^{-1} \a_{\ell m}}}{\sqrt{|\C_{\ell}|}}
  P(\C_{\ell})\\
  &=
  \frac{\e^{-\frac{2\ell+1}{2}\mathrm{tr}(\upsigma_{\ell}\C_{\ell}^{-1})}}{|\C_{\ell}|^{\frac{2\ell+1}{2}}}
  P(\C_{\ell}),
  \label{eq:invwishart}
\end{align}
where $\a_{\ell m} = \{a_{\ell m}^{T}, a_{\ell m}^{E}, a_{\ell
  m}^{B}\}$ and
\begin{align}
  \C_{\ell}
&\equiv
  \left[
    \begin{array}{ccc}
      C_{\ell}^{TT} & C_{\ell}^{TE} & C_{\ell}^{TB} \\
      C_{\ell}^{TE} & C_{\ell}^{EE} & C_{\ell}^{EB} \\
      C_{\ell}^{TB} & C_{\ell}^{EB} & C_{\ell}^{BB} 
    \end{array}
    \right];\\
  \upsigma_{\ell}
&\equiv \frac{1}{2\ell+1}\sum_{m=-\ell}^{\ell}
  \left[
    \begin{array}{ccc}
      (a^T_{\ell m})^* a^T_{\ell m} & (a^T_{\ell m})^* a^E_{\ell m} &
      (a^T_{\ell m})^* a^B_{\ell m} \\
      (a^E_{\ell m})^* a^T_{\ell m} & (a^E_{\ell m})^* a^E_{\ell m} &
      (a^E_{\ell m})^* a^B_{\ell m} \\
      (a^B_{\ell m})^* a^T_{\ell m} & (a^B_{\ell m})^* a^E_{\ell m} &
      (a^B_{\ell m})^* a^B_{\ell m}
    \end{array}
    \right].
\end{align}
We typically adopt uniform priors on $C_{\ell}$ (although for a
discussion of non-uniform priors, see \citealp{larson:2006}), and the
distribution in Eq.~\eqref{eq:invwishart} is then known as the inverse
Wishart distribution, which has a very simple sampling algorithm:
\begin{enumerate}
  \item Draw $2\ell-n$ Gaussian random vectors, $\eta_i$, from the
    empirical covariance matrix $(2\ell+1)\upsigma_{\ell}$, each of
    length $n$, where $n$ is the dimension of $\C_{\ell}$;
  \item Compute the outer product of these vectors, $\uprho_{\ell} =
    \sum_{i=1}^{2\ell-n} \eta_i\eta_i^t$;
  \item Set $\C_{\ell} = \upsigma_\ell/\uprho_{\ell}$.
\end{enumerate}
Note that if $\C$ is block-diagonal, as for instance is the case if
$C_{\ell}^{TB} = C_{\ell}^{EB}=0$, then this algorithm should be
applied separately block-by-block. Also, if binning is desired, for
instance to increase the effective signal-to-noise ratio of a given
power spectrum coefficient, this is most conveniently done in terms of
$D_{\ell} = C_{\ell}\,\ell(\ell+1)/2\pi$; for details, see
\citet{larson:2006}.

\begin{table*}[t]
  \begingroup
  \newdimen\tblskip \tblskip=5pt
  \caption{Computational resources required for end-to-end
    \BP\ processing. All times correspond to CPU hours. All reported times are
    averaged over more than 100 samples, and vary by $\lesssim\,5\,\%$ from sample to
    sample. Reproduced from \citet{bp03}.}
  \label{tab:resources}
  \nointerlineskip
  \vskip -3mm
  \footnotesize
  \setbox\tablebox=\vbox{
    \newdimen\digitwidth
    \setbox0=\hbox{\rm 0}
    \digitwidth=\wd0
    \catcode`*=\active
    \def*{\kern\digitwidth}
    \newdimen\signwidth
    \setbox0=\hbox{-}
    \signwidth=\wd0
    \catcode`!=\active
    \def!{\kern\signwidth}
 \halign{
      \hbox to 7.5cm{#\leaderfil}\tabskip 1em&
      \hfil#\tabskip 1em&
      \hfil#\tabskip 1em&
      \hfil#\tabskip 1em&
      \hfil#\tabskip 2em&
      #\tabskip 0em\hfil\cr
    \noalign{\doubleline}
      \omit\textsc{Item}\hfil&
      \omit\hfil\textsc{30 GHz}\hfil&
      \omit\hfil\textsc{44 GHz}\hfil&
      \omit\hfil\textsc{70 GHz}\hfil&
      \omit\hfil\textsc{Sum}\hfil&
      \omit\hfil\textsc{Reference}\hfil\cr
      \noalign{\vskip 4pt\hrule\vskip 4pt}
      \multispan5\textit{Data volume}\hfil\cr
      \noalign{\vskip 2pt}
      \hskip 10pt Uncompressed TOD volume                 & {\gray 761 GB} &
      {\gray 1\,633 GB} & {\gray 5\,522 GB} & {\gray 7\,915 GB}& \cr
      \hskip 10pt Compressed TOD volume & **86
      GB & *178 GB & **597 GB & ***861 GB& \cr
      \hskip 10pt Non-TOD-related RAM usage &   &  &  & ***659 GB& \cr
      \hskip 10pt {\bf Total RAM requirements} &  &  &  & **{\bf1\,520 GB}& \cr      
      \noalign{\vskip 2pt}      
      \multispan5\textit{Processing time (cost per run)}\hfil\cr
      \hskip 10pt TOD initialization/IO time                    & 3.8\,h & 4.3\,h & 12.5\,h & 20.6\,h& \cr
      \hskip 10pt Other initialization                          &  &  &  &  43.4\,h& \cr
      \hskip 10pt {\bf Total initialization}                          &  &  &  &  {\bf 64.0\,h}& \cr
      \noalign{\vskip 2pt}      
      \multispan5\textit{Gibbs sampling steps (cost per sample)}\hfil\cr
      \hskip 10pt Huffman decompression                            & 1.1\,h& 1.8\,h& 7.1\,h& 10.0\,h& \citet{bp03}\cr
      \hskip 10pt TOD projection ($\P$ operation)               & 0.3\,h& 0.7\,h& 3.1\,h&  4.1\,h& \citet{bp01}\cr
      \hskip 10pt Sidelobe evaluation ($\s_{\mathrm{sl}}$)         & 1.1\,h& 2.1\,h& 6.5\,h&  9.7\,h& \citet{bp08}\cr
      \hskip 10pt Orbital dipole ($\s_{\mathrm{orb}}$)             & 0.5\,h& 1.1\,h& 4.6\,h& 6.2\,h& \citet{bp07}\cr
      \hskip 10pt Gain sampling ($g$)                           & 0.6\,h& 0.7\,h& 4.7\,h& 6.0\,h& \citet{bp07}\cr
      \hskip 10pt 1\,Hz spike sampling ($s_{\mathrm{1hz}}$)      &
      0.2\,h& 0.3\,h& 1.9\,h& 2.4\,h& \citet{bp01}\cr      
      \hskip 10pt Correlated noise sampling ($\n_{\mathrm{corr}}$) & 1.7\,h& 3.6\,h& 24.8\,h& 30.1\,h& \citet{bp06}\cr
      \hskip 10pt Correlated noise PSD sampling ($\xi_{\mathrm{n}}$) & 3.3\,h& 4.0\,h& 1.1\,h& 8.4\,h& \citet{bp06}\cr
      \hskip 10pt TOD binning ($\P^t$ operation)                &
      0.2\,h& 0.5\,h& 4.1\,h& 4.8\,h& \citet{bp10}\cr
      \hskip 10pt Sum of other TOD processing                   & 1.3\,h& 2.5\,h& 10.9\,h& 14.7\,h& \citet{bp03}\cr
      \hskip 10pt {\bf TOD processing cost per sample}          & {\bf
        10.4\,h}& {\bf 17.4\,h}& {\bf 69.1\,h}&  {\bf 96.9\,h}& \cr
      \noalign{\vskip 2pt}
      \hskip 10pt Amplitude sampling, $P(\a\mid \d, \omega\setminus\a)$  &   &  &  & 23.9\,h& \citet{bp13}\cr
      \hskip 10pt Spectral index sampling, $P(\beta\mid \d, \omega\setminus\beta)$  &   &  &  & 40.3\,h& \citet{bp14}\cr
      \hskip 10pt Other steps                                   &      &  &  &  0.6\,h& \citet{bp01}\cr
      \noalign{\vskip 2pt}
      \hskip 10pt {\bf Total cost per sample}                   &   &  &  &  {\bf 163.9\,h}& \cr
      \noalign{\vskip 4pt\hrule\vskip 5pt} } }
  \endPlancktablewide \endgroup
\end{table*}


The above algorithm describes a self-consistent Gibbs-based approach to CMB
power spectrum sampling, as originally suggested by
\citet{wandelt2004}. The product from this procedure is a set of joint
samples $(\s_{\mathrm{CMB}}, C_{\ell})_i$. However, the algorithm does
not specify how to constrain cosmological parameters from these
samples. Indeed, many different approaches may be adopted for this
purpose, each making different assumptions and choices with regard to
computational cost and robustness to systematic errors. Some
approaches presented in the literature include
\begin{itemize}
\item \emph{the Blackwell-Rao estimator} \citep{chu2005}: Direct
  averaging over $\sigma_{\ell}$ samples given the analytic smoothing
  kernel in Eq.~\eqref{eq:invwishart}. Exact, but converges slowly in
  low signal-to-noise regime. Used by \WMAP\ low-$\ell$ $TT$
  likelihood \citep{hinshaw2012}.
\item \emph{the Gaussianized Blackwell-Rao estimator}
  \citep{rudjord:2009}: Multivariate Gaussian approximation to the
  above, following a Gaussian change-of-variable defined by univariate
  marginal distribution. Converges much faster than direct
  Blackwell-Rao estimator, and is highly accurate for typical
  masks. Used by \Planck\ low-$\ell$ $TT$ likelihood
  \citep[e.g.,][]{planck2016-l05}.
\item \emph{joint Metropolis-Hastings sampling} of $\{\a_{\ell m},
  \C_{\ell}\}$ \citep{jewell:2009,racine:2016}: Efficient in both low
  and high signal-to-noise regimes; may be applied to both $C_{\ell}$
  and cosmological parameter estimation. 
\end{itemize}
The first two of these methods define a CMB power spectrum likelihood
function, $\mathcal{L}(\C_{\ell})$, which then must be coupled to a
cosmological parameter estimation code. We employ the widely employed
\texttt{CosmoMC} \citep{cosmomc} code for this purpose, as detailed in
\citet{bp12}. In contrast, when applied to cosmological parameter
estimation, the third method requires a means to convert between
cosmological parameters and angular power spectra, such as \texttt{CAMB}
\citep{Lewis:1999bs}. In this paper, we adopt the
Gaussianized Blackwell-Rao estimator as our default solution, and
leave the full integrated MCMC sampling approach for future work.

\subsection{Computational requirements and optimization}
\label{sec:compcost}

The end-to-end algorithm summarized in the last few sections
represents a significant computational challenge, both in terms of
fundamental hardware requirements and in terms of software
optimization. In this section we briefly review some critical
computational features implemented in the current code, while in-depth
presentations are provided by \citet{bp03} and \citet{bp05}. In
addition, we highly recommend the interested reader to consult the
source code.\footnote{The \BP\ software is available under a GNU
  Public Library (GPL) open-source license at
  \url{https://github.com/Cosmoglobe/Commander}.} At the same time, we
emphasize that these codes are most definitely works in progress, and
still undergo rapid development. Nearly every single component and
function have room for further improvement and optimization. However,
it is our hope and intention that by providing all codes to the
general community under an open-source license, new collaborations,
efforts and ideas will emerge, and this will leading to more mature,
efficient and generally applicable code.

With these caveats in mind, Table~\ref{tab:resources} summarize the
overall computational cost of the current implementation, both in
terms of initialization and cost per sample. These benchmarks were
obtained by running the pipeline on a single compute node with 128
AMD EPYC 7H12 2.6\,GHz cores and 2\,TB of RAM. All time
related costs are provided in units of wall-time, and must therefore
be multiplied with 128 to convert to CPU time.

Overall, the computational complexity of the \BP\ Gibbs sampler is
determined by three fundamentally different types of
operations. First, the low-level analysis is dominated by TOD memory
management. Second, the high-level amplitude sampling step is
dominated by spherical harmonic transforms. Third, the spectral index
sampling step is dominated by map-based operations, typically either
spherical harmonic transforms or $\chi^2$ evaluations. Efficient
parallelization of each of these three types of operations is
therefore the critical design driver for the current implementation. We
now briefly review how the \BP\ pipeline optimizes each of these
aspects, and refer the interested reader to \citet{bp03} for further
details.

\subsubsection{Low-level optimization}

Starting with the low-level TOD-oriented operations, we first note in
Table~\ref{tab:resources} that the full data volume of four years of
\Planck\ LFI observations is 8\,TB. This number includes all science
and housekeeping data. A single read of the full data set from
spinning disks on a typical intermediate-sized high-performance
computing (HPC) cluster therefore requires a few hours of wall time,
assuming $\mathcal{O}(1\,\mathrm{GB}\,\mathrm{s^{-1}})$ read
speed. While acceptable as a one-time initialization cost, integrating
such expenses into the Gibbs loop clearly leads to impractical run
times. A first requirement for efficient end-to-end TOD processing is
thus that the entire data set may be stored in RAM. Likewise, noting
that the memory bus from the RAM to the CPU is relatively slow
compared to CPU operations, a corollary requirement is that the
overall memory footprint should be aggressively minimized.

With these observations in mind, we first choose to read only those
parts of the data that are strictly required for the analysis in
question; all unnecessary housekeeping data are omitted. For each
\Planck\ LFI radiometer the only retained quantities therefore include
1) differenced detector voltages, $\d_t$ (one float per sample); 2)
pointing, $\P_t$ (three double precision values per sample); and 3)
flags, $f_t$ (one integer per sample). Nominally, a total of 32
bytes/sample/radiometer are required to store the TOD information.

However, as detailed by \citet{bp03}, because the pointing and flags
are both very smooth functions of time, they lend themselves to highly
efficient compression. We exploit this by transforming and
discretizing each relevant quantity into integers; taking the
difference between consecutive samples to minimize their dynamic
range; and finally Huffman compressing \citep{huffman:1952} the
resulting time streams, i.e., we assign bit patterns of variable
lengths to each integer according to their relative frequency. The
average number of bits per sample is thus reduced by a factor of
5--6. These compressed TOD arrays are then stored in memory
PID-by-PID, and only decompressed when needed. The total data volume
is in this way reduced from 8\,TB to 861~GB, which fits into the RAM
of a single modern compute node. The decompression cost accounts for
about 5\,\% of the total analysis wall time, which we consider well
worth the memory savings. However, as discussed by \citet{bp03}, this
compression does have notable implications in terms of the overall
Gibbs sampling structure, as the full decompressed TOD set can never
be stored in memory at once, nor is it possible to store multiple
copies of the TOD. Accordingly, careful relative ordering of the
various Gibbs sampling steps is necessary. In practice, four full
scans are made through the entire TOD within each Gibbs iteration,
where each scan corresponds to sampling one global TOD-related
parameter, namely three gain components (see
Sect.~\ref{sec:gain}) and the bandpass correction parameter; none of
these can be sampled simultaneously without breaking the Gibbs chain.

Next, the low-level parallelization scheme for TOD processing is
organized according to PIDs, such that each computing core processes a
distinct subset of PIDs. Load balancing is achieved by first measuring
the effective computing time for each PID, and then distributing them
according to cost in a round-robin manner among the various computing
cores.

Inspecting the costs of individual steps in Table~\ref{tab:resources},
we see that the dominant TOD operation is associated with sampling
$\n_{\mathrm{corr}}$, which makes intuitive sense: While most
operations scale linearly in the number of samples,
$\mathcal{O}(N_{\mathrm{tod}})$, the correlated noise step requires
two Fourier transforms, and therefore scales as
$\mathcal{O}(N_{\mathrm{tod}}\log N_{\mathrm{tod}})$. To optimize this
step, we first of all employ the \texttt{FFTW} library \citep{FFTW05} for all
FFT operations. Second, we note that the speed required for a single
FFT transform depends sensitively and highly non-linearly on
$N_{\mathrm{tod}}$. Values of $N_{\mathrm{tod}}$ that happen to
factorize into particularly favorable combinations of primes may
happen to be, say, three to five times faster than neighboring
values. We exploit this by first measuring the time required per FFT
for every length between 1 and $10^6$, and construct a table of
optimal lengths, with at least one value per 100th sample. At read
time, we then truncate the length of each PID until it equals the
closest lower optimal length. As such, we lose on average one second of
data per PID, corresponding to about 0.03\,\% of the total data
volume, while gaining a factor of three or more in overall TOD
processing time.

\begin{figure*}[t]
  \center
  \includegraphics[width=\linewidth]{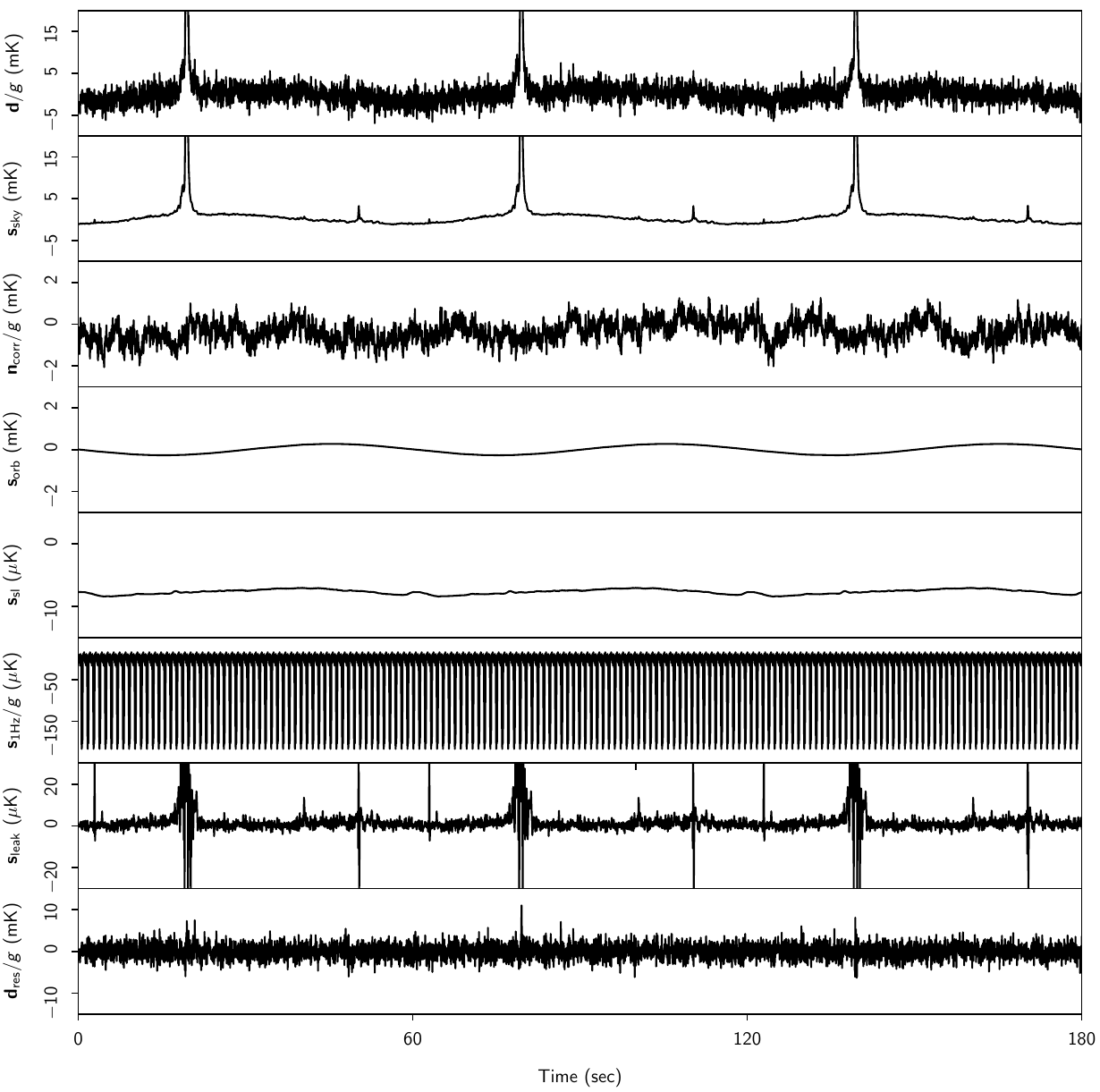}
  \caption{Time-ordered data segment for the 30\,GHz LFI 27M
    radiometer. From top to bottom, the panels show 1) raw calibrated
    TOD, $\d/g$; 2) sky signal, $\s_{\mathrm{sky}}$; 3) calibrated
    correlated noise, $\n_{\mathrm{corr}}/g$; 4) orbital CMB dipole
    signal, $\s_{\mathrm{orb}}$; 5) sidelobe correction,
    $\s_{\mathrm{sl}}$; 6) electronic 1\,Hz spike correction,
    $\s_{\mathrm{1Hz}}$; 7) leakage mismatch correction,
    $\s_{\mathrm{leak}}$; and 8) residual TOD, $\d_{\mathrm{res}} =
    (\d-\n_{\mathrm{corr}} - \s_{\mathrm{1Hz}})/g - \s_{\mathrm{sky}}
    - \s_{\mathrm{orb}} - \s_{\mathrm{leak}} - \s_{\mathrm{sl}}$. Note
    that the vertical range vary significantly from panel to
    panel. Reproduced from \citet{bp10}.}
  \label{fig:todplot}
\end{figure*}

\begin{figure}[t]
  \center
  \includegraphics[width=\linewidth]{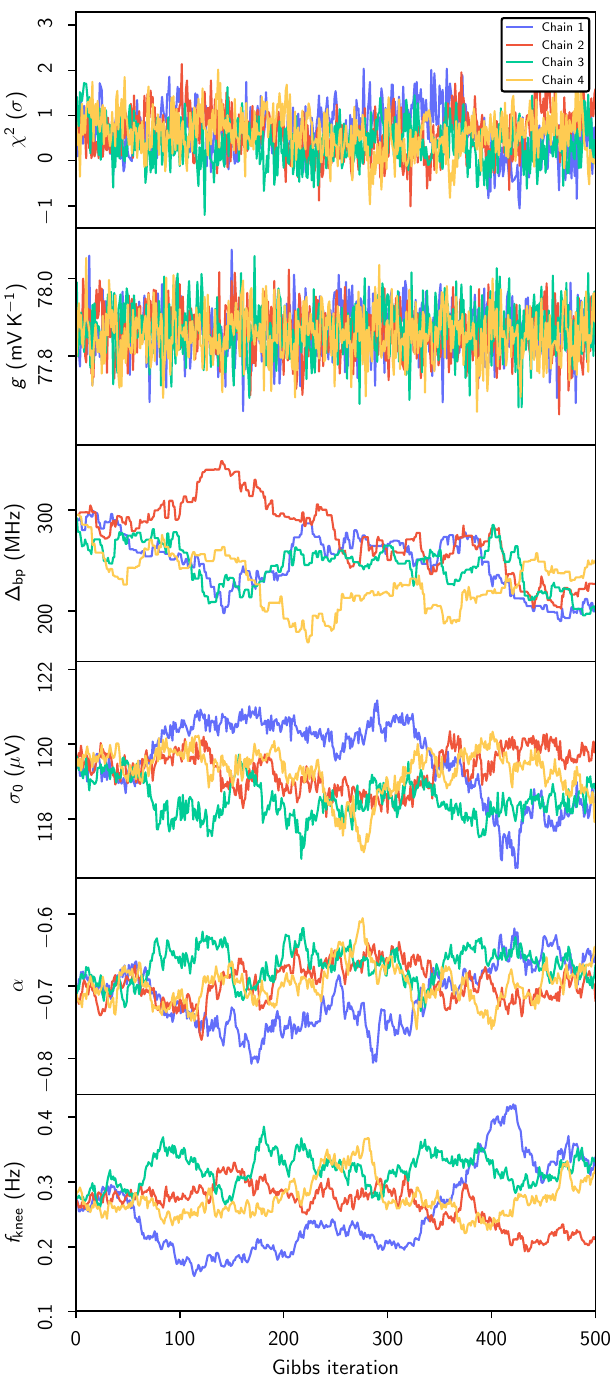}
  \caption{Example of TOD parameters Gibbs chain for the 30\,GHz LFI 27M
    radiometer. From top to bottom, the panels show normalized reduced
    $\chi^2$ for a single Pointing Period; gain for the same PID in
    units of mV\,K$^{-1}$; bandpass correction in MHz; white noise
    level, $\sigma_0$; correlated noise slope, $\alpha$; and
    correlated noise knee frequency, $f_{\mathrm{knee}}$. The four
    different colored curves correspond to independent Gibbs
    chains.}\label{fig:traceplot}
\end{figure}

After the FFT-based operations, the dominant TOD operations are the
sidelobe and orbital dipole evaluations, as well as the pointing
projections operators, $\P$ and $\P^t$. Here it is worth noting that
the TOD analysis is currently memory-bus limited. That is, the cost is
associated simply with transferring data from RAM into the CPU. As
such, the specific algorithmic details of each step are largely
irrelevant, and the important factor is simply the total data
volume. To improve the performance of these steps, the best approach
would be to run across multiple nodes, which thereby increase the
number of memory buses available. On the other hand, this also leads
to lower performance for the CPU dominated operations, and most
notably the spherical harmonics transforms. A future optimal solution
should implement a better tuned parallelization strategy where SHTs
are parallelized within nodes, while TOD operations are parallelized
across nodes; this is left for future development.

Next, the two TOD projection operators warrant a few comments. First,
we recall that $\P$ converts a map into a time stream. This represents
a computational challenge in itself, because each core then needs
access to all pixels in the map. However, actually storing the full
map per core would require substantial amounts of memory. To solve
this, we exploit a MPI-3 shared memory feature, and only store one
copy of the map per compute node, rather than one per core. However,
we do observe that the memory access speed associated with these
shared-memory arrays is typically five times slower than for local
arrays, and further optimizations are therefore possible.

In contrast, the $\P^t$ operation co-adds samples in a time-stream
into a map. In terms of practical code, this is a more complex
operation than $\P$, since all cores need to update the values stored
in each sky map pixel, not only read them. This can easily lead to
race conditions in which different cores simulatenously write to the
same parts of memory, resulting in corrupt data, and a direct shared
array approach is therefore impractical. At the same time, allocating
a full sky map per core is not an option due to the same memory
constraints discussed above. As a compromise, we instead first scan
the full pointing stored by each core, and accumulate a list of all
locally observed pixels. Due to the sparse \Planck\ scanning strategy,
this typically amounts to only 5--10\,\% of all pixels for each
core. Allocating and maintaining a sub-map of this limited size is
acceptable in terms of total memory footprint. Co-addition over cores
is then achieved using a combination of shared arrays within each
computing node, and a single \texttt{MPI\_ALLREDUCE} operation between
nodes. Clearly, further optimization is very likely possible also with
respect to this operation.

\subsubsection{High-level parallelization and optimization}

Next, we consider optimization of the high-level routines, and in
particular of the amplitude and spectral index sampling steps. These
are largely overlapping in terms of essential low-level routines, and
so we will also discuss them jointly.

The single most important computational routine involved in these
operations is the spherical harmonics transform, needed both for
solving the Wiener filter defined by Eq.~\eqref{eq:wiener} and for
smoothing maps to a common angular resolution as required for
Eq.~\eqref{eq:beta_marg}. Indeed, the importance of this operation is so
critically important that we base our entire map parallelization
strategy of our codes around it. With this in mind, we adopt the
\texttt{libsharp2} \citep{reinecke2013} spherical harmonics library
for all harmonic transforms, which is the most efficient library for
this purpose available today. This library is based on a deep
parallelization level in both pixel and harmonic space, distributing
both constant-latitude rings and constant-$m$ harmonics across
different cores. We adopt these parallelization conventions without
modification. 

The second most important operation involved in these operations is
multiplication with the mixing matrix, $\M(\beta;
\Delta_{\mathrm{bp}})$. As described in Sect.~\ref{sec:bandpass}, this
expression involves integration of an ideal parametric SED with the
bandpass of each instrumental detector. It also varies from
pixel-to-pixel, depending on the local properties of the spectral
parameters, $\beta$. For this reason, we pre-compute the full mixing
matrix prior to each full amplitude sampling step,
pixel-by-pixel. Taking advantage of the \texttt{libsharp} parallelization
scheme, which distributes rings across all available cores, the memory
requirements for this is fairly limited. Furthermore, employing the
spline-based library discussed in Sect.~\ref{sec:bandpass}, the actual
evaluation of this matrix only carries a cost equal to a polynomial
evaluation per pixel. However, it is important to note that actually
changing the bandpass correction parameters, $\Delta_{\mathrm{bp}}$, requires
a full re-evaluation of the underlying splines, as well as all
higher-level mixing matrices, and this particular operation is
therefore very computationally intensive. As a result, it is done as
infrequently as possible.

\begin{figure}[t]
  \center
  \includegraphics[width=\linewidth]{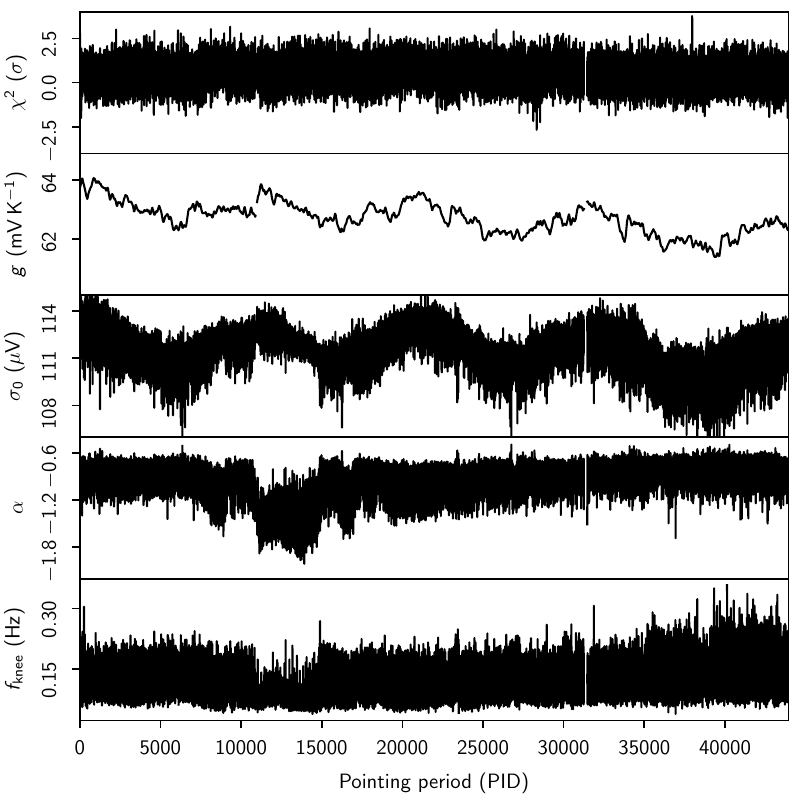}
  \caption{A single Gibbs sample for the 30\,GHz LFI 28M radiometer as a
    function of PID. From top to bottom the panels show 1)
    normalized reduced $\chi^2$; 2) gain, $g$; 3) white noise level,
    $\sigma_0$; 4) correlated noise slope, $\alpha$; and 5) correlated
    noise knee frequency, $f_{\mathrm{knee}}$. }
  \label{fig:pidpar}
\end{figure}

Finally, as described above, many of the various sampling steps are
carried out with a standard Metropolis sampler. Although
conceptionally and implementationally straightforward, this sampler
does have the drawback of requiring specific tuning of the step size
to be efficient. For most of these samplers, we therefore typically
run a short tuning chain during the first iteration, if the
computational cost of the sampler is limited (which, for instance, is
the case for the point source sampler), or insert a pre-calculated
proposal matrix into the run through a parameter file (which, for
instance, is the case for the bandpass correction sampler). Such
tuning is essential to achieve acceptable mixing for the overall
chain.

\section{Results}
\label{sec:results}

We are now finally ready to present the main results resulting from
applying the algorithms summarized in
Sects.~\ref{sec:model}--\ref{sec:posterior} to the data combination
described in Sect.~\ref{sec:data}. For the analysis shown here, we
have produced a total of four independent Monte Carlo Markov chains of
samples drawn from the posterior distribution $P(\omega\mid\d)$, as
described in Sect.~\ref{sec:gibbschain}. Each chain has 1000 samples,
and we conservatively discard the first 200 samples for burn-in. We
thus retain a total of 3200 accepted samples for final
processing. With a computational cost of 1.3 wall hours/sample
\citep{bp03}, this set took about three months of continuous run time
to produce on two nodes, for a total computational cost of 670\,000
CPU hours. Although not directly comparable, it is still interesting
to note that the production of the \Planck\ FFP8 simulation set
required a total of 25 million CPU hours, and the cost of constructing
only a single component of a single Monte Carlo realization of the
70\,GHz channel cost 9360\,CPU-hours \citep{planck2014-a14}. The full
analysis shown in the following thus carries a total computational
cost that is equivalent to about 70 \Planck\ FFP8 70\,GHz
simulations. This clearly demonstrates the computational feasibility
of the Bayesian end-to-end approach, and the algorithms shown here do
not require the use of a massive super-computer center to be
useful. At the same time, it is also clear that future work should
concentrate on increasing the concurrency of the current
implementation through better parallelization schemes, such that the
wall time can be reduced to hours or days, as opposed to months, when
more resources are available.

\begin{figure}[t]
  \center
  \includegraphics[width=\linewidth]{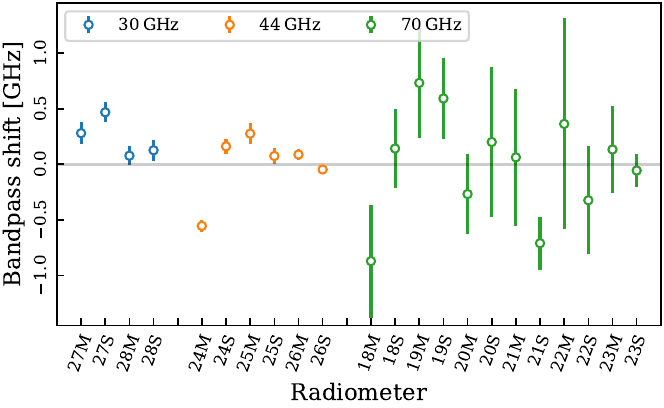}
  \caption{Estimated bandpass corrections for each LFI radiometer. Error bars indicate $\pm3\sigma$ uncertainties. Note these parameters are constrained to have vanishing mean within each frequency, and are as such strongly correlated. Reproduced from \citet{bp09}. }\label{fig:bpshift}
\end{figure}

\subsection{Instrumental parameters}
\label{sec:tod_params}

We start our review by inspecting the data and model at the lowest
level, and for this purpose we select the 30\,GHz channel as a
reference test case, for which the foreground contamination is the
largest, and therefore the calibration challenge the hardest. The top
panel of Fig.~\ref{fig:todplot} (reproduced from \citealp{bp10}) shows
a 3~minute chunk of the 30\,GHz LFI 27M TOD, in which the only
pre-processing steps are differencing with the 4\,K load signal and
ADC corrections (see Sect.~\ref{sec:data} for details).

The top panel shows the raw measurements, which are visually dominated
by the CMB dipole, as seen by the slow sinusoidal oscillations; the
Galactic plane, as seen by the sharp spikes; and instrumental
noise. The second panel shows the estimated sky signal for one random
sample; here we see small-amplitude perturbations in addition to the
large dipole and galactic contributions, and these are dominated by
the CMB temperature fluctuations, which are the main scientific target
of the entire analysis. Rows 3--7 show various corrections arising
from correlated noise, the orbital CMB dipole, sidelobes, the 1\,Hz
spike signal, and bandpass and beam leakage. The last panel shows the
residual after subtracting all the above terms from the raw data,
and this highlights anything that is not explicitly captured by the
parametric model; overall, this is largely consistent with white
noise, except for a few spikes near the Galactic plane crossings,
which are masked both in low-level and high-level processing. For
further discussion of this plot, see \citet{bp10}.

Figure~\ref{fig:todplot} represents one single Gibbs sample in the
full chain. In contrast, Fig.~\ref{fig:traceplot} shows samples from
all four Gibbs chains for the instrumental parameters for one PID, but
this time plotted as a function of Gibbs iteration. For perfect Markov
chain mixing, these should all scatter around a well-defined mean
value with a short correlation length.

The top panel shows the normalized reduced $\chi^2$ as defined by
\begin{equation}
\chi^2 \equiv \frac{\sum_{t=1}^{N_{\mathrm{tod}}} \left(\frac{d_t - s_t^{\mathrm{tot}}}{\sigma_0}
  \right)^2 - N_{\mathrm{tod}}}{\sqrt{2N_{\mathrm{tod}}}}.
\label{eq:redchisq}
\end{equation}
Recalling that the $\chi^2$ distribution with $n$ degrees of freedom
converges towards a Gaussian with mean equal to $n$ and variance equal
to $2n$, this quantity should be approximately distributed as $N(0,1)$
for ideal data, with deviations measured in units of $\sigma$. We
adopt this $\chi^2$ as a convenient goodness-of-fit measure. We see
that the mean value is $\chi^2 \approx 0.5\sigma$, which indicates
good a good fit overall. 

\begin{figure*}[p]
  \center
  \includegraphics[width=0.265\linewidth]{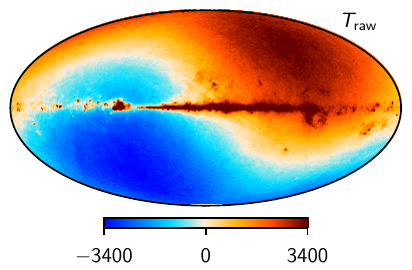}
  \includegraphics[width=0.265\linewidth]{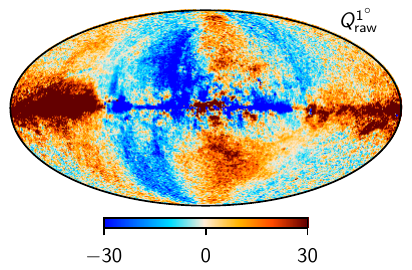}
  \includegraphics[width=0.265\linewidth]{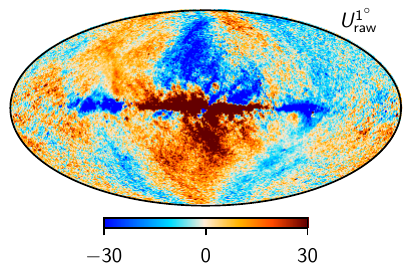}\\\vspace*{3mm}
  \includegraphics[width=0.265\linewidth]{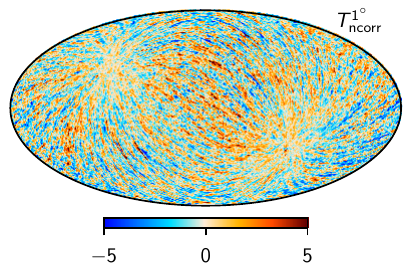}
  \includegraphics[width=0.265\linewidth]{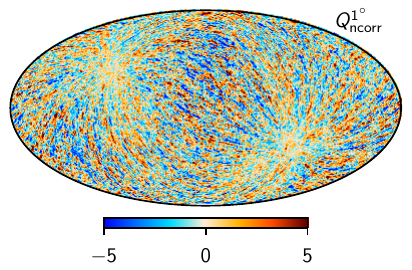}
  \includegraphics[width=0.265\linewidth]{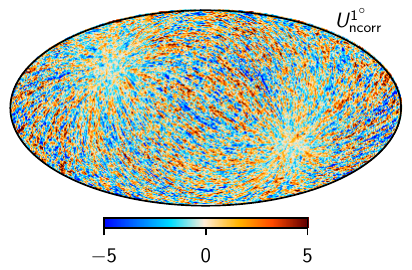}\\
  \includegraphics[width=0.265\linewidth]{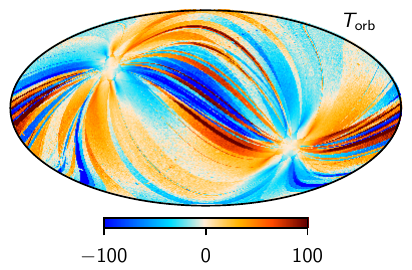}
  \includegraphics[width=0.265\linewidth]{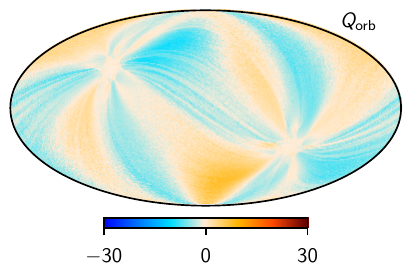}
  \includegraphics[width=0.265\linewidth]{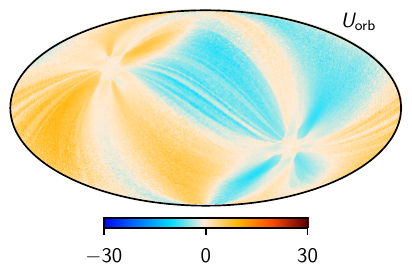}\\
  \includegraphics[width=0.265\linewidth]{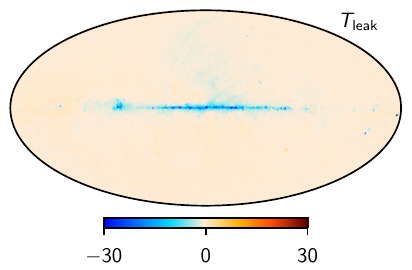}
  \includegraphics[width=0.265\linewidth]{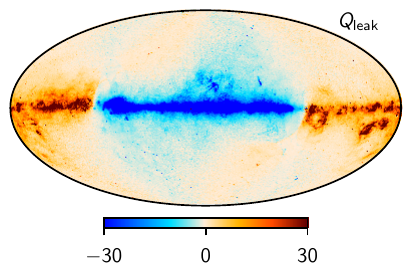}
  \includegraphics[width=0.265\linewidth]{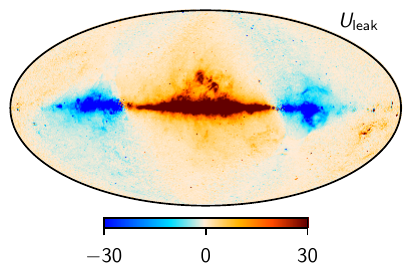}\\
  \includegraphics[width=0.265\linewidth]{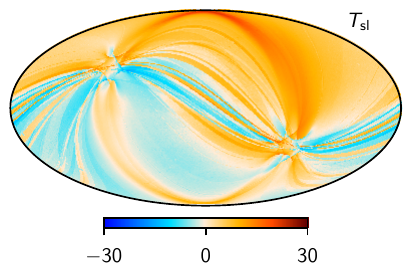}
  \includegraphics[width=0.265\linewidth]{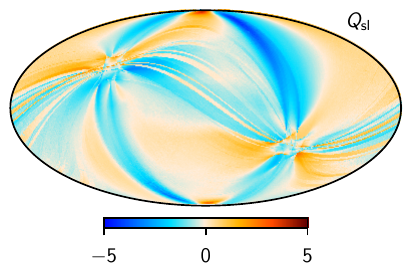}
  \includegraphics[width=0.265\linewidth]{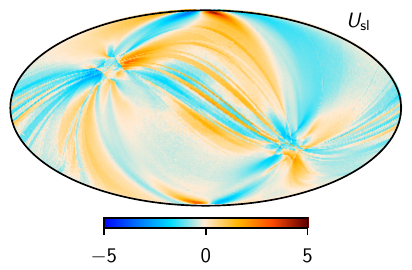}\\
  \includegraphics[width=0.265\linewidth]{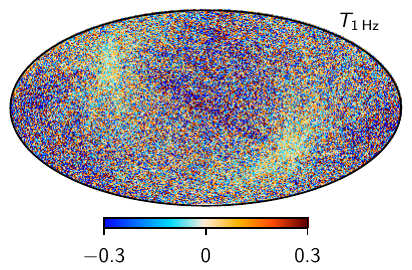}
  \includegraphics[width=0.265\linewidth]{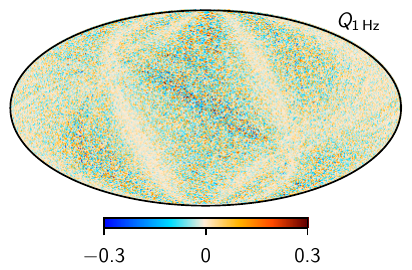}
  \includegraphics[width=0.265\linewidth]{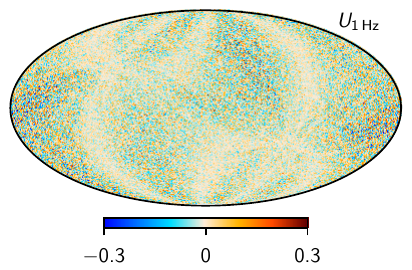}\\
  \includegraphics[width=0.265\linewidth]{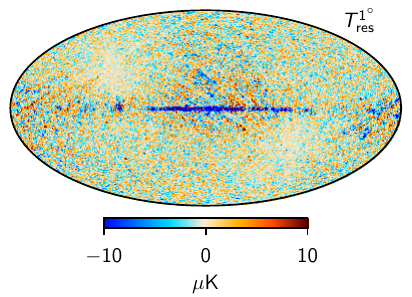}
  \includegraphics[width=0.265\linewidth]{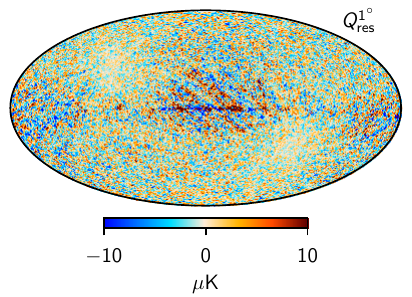}
  \includegraphics[width=0.265\linewidth]{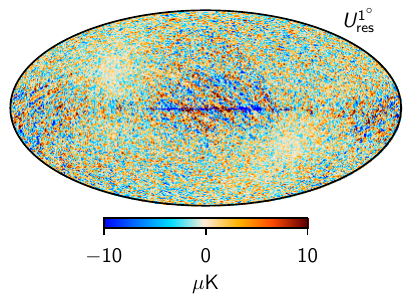}
  \caption{Comparison between TOD corrections for the 30\,GHz channel
    for a single Gibbs sample, projected into sky maps. Columns show
    Stokes $T$, $Q$, and $U$ parameters. Rows show, from top to
    bottom, 1) raw TOD; 2) correlated noise; 3) the orbital dipole; 4)
    bandpass and beam mismatch leakage; 5) sidelobe corrections; and
    6) 1\,Hz electronic spike correction. The bottom row shows the
    shows the residual obtained when binning the sky and systematics
    subtracted TOD into a sky map. Note that some components have been
    smoothed to an angular resolution of $1^{\circ}$ FWHM.  }\label{fig:corrmaps}
\end{figure*}

The second panel shows the gain $g$ for the same PID. In this case,
the Markov correlation length appears very short. Considering that we
have 3200 full Gibbs samples available, this implies that the number
of independent gain samples per PID is quite high, and their histogram
provides a useful estimate of the true underlying distribution; in
this case the marginal posterior may be summarized as
$g=77.85\pm0.02\,\textrm{mV}\,\mathrm{K}^{-1}$. For further discussion
of the gain posteriors, see \citet{bp07}.

The third panel shows the bandpass shift, $\Dbp$, for the 30\,GHz LFI
27M radiometer. As already noted in Sect.~\ref{sec:bandpass}, this
parameter is the single most difficult quantity to estimate in the
entire framework, because of the highly non-linear, non-Gaussian and
global nature of its impact; virtually \emph{all} stochastic variables
in the entire model depend on the instrumental bandpass in one form or
another, and changes in this parameter therefore take a substantial
amount of time to propagate throughout the model. Furthermore, the
sampling algorithm used for this parameter is a basic Metropolis
sampler, simply because of a lack of better alternatives. The result
is a long correlation length of about 100 samples, resulting in
perhaps as few as 32 uncorrelated samples in the full sample
set. Still, even with this crude sampler, we do see that the four
chains mix reasonably well, and it is possible to establish a useful
estimate for the marginal posterior, which in this case may be
summarized as $\Dbp = 240\pm30\,\textrm{MHz}$. However, in this case
the sampling uncertainty accounts for a large fraction of the error
bar. For further discussion of the bandpass posteriors, see
\citet{bp09}.

The three last panels show the three noise PSD parameters, $\sigma_0$,
$\alpha$ and $f_{\mathrm{knee}}$, for the same radiometer. These also
show long correlation lengths. However, as discussed by
\citet{bp04,bp06}, this long correlation length is due to internal
degeneracies among the correlated noise parameters, $\xi_n$, and since
all other parameters, such as the gain or CMB component, only care
about the total noise power spectral density, $P_n(f)$, and not the
individual $\xi_n$ parameters, this poor mixing does not represent a
significant limitation for the analysis; see Fig.~15 in
\citet{bp04}. For further discussion of the noise posteriors, see
\citet{bp06}.

\begin{figure}[t]
  \center
  \includegraphics[width=\linewidth]{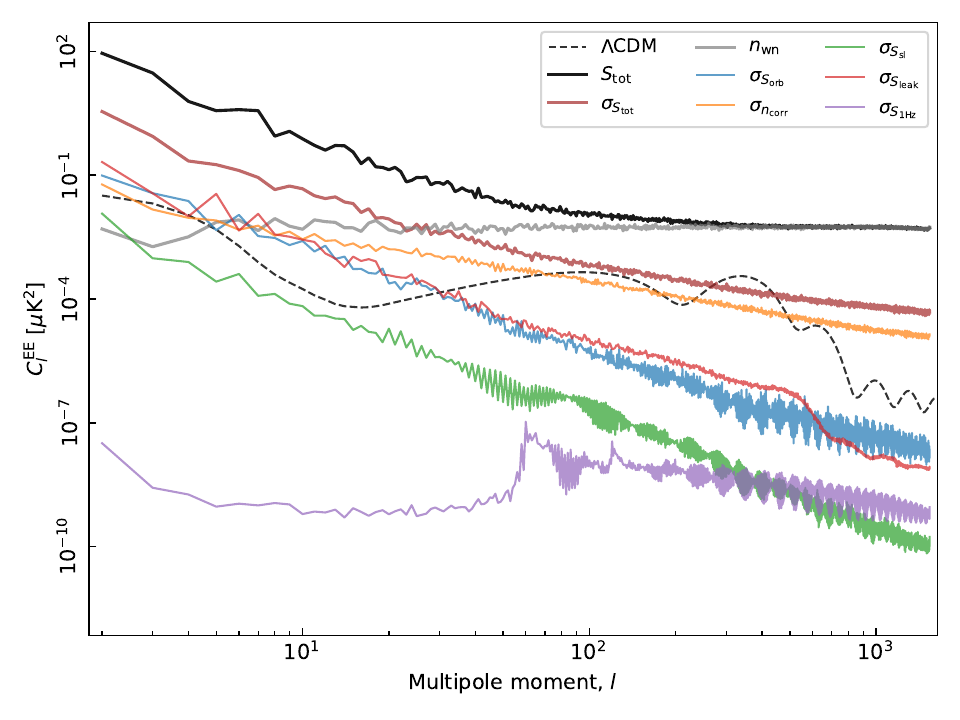}
  \caption{$EE$ pseudo-spectrum standard deviation for each
    instrumental systematic correction at 30\,GHz shown (\emph{thin
      colored lines}). For comparison, the thick black line shows the
    spectrum for the full co-added frequency map; the thick red line
    shows the standard deviation of the same (i.e., the full
    systematic uncertainty); the gray line shows white noise; and the
    dashed black line shows the best-fit \Planck\ 2018 $\Lambda$CDM
    power spectrum convolved with the instrument beam. All spectra
    have been derived outside the CMB confidence mask \citep{bp13}
    using the HEALPix \texttt{anafast} utility, correcting only for
    sky fraction and not for mask mode coupling. Reproduced from
    \citep{bp10}. \label{fig:corrmap_powspec_stddev} }
\end{figure}

\begin{figure*}{t}
  \center
  \includegraphics[width=0.33\linewidth]{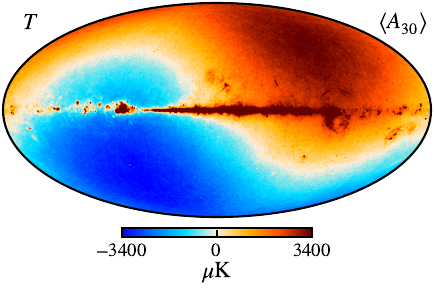}
  \includegraphics[width=0.33\linewidth]{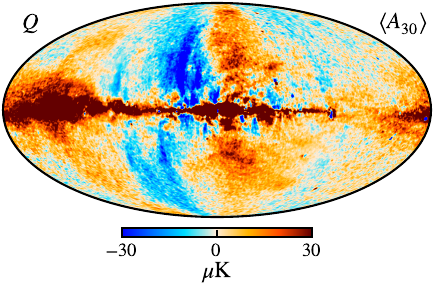}
  \includegraphics[width=0.33\linewidth]{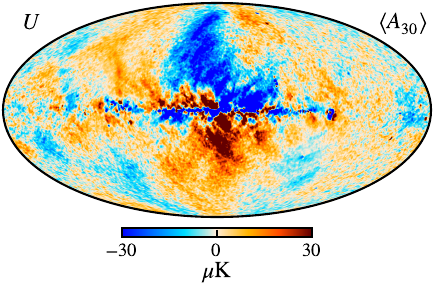}\\
  \includegraphics[width=0.33\linewidth]{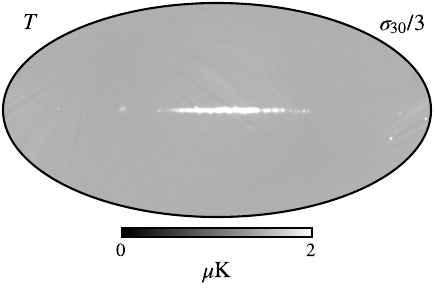}
  \includegraphics[width=0.33\linewidth]{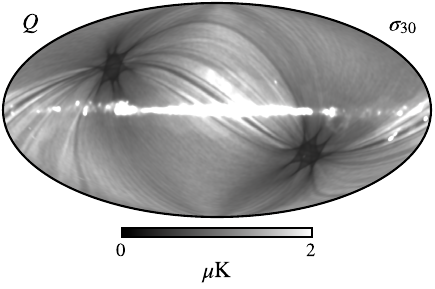}
  \includegraphics[width=0.33\linewidth]{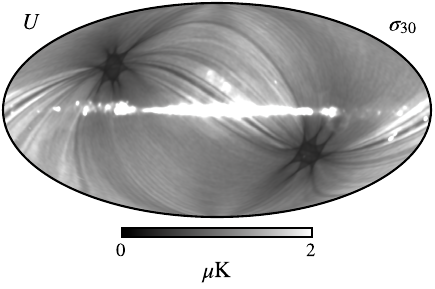}\\
  \includegraphics[width=0.33\linewidth]{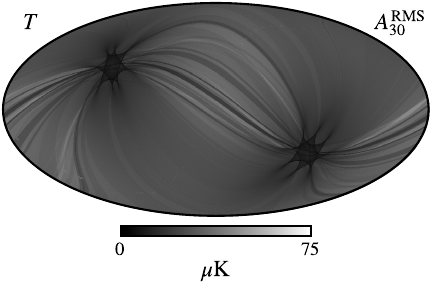}
  \includegraphics[width=0.33\linewidth]{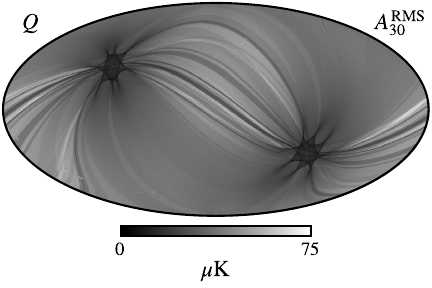}
  \includegraphics[width=0.33\linewidth]{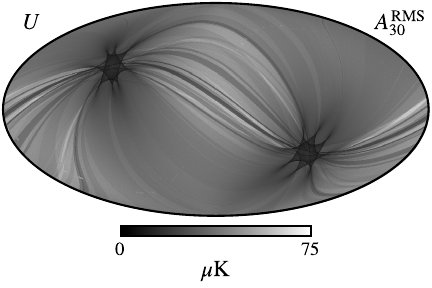}\\
  \includegraphics[width=0.33\linewidth]{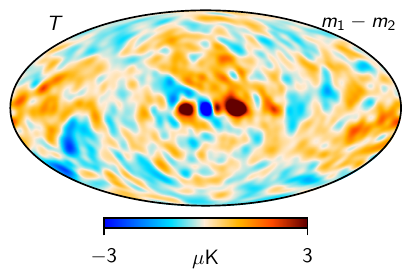}
  \includegraphics[width=0.33\linewidth]{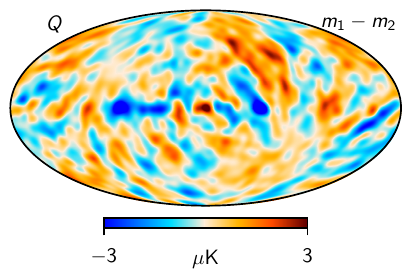}
  \includegraphics[width=0.33\linewidth]{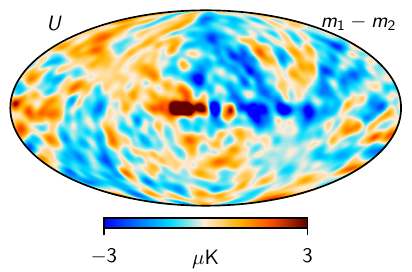}  
  \caption{Posterior summary maps for the LFI 30\,GHz channel. Columns
    show the temperature and Stokes $Q$ and $U$ parameters, while rows
    show, from top to bottom, the posterior mean, the posterior
    standard deviation, the white noise rms per pixel, and the
    difference between two random individual Gibbs samples. The
    polarization maps in the two two rows have been smoothed to an
    angular resolution of $1^{\circ}$ FWHM to visually reduce the
    noise level, while the bottom row is smoothed to $7^{\circ}$
    FWHM.}
  \label{fig:freq_postmean}
\end{figure*}

Figure~\ref{fig:pidpar} shows corresponding values as a function of
PID for one single Gibbs sample, this time for the 30\,GHz LFI 28M
radiometer. The top panel shows the normalized reduced $\chi^2$, as
defined by Eq.~\eqref{eq:redchisq}. As discussed above, this function
should ideally be independent between PIDs, and distributed according
to $N(0,1)$. This plot is therefore a powerful monitor for identifying
unmitigated and non-stationary instrumental systematic effects in a
given radiometer. In this case, the distribution scatters around zero
with roughly the expected variations, although there is a slight shift
towards positive values of about 0.5--$1\,\sigma$; overall, the model
appears to perform wellx.


The second panel shows the gain for the same 28M radiometer as a
function of PID. Here we see clear evidence of a systematic
oscillation with a period of one year, and a maximum variation of
about 1--2\,\% throughout the mission. The oscillatory behaviour is
primarily due to variations in the incoming solar radiation during the
year, effectively changing the heating of the instrument depending on its
precise orientation with respect to the Sun.

The three bottom panels of Fig.~\ref{fig:pidpar} show corresponding
plots of the three $1/f$ noise parameters as a function of
PID. Similar features as observed in the gain are seen also here,
although with lower signal-to-noise ratio. Overall, it is visually
obvious that the noise properties of this channel are not stationary
throughout the mission, but rather vary significantly in time. In
particular, the white noise level varies by 3--4\,\% throughout the
mission, and mirrors the gain variations seen above. For the slope,
$\alpha$, the most noteworthy feature are overall steeper values
between PIDs 11\,000 and 15\,000; as shown by \citet{bp06}, these can
be traced to changes in the thermal environment of the satellite using
house-keeping data.



Next, Fig.~\ref{fig:bpshift} shows the marginal posterior mean and
standard deviation for the bandpass correction,
$\Delta_{\mathrm{bp}}$, of each radiometer \citep{bp09}. Recalling
that we fix the absolute corrections for the 44 and 70\,GHz at zero
and only fit the 30\,GHz offset, we find an overall 30\,GHz frequency
shift of $0.24\pm0.03\,\mathrm{GHz}$, in agreement with
\citet{planck2014-a12}. Regarding the relative corrections, we note
that these are depend sensitively on the foreground amplitude at a
given frequency, and as a result, the relative bandpass uncertainties
are small at 30\,GHz, where the foregrounds are bright, while they are
large at 70\,GHz, where the foregrounds are weak.

Next, we consider the spatial structure of each of the various TOD
model terms in pixel space, and Fig.~\ref{fig:corrmaps} shows each of
the TOD objects binned into a 3-component Stokes $IQU$ sky map for the
30\,GHz channel and one arbitrarily selected Gibbs sample. This plot
corresponds essentially to a binned version of Fig.~\ref{fig:todplot},
and shows, from top to bottom, 1) the raw data; 2) correlated noise;
3) the orbital dipole; 4) bandpass and beam leakage corrections; 5)
far sidelobe corrections; 6) 1\,Hz spike corrections; and 7) the total
unmodelled residual. Note that the various terms are plotted with very
different color ranges, and, for instance, the range used for the
1\,Hz correction is only 0.3\muK, and this correction is therefore for
most practical purposes negligible. On the other hand, the range of
the polarized bandpass leakage correction is 30\,\muK, and this
represents as such the single most important large-scale polarization
correction. 

Figure~\ref{fig:corrmap_powspec_stddev} summarizes the uncertainty of
each systematic effect for the 30\,GHz channel and for the $EE$
angular power spectrum. The black line shows the power spectrum of the
full amplitude map, while the thick red line shows the corresponding
posterior standard deviation, which essentially summarizes the total
systematic uncertainty. The thin colored lines break down this into
individual contributions. Here we see that for $EE$ polarization at
the \Planck\ 30\,GHz channel, the three dominant uncertainty
contributions are correlated noise (orange), gain fluctuations
(through the orbital dipole; blue), and bandpass corrections (thin
red). All these are roughly of the same order of magnitude at
$\ell\lesssim10$, and jointly accounting for all three is therefore
essential in order to properly capture the full uncertainties. For
reference, we note that only the correlated noise contribution was
fully accounted in the official LFI low-resolution covariance
matrices. A complete survey of similar maps and power spectra for all
three LFI frequencies is provided by \citet{bp10}.

\subsection{Frequency maps}
\label{sec:freqmaps}

We now turn our attention to co-added frequency maps, as solved for
deterministically through Eq.~\eqref{eq:binmap}. For many users, these
represent the most convenient form of the \BP\ products, and we
provide these maps both in the form of individual samples, each
corresponding to one possible realization of all modelled systematic
effects, and as more traditional posterior mean and standard deviation
maps,
\begin{align}
  \hat{\m}_{\nu} &= \left<\m_{\nu}^i \right>\\
  \sigma_{\nu}(p) &= \sqrt{\left<\left(m_{\nu}^i(p)-\hat{m}_{\nu}(p)\right)^2 \right>},
\end{align}
where brackets indicate averaging over Monte Carlo samples. Note that
$\sigma_{\nu}$, as defined here, only accounts for systematic uncertainties
per pixel, not white noise uncertainties as defined by the diagonal of
the inverse coupling matrix in Eq.~\eqref{eq:binmap},
$\sigma_{\nu}^{\mathrm{wn}}(p)$. To obtain the full uncertainty, these
two terms must be added in quadrature,
\begin{equation}
  \sigma_{\nu}^{\mathrm{tot}}(p) = \sqrt{\sigma_{\nu}(p)^2 + \sigma_{\nu}^{\mathrm{wn}}(p)^2}.
\end{equation}
We stress, however, that analysis of these posterior mean maps is
likely to be sub-optimal for most scientific applications, and will
not exploit the full power of the \BP\ framework. Instead, we highly
recommend users to analyze the full ensemble of individual posterior
samples; that is by far the most robust and statistically correct
method for propagating \BP\ uncertainties into any higher-level
analysis.

\begin{figure*}[p]
    \vspace*{1cm}
  
  \center
  \includegraphics[width=0.33\linewidth]{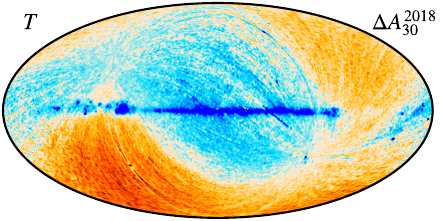}
  \includegraphics[width=0.33\linewidth]{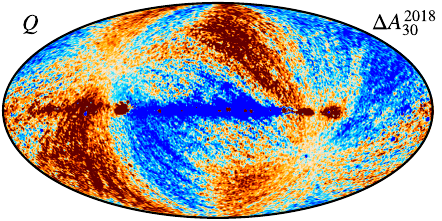}
  \includegraphics[width=0.33\linewidth]{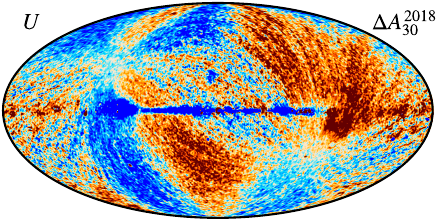}\\
  \includegraphics[width=0.33\linewidth]{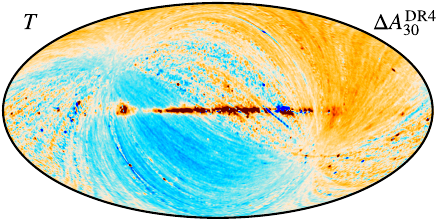}
  \includegraphics[width=0.33\linewidth]{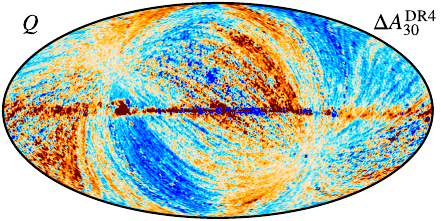}
  \includegraphics[width=0.33\linewidth]{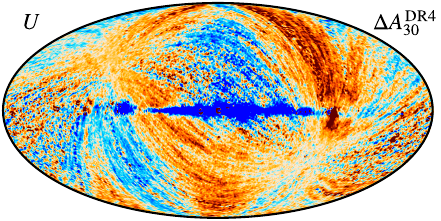}\\
  \vspace*{5mm}
  \includegraphics[width=0.33\linewidth]{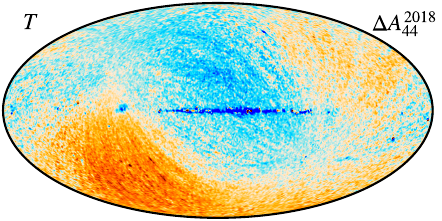}
  \includegraphics[width=0.33\linewidth]{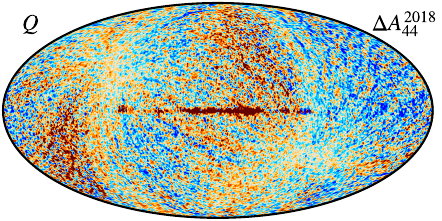}
  \includegraphics[width=0.33\linewidth]{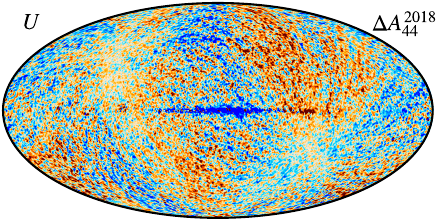}\\
  \includegraphics[width=0.33\linewidth]{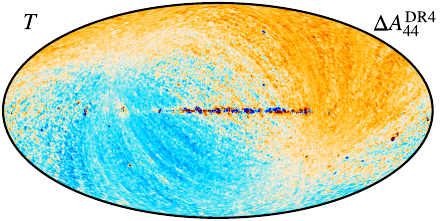}
  \includegraphics[width=0.33\linewidth]{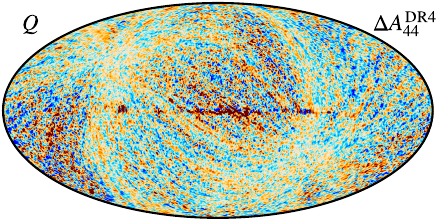}
  \includegraphics[width=0.33\linewidth]{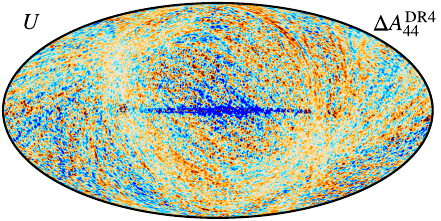}\\
  \vspace*{5mm}
  \includegraphics[width=0.33\linewidth]{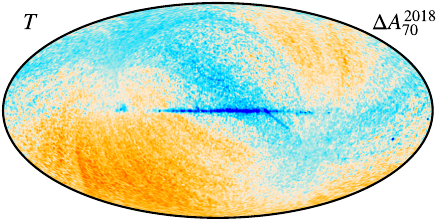}
  \includegraphics[width=0.33\linewidth]{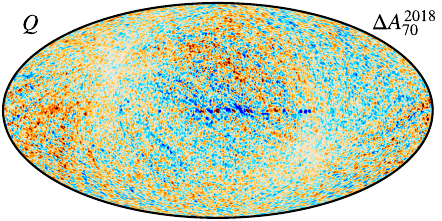}
  \includegraphics[width=0.33\linewidth]{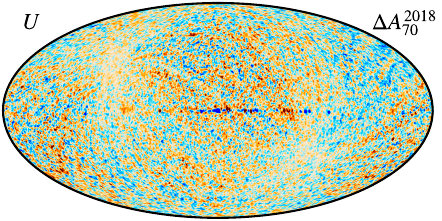}\\
  \includegraphics[width=0.33\linewidth]{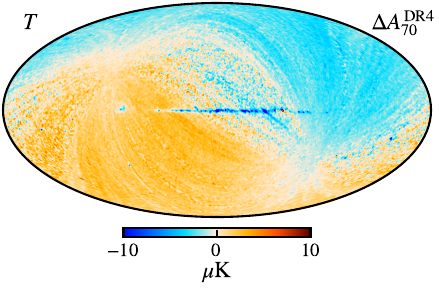}
  \includegraphics[width=0.33\linewidth]{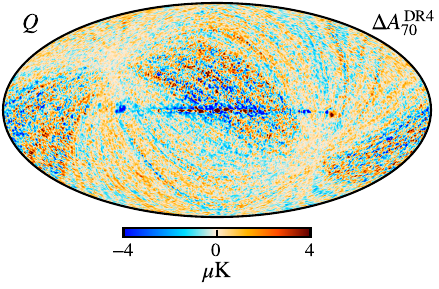}
  \includegraphics[width=0.33\linewidth]{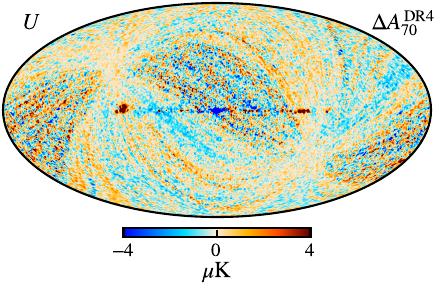}\\
  \caption{Differences between \BP\ and 2018 or \npipe\ frequency
    maps, smoothed to a common angular resolution of $2^{\circ}$
    FWHM. Columns show Stokes $T$, $Q$ and $U$ parameters,
    respectively, while rows show pair-wise differences with respect
    to the pipeline indicated in the panel labels. A constant offset
    has been removed from the temperature maps, while all other modes
    are retained. The 2018 maps have been scaled by their respective
    beam normalization prior to subtraction.
  }\label{fig:freqdiff}
\end{figure*}

\begin{figure*}[t]
  \center
  \includegraphics[width=\linewidth]{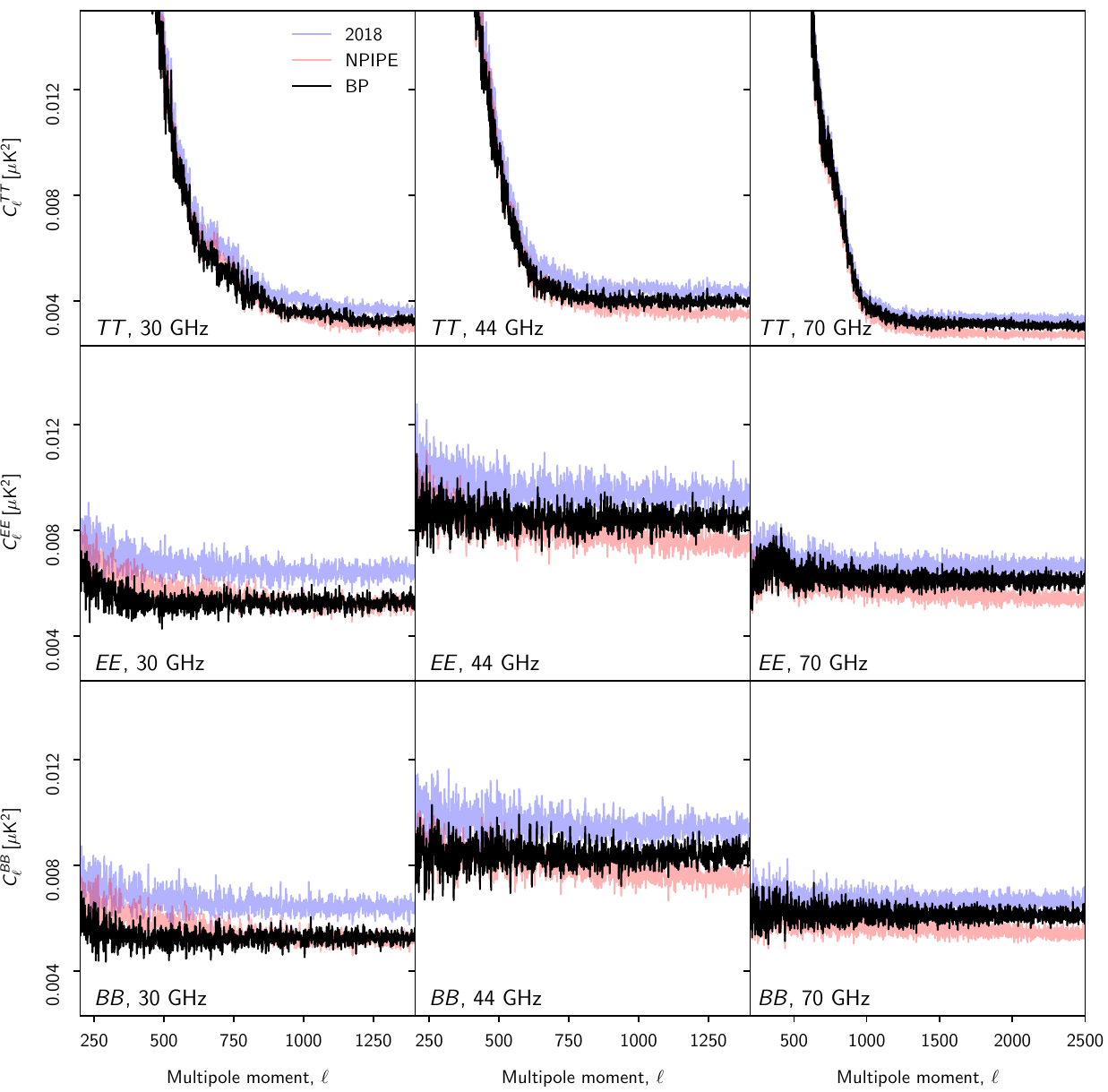}
  \caption{Comparison between angular auto-spectra computed from the \BP\
    (\emph{black}), \Planck\ 2018 (\emph{blue}), and \npipe\ (\emph{red})
    full-frequency maps. Rows show different frequencies, while columns show
    $TT$, $EE$, and $BB$ spectra. All spectra have been estimated with
    \texttt{PolSpice} using the \Planck\ 2018 common component separation confidence mask \citep{planck2016-l04}. Reproduced from \citet{bp10}. \label{fig:powspec_full}}
\end{figure*}

With these caveats in mind, Fig.~\ref{fig:freq_postmean} shows various
posterior summary maps for the 30\,GHz channel; the full set of maps
are shown and discussed by \citet{bp10}. From top to bottom, the
various rows show the posterior mean, the posterior standard
deviation, the white noise rms, and a straight difference map between
two independent Gibbs samples. The polarization maps in the top two
rows have been smoothed to an angular resolution of $1^{\circ}$ FWHM
to reduce noise, while the bottom row has been smoothed to $7^{\circ}$
FWHM.

Regarding the posterior mean map, perhaps the most striking feature is
that the \BP\ temperature map retains the CMB dipole, similar to
\Planck\ PR4 \citet{planck2020-LVII}, but contrary to the
\Planck\ 2018 and \WMAP\ frequency maps. Leaving this component in the
maps ensures that the full information content of the data is
available for subsequent component separation and calibration
applications.

The posterior standard deviation maps in the second row summarize the
combined effect of all the various systematic corrections made to the
frequency map. The most striking features include:
\begin{enumerate}
  \item a large monopole variation in the 30\,GHz temperature map,
    resulting in a nearly uniform morphology dipole variations;
  \item excess variance for rings aligned with the Galactic plane in
    polarization, reflecting the higher uncertainties in the
    time-variable gain resulting from the processing mask;
  \item excess variance along the Galactic plane, reflecting the
    higher uncertainties here due to gain and bandpass variations;
\end{enumerate}
all of which are super-imposed on the general \Planck\ scanning
pattern, which itself reflects correlated noise variations. We also
note that the upper limit of the temperature color scale is only
2\muK, which indicates that these variations are much lower than the
intrinsic variance of the CMB temperature fluctuations, which is about
30\,muK on these angular scales, and minor details in the systematic
model are therefore unlikely to affect final cosmological results. In
contrast, the standard deviation of the polarization maps at high
Galactic latitudes is typically about 0.5\muK, which is of the same
order of magnitude as the expected polarization imprint from cosmic
reionization.

The third row shows the corresponding white noise standard deviation
maps. These maps are fully specified by the detector white noise level
$\sigma_0$, the time-variable gains $g_t$, and the number of
observations per pixel. An important point regarding these white noise
rms maps is that their amplitude scales directly with the pixel size,
while the posterior standard deviation maps in the second row do
not. When smoothed over sufficiently large angular scales, the
systematic uncertainties will therefore eventually start to dominate
over the white noise level. 

The bottom row shows the difference between two frequency map samples,
smoothed to a common angular resolution of $7^{\circ}$ FWHM. Here we
clearly see correlated noise stripes along the \Planck\ scan direction
in all three frequency channels, but significantly more pronounced in
the 30\,GHz channel than in the other two frequencies. We also see
fluctuations along the Galactic plane, which are dominated by
uncertainties in the bandpass correction parameters,
$\Delta_{\mathrm{bp}}$. Clearly, modelling such correlated
fluctuations in terms of a single standard deviation per pixel is
unlikely to be adequate for any high-precision analysis, and the only
way to robustly propagate uncertainties is through analysis of the
full ensemble of Gibbs samples.

Figure~\ref{fig:freqdiff} shows differences between the \BP\ frequency
maps and those presented in the \Planck\ 2018 and \npipe\ data
releases. To ensure that this comparison is well defined, the 2018
maps have been scaled by the uncorrected beam efficiencies, and the
best-fit \Planck\ 2018 solar dipole has been added to each map, before
computing the differences. Overall, we see that the \BP\ maps agree
with the other two pipelines to $\lesssim 10\muK$ in temperature, and
to $\lesssim 4\muK$ in polarization. In temperature, we see that the
main difference between \npipe\ and \BP\ is an overall dipole, while
differences with respect to the 2018 maps show greater morphological
differences. The sign of the \npipe\ dipole differences changes with
frequency. This result is consistent with the original
characterization of the \npipe\ maps derived through multi-frequency
component separation in \citet{planck2020-LVII}; that paper reports a relative
calibration difference between the 44 and 70\,GHz channel of 0.31\,\%,
which corresponds to 10\muK\ in the map-domain. Overall, in temperature
\BP\ is thus morphologically similar to \npipe, but it improves a
previously reported relative calibration uncertainty between the
various channels by performing joint analysis.

In polarization, the dominant large-scale structures appear to be
dominated by effectively different offset determinations per PID,
which may originate from different gain or correlated noise
solutions. It is worth noting that the overall morphology of these
difference maps is structurally similar between frequencies, and that
\ the apparent amplitude of the differences falls with frequency. This
strongly suggests that different foreground modelling plays a crucial
role. In this respect, two observations are particularly noteworthy:
First, while both the \Planck\ 2018 and \npipe\ pipelines incorporate
component separation as an external input as defined by the
\Planck\ 2015 data release \citep{planck2014-a12}, \BP\ performs a
joint fit of both astrophysical foregrounds and instrumental
parameters. Second, both the LFI DPC and the \npipe\ pipeline consider
only \Planck\ observations alone, while \BP\ also exploits
\WMAP\ information to establish the sky model, which is particularly
important to break scanning-induced degeneracies in polarization.

In Fig.~\ref{fig:powspec_full} we show the high-$\ell$
auto-correlation spectra for each of the three generations of
\Planck\ LFI maps (2018, PR4, and \BP) as computed with
\texttt{PolSpice} \citep{chon2004}. In all cases, we apply the
\Planck\ 2018 common component separation confidence mask
\citep{planck2016-l04}, which accepts a sky fraction of 80\,\%.  The
most notable features here are, first, that the overall noise levels
of the \BP\ maps are slightly lower than in the \Planck\ 2018 maps,
but also somewhat higher than PR4. Secondly, we also note that the
\BP\ spectra are notably flatter than the other two pipelines, and in
particular than \npipe, which shows a clearly decreasing trend toward
high multipoles.

As discussed by \citet{planck2020-LVII} and \citet{bp10}, \Planck\ PR4
achieves lower noise than \Planck\ 2018 primarily through three
modifications, namely use of the \Planck\ repointing periods (which
accounts for 8\,\% of the total data volume); use of the so-called
``ninth survey'' (which accounts for 3\,\% of the total data volume);
and by smoothing the LFI reference load data prior to diode
differencing. Among these, \BP\ only implements the repointing period
extension, for which we find no measurable issues. For the ninth
survey data, on the other hand, we find that the TOD $\chi^2$
statistics vary more strongly from PID to PID, which suggests poorer
instrumental stability. We therefore exclude these data, following
\Planck\ 2018. Finally, we note that the decreased white noise floor
that results from load smoothing also comes at a cost of increased
colored (or correlated) noise at high multipoles, and this is both
computationally expensive to model properly. For further discussion
regarding these frequency maps and power spectra, see \citet{bp10}.

\begin{figure*}[t]
  \center
  \includegraphics[width=0.49\linewidth]{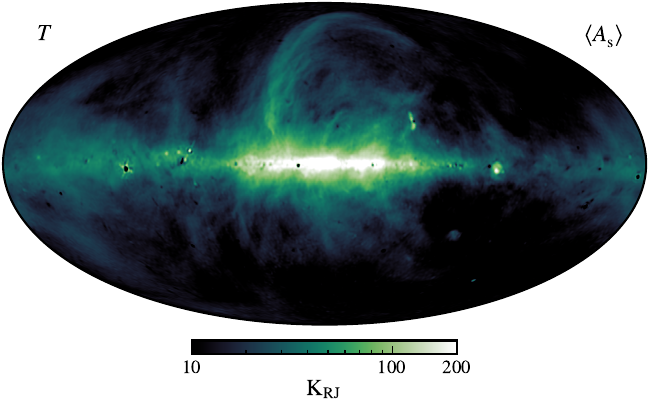}
  \includegraphics[width=0.49\linewidth]{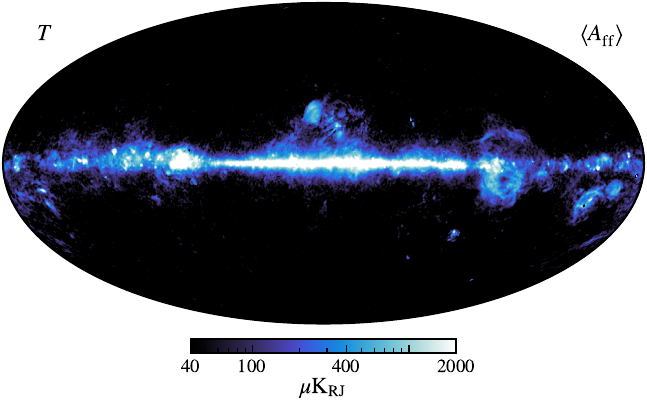}\\
  \includegraphics[width=0.49\linewidth]{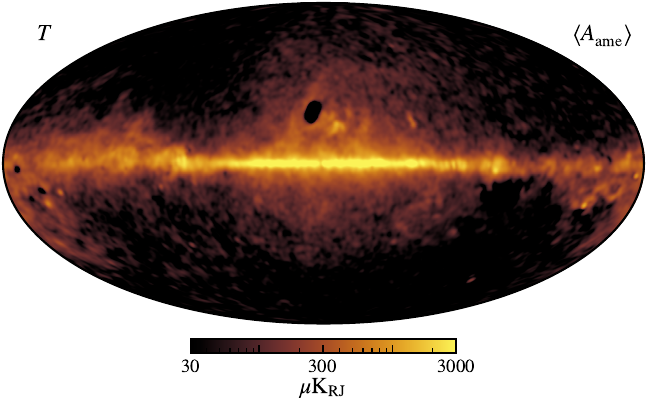}
  \includegraphics[width=0.49\linewidth]{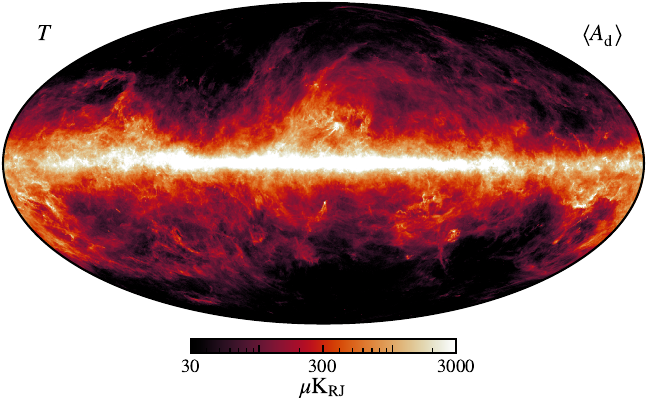}
  \caption{
    Posterior mean maps of the amplitude of each of the four intensity
    foreground components included in the \BP\ analysis. (\emph{Top left:})
    Synchrotron amplitude, evaluated at 30~GHz and smoothed to $2^{\circ}$ FWHM
    resolution.  (\emph{Top right:}) Free-free amplitude, evaluated at 40~GHz
    and smoothed to $30\arcmin$ FWHM resolution. (\emph{Bottom left:}) AME
    amplitude, evaluated at 22~GHz and smoothed to $2^{\circ}$ FWHM resolution.
    (\emph{Bottom right:}) Thermal dust amplitude, evaluated at 545~GHz and smoothed to $10\arcmin$ FWHM resolution. Note that the color bars vary between panels. See \citet{bp13} for further discussion of these maps.
  }\label{fig:fg_temp}
\end{figure*}

\subsection{Astrophysical component posteriors}
\label{sec:sky_params}

We now turn our attention to the astrophysical component posteriors. However,
before presenting the results, we recall that a main design feature of the
current analysis was to let the LFI data play the main role in the CMB
reconstruction. In practice, this means that neither the CMB-dominated HFI
frequencies, nor the \WMAP\ \emph K-band observations, are included in the
analysis. As a result, we note that the derived foreground posterior constraints
shown here are significantly weaker than those presented by the \Planck\ team in
\citet{planck2014-a12}, \citet{planck2016-l04}, and \citet{planck2020-LVII}. Full joint
analysis of all data sets is left for future work.

\begin{figure*}[t]
  \center
  \includegraphics[width=0.49\linewidth]{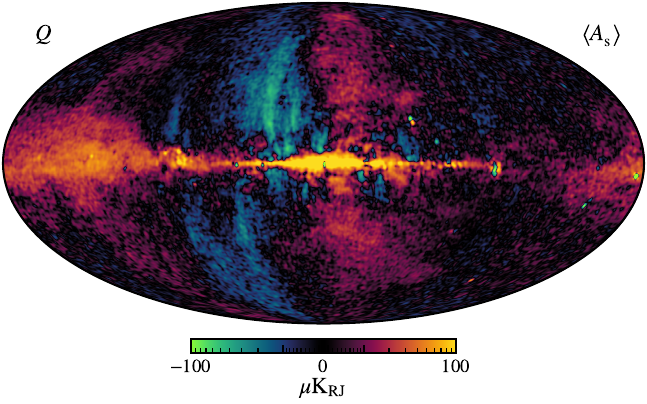}
  \includegraphics[width=0.49\linewidth]{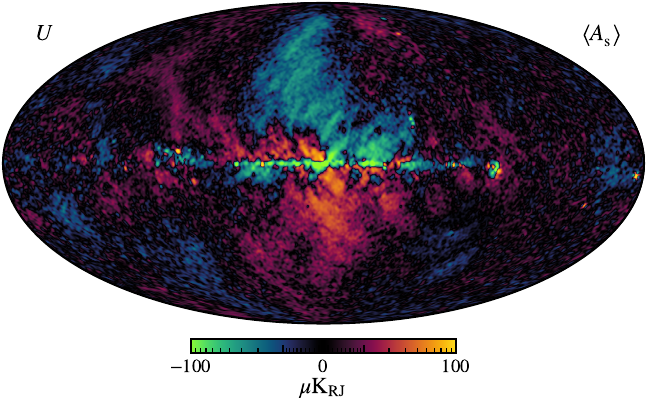}
  \caption{Posterior mean maps of polarized synchrotron amplitude derived from \BP,
    evaluated at 30\,GHz and smoothed to an angular resolution of $1^{\circ}$ FWHM. The two columns show Stokes $Q$ and $U$ parameters, respectively; see \citet{bp14} for further discussion of these maps.}\label{fig:synch_pol}
\end{figure*}

\begin{figure*}[t]
  \center
  \includegraphics[width=0.49\linewidth]{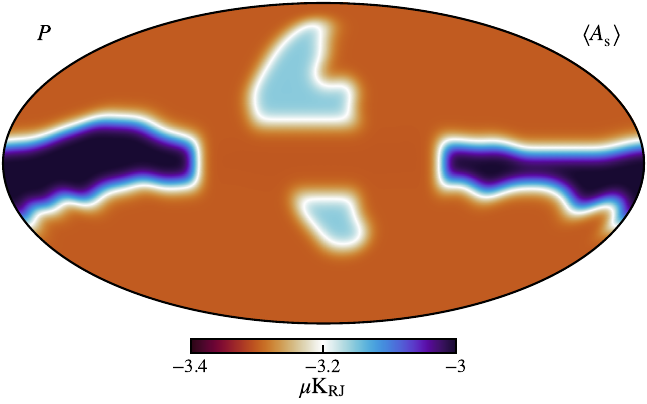}
  \includegraphics[width=0.49\linewidth]{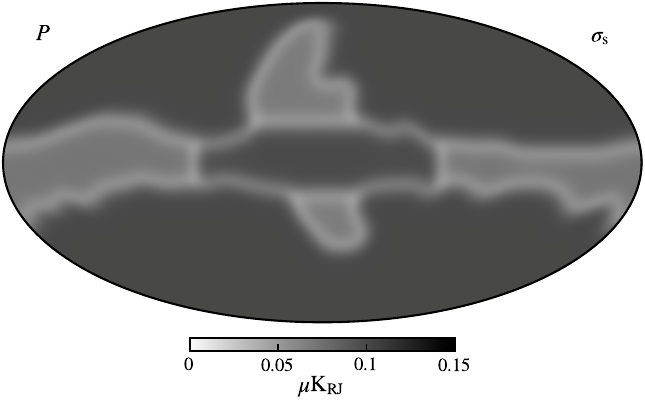}
  \caption{Posterior mean and standard deviation maps for the spectral
    index of polarized synchrotron emission,
    $\beta_{\mathrm{s}}$. Note that $\beta_{\mathrm{s}}$ is fitted in
    terms of four disjoint regions, each with a constant value but
    smoothed with a $10^{\circ}$ FWHM Gaussian beam to avoid edge
    effects. The effect of this smoothing is seen in both the mean and
    standard deviation maps. Reproduced from \citet{bp14}.}\label{fig:synch_beta_pol}

  \center
  \includegraphics[width=0.9\linewidth]{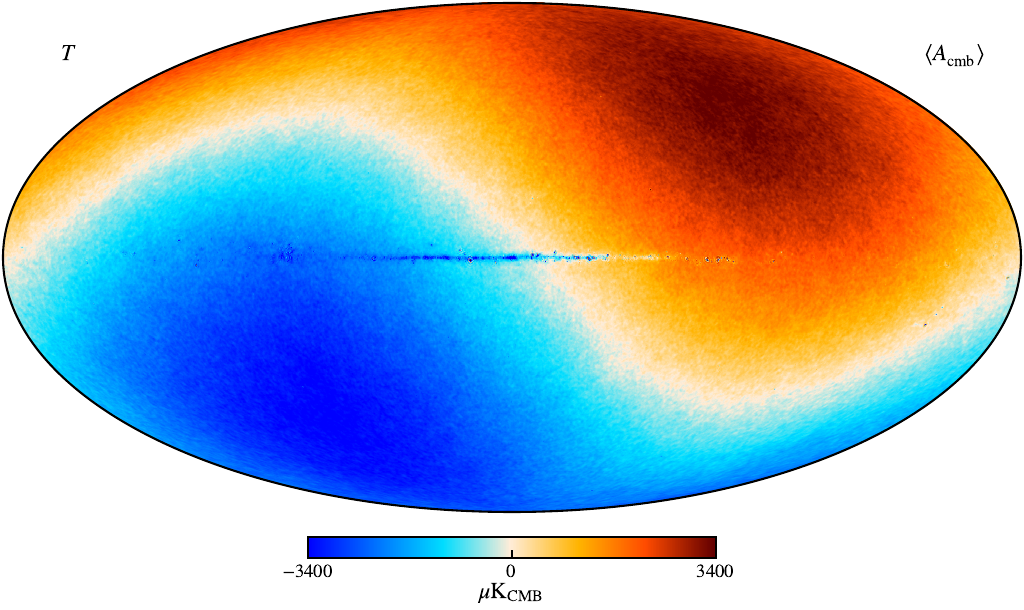}
  \caption{Posterior mean CMB \BP\ temperature map, smoothed to an
    angular resolution of $14\arcm$ FWHM. Reproduced from \citet{bp11}.}\label{fig:cmb_with_dipole}
\end{figure*}

\begin{figure}[t]
  \center
  \includegraphics[width=\linewidth]{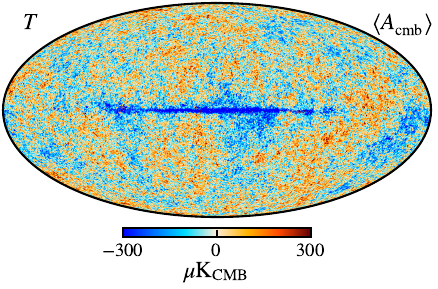}\\
  \includegraphics[width=\linewidth]{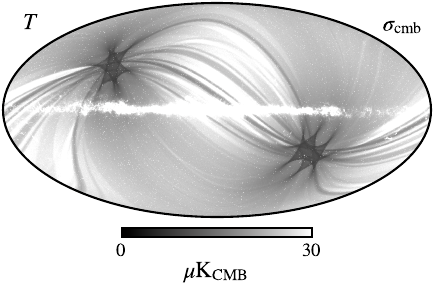}\\  
  \includegraphics[width=\linewidth]{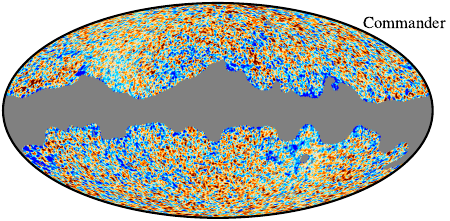}
  \includegraphics[width=0.68\linewidth]{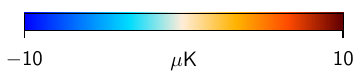}
  \caption{Difference map between the \BP\ and the \Planck\ 2018
    \commander\ CMB temperature map \citep{planck2016-l04}, smoothed
    to a common angular resolution of $1^{\circ}$
    FWHM.}\label{fig:cmb_diff}
\end{figure}

With that caveat in mind, Fig.~\ref{fig:fg_temp} shows the posterior
mean maps for each of the four modelled temperature foregrounds,
namely synchrotron, free-free, AME, and thermal dust emission. As
discussed by \citet{bp13}, these are consistent with earlier results
of the same type \citep{planck2014-a12}, but with notably higher
uncertainties, because of the more limited data set employed here.

Similarly, Fig.~\ref{fig:synch_pol} shows the posterior mean amplitude
for polarized synchrotron emission, and Fig.~\ref{fig:synch_beta_pol}
summarizes the posterior mean (left panel) and standard deviation
(right panel) for the power-law index of polarized synchrotron
emission. In this case, it is worth pointing out that the
\Planck\ team never published a joint polarized synchrotron solution
that included both \Planck\ and \WMAP\ observations, for the simple
reason that these data sets could never made to agree statistically to
a satisfactory degree when analyzed separately; when attempting to fit
a single synchrotron spectral index across both data sets, the
resulting constraints were clearly nonphysical, and led to large
$\chi^2$ excesses.

\begin{figure}[t]
  \center
  \includegraphics[width=\linewidth]{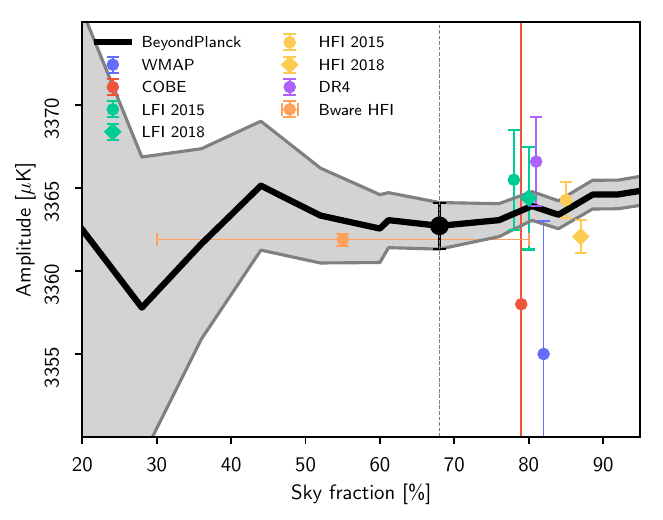}
  \caption{CMB dipole amplitude as a function of sky fraction, reproduced from \citet{bp11}.
    The gray band indicates the 68\,\% posterior confidence region.}\label{fig:cmb_dipole}
\end{figure}

Thus, the \BP\ analysis represents the first reduction of the \Planck\ LFI
data set for which a joint foreground polarization analysis
with \WMAP\ yields statistically meaningful results. However, as shown
by \citet{bp14}, even the combination of the two data sets does not
constrain the spectral index very strongly, and for this reason we
choose to fit only a small number of independent spectral indices
across the sky. Specifically, we partition the sky into four disjoint
regions, corresponding to the Galactic Center (GC), the Galactic Plane
(GP), the North Galactic Spur (NGS), and High Galactic Latitudes
(HGL), and treat each region separately. Adopting
\citet{planck2014-a12} as a reference, we enforce a Gaussian prior of
$\beta_{\mathrm{s}}\sim N(-3.1,0.1^2)$. Finally, each spectral index
sample is smoothed with a Gaussian beam of $10^{\circ}$ FWHM to avoid
edge effects.

For the GP and NGS regions, which both have significant
signal-to-noise ratio with respect to polarized synchrotron emission
and low systematic effects, we fit $\beta_{\mathrm{s}}$ using the full
posterior distribution as described in Sect.~\ref{sec:beta}. However,
for the HGL region, in which the effective synchrotron signal-to-noise
ratio is very low, we simply marginalize over the prior, and exclude
the likelihood term. The reason for this is simply that unconstrained
degeneracies with other parameters, such as the gain, tend to bias
$\beta_{\mathrm{s}}$ toward high values
($\beta_{\mathrm{s}}^{\mathrm{HGL}}\approx-2.5$; see \citealp{bp14})
when fitted freely. 

We also do the same for the GC region, for which
temperature-to-polarization leakage and bandpass effects are
particularly important, and the synchrotron signal may also be biased
by Faraday rotation. When fitting this region freely, we find an
effective spectral index of
$\beta_{\mathrm{s}}^{\mathrm{GC}}\approx-4$, which is also clearly
unphysical. Rather than letting these unmodelled systematic effects
feed into the other components, we marginalize over the physically
motivated prior.

This leaves us with two main regions usable for scientific
interpretation, and these may be seen as blue regions in the standard
deviation map in Fig.~\ref{fig:synch_beta_pol}. Specifically, we find
${\beta_{\mathrm{s}}^{\mathrm{GP}}=-3.03\pm0.07}$ and
${\beta_{\mathrm{s}}^{\mathrm{NGS}}=-3.17\pm0.06}$, respectively
\citep{bp14}. We note that these values are broadly consistent with
previous temperature-only constraints, such as those reported by
\citet{planck2014-a12}, who found $\beta_{\mathrm{s}}=-3.1$. They are
also consistent with the observation that the spectral index is
flatter towards the Galactic plane than in the North Galactic Spur
\citep[e.g.,][]{kogut:2012,fuskeland2014,fuskeland:2019}, although the
statistical significance of this observation is marginal. In this
respect, it is worth noting that the low Galactic latitudes are
particularly sensitive to both systematic and astrophysical modelling
errors, both in temperature and polarization.  For a full discussion
of these results, we refer the interested reader to \citet{bp14}.

\subsection{CMB temperature posteriors}
\label{sec:cmb_params}

Finally, we arrive at the main scientific target application of the
paper, the CMB posteriors. For a full discussion of the following
results, we refer the interested reader to \citet{bp11,bp12}, and here
we only provide a brief overview of the main points.

First, Fig.~\ref{fig:cmb_with_dipole} shows the marginal CMB
temperature fluctuation posterior mean map as derived in \BP, given
the data, model and priors described above. This map is massively
dominated by the CMB solar dipole, with only a small imprint of the
Galactic plane being visible in the very center. At high latitudes,
CMB temperature fluctuations may be seen as tiny ripples superimposed
on the dipole.

\begin{table*}
\newdimen\tblskip \tblskip=5pt
\caption{Comparison of Solar dipole measurements from \COBE, \WMAP, and \Planck. }
\label{tab:dipole}
\vskip -4mm
\footnotesize
\setbox\tablebox=\vbox{
 \newdimen\digitwidth
 \setbox0=\hbox{\rm 0}
 \digitwidth=\wd0
 \catcode`*=\active
 \def*{\kern\digitwidth}
  \newdimen\dpwidth
  \setbox0=\hbox{.}
  \dpwidth=\wd0
  \catcode`!=\active
  \def!{\kern\dpwidth}
  \halign{\hbox to 2.5cm{#\leaderfil}\tabskip 2em&
    \hfil$#$\hfil \tabskip 2em&
    \hfil$#$\hfil \tabskip 2em&
    \hfil$#$\hfil \tabskip 2em&
    #\hfil \tabskip 0em\cr
\noalign{\doubleline}
\omit&&\multispan2\hfil\sc Galactic coordinates\hfil\cr
\noalign{\vskip -3pt}
\omit&\omit&\multispan2\hrulefill\cr
\noalign{\vskip 3pt} 
\omit&\omit\hfil\sc Amplitude\hfil&l&b\cr
\omit\hfil\sc Experiment\hfil&[\muK_{\rm
CMB}]&\omit\hfil[deg]\hfil&\omit\hfil[deg]\hfil&\hfil\sc Reference\hfil\cr
\noalign{\vskip 3pt\hrule\vskip 5pt}
\COBE \rlap{$^{\rm a,b}$}&                  3358!**\pm23!**&     264.31*\pm0.16*&
     48.05*\pm0.09*&\citet{lineweaver1996}\cr
\WMAP\ \rlap{$^{\rm c}$}&                  3355!**\pm*8!**&     263.99*\pm0.14*&
     48.26*\pm0.03*&\citet{hinshaw2009}\cr
\noalign{\vskip 3pt}
LFI 2015 \rlap{$^{\rm b}$}&              3365.5*\pm*3.0*&     264.01*\pm0.05*&
     48.26*\pm0.02*&\citet{planck2014-a03}\cr
HFI 2015 \rlap{$^{\rm d}$}&              3364.29\pm*1.1*&     263.914\pm0.013&
     48.265\pm0.002&\citet{planck2014-a09}\cr
\noalign{\vskip 3pt}
LFI 2018 \rlap{$^{\rm b}$}&              3364.4*\pm*3.1*&     263.998\pm0.051&
     48.265\pm0.015&\citet{planck2016-l02}\cr
HFI 2018 \rlap{$^{\rm d}$}&              3362.08\pm*0.99&     264.021\pm0.011&
     48.253\pm0.005&\citet{planck2016-l03}\cr
\noalign{\vskip 3pt}
Bware & 3361.90\pm*0.40 & 263.959\pm0.019 & 48.260\pm0.008  & \citet{delouis:2021}  \cr
\Planck\ PR4\ \rlap{$^{\rm a,c}$}& 3366.6*\pm*2.6*& 263.986\pm0.035&
48.247\pm0.023&\citet{planck2020-LVII}\cr
\noalign{\vskip 3pt}
\bf\BP\ \rlap{$^{\rm e}$} & \bf3362.7*\pm*1.4*& \bf264.11*\pm0.07*&
 \bf48.279\pm0.026&\citet{bp11}\cr
\noalign{\vskip 5pt\hrule\vskip 5pt}}}
\endPlancktablewide
\tablenote {{\rm a}} Statistical and systematic uncertainty estimates are added in quadrature.\par
\tablenote {{\rm b}} Computed with a naive dipole estimator that does not account for higher-order CMB fluctuations.\par
\tablenote {{\rm c}} Computed with a Wiener-filter estimator that estimates, and marginalizes over, higher-order CMB fluctuations jointly with the dipole.\par
\tablenote {{\rm d}} Higher-order fluctuations as estimated by subtracting a dipole-adjusted CMB-fluctuation map from frequency maps prior to dipole evaluation. \par
\tablenote {{\rm e}} Estimated with a sky fraction of 68\,\%. Error bars include only statistical uncertainties, as defined by the global \BP\ posterior framework, and they thus account for instrumental noise, gain fluctuations, parametric foreground variations etc. 
\par
\end{table*}

\begin{figure}[t]
  \center
  \includegraphics[width=\linewidth]{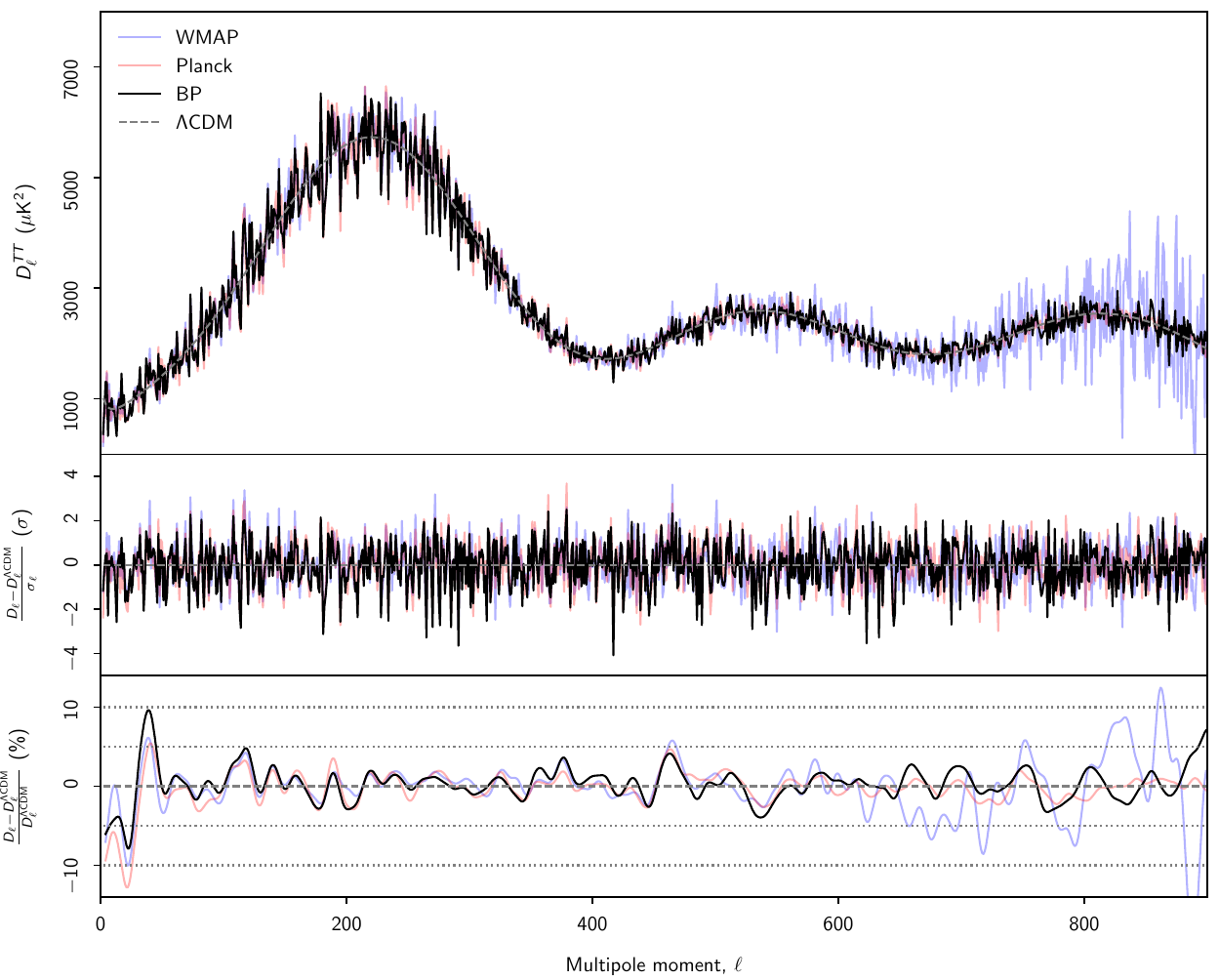}
  \caption{(\emph{Top}:) Angular CMB temperature power spectrum,
    $D_{\ell}^{TT}$, as derived by \BP\ (black), \Planck\ (red), and
    \WMAP\ (blue). The best-fit \Planck\ 2018 $\Lambda$CDM power
    spectrum is shown in dashed gray. (\emph{Middle}:) Residual power
    spectrum relative to $\Lambda$CDM, measured relative to full
    quoted error bars, $r_{\ell} \equiv
    (D_{\ell}-D_{\ell}^{\Lambda\mathrm{CDM}})/\sigma_{\ell}$. For
    pipelines that report asymmetric error bars, $\sigma_{\ell}$ is
    taken to be the average of the upper and lower error
    bar. (\emph{Bottom}:) Fractional difference with respect to the
    \Planck\ $\Lambda$CDM spectrum. In this panel, each curve has been
    boxcar averaged with a window of $\Delta\ell=100$ to suppress
    random fluctuations. }\label{fig:cl_TT}
\end{figure}

Figure~\ref{fig:cmb_diff} shows, from top to bottom, the \BP\ posterior
mean temperature map (after subtracting the CMB solar dipole), the
posterior standard deviation, and the difference with respect to the
\Planck\ 2018 \commander\ CMB map. Both the posterior mean and
standard deviation maps show clear evidence for foreground residuals
near the Galactic plane, but these are well covered by the confidence
mask. The differences seen in the bottom panel range between
$\pm10\muK$, which is the same order of magnitude as observed between
the various component separation algorithms used internally in the
\Planck\ collaboration \citep{planck2016-l04}. 

Next, Fig.~\ref{fig:cmb_dipole} shows the CMB Solar dipole amplitude
as derived from masks with different sky fractions, ranging between 20
and 95\,\%, and the value corresponding to the fiducial analysis mask
of a 68\,\% sky fraction is tabulated in
Table~\ref{tab:dipole}. Overall, we see that the \BP\ estimates agree
well with previous estimates, both in terms of amplitude and
direction, although there is a $1.5\,\sigma$ shift in the . The main difference between the \BP\ and previous results
lies thus not in central values, but rather in their uncertainties. In
particular, it is important to note that the CMB dipole is within the
\BP\ framework estimated on completely the same footing as any other
mode in the CMB sky, and is represented in terms of three spherical
harmonic coefficients in $\s_{\mathrm{CMB}}$. No special-purpose
component separation algorithms are applied to derive the CMB dipole,
nor are there any \emph{ad-hoc} instrumental uncertainty corrections
involved in the estimation of the error bars; both the posterior mean
and standard devation are evaluated directly from the Gibbs
samples. 

Figure~\ref{fig:cl_TT} shows the angular temperature power spectrum
derived from the sample set with a Blackwell-Rao estimator
\citep{chu2005,bp11}, and compared with the similar spectra presented
by \Planck\ \citep{planck2016-l05} and \WMAP\ \citep{hinshaw2012}. For
reference, the gray dashed line shows the best-fit \Planck\ 2018
$\Lambda$CDM spectrum. The middle panel shows the difference of each
measured spectrum with respect to the model spectrum in units of each
pipeline's respective error bars, while the bottom panel shows the
corresponding fractional difference with respect to the best-fit
\Planck\ 2018 $\Lambda$CDM spectrum in units of percent. At
$\ell\lesssim 500$, where these data sets are all signal-dominated,
the three spectra follow each other almost $\ell$-by-$\ell$, while at
higher multipoles, where \WMAP\ becomes noise-dominated, larger
variations are seen within multipoles. Overall, the agreement between
the three estimates is very good, both as measured by fractional
differences and in units of $\sigma$. Correspondingly, standard
$\Lambda$CDM parameter constraints from the \BP\ analysis are fully
compatible with those derived from the standard \Planck\ processing
when considering the same multipole ranges \citep{bp12}.

\subsection{CMB polarization posteriors}

\begin{table*}[t]       
  \begingroup                                                                            
  \newdimen\tblskip \tblskip=5pt
  \caption{Summary of cosmological parameters dominated by large-scale
    polarization and goodness-of-fit statistics. Columns list, from
    left to right, 1) analysis name; 2) basic data sets included in
    the analysis; 3) effective accepted sky fraction; 4) posterior
    mean estimate of the optical depth of reionization with 68\,\%
    error bars; 5) upper limit on tensor-to-scalar ratio at 95,\%
    confidence; 6) $\chi^2$ goodness-of-fit statistic as measured in
    terms of probability-to-exceed; and 7) primary reference. Reproduced from \citet{bp12}.
    \label{tab:cospar_pol}}
  \nointerlineskip                                                                                                                                                                                     
  \vskip -2mm
  \footnotesize                                                                                                                                      
  \setbox\tablebox=\vbox{ %
  \newdimen\digitwidth                                                                                                                          
  \setbox0=\hbox{\rm 0}
  \digitwidth=\wd0
  \catcode`*=\active
  \def*{\kern\digitwidth}
  \newdimen\signwidth
  \setbox0=\hbox{+}
  \signwidth=\wd0
  \catcode`!=\active
  \def!{\kern\signwidth}
  \newdimen\decimalwidth
  \setbox0=\hbox{.}
  \decimalwidth=\wd0
  \catcode`@=\active
  \def@{\kern\signwidth}
  \halign{ \hbox to 1.8in{#\leaderfil}\tabskip=1.0em&
    \hfil$#$\tabskip=1.5em&
    \hfil$#$\hfil\tabskip=1em&
    \hfil$#$\hfil\tabskip=1em&
    \hfil$#$\hfil\tabskip=0.5em&
    \hfil$#$\hfil\tabskip=1em&
    $#$\hfil\tabskip=0em\cr
  \noalign{\doubleline}
  \omit{\sc Analysis Name}\hfil& \omit{\sc Data Sets}\hfil&\hfil f^{\mathrm{pol}}_{\mathrm{sky}}\hfil&\hfil \tau\hfil&\hfil r^{BB}_{95\,\%}\hfil&\omit\hfil $\chi^2$ {\sc PTE}\hfil&\omit{\sc Reference}\hfil\cr
  \noalign{\vskip 3pt\hrule\vskip 5pt}
  \WMAP\ 9-yr& \omit \WMAP\ \emph{Ka}--\emph V\hfil& 0.76&  0.089\pm0.014& & & \omit\citet{hinshaw2012}\hfil\cr
    Natale et al.& \omit LFI 70, \WMAP\ \emph{Ka}--\emph V\hfil& 0.54&
    0.069\pm0.011& <0.79 & & \omit\citet{natale:2020}\hfil\cr
  \Planck\ 2018& \omit HFI 100$\times$143\hfil& 0.50&  0.050\pm0.009& <0.41 & & \omit\citet{planck2016-l05}\hfil\cr
  \srollTwo & \omit HFI 100$\times$143\hfil& 0.50&  0.059\pm0.006& & & \omit\citet{pagano:2020}\hfil\cr    
  \npipe\ (\commander\ CMB)& \omit LFI+HFI\hfil& 0.50&  0.058\pm0.006&
  < 0.16 & & \omit\citet{tristram:2020}\hfil\cr
  \noalign{\vskip 3pt}   
  {\bf \BP}, $\ell=2$--8 & \omit \bf LFI, \WMAP\ \emph{Ka}--\emph V\hfil&
  \bf 0.68&  \bf 0.066\pm{0.013}& \bf <0.84& \bf 0.32& \textbf{\citet{bp12}}\cr 
  \BP, $\ell=3$--8 & \omit LFI, \WMAP\ \emph{Ka}--\emph V\hfil& 0.68&  0.066\pm{0.014}& <1.0& 0.32& \textrm{\citet{bp12}}\cr 
    \noalign{\vskip 3pt\hrule\vskip 5pt}   
  }}
  \endPlancktablewide                                                                                                                                            
  \endgroup
\end{table*}

\begin{figure}[t]
  \center
  \includegraphics[width=\linewidth]{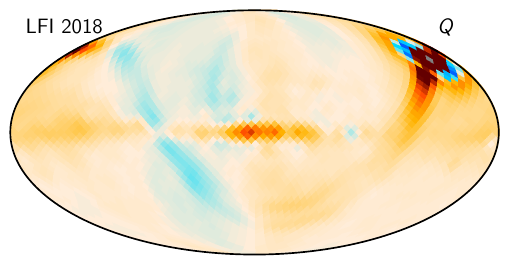}
  \includegraphics[width=\linewidth]{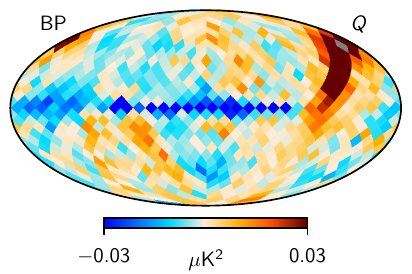}
  \caption{Single column of the low-resolution
    CMB noise covariance matrix, as estimated by the LFI DPC
    (\emph{top row}) and \BP\ (\emph{bottom row}). The column
    corresponds to the Stokes $Q$ pixel marked in gray, which is
    located in the top right quadrant near the `$Q$' label. Note that
    the DPC covariance matrix is constructed at $N_{\mathrm{side}}=16$
    and includes a cosine apodization filter, while the \BP\ matrix is
    constructed at $N_{\mathrm{side}}=8$ with no additional
    filter.}\label{fig:ncov}
\end{figure}

\begin{figure}[t]
  \center
  \includegraphics[width=\linewidth]{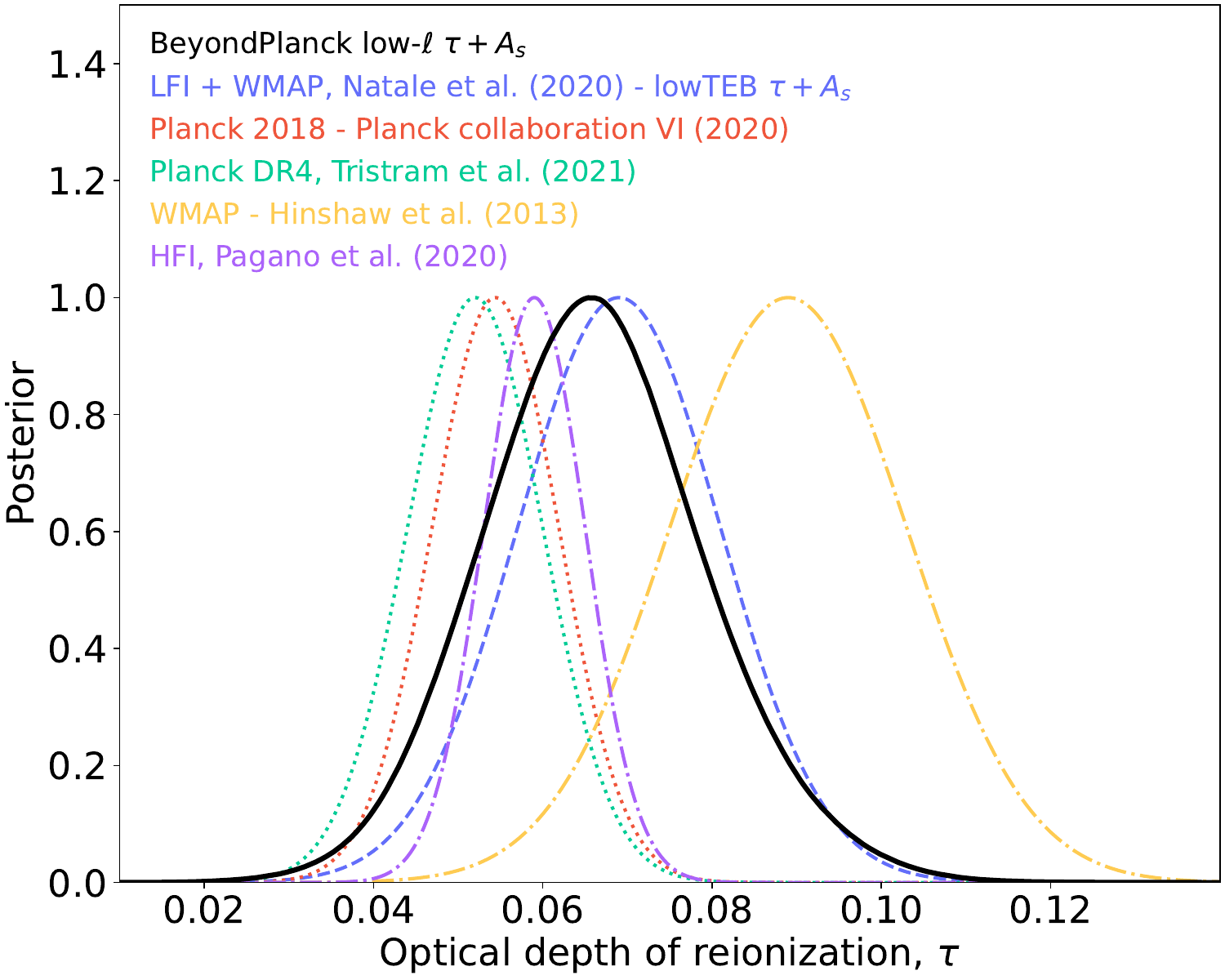}
  \caption{Comparison of (unnormalized) marginal
    posterior distributions of the reionization optical depth from
    \BP\ (solid black line) and various previously published estimates
    (colored dotted/dashed curves). Reproduced from \citet{bp12}.}\label{fig:tau}
\end{figure}

\begin{figure}[t]
	\center
	\includegraphics[width=\linewidth]{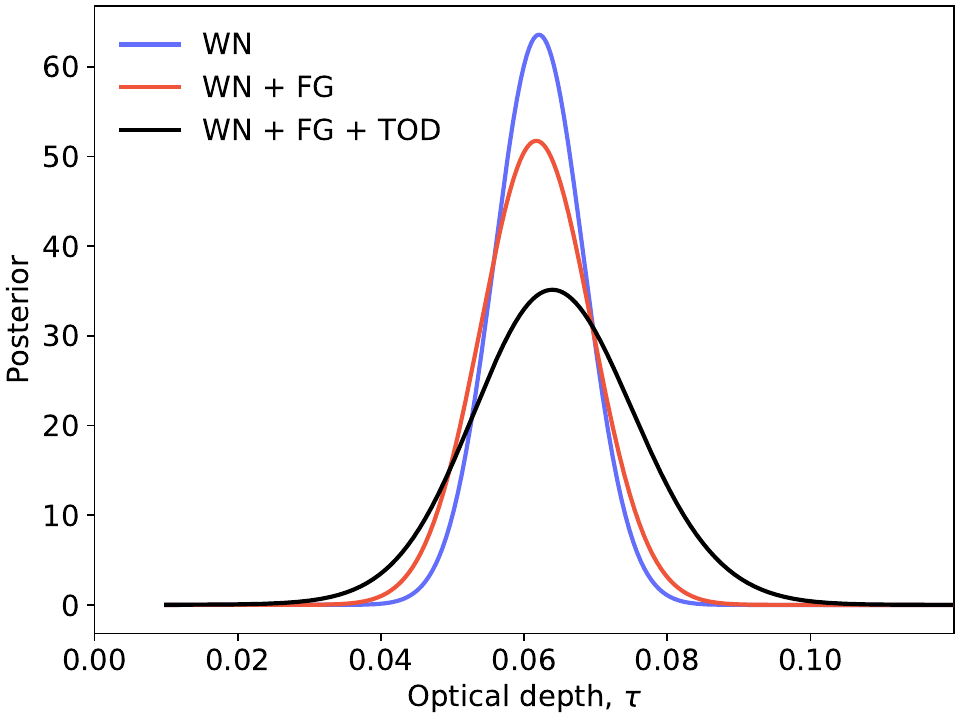}
        \caption{Estimates of $\tau$ under different uncertainty
          assumptions. The blue curve shows marginalization over white
          noise only; the red 
          curve shows marginalization over white noise and
          astrophysical uncertainties; and, finally, the black curve
          shows marginalization over all contributions, including
          low-level instrumental uncertainties, as in the
          final \BP\ analysis.}
	\label{fig:tau_assess}
\end{figure}

Turning to low-resolution polarization estimation, which we stated
served as the main scientific target of the entire \BP\ processing, we
start by showing a slice through the CMB Stokes $Q$ covariance matrix
in the bottom panel Fig.~\ref{fig:ncov}; for comparison, the top panel
shows the corresponding DPC 70\,GHz covariance matrix that takes into
account correlated noise and foreground template subtraction. These
matrices are sliced through the pixel marked in gray in the upper
right quadrant. As discussed by \citet{bp11}, the dominant features in
the \BP\ matrix are 1) correlated noise at high latitudes, tracking
the same scanning path as seen in the DPC matrix; 2) bandpass and
foreground uncertainties along the Galactic plane; and 3) gain and
calibration uncertainties at high latitudes. For the LFI 30\,GHz, the
latter effect is in fact slightly stronger than the correlated noise
contribution, as already pointed out in
Fig.~\ref{fig:corrmap_powspec_stddev}, and neglecting this
contribution significantly underestimates the total uncertainties.

Figure~\ref{fig:tau} compares the posterior distribution for the
optical depth of reionization, $\tau$, with similar estimates derived
using different data sets and methods, while
Table~\ref{tab:cospar_pol} compares numerically $\tau$, $r$, and the
reduced $\chi^2$ goodness-of-fit statistics for the same
analyses. First and foremost, we see in Fig.~\ref{fig:tau} that the
\BP\ estimates of $\tau$ agree very well with previous analyses in
terms of the posterior mean value for $\tau$, with a central value of
$\tau=0.066\pm0.013$. We also see that the tensor-to-scalar ratio is
consistent with zero, and the $\chi^2$ statistic has a
probability-to-exceed of 32\,\%, both of which suggests that the
adopted model performs well.

In particular, the agreement in terms of $\tau$ is very close with
respect to \citet{natale:2020}, who reported a value of
$\tau=0.069\pm0.011$ using an almost identical data
selection. However, as discussed by \citet{bp12}, it is important to
note that the \BP\ uncertainty is in fact significantly larger than
that reported by \citet{natale:2020}. Not only is the actually
reported values 18\,\% larger in itself, but the \BP\ analysis mask is
also more permissive, allowing 68\,\% of the sky to be used, which is
to be compared with a sky fraction of 54\,\% in the previous
analysis. When accounting for all effects, the \BP\ uncertainty is
about 30\,\% larger than the Natale et al.\ result, and we interpret
this as being caused by marginalizing over a more complete set of
uncertainties.

This effect is explicitly demonstrated in Fig.~\ref{fig:tau_assess} by
analyzing different models within the \BP\ framework alone: The blue
curve shows the posterior distribution when only accounting for white
noise; the red curve shows the same when additionally marginalizing
over foreground uncertainties; and, finally, the black curve shows the
posterior distribution with marginalizing over all sources of
uncertainty; white noise, foregrounds, and instrumental
parameters. Neglecting the instrumental effects, such as gain and
bandpass uncertainties, significantly under-estimates the true
uncertainty, and properly accounting for all these effects is a main
goal of the \BP\ framework.

\section{Reproducibility and Open Science}
\label{sec:software}

As discussed in Sect.~\ref{sec:introduction}, the main long-term
scientific goal and motivation of the \BP\ program is to establish an
end-to-end analysis framework for CMB observations that, we hope, will
be useful for the general community. This framework is designed to be
sufficiently flexible to allow analysis of different and complementary
experiments, and thereby exploit the strengths of one instrument to
break degeneracies in another. Organizing and promoting this work is a
key goal for the \cosmoglobe\ project, which aims at performing a
similar analysis as \BP, but for a much wider selection of datasets,
and eventually build a single statistically consistent and
multi-experiment model of the radio, microwave, and sub-millimeter
sky. An early example of this work has already been demonstrated in
the current paper suite by \citet{bp17}, who have extended the current
framework to re-analyze the \WMAP\ $Q$-band data with very promising
results.

For this project to succeed, substantial efforts have been spent
within the \BP\ program on the issue of \emph{reproducibility}. These
efforts are summarized by \citet{bp05}, both in terms of the internal
process itself and some lessons learned, and also in terms of the
final practical solutions that have been implemented. Here we provide
a brief summary of the main points.

\subsection{Reproducibility}

For the \BP\ and \cosmoglobe\ framework to be useful for other
experiments it must be \emph{reproducible}: Researchers outside of the
current collaboration must be able to repeat our analysis, before
improving and extending it. To support this, we have focused on four
main items:
\begin{enumerate}
\item \emph{Documented open-source code} -- the full
  \commander\footnote{\url{http://beyondplanck.science}} source code,
  as well as various pre- and post-processing
  tools,\footnote{\url{https://github.com/cosmoglobe/c3pp}} are made
  publicly available in a GitHub repository under a GPL license, and
  may be freely downloaded and extended within the general restriction
  of that license. Preliminary documentation is
  provided,\footnote{\url{https://docs.beyondplanck.science}} although
  it is under continuous development, as is the source code itself.
\item \emph{Cmake compilation} -- easy compilation is supported
  through the Cmake environment; required external libraries are
  automatically downloaded and compiled.
\item \emph{Data downloader} -- a Python-based tool is provided that
  automatically downloads all \BP\ input data to a user-specified
  directory, together with the parameter files that are needed to run
  the code.
\end{enumerate}
In addition, all main results (both full chain files and selected
post-processed posterior mean and standard deviation maps) are
available from the
\cosmoglobe\ homepage,\footnote{\url{https://cosmoglobe.uio.no}} and
eventually through the \Planck\ Legacy
Archive.\footnote{\url{https://pla.esac.esa.int/}} For further details
regarding the reproducibility aspects of the work, we refer the
interested reader to \citet{bp05}.

\subsection{Software}

A second requirement for the \BP\ framework to be useful for other
users is that the software is computationally efficient so that it can
be run on generally available hardware, and also that the source code
is extendable without expert knowledge. Regarding the former point, we
note that great emphasis has been put on minimizing the required
computational resources throughout the implementation. This appears to
be at least partially successful, as summarized in
Sect.~\ref{sec:compcost} and by \citet{bp03}: The full \BP\ analysis,
as presented here, has a computational cost of 670\,000 CPU hours,
which is roughly equivalent to the cost of producing 70 end-to-end
\Planck\ FFP8 70\,GHz realizations using the traditional pipeline
\citep{planck2014-a14}. Furthermore, by compressing the TOD inputs the
memory footprint of the LFI data set has been reduced by about an
order of magnitude (see Table~\ref{tab:resources} and \citealp{bp03}),
and now requires only about 1.5\,TB of RAM to run. Computers with this
amount of memory and clock cycles are now available outside
super-computing centers, and a full \Planck\ LFI analysis therefore no
longer requires the use of expensive super-computers -- although they
will of course be beneficial when available.

Regarding the software itself, the current main code base is written
in object-oriented Fortran 2003. Clearly, this may represent a
significant hurdle for many users, as most astrophysics students today
are typically more exposed to languages like Python or C than
Fortran. This choice of language is primarily historical, and due to
the fact that a large part of the legacy code base was originally
written in Fortran, most notably \HEALPix\ \citep{gorski2005} and
\commander\ \citep{eriksen:2004,eriksen2008}. However, a second
important motivation for adopting Fortran is that it remains one of
the fastest languages even today in terms of computational speed
and memory management. As far as readability and extendability goes,
the code has been designed with a strong focus on object-orientation,
and we believe that adding support for several types of new
sub-classes is relatively straight-forward. This includes classes for
new signal components; noise or beam representations; or TOD
models. On the other hand, modifying the underlying memory management,
component separation infrastructure, or parallelization paradigm, is
likely to be difficult without expert knowledge. A guide to the
current software is provided by \citet{bp03}. As a real-world
demonstration of the extendability of the framework, we present
a preliminary application to \WMAP\ in \citep{bp17}.

As useful as we hope the current version will be, we do believe that
developing a massively parallel version of \commander\ in Python would
be a useful, interesting and intellectually challenging task, and we
would encourage (and support!) work in this direction. For reference,
the current \commander\ Fortran source code spans 45\,000
lines,\footnote{Interestingly, only about 6000 lines are directly
  associated with TOD processing, while 14\,000 lines are directly
  associated with component separation; the rest is spent on general
  data infrastructure and tools.} which can likely be reduced by a
significant factor if written in a less verbose language; porting this
to Python would obviously be a major undertaking, but certainly
feasible for even just a small team of talented and motivated
researchers.

\begin{figure*}[t]
  \center
  \includegraphics[width=0.9\linewidth]{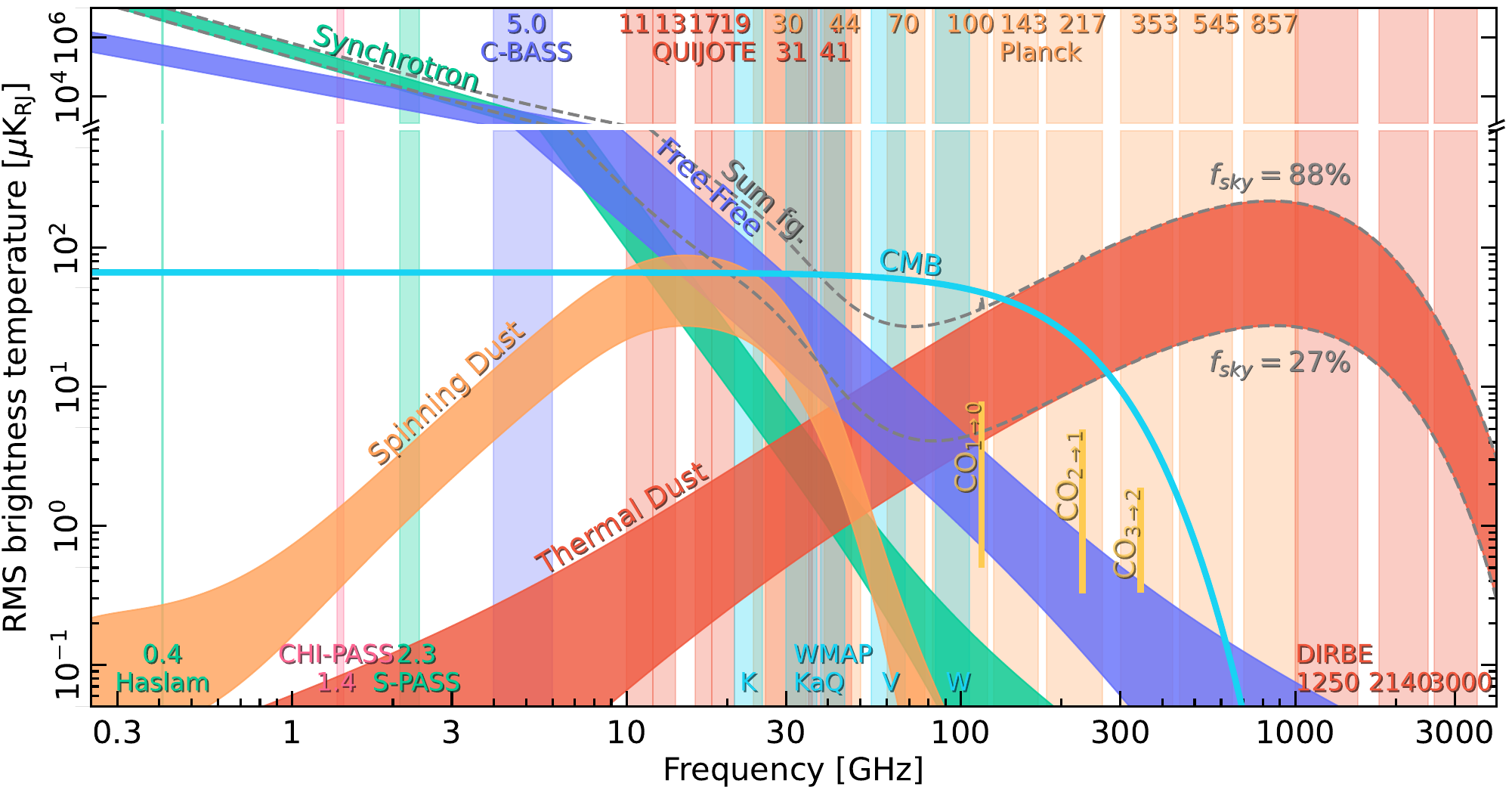}
  \includegraphics[width=0.9\linewidth]{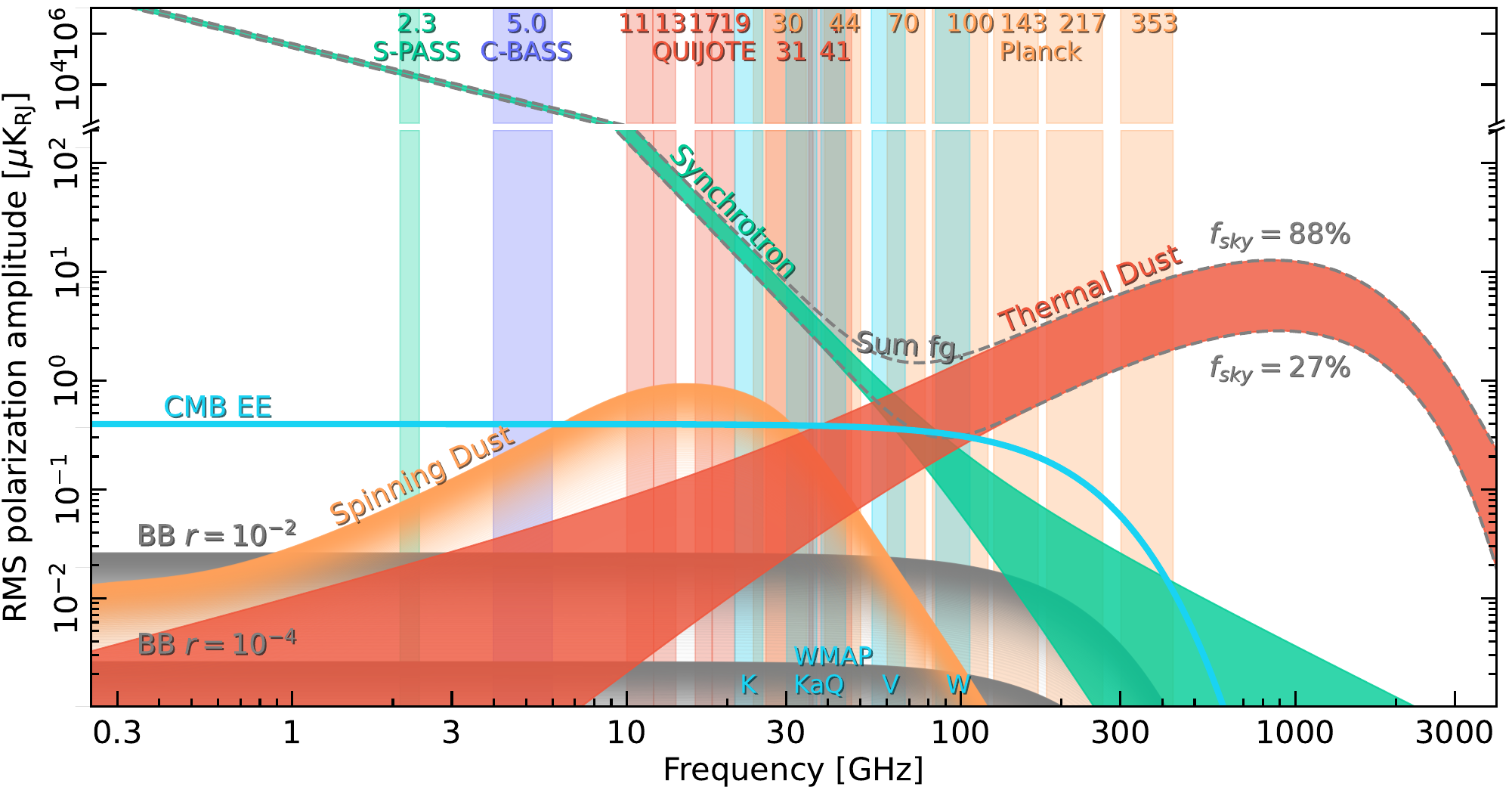}
  \caption{Brightness temperature (\emph{top panel}) and polarization
    amplitude (\emph{bottom panel}) RMS as a function of frequency and
    astrophysical component, and as derived from the \BP\ and
    \Planck\ sky models. Vertical bands indicate the frequency ranges
    of various experiment. All components have been evaluated at a
    common angular resolution of $1^{\circ}$ FWHM. The widths of each
    confidence region correspond to the spread when evaluating the RMS
    over two different masks with sky fractions of 88 and 27\,\%,
    respectively.  The cyan curve shows the level of CMB fluctuations
    evaluated for the best-fit \Planck\ $\Lambda$CDM spectrum. In the
    current \BP\ analysis, only the three LFI channels are modelled in
    the time-domain. For polarization, the spinning dust component
    (orange curve) indicates an upper limit as presented by \citet{bp15},
    not a detection. A long-term future goal is to include all
    publicly available and relevant data (for instance \WMAP\ and
    \Planck\ HFI) into this model; preferably in the form of
    time-ordered data, but if this is not technically or financially
    possible, then at least in the form of pre-processed sky
    maps. This work will be organized within the
    \cosmoglobe\ project.}
  \label{fig:overview_fg}
\end{figure*}

\section{Conclusions, summary and outlook}
\label{sec:conclusions}

The \Planck\ project represents a landmark achievement in
international cosmology, mapping out the primary temperature
fluctuations in the CMB to a precision determined by astrophysical
constraints. This achievement was made possible by the dedication and
long-term contributions from ESA and NASA; from tens of national
funding agencies; and many hundreds of scientists and engineers
working for more than two decades. At the end of the mission, a
massive amount of knowledge and expertise regarding optimal analysis
of CMB experiments had been generated within the collaboration, as
clearly demonstrated through more than 150 refereed scientific publications.

A central goal of the \BP\ project was to translate a significant part
of this aggregated experience into a practical computer code that can
analyse \Planck\ data from end-to-end, and to make this code available
to the community in general. Main advantages of an integrated
end-to-end Bayesian approach include:
\begin{enumerate}
\item \emph{Faithful error propagation}: \BP\ implements global
  end-to-end Bayesian CMB analysis framework. The single most
  important advantage of this is faithful propagation of uncertainties
  from raw TOD to final cosmological parameters. Instrumental and
  astrophysical systematic errors are propagated to the final CMB
  likelihood on the same footing as any other nuisance
  parameter. While already important for \Planck, this issue will
  become absolutely critical for future planned high-precision
  $B$-mode experiments, such as \emph{LiteBIRD} or \emph{PICO}.
\item \emph{Breaking degeneracies and saving costs by exploiting
  synergistic observations}: Combining data from complementary sources is
  essential to break fundamental degeneracies within a given
  experiment. For instance, both \Planck\ and \WMAP\ have degenerate
  polarization modes that they cannot measure well on their own, due
  to peculiarities in their respective scanning strategies---but there
  are no degenerate modes in the combined data set. In general,
  however, the usefulness of joint analysis with external data is
  often limited by systematic errors. The \BP\ framework addresses
  this by providing a common platform for performing joint low-level
  analysis of different experiments. Also noting that the lion's share
  of the analysis cost of any real-world CMB experiment is associated
  with understanding degeneracies and systematic errors, we believe that a global approach
  will lead to better and cheaper science for each experiment.
\item \emph{Fewer human errors}: Tight analysis integration also leads
  to many important practical advantages, including less room for
  human errors or miscommunication; greater transparency of both
  explicit and implicit priors; better optimization of computing
  resources; and significantly reduced end-to-end wall-clock time by
  eliminating intermediate human interaction. 
\item \emph{``Faster, better and cheaper'' through open-source 
  science}: True inter-experiment global analysis will clearly not
  succeed without active contributions and support from a large part
  of the general community. For this reason, we make our source codes
  publicly available under a GPL open-source license to ensure
  long-term stability of the currently released software.  It
  also means that future improvements must be released under a
  similarly generous license, in recognition of the fact that this
  project is intended to be collaborative, open, and inclusive. The
  use of stakeholder knowledge is critically important---and we hope
  that many stakeholders will indeed be interested in actively
  contributing to the program, ultimately leading to ``faster, better,
  and cheaper'' science for everyone.
\end{enumerate}

As discussed above, the \BP\ program has focused on the \Planck\ LFI
data. The reasons for doing so were three-fold. First and foremost,
many \BP\ collaborators have been working with the LFI data for one or
two decades, and the aggregated experience with this data set within
the collaboration implied a low start-up cost; results could be
produced quickly. Second, the full LFI data volume is fairly limited
in size, comprising less than 1\,TB after compression, which is good
for fast debugging and testing. Third, the LFI instrument is based on
HEMT radiometers, which generally both have a relatively high noise
contribution and low systematic errors per sample. The combination of
these three points made LFI a natural starting point for the work.

However, now that the computational framework already exists, it will
require substantially less effort to generalize it to other and
complementary data sets. Currently on-going generalizations by various
teams involve \WMAP, \emph{LiteBIRD}, CHIPASS, and COBE-DIRBE, and we
warmly welcome initiatives targeting any other experiment as well. In
this respect, it may be useful to distinguish between four types of
experiments, each with their own set of algorithmic complexities.

First, many radio, microwave and sub-millimeter experiments may be
modelled within nearly the same sky and instrument model as
\BP. Examples include C-BASS, QUIET and QUIJOTE, all of which simply
provide additional signal-to-noise and/or frequency coverage, as far
as the underlying algorithms are concerned. For these, analysis within
the \BP\ framework may turn out to amount simply to writing one or
more TOD processing modules (for instance using the current LFI module
as a template) to take into account the various instrument-specific
systematic effects of the experiment in question. These experiments
should be, relatively speaking, the easiest to integrate into the
current framework.

Other experiments may build on the same sky model and component
separation procedures as \BP, but require a different mapmaking
algorithm. One prominent example of this is \WMAP, which is
differential in nature, and therefore requires a different Conjugate
Gradient mapmaking algorithm to translate cleaned TOD into pixelized
maps; this work is already on-going \citep{bp17}. Experiments of this type should
also be relatively straighforward to integrate.

A third class of experiments are those that can use the same type of
sky models, but requires a significantly different instrumental
model. The most prominent example of such are TES bolometer-based
instruments. These often have both higher signal-to-noise ratios and
systematic errors per sample, and therefore require a richer set of
systematics corrections. They also typically have a significant
multiplicative transfer function, which means that unbiased maps
cannot be produced simply by introducing additive TOD corrections, as
is done in the current implementation.  Instead, they will also require
a dedicated Conjugate Gradient mapmaker to take into account the
multiplicative effects. Examples of potentially relevant experiments
include for instance BICEP2, CLASS, SPIDER, and \emph{LiteBIRD}. Integrating
these will thus be more challenging than HEMT-based experiments like
LFI or \WMAP, but it should certainly be feasible, and the scientific
rewards will be massive.

The fourth and final group of experiments are those that either
produce massive amounts of time-ordered data, or very high-resolution
data. Important examples are ACT , SPT, Simons Observatory, and CMB-S4. 
These will all require a fundamental redesign of the existing
code base, simply to handle the massive amounts of
memory and network communication efficiently. Additionally,
experiments that observe only a fraction of the sky, but at high
angular resolution, cannot employ the spherical harmonics basis that
we currently use for component separation without introducing large
degeneracies and singular modes; all spatial modes need to be
constrained by at least one experiment for the current implementation
to work properly. Developing a new version of the Bayesian framework
that can handle higher levels of parallelization, and also use more
general basis sets, is thus an important goal for future work.

Returning to the specific scientific results derived by the
\BP\ project, we note that cosmological constraints derived from LFI
and \WMAP\ alone will never be competitive in terms of overall
uncertainties as compared to an HFI-based analysis. Nevertheless, many
interesting results have been established during the course of the
project. Some of the most noteworthy among these are the
following:
\begin{enumerate}
  \item We have succeeded in integrating the LFI 44\,GHz channel into
    a statistically viable low-$\ell$ CMB likelihood. Two key
    algorithmic steps that were required for this were, firstly, to
    implement a more flexible gain smoothing algorithm stiff than a
    hard boxcar average, and, secondly, to model a previously unknown
    additional correlated noise contribution on 0.01--1\,sec time
    scales. Less important, but still notable, improvements include
    more efficient ADC corrections and removal of known bandpass
    artefacts. After making all these corrections, we are not
    currently aware of any significant outstanding systematic effects
    in the \Planck\ LFI data.
  \item We have for the first time constructed full dense
    low-resolution CMB and frequency map covariance matrices that
    account for marginalization over a wide range of important
    systematic time-ordered effects, including gain, bandpass, and
    foreground corrections, and correlated noise. This results in a
    low-$\ell$ polarization likelihood that yields results consistent
    with the latest HFI analyses, and a best-fit value of the
    reionization optical depth of
    $\tau=0.066\pm0.013$. The associated $\chi^2$
    goodness-of-fit statistics are statistically acceptable, although
    there might be weak hints of excess power, possibly due to the
    break-down of the $1/f$ noise model.
  \item We have produced a statistically consistent and joint estimate
    of the CMB dipole using both \Planck\ and \WMAP\ data. The
    best-fit dipole amplitude of $3362.7\pm1.4\muK$ is consistent with
    all published results, including the latest HFI-based
    measurements, and the quoted error estimate is derived strictly
    within the well-defined Bayesian statistical framework.
  \item We are for the first time able to fit a physically
    meaningful spectral index of polarized synchrotron emission using
    both \WMAP\ and \Planck. This is the direct result of performing a
    truly joint analysis with LFI and \WMAP\, as described above,
    using information from one experiment to break degeneracies within
    the other.
\end{enumerate}

While the \BP\ project itself contractually ended on November 30th,
2020, the work will in general continue with various alternative
funding sources, and, we hope, also with the help of a continuously
growing community of supporting collaborators and
experiments. Figure~\ref{fig:overview_fg} shows a compilation of the
current \BP\ sky model and data sets in both temperature (top panel)
and polarization (bottom panel) together with selected external
products. The long-term goal of this work is to populate this plot
with all available experimental data, and thereby gradually refine the
sky model. The ERC-funded \cosmoglobe\ project aims to coordinate
these efforts, and will serve as a stable platform for all parties
interested in global Bayesian CMB analysis. \cosmoglobe\ will also
serve as the long-term home for all \BP\ material and products, long
after the current \BP\ web portal vanishes.

Finally, we end with an important \emph{caveat emptor}, and emphasize
that \commander\ is very much a work-in-progress---and it will
remain so for all foreseeable future. Essentially every single step in
the pipeline can and will be replaced by smarter and more capable
sampling algorithms; the user-interface could most certainly be made
more intuitive; and so on. This is an unavoidable side-effect of being
at the cutting edge of algorithmic research, where new ideas are
continuously being explored, implemented and tested. However, at the
same time, it is also our belief that the current platform is now
finally sufficiently mature to allow external users and developers to
use it productively for their own analyses, and to extend it as they
see fit. In other words, we believe that now is the right time for
Bayesian end-to-end CMB analysis to go OpenSource, and we invite all
interested parties to participate in this work.

\begin{acknowledgements}
  We thank Prof.\ Pedro Ferreira and Dr.\ Charles Lawrence for useful suggestions, comments and 
  discussions. We also thank the entire \Planck\ and \WMAP\ teams for
  invaluable support and discussions, and for their dedicated efforts
  through several decades without which this work would not be
  possible. The current work has received funding from the European
  Union’s Horizon 2020 research and innovation programme under grant
  agreement numbers 776282 (COMPET-4; \BP), 772253 (ERC;
  \textsc{bits2cosmology}), and 819478 (ERC; \textsc{Cosmoglobe}). In
  addition, the collaboration acknowledges support from ESA; ASI and
  INAF (Italy); NASA and DoE (USA); Tekes, Academy of Finland (grant
   no.\ 295113), CSC, and Magnus Ehrnrooth foundation (Finland); RCN
  (Norway; grant nos.\ 263011, 274990); and PRACE (EU).
\end{acknowledgements}

\bibliographystyle{aa}

\bibliography{Planck_bib,BP_bibliography}

\appendix

\section{Review of frequently used textbook sampling algorithms}
\label{app:samplers}

As described in Sect.~\ref{sec:gibbs}, the \BP\ pipeline is designed
in the form of a Gibbs sampler in which each parameter is sampled
conditionally on all other parameters. Each parameter must therefore
be associated with a specific sampling algorithm that samples from the
correct distribution. In this appendix, we therefore review some of
the most common sampling techniques that are used in the
\BP\ framework, while noting that all of this is textbook material;
this is just provided for reference purposes. In all cases below, we
assume that samplers for both the uniform distribution, $U[0,1]$, and
the standard univariate normal (Gaussian) distribution, $N(0,1)$, are
already available through some numerical library; we use routines
provided in \HEALPix.

\subsection{Univariate and low-dimensional Gaussian sampling}
\label{sec:gauss_lowdim}

Perhaps the single most common distribution in any Bayesian pipeline
is the univariate Gaussian distribution $N(\mu,\sigma^2)$ with mean
$\mu$ and standard deviation $\sigma$. A sample $x$ from this
distribution can be trivially generated by
\begin{equation}
  x = \mu + \sigma \eta,
\end{equation}
where $\eta\sim N(0,1)$ is a standard normal variate. Note that
$\left<x\right> = \mu$, because $\left<\eta\right> = 0$ and
$\left<(x-\mu)^2\right> = \sigma^2$ because $\left<\eta^2\right> = 1$.

A sample from a multivariate normal distribution $N(\boldsymbol{\mu}, \C)$
with mean vector $\boldsymbol{\mu}$ and covariance matrix $\C$ may be
produced in a fully analogous manner,
\begin{equation}
  \x = \boldsymbol{\mu} + \C^{\frac{1}{2}} \boldsymbol{\eta},
\end{equation}
where now $\boldsymbol{\eta}$ is a vector of independent $N(0,1)$ variates,
and $\C^{\frac{1}{2}}$ denotes some matrix for which $\C =
\C^{\frac{1}{2}}(\C^{\frac{1}{2}})^t$. The two most typical examples
are the Cholesky decomposition ($\C = \L\L^t$ for positive definite
matrices, where $\C^{\frac{1}{2}} = \L$) and singular-value
decomposition ($\C = \V\tens\Sigma\V^t$ for singular matrices, where
$\C^{\frac{1}{2}} = \V\tens\Sigma^{\frac{1}{2}}\V^t$). A notable advantage
regarding the latter is that it is symmetric, and therefore less
bug-prone than the Cholesky factor; on the other hand, it is slightly
more computationally expensive.

\subsection{High-dimensional Gaussian sampling}
\label{sec:gauss_highdim}

It is important to note that evaluating a ``square root'' of a matrix,
whether it is through Cholesky or eigenvector decomposition, is an
$\mathcal{O}(n^3)$ operation, where $n$ is the dimension of the
matrix. As such, the direct approach is only computationally
practical for relatively low-dimensional distributions, and just with
a few thousand elements or less. For distributions with millions of
correlated variables, the above prescription is entirely
impractical. In the following, we therefore describe a widely used
method to sample from high-dimensional distributions, effectively by
inverting the covariance matrix iteratively by Conjugate Gradients.

Again, let $\x$ be a random Gaussian field of $n$ elements with an
$n\times n$ covariance matrix $\S$, i.e., $\x \sim N(\vec 0,
\S)$. Further, to put the notation into a familiar context, we assume
we have some observations $\d$ that can be modeled as
\begin{equation}
    \d = \T\x + \n,
    \label{eq:gauss_highdim_datadef}
\end{equation}
where $\n$ is a stochastic noise vector of size $n_d$ (which in
general is different from $n$) which is drawn from a Gaussian
distribution with zero mean and covariance $\N$, and $\T$ is a matrix
of size $n_d\times n$, which effectively translates $\x$ into the
vector space of $\d$. In other words, we assume that the data may be
modelled as a linear combination of $\x$ plus a well-defined noise
contribution.

Note that this assumption about $\d$ does not preclude the cases where we have observations that can be written as $\d = \T\x + \n + \b$, where $\b$ is known and independent of $\x$ - in this case, we are free to redefine $\d$: $\d' \rightarrow \d - \b$, in which case our assumption in Eq.~\eqref{eq:gauss_highdim_datadef} would be met for $\d'$.

In general, $\T$ will not depend on $\x$. In the context of the Gibbs
framework of this paper, however, $\T$ typically \emph{will} depend on
other quantities that we do sample, but which we assume to be known
with respect to the current conditional of the Gibbs chain.

Our goal is then to draw a sample from $P(\x \mid  \d, \T, \S, \N)$, the
posterior of $\x$, given $\d$ and the other quantities, denoted
$P(\x\mid \d)$ as a shorthand. Using Bayes' theorem, we can write this as
\begin{equation}
    P(\x \mid  \d) \propto P(\d \mid  \x) P(\x).
\end{equation}
Here $P(\x)$ is a prior for $x$, which we assume takes the form
$\mathcal{N}(\vec 0, \S)$, whereas the likelihood term, $P(\d \mid  \x)$, is
simply given by a Gaussian distribution with covariance $\N$ and mean $\T\x$.
This gives (neglecting the pre-factors of the exponentials, as they are independent of $\x$ and end up as normalization constants)
\begin{align}
    -2\ln P(\x\mid \d) & = \x^t\S^{-1}\x + (\d-\T\x)^t\N^{-1}(\d - \T\x) \nonumber \\
             & = \x^t\S^{-1}\x + \d^t\N^{-1}\d + \x^t\T^t\N^{-1}\T\x -\nonumber
    \\ &\quad\quad\quad\d^t\N^{-1}\T\x - \x^t\T^t\N^{-1}\d \nonumber \\ 
             & = \x^t(\S^{-1} + \T^t\N^{-1}\T)\x - 2\x^t\T^t\N^{-1}\d,
    \label{eq:gauss_highdim_posterior}
\end{align}
where, in the last transition, we neglect also the terms that do not
include $\x$. We also use the identity $\a^t\C \b = \b^t\C \a$,
which is valid for a symmetric matrix $\C$, in order to gather the
terms that are linear in $\x$.

This expression for $P(\x\mid \d)$ can be written as a Gaussian
distribution by ``completing the square'': We are looking for a matrix
$\F$ and a vector $\vec{c}$ such that
\begin{align}
    P(\x\mid \d) &= \exp\biggl[-\frac{1}{2}(\x - \vec{c})^t\F^{-1}(\x - \vec{c})\biggr] \nonumber \\
             &\propto \exp\biggl[-\frac{1}{2}\biggl(\x^t\F^{-1}\x - 2\x^t\F^{-1}\vec{c}\biggr)\biggr].
    \label{eq:gauss_highdim_completesquares}
\end{align}
Comparing terms in Eqs.~\eqref{eq:gauss_highdim_posterior} and \eqref{eq:gauss_highdim_completesquares}, we find that the terms that are quadratic in $\x$ enforce
\begin{equation}
    \F^{-1} = \S^{-1} + \T^t\N^{-1}\T.
    \label{eq:gauss_highdim_Fdef}
\end{equation}
Inserting this into the terms that are linear in $\x$, we find
\begin{equation}
                       \vec{c} = (\S^{-1}+ \T^t\N^{-1}\T)^{-1}\T^t\N^{-1}\d.
    \label{eq:gauss_highdim_cdef}
\end{equation}
Thus, the posterior of $\x$ is a Gaussian distribution with covariance
given by Eq.~\eqref{eq:gauss_highdim_Fdef} and mean (and mode) given by
Eq.~\eqref{eq:gauss_highdim_cdef}.

In order to draw a sample, $\tilde{\x}$, from this distribution, we
can in principle use the standard prescription for sampling from
multivariate Gaussian distributions, 
\begin{equation}
	\boldsymbol x=(\mathsf S^{-1}+\mathsf T^t\mathsf N^{-1}\mathsf T)^{-1}
	\mathsf T^t\mathsf N^{-1}\boldsymbol d+(\mathsf S^{-1}+\mathsf T^t\mathsf N^{-1}\mathsf T)^{-1/2}\boldsymbol\eta,
\end{equation}
as summarized in the previous
section. However, inverting the covariance matrix, $\S^{-1}+
\T^t\N^{-1}\T$, is once again an $\mathcal{O}(n^3)$ operation. To
circumvent this problem, we instead consider the same equation in the
form
\begin{equation}
  (\S^{-1} + \T^t\N^{-1}\T)\x = \T^t\N^{-1}\d
	+(\mathsf S^{-1}+\mathsf T^t\mathsf N^{-1}\mathsf T)^{1/2}\boldsymbol\eta.
  \label{eq:gauss_highdim_cdef_linsys}
\end{equation}
Since the matrix on the left-hand side is both symmetric and
semi-positive definite, this equation can be solved iteratively by
Conjugate Gradients; for a brilliant review of this algorithm, see
\citet{shewchuk:1994}. Additionally, to obtain the correct covariance
structure, one can instead add one random zero-mean covariance term for
each element in the covariance matrix to the right-hand side of the
equation,
\begin{equation}
    (\S^{-1} + \T^t\N^{-1}\T)\x = \T^t\N^{-1}\d + \T^t\N^{-1/2}\vec{\eta}_1 +
    \S^{-1/2}\vec{\eta}_2. 
    \label{eq:multigauss}
\end{equation}
With this definition, $\left<\x\right> = \vec{c}$,
and $\left<\x\x^t\right> = (\S^{-1} + \T^t\N^{-1}\T)^{-1} = \F$, as desired.

A fully analogous calculation may be done also with a non-zero prior
mean, $\m$, in which case an additional term is introduced on
the right-hand side of Eq.~\eqref{eq:gauss_highdim_cdef_linsys},
\begin{equation}
  (\S^{-1} + \T^t\N^{-1}\T)\x = \T^t\N^{-1}\d + \S^{-1}\m + \T^t\N^{-1/2}\vec{\eta}_1 +
    \S^{-1/2}\vec{\eta}_2. 
  \label{eq:gauss_highdim_prior}
\end{equation}
The relative strength of the data and prior terms is thus effectively
determined by the overall signal-to-noise ratio of the data as
measured by $\S$ and $\N$, and in the limit of vanishing
signal-to-noise (i.e., $\N^{-1}\rightarrow0$), $\left<\x\right>=\m$,
as desired. Note, also, that $\S$ quantifies the covariance of the
\emph{fluctuations around the mean}, not the co-variance of the entire
field $\x$ itself. In the limit of $\S\rightarrow0$ (or, equivalently,
$\S^{-1}\rightarrow\infty$), we therefore also have
$\left<\x\right>=\m$. Thus, the magnitude of $\S$ represents a direct
handle for adjusting the strength of the prior.

\subsection{Inversion sampling}
\label{sec:inversion}

The samplers discussed in the two previous sections only concerns
Gaussian distributions. In contrast, the so-called \emph{inversion}
sampler is a completely general sampler that works for all univariate
distributions. 

Let $P(x)$ be a general probability distribution for some random
variable $x$. The inversion sampler is then defined as follows:
\begin{enumerate}
  \item Compute $P(x)$ over a grid in $x$, making sure to probe the
    tails to sufficient accuracy.
  \item Compute the cumulative probability distribution, ${F(x) =
    \int_{-\infty}^{x} P(x')\,\mathrm dx'}$.
  \item Draw a random uniform variate, $\eta \sim U[0,1]$.
  \item Solve the nonlinear equation $\eta = F(x)$ for $x$.
\end{enumerate}

Clearly, this is a computationally very expensive algorithm, noting
that it actually requires the user to map the full distribution,
$P(x)$, in the first step. This typically requires a preliminary
bisection search to first identify a sufficiently wide region in $x$
to cover all significant parts of $P$. Then another 50--100
evaluations are required to grid the (log-)probability
distribution. 

However, the facts that this sampler requires no manual tuning, and
that it produces independent samples, make it an attractive component
in many Gibbs samplers; typically, the overall computational cost of
the entire Gibbs chain is dominated by completely different operations.





\end{document}